\documentclass{elsart}

\usepackage{graphicx}
\usepackage{epsfig}

\usepackage{amssymb}

\def\gsim{\raise0.3ex\hbox{$>$\kern-0.75em\raise-1.1ex\hbox{$\sim$}}}

\newcommand {\AuAu}      {\mbox{Au+Au}}

\newcommand {\dAu}       {\mbox{$d+$Au}}

\newcommand {\Npart}{\mbox{$N_{\rm part}$}}

\newcommand {\cosmod} {\mbox{$\cos{2\Delta \phi}$}}

\newcommand {\dEdx} {\mbox{$\Delta E/\Delta x$}}

\newcommand {\dphi} {\mbox{$\Delta \phi$}}

\newcommand {\dngdy} {\mbox{$dn_g/dy$}}

\newcommand {\piz} {\mbox{$\pi^0$}}
\newcommand {\pp} {\mbox{$p+p$}}
\newcommand {\pperp} {\mbox{$p_T$}}
\newcommand {\psy}   {\mbox{$\eta$}}

\newcommand {\RAApt} {\mbox{$R_{AA}(\pperp)$}}
\newcommand {\RAAAvgnpart} {\mbox{$R_{AA}$}}
\newcommand {\RAAnpart} {\mbox{$R_{AA}^{N_{\rm part}}$}}

\newcommand {\RAA} {\mbox{$R_{AA}$}}
\newcommand {\RdA} {\mbox{$R_{dA}$}}
\newcommand {\Rcp} {\mbox{$R_{CP}$}}

\newcommand {\Sloss}{\mbox{$S_{\rm loss}$}}

\newcommand {\TAB} {\mbox{$T_{AB}$}}
\newcommand {\xt} {\mbox{$x_T$}}

\begin{document}
\begin{frontmatter}
 
 
 
\title{Formation of dense partonic matter in relativistic nucleus-nucleus collisions
at RHIC:\\  Experimental evaluation by the PHENIX collaboration}
 
\author[vandy]{K.~Adcox}
\author[bnl]{S.S.~Adler}
\author[jinrdubna]{S.~Afanasiev}
\author[bnl,columbia]{C.~Aidala}
\author[stonybrkc]{N.N.~Ajitanand}
\author[kek,riken,rikjrbrc]{Y.~Akiba}
\author[nmsu]{A.~Al-Jamel}
\author[stonybrkc]{J.~Alexander}
\author[fsu]{R.~Amirikas}
\author[kyoto,riken]{K.~Aoki}
% ...
\author{(PHENIX Collaboration)}
% ...
\address[bnl]{Brookhaven National Laboratory, Upton, NY 11973-5000, USA}
% ...
\address[columbia]{Columbia University, New York, NY 10027 and Nevis Laboratories, Irvington, NY 10533, USA}
% ...
\address[fsu]{Florida State University, Tallahassee, FL 32306, USA}
% ...
\address[jinrdubna]{Joint Institute for Nuclear Research, 141980 Dubna, Moscow Region, Russia}
% ...
\address[kek]{KEK, High Energy Accelerator Research Organization, Tsukuba-shi, Ibaraki-ken 305-0801, Japan}
% ...
\address[kyoto]{Kyoto University, Kyoto 606, Japan}
% ...
\address[nmsu]{New Mexico State University, Las Cruces, NM 88003, USA}
% ...
\address[riken]{RIKEN (The Institute of Physical and Chemical Research), Wako, Saitama 351-0198, JAPAN}
\address[rikjrbrc]{RIKEN BNL Research Center, Brookhaven National Laboratory, Upton, NY 11973-5000, USA}
% ...
\address[stonybrkc]{Chemistry Department, Stony Brook University, Stony Brook, SUNY, NY 11794-3400, USA}
% ...
\address[vandy]{Vanderbilt University, Nashville, TN 37235, USA}
% ...
\thanks[deceased]{Deceased}
 
\begin{abstract}
Extensive experimental data from high-energy nucleus-nucleus collisions
were recorded using the PHENIX detector at the Relativistic Heavy Ion
Collider (RHIC).  The comprehensive set of measurements from the first three
years of RHIC operation includes charged particle multiplicities, 
transverse energy, yield ratios and spectra of identified hadrons in a wide range of
transverse momenta ($p_T$), elliptic flow, two-particle correlations,
non-statistical fluctuations, and suppression of particle production at
high $p_T$.  The results are examined with an emphasis on implications for the
formation of a new state of dense matter.  We find that the state of
matter created at RHIC cannot be described in terms of ordinary color
neutral hadrons.
\end{abstract}
 
\end{frontmatter}
 
\section{INTRODUCTION}
\label{Sec:intro}
\subsection{Historical Introduction}
A recurring theme in the history of physics is the desire to study
matter under extreme conditions. The latter half of the twentieth
century saw this quest extended from 'ordinary' atomic systems to
those composed of nuclear matter. Even prior to the identification
of Quantum Chromodynamics (QCD) as the underlying theory of the
strong interaction, there was considerable interest in the fate of
nuclear matter when subjected to density and temperature \
extremes \cite{Chapline:1973zf,Lee:1974ma,Lee:1975kn}. Particularly
intriguing was the suggestion that new phases of nuclear matter
could be associated with a corresponding change in the structure of
the vacuum \cite{Lee:1979nz}. These considerations gained additional
impetus with the realizations that a) QCD was the correct theory of
the strong interaction, b) the phenomena of quark confinement was a
consequence of the nonperturbative structure of the vacuum and c)
this vacuum structure is modified at high temperatures and/or
densities, suggesting that quarks and gluons under such conditions
would be deconfined. Taken together, these facts suggest that QCD is
a fundamental theory of nature containing a phase transition {\em
that is accessible to experimental investigation}.

It is quite remarkable that this understanding was achieved very
early in the development of QCD.  Collins and Perry noted in
1975 \cite{Collins:1975ky} that the reduction of the coupling
constant at small distances indicated that the dense nuclear matter
at the center of neutron stars would consist of deconfined quarks
and gluons\footnote{In fact, prior to the development of QCD the
quark hypothesis raised serious issues concerning the
stability of neutron stars \cite{Itoh:1970}.}. Their treatment
focused on the high-density, low-temperature regime of QCD, but they
did note that similar arguments might apply to the high temperatures
present in the early universe. An extensive review by Shuryak in
1980 \cite{Shuryak:1980tp} is the first to have examined the
high-temperature phase in detail, and is also notable for proposing
the phrase ``quark-gluon plasma'' (QGP) to describe the deconfined
state:
\begin{quotation}
{\em When the {\em energy} density $\varepsilon$ exceeds some typical
hadronic value ($\sim 1$ \ GeV/fm$^3$), matter no longer
consists of separate hadrons (protons, neutrons, etc.), but as their
fundamental constituents, quarks and gluons. Because of the apparent
analogy with similar phenomena in atomic physics we may call this
phase of matter the QCD (or quark-gluon) plasma. }
\end{quotation}


Developing a quantitative understanding of the deconfining phase
transition in hadronic matter and of QGP properties has proven to be
a challenging task. While simple dimensional arguments suffice to
identify both the critical energy density $\varepsilon_C \sim $
1 GeV/fm$^3$ and the associated critical temperature $T_C \sim
170$ MeV, these values also imply that the transition occurs
in a regime where the coupling constant is of order unity, thereby
making perturbative descriptions highly suspect.


Progress in understanding QCD in the extremely non-perturbative
domain  near the critical temperature has relied on an essential
contribution by Creutz \cite{Creutz:1977ch}, who showed that
numerical implementations of Wilson's lattice
formulation \cite{Wilson:1974sk} could be used to study phase transition
phenomena. This work, together with the continued exponential
increases in computing power, stimulated the development of lattice
QCD, which in turn has led to detailed investigations of the
thermodynamic properties of quarks and gluons \cite{Laermann:2003cv}.

\begin{figure}[tbhp]
\includegraphics[width=1.0\linewidth]{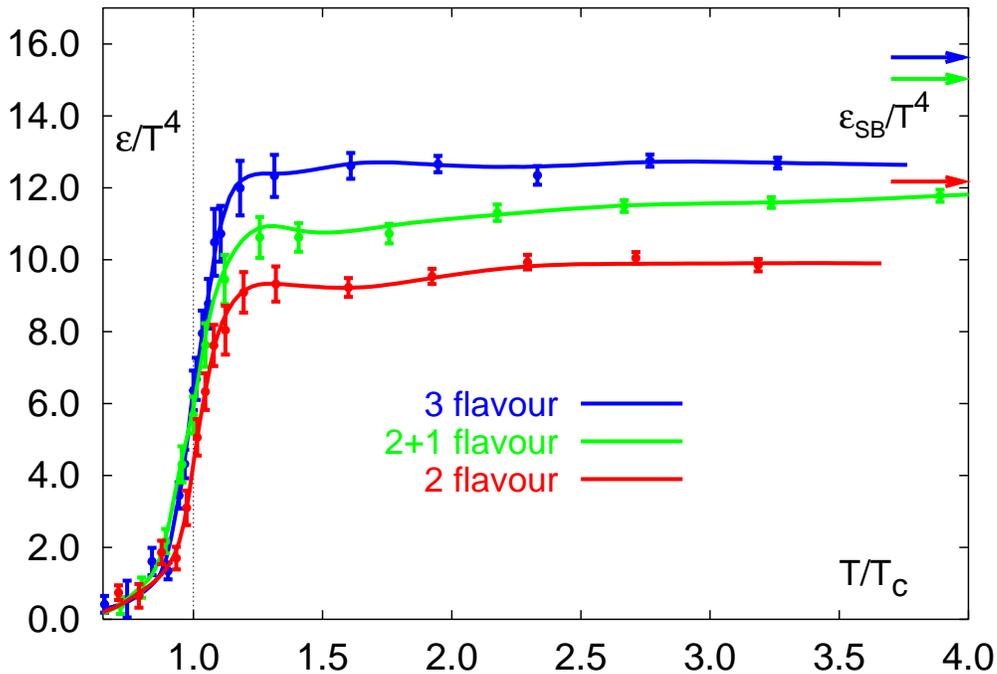}
\caption{Lattice QCD results \cite{Karsch:2001cy} for the energy
density / $T^4$ as a function of the temperature scaled by the
critical temperature $T_C$. Note the arrows on the right side
indicating the values for the Stefan-Boltzmann limit.
\label{fig_lattice}}
\end{figure}

Lattice QCD predicts a phase transformation to a quark-gluon plasma
at a temperature of approximately $T \approx 170$  MeV $\approx
10^{12}$ K, as shown in Fig. \ref{fig_lattice} \cite{Karsch:2001cy}.
This transition temperature corresponds to an energy density
$\varepsilon \approx 1 {\rm GeV/fm}^{3}$, nearly an order of magnitude
larger than that of normal nuclear matter. As noted above, this
value is plausible based on dimensional grounds, since such
densities correspond to the total overlap of several (light) hadrons
within a typical hadron volume of 1--3 ${\rm fm}^3$. No plausible
mechanism exists under which hadrons could retain their {\em in
vacuo} properties under these conditions. Lattice calculations also
indicate that this significant change in the behavior of the system
occurs over a small range in temperature ($\sim$20 MeV), and suggest
that the change of phase includes the restoration of approximate
chiral symmetry resulting from greatly reduced or vanishing quark
constituent masses.

\begin{figure}[tbhp]
\includegraphics[width=1.0\linewidth]{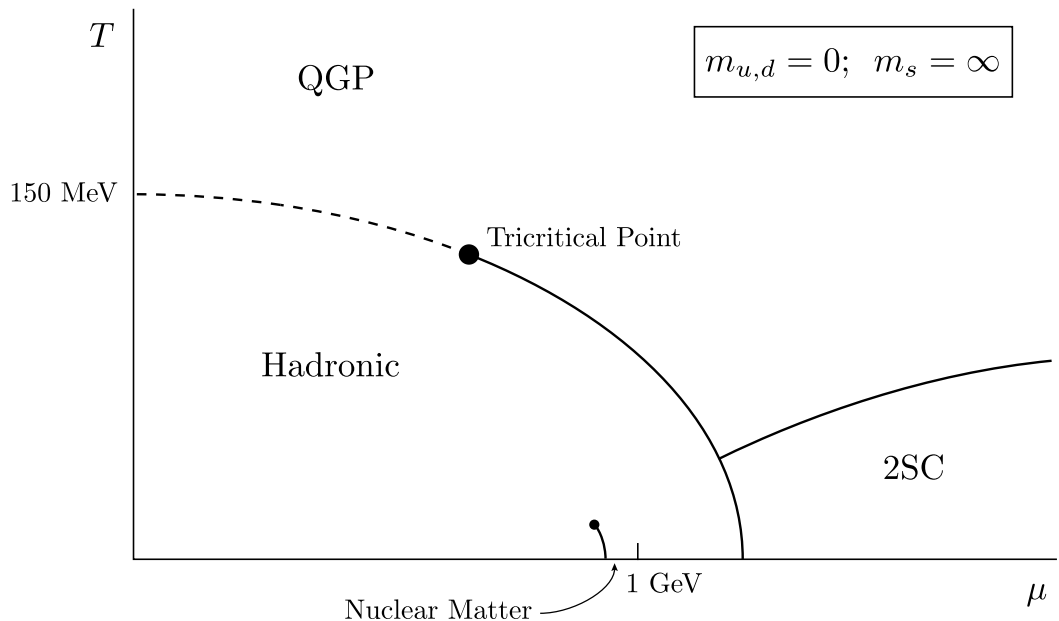}
\caption{Theoretical phase diagram of nuclear matter for two
massless quarks as a function of temperature $T$ and baryon chemical
potential $\mu$ \cite{Rajagopal:2000uu}. \label{fig_phase}}
\end{figure}

In the limit of massless noninteracting particles, each bosonic
degree of freedom contributes ${{\pi^{2}} \over {30}}T^{4}$  to the
energy density; each fermionic degree of freedom contributes ${{7}
\over {8}}$ this value. The corresponding ``Stefan-Boltzmann''
limits of the energy density $\varepsilon_{SB}$ for the case of 2(3)
active flavor quark-gluon plasma is then
\begin{eqnarray}
&  \{ 2_f \cdot 2_s \cdot 2_q \cdot 3_c \frac{7}{8} + 2_s \cdot 8_c
\}
           \frac{\pi^2}{30}T^4 = 37 \  \frac{\pi^2}{30}T^4
           \\ \nonumber
\varepsilon_{SB} = & \\
&  \{ 3_f \cdot 2_s \cdot 2_q \cdot 3_c \frac{7}{8} + 2_s \cdot 8_c
\}
           \frac{\pi^2}{30}T^4 = 47.5 \ \frac{\pi^2}{30}T^4
           \\ \nonumber
\end{eqnarray}
after summing over the appropriate flavor, spin, quark/antiquark and
color factors for quarks and spin times color factors for gluons.
The large numerical coefficients (37 and 47.5) stand in stark
contrast to the value of $\sim$3 expected for a hadron gas with
temperature $T < T_C$, in which case the degrees of freedom are
dominated by the three pion species $\pi^-,\pi^0,\pi^+$.

The exact order of this phase transition is not known.  In a pure
gauge theory containing only gluons the transition appears to be
first order. However, inclusion of two light quarks (up and down) or
three light quarks (adding the strange quark) can change the
transition from first order to second order to a smooth crossover.
These results are obtained at zero net baryon density; dramatic
changes in the nature of the transition and in the medium itself are
expected when the net baryon density becomes significant. A
schematic version of the phase diagram for an idealized form of
nuclear matter with vanishing light quark (up and down) masses and
infinite strange quark mass is presented in
Fig. \ref{fig_phase} \cite{Rajagopal:2000uu}. For sufficiently large
values of the baryon chemical potential $\mu$ this system exhibits a
first order phase transition between hadronic matter and QGP, along
with a tricritical point below which the transition becomes second
order.   However, non-zero values of the light quark masses
dramatically alter this simple picture: The second order phase
transition denoted by the dashed line in Fig. \ref{fig_phase}
becomes a smooth crossover, and the tricritical point
correspondingly becomes a critical point designating the end of the
first order transition found at higher values of $\mu$. For example,
recent calculations \cite{Fodor:2004nz,Ejiri:2003dc} indicate that
the transition is a crossover for values of $\mu < \sim 400$ MeV.
Given that both theoretical arguments and experimental data suggest
that nucleus-nucleus collisions at RHIC (at least near mid-rapidity)
are characterized by low net baryon density, we will restrict our
attention to this regime, while noting that the predicted smooth
nature of the transition in this region increases the experimental
challenges of unambiguously establishing that such a transition has
occurred. We also note that while Fig. \ref{fig_phase} shows that
the region of low temperature and high baryon density is expected to
show a transition to a color superconducting phase of matter, this
regime is not accessible to RHIC collisions and will not be
discussed further.

While the lattice results plotted in Fig. \ref{fig_lattice} show
that the energy density reaches a significant fraction ($\sim 0.8$)
of the Stefan-Boltzmann values in the deconfined phase, the
deviation from $\varepsilon_{SB}$, and the reason for the persistence
of that deviation  to the highest  studied values of $T/T_C$, are of
great interest. For instance, Greiner has noted \cite{greiner} that
``in order to allow for simple calculations the QGP is usually
described as a free gas consisting of quarks and gluons.  This is
theoretically not well founded at $T \approx T_{c}$''. In fact,
analysis of the gluon propagator in a thermal
system \cite{Klimov:1982bv,Weldon:1982aq} has demonstrated that
effective masses of order $g(T) T$ are generated, suggesting that
the relevant degrees of freedom are in fact massive near $T_C$.
$m_{g} \approx T_{c}$ could be generated by gluons.
Especially interesting is recent work which indicates that both
heavy \cite{Datta:2002ck,Asakawa:2003re,Datta:2003ww} and
light \cite{Karsch:2002wv} flavor states may remain bound above
$T_C$, calling into question the naive interpretation of
$\varepsilon(T)$ as an indicator of the explicit  appearance of quark
and gluon degrees of freedom. This is supported by explicit
calculations of the spectrum of bound states above
$T_C$ \cite{Shuryak:2004tx} which predict a rich structure of states
that belies a description as a weakly interacting parton gas.

To emphasize this point, consider the standard measure of the degree
of coupling in a classical plasma, obtained by comparing the
relative magnitudes of the average kinetic and potential energies:
\begin{equation}
\Gamma \equiv { {\langle V(r) \rangle}
                 \over
                {\langle E_{kin} \rangle}
              }.
\end{equation}
In the case of the QCD plasma the mean inter-particle spacing should
scale as some numerical coefficient times $1/T $. Naively, this
gives a mean potential energy $\langle V(r) \rangle \sim
\alpha_s(T)\langle 1/r \rangle \sim \alpha_s(T) T$, leading to
\begin{equation}
\Gamma \sim { {\alpha_s(T) T}
              \over
              { 3 T }
             }
             \sim \alpha_s(T).
\end{equation}
Any reasonable estimate for the numerical coefficients leads to
$\Gamma > 1$, which is the condition for a ``strongly-coupled''
plasma. In reality, the screening present at such densities (or
equivalently, the generation of effective gluon masses) modifies the
mean potential energy to $\langle V(r) \rangle \sim g(T) T$, which
only increases the estimated value of $\Gamma$\ \cite{BerndtPrivate}.
Considerations such as these have led some
authors \cite{Shuryak:2003xe,Gyulassy:2004zy} to denote quark-gluon
plasma in this regime as ``sQGP" for ``strongly interacting QGP".

It is worth noting that this state of affairs has been anticipated
by many authors. Whether the argument was based on the divergence of
perturbative expansions \cite{Baym:1982sn}, on phenomenological
descriptions of confinement \cite{Plumer:1984tq}, on the development
of effective gluon masses from plasmon modes \cite{Muller:1992tb} or
on general principles \cite{greiner}, it is clear that the QGP near
$T_C$ should not be regarded as an ideal gas of quarks and gluons.

\begin{figure}[tbhp]
\includegraphics[width=1.0\linewidth]{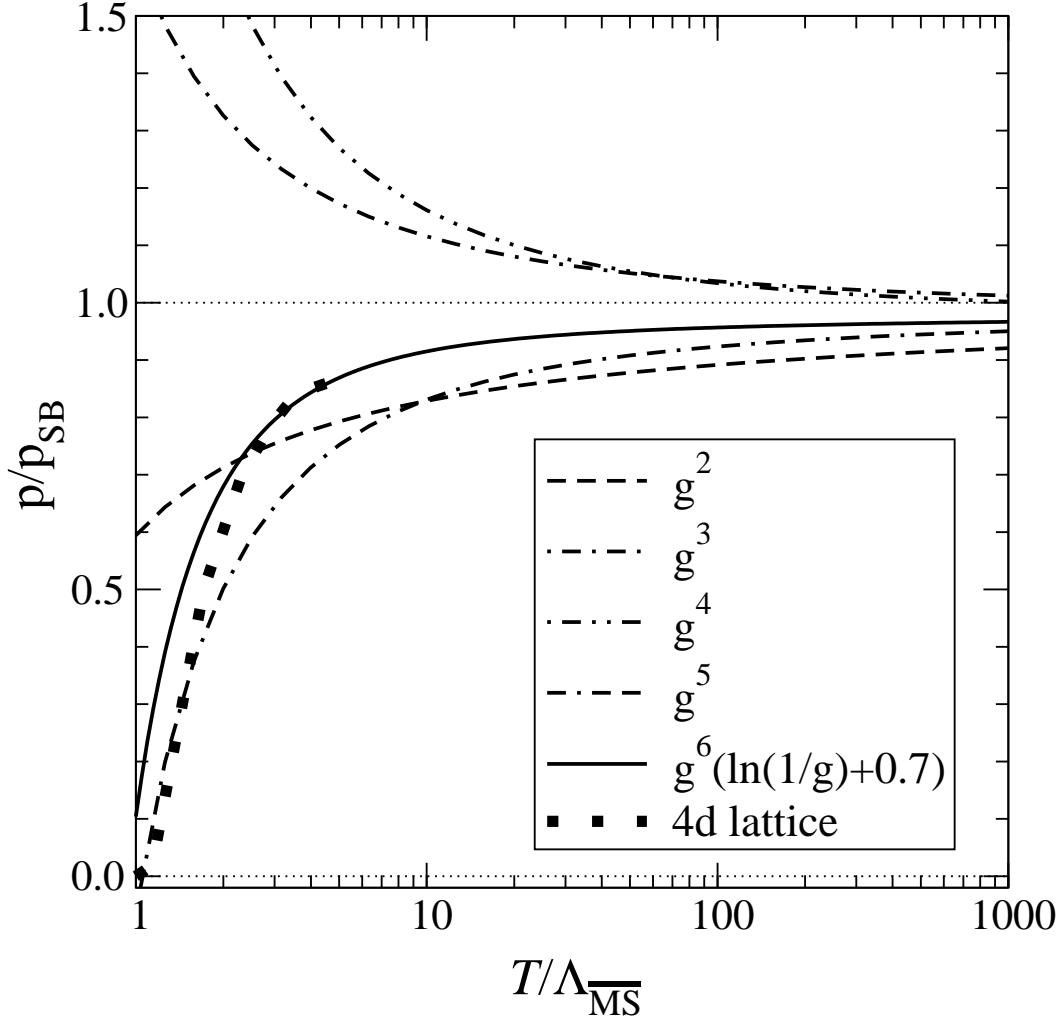}
\caption{Perturbative QCD results for the pressure as a function of
temperature at various orders normalized to the Stefan-Boltzmann
value $p_{SB}$ \cite{Kajantie:2002wa}.\label{fig_hotqcd}}
\end{figure}

How high a temperature is needed not just to form a quark-gluon
plasma, but to approach this ``weakly'' interacting plasma? A
calculation of the pressure of hot matter within perturbative
QCD \cite{Kajantie:2002wa} is shown in Fig. \ref{fig_hotqcd}. The
pressure result oscillates significantly as one considers
contributions of different orders. These oscillations are an
indication that the expansion is not yielding reliable results.
However at temperatures approaching 1000 times of $T_{C}$ ($\approx
\Lambda_{\bar{MS}}$), they appear to be converging toward the
Stefan-Boltzmann limit (asymptotically free partons).  It is
interesting that in considering the highest-order term, the results
are still nonconvergent though one seems to approach the lattice
calculated pressure.  Unlike the case of single parton-parton
scattering at zero temperature, the infrared problems of
finite-temperature field theory prevent further analytic progress
even for very small values of the coupling
constant \cite{Kajantie:2002wa,Linde:1980ts,Gross:1980br}.

The goal of relativistic heavy ion physics is the experimental study
of the nature of QCD matter under conditions of extreme temperature.
A great emphasis has been placed on ``the discovery of the
quark-gluon plasma'', where the terminology ``quark-gluon plasma''
is used as a generic descriptor for a system in which the degrees of
freedom are no longer the color neutral hadron states observed as
isolated particles and resonances. This definition is limited since
high-energy proton-proton reactions cannot be described purely in
terms of color-neutral hadrons, but rather require analysis of the
underlying partonic interactions. The hoped-for essential difference
in heavy ion collisions is the dominance of the partonic-level
description for essentially all momentum scales and over nuclear
size distances. Beyond this simple criterion, in order to
characterize the produced system as a state of matter it is
necessary to establish that these non-hadronic degrees of freedom
form a statistical ensemble, so that concepts such as temperature,
chemical potential and flow velocity apply and the system can be
characterized by an experimentally determined equation of state.
Additionally, experiments eventually should be able to determine the
physical characteristics of the transition, for example the critical
temperature, the order of the phase transition, and the speed of
sound along with the nature of the underlying quasi-particles. While
at (currently unobtainable) very high temperatures $T \gg T_{c}$ the
quark-gluon plasma may act as a weakly interacting gas of quarks and
gluons, in the transition region near $T_{c}$ the fundamental
degrees of freedom may be considerably more complex.  It is
therefore appropriate to argue that the quark-gluon plasma must be
defined in terms of its unique properties {\em at a given
temperature}. To date the definition is provided by lattice QCD
calculations. Ultimately we would expect to validate this by
characterizing the quark-gluon plasma in terms of its experimentally
observed properties.  However, the real discoveries will be of the
fascinating properties of high temperature nuclear matter, and not
the naming of that matter.


\subsection{Experimental Program}

The theoretical discussion of the nature of
hadronic matter at extreme densities has been greatly stimulated by
the realization that such conditions could be studied via
relativistic heavy ion collisions \cite{ExpHistory}. Early
investigations at the Berkeley Bevalac (c. 1975--1985), the BNL AGS
(c. 1987--1995) and the CERN SPS (c. 1987--present) have reached their
culmination with the commissioning of BNL's Relativistic Heavy Ion
Collider (RHIC), a dedicated facility for the study of nuclear
collisions at ultra-relativistic energies \cite{Baym:2001in}.

The  primary goal of RHIC is the experimental study of the QCD phase transition.
The 2002 Long-Range Plan for Nuclear Science \cite{LRP02} clearly enunciates this objective:
\begin{quotation}
{\em
...the completion
of RHIC at Brookhaven has ushered in a new era. Studies
are now possible of the most basic interactions predicted by
QCD in bulk nuclear matter at temperatures and densities
great enough to excite the expected phase transition to a
quark-gluon plasma. As the RHIC program matures, experiments
will provide a unique window into the hot QCD vacuum,
with opportunities for fundamental advances in the
understanding of quark confinement, chiral symmetry breaking,
and, very possibly, new and unexpected phenomena in
the realm of nuclear matter at the highest densities.}
\end{quotation}
The RHIC accelerator and its four experiments were commissioned and
brought online in the summer of 2000. The initial operation of both
RHIC and the experiments has been remarkably successful. In these
first three years the accelerator has collided, and the experiments
have acquired data on, Au+Au collisions at five energies, an
essential $p+p$ baseline data set, and a critical $d$+Au comparison.
The analyses of these various systems have resulted in a
correspondingly rich abundance of results, with over 90 publications
in the refereed literature.

It is therefore appropriate to reflect on the physics
accomplishments to date, with a particular emphasis on their
implications for the discovery of a new state of matter. At the same
time, it is essential to identify those features of the data (if
any) that are at odds with canonical descriptions of the produced
matter, to specify those crucial measurements which remain to be
made, and to outline a program for continued exploration and
characterization of strongly interacting matter at RHIC. The PHENIX
collaboration \cite{PHENIX} has performed such an assessment; this
document represents a summary of its findings.

The PHENIX Conceptual Design Report \cite{phenix:CDR}, submitted to
BNL/RHIC management on January 29th, 1993, outlined a comprehensive
physics program focused on the search for and characterization of
new states of nuclear matter. The measurement of electromagnetic
probes and high-transverse-momentum phenomena formed a major thrust
of the proposed program. It was also realized that the measurement
of global variables and soft identified hadron spectra in the same
apparatus was essential to the goal of understanding the evolution
of the produced matter over all relevant timescales. These diverse
criteria required combining an unprecedented number of subsystems
together with a high-bandwidth trigger and data-acquisition system
into an integrated detector design. Particular attention was given
to minimizing the conflicting design criteria of the central arm
spectrometers, with their requirement for minimal mass in the
aperture, and those of the muon spectrometers which require maximal
absorption of the incident hadron flux. The data acquisition and
trigger system was designed to accommodate the great variety of
interaction rates and event sizes provided by RHIC. Every effort was
made to provide for future upgrades, both in the geometry of the
experiment and in the architecture and design parameters of the
read-out system.

%
The published PHENIX results of Au+Au collision
at a center-of-mass energy per nucleon
pair, $\sqrt{s_{NN}}$, of 130 GeV
\cite{Adcox:2000sp,Adcox:2001ry,Adcox:2001jp,Adcox:2001mf,Adcox:2002uc,Adcox:2002cg,Adcox:2002au,Adcox:2002mm,Adcox:2002pa,Adcox:2002ms,Adcox:2002pe,Adcox:2003nr} and at
$\sqrt{s_{NN}} = 200$ GeV
\cite{Adler:2003qi,Adler:2003kt,Adler:2003rc,Adler:2003kg,Adler:2003au,Adler:2003cb,Adler:2003xq,Adler:2004rq,Adler:2004ta,Adler:2004uy,Adler:2004zn},
$p+p$ collisions at $\sqrt{s} = 200$ GeV \cite{Adler:2003pb,Adler:2003qs,Adler:2004ps,Adler:2005qk},
and $d$+Au at $\sqrt{s_{NN}} = 200$ GeV \cite{Adler:2003ii,Adler:2004eh}
clearly demonstrate that PHENIX's
goal to make high-quality measurements in both hadronic and leptonic
channels for collisions ranging from $p+p$ to Au+Au has been realized.
A summary of these results illustrates this point:
\begin{itemize}

\item First measurement of the dependence of the charged particle pseudo-rapidity density \cite{Adcox:2000sp} and the
      transverse energy \cite{Adcox:2001ry} on the number of participants in Au+Au collisions at $\sqrt{s_{NN}}$ = 130 GeV;
      systematic study of the centrality and $\sqrt{s_{NN}}$ dependence of $dE_T/d\eta$ and $dN_{ch}/d\eta$ \cite{Adler:2004zn}.

\item Discovery of suppressed production for $\pi^0$'s and charged particles at high $p_T$
      in Au+Au collisions at $\sqrt{s_{NN}}$ = 130 GeV \cite{Adcox:2001jp} and a systematic study
      of the scaling properties of the suppression \cite{Adcox:2002pe};
      extension of these results to much higher transverse momenta in
      Au+Au collisions
      at $\sqrt{s_{NN}}$ = 200 GeV \cite{Adler:2003qi,Adler:2003au}.

\item Co-discovery (together with BRAHMS \cite{Arsene:2003yk}, PHOBOS \cite{Back:2003ns} and STAR \cite{Adams:2003im}) of absence of high-$p_T$ suppression in
      $d$+Au collisions at $\sqrt{s_{NN}}$ = 200 GeV \cite{Adler:2003ii}.

\item Discovery of the anomalously large proton and anti-proton yields
      at intermediate transverse momentum in
      Au+Au collisions at $\sqrt{s_{NN}}$ = 130 GeV through the systematic study of
      $\pi^\pm$, $K^\pm$, $p$ and ${\bar p}$ spectra \cite{Adcox:2001mf};
      study of the scaling properties of the proton and anti-proton yields
      in Au+Au collisions at
      $\sqrt{s_{NN}}$ = 200 GeV \cite{Adler:2003kg};

\item Measurement of $\Lambda$'s and ${\bar \Lambda}$'s in
      Au+Au collisions at $\sqrt{s_{NN}}$ = 130 GeV \cite{Adcox:2002au};
      measurement of $\phi$'s at $\sqrt{s_{NN}}$ = 200 GeV \cite{Adler:2004hv};
      measurement of deuteron and anti-deuteron spectra
      at $\sqrt{s_{NN}}$ = 200 GeV \cite{Adler:2004uy}.

\item Measurement of Hanbury-Brown-Twiss (HBT) correlations in $\pi^+\pi^+$ and $\pi^-\pi^-$ pairs in
       Au+Au collisions at $\sqrt{s_{NN}}$ = 130 GeV \cite{Adcox:2002uc} and 200 GeV \cite{Adler:2004rq},
       establishing that the ``HBT puzzle'' of
       $R_{\mbox{\rm out}} \approx R_{\mbox{\rm side}}$ extends to high pair momentum.

\item First measurement of single electron spectra in
       Au+Au collisions at $\sqrt{s_{NN}}$ = 130 GeV, suggesting that charm production
       scales with the number of binary collisions \cite{Adcox:2002cg};
       measurement of centralty dependence of charm production in
       Au+Au collisions at $\sqrt{s_{NN}}$ = 200 GeV \cite{Adler:2004ta}.

\item Sensitive measures of charge fluctuations \cite{Adcox:2002mm}
      and fluctuations in mean $p_T$ and transverse energy per particle \cite{Adcox:2002pa,Adler:2003xq}
      in Au+Au collisions at $\sqrt{s_{NN}}$ = 130 GeV and 200 GeV.

\item Measurements of elliptic flow for charged particles
      from Au+Au collisions at $\sqrt{s_{NN}}$ = 130 GeV \cite{Adcox:2002ms}
      and 62 GeV to 200 GeV \cite{Adler:2004cj}
      and identified charged hadrons
      from Au+Au collisions at $\sqrt{s_{NN}}$ = 200 GeV \cite{Adler:2003kt}.

\item Extensive study of hydrodynamic flow, particle yields, ratios and spectra
      from Au+Au collisions at $\sqrt{s_{NN}}$ = 130 GeV \cite{Adcox:2003nr}
      and 200 GeV \cite{Adler:2003cb}.

\item First observation of $J/\psi$ production in
      Au+Au collisions at $\sqrt{s_{NN}}$ = 200 GeV \cite{Adler:2003rc}.

\item Measurement of the nuclear modification factor for hadrons at forward and
      backward rapidities in
      $d$+Au collisions at $\sqrt{s_{NN}}$ = 200 GeV \cite{Adler:2004eh}.

\item First measurement of  the jet structure of baryon excess in
      Au+Au collisions at $\sqrt{s_{NN}}$ = 200 GeV \cite{Adler:2004zd}.

\item First measurement of elliptic flow of single electrons from charm decay in
      Au+Au collisions at $\sqrt{s_{NN}}$ = 200 GeV \cite{Adler:2005ab}.

\item First measurement of direct photons in
      Au+Au collisions at $\sqrt{s_{NN}}$ = 200 GeV \cite{Adler:2005ig}.

\item Measurement of crucial baseline data on $\pi^0$ spectra \cite{Adler:2003pb},
      direct photon producion \cite{Adler:2005qk},
      and $J/\psi$ production \cite{Adler:2003qs} in
      $p+p$ collisions at $\sqrt{s}$ = 200 GeV.

\item First measurement of the double longitudinal spin asymmetry
      $A_{LL}$ in $\pi^0$ production for polarized $p+p$ collisions at
      $\sqrt{s}$ = 200 GeV \cite{Adler:2004ps}.

\end{itemize}

These publications encompass physics from the barn to the picobarn
level; their very breadth precludes a detailed presentation here.
These data, together with a rich program of future RHIC
measurements, will allow us to address many of the features that
would characterize a quark-gluon plasma:
\begin{itemize}
 \item Temperature
 \item Parton number density
 \item Energy density
 \item Opacity
 \item Collective behavior
 \item Thermalization leading to the quark-gluon phase
 \item Deconfinement
 \item Number and nature of degrees of freedom
 \item Recombination of quarks and gluons to form final-state hadrons
 \item Chiral symmetry restoration
 \item Time evolution of system parameters
 \item Equation of state
 \item Color and thermal transport properties
 \item Critical behavior
\end{itemize}
As emphasized above, the present PHENIX data set from RHIC 
runs in year 2000 to 2003 already provides an extensive set of
measurements on global variables: (transverse energy and
multiplicity, elliptic flow); correlations and fluctuations:
(fluctuations in charge and $\langle p_T \rangle$, HBT
measurements), hadron spectra: (low-$p_T$ single-hadron spectra and
radial flow, particle ratios, resonances, anomalous $p/\pi$ ratio at
intermediate $p_T$); high-$p_T$ physics: (high-$p_T$ singles
spectra, suppression phenomena in $A+A$, nonsuppression in $d+A$,
high-$p_T$ two-particle correlations, nuclear
suppression/enhancement in forward/backward directions), heavy
flavor production: (charm, $J/\psi$), and electromagnetic probes:
(direct photons).  However, an important conclusion of this report
is that systematic studies of these observables (vs. collision
species and energy) are needed to extract unambiguous information on
most of these features.
\subsection{Organization of this Document}
As a result, this paper concentrates on those aspects of the
present data that address the broad features of energy density,
thermalization, deconfinement and critical behavior. The focus in
most cases will be on the data of the PHENIX experiment, but the
data of the other RHIC experiments will be cited to support and to
extend the discussion\footnote{An underappreciated aspect of the
RHIC program is the excellent agreement between the various
experiments in almost all measured channels.}. The experimental
tools that allow the systematic study of all phenomena as a function
of the inferred impact parameter are presented in the context of
hard-scattering phenomena. These methods and the associated data are
then used to discuss the experimental evidence for the formation of
a state of high-density matter. The measured abundances, spectra and
flow patterns are used to analyze the degree of thermalization and
collectivity in the produced matter. These results are then examined
for evidence establishing the role of deconfined quarks and gluons
in the produced system, along with the implications for its
description as a quark-gluon plasma. A concluding section summarizes
the findings and identifies key future measurements required to
further refine our observations.
 
\section{ENERGY DENSITY AND $E_T$, $N_{CH}$}
\label{Sec:denseE}
%
%
%
%


A prerequisite for creating  a quark-gluon plasma is producing a
system with sufficiently large energy density.  From both elementary
estimates \cite{Shuryak:2004tx} and from extensive numerical studies
in lattice QCD \cite{Karsch:2001cy,Laermann:2003cv}, the required
density is known to be on the order of 1 GeV/fm$^3$. Establishing
that this energy density is created in RHIC collisions is a basic
ingredient in establishing the creation of a QGP at RHIC.

In this section we explore what can be deduced about the energy
densities achieved in RHIC $A+A$ collisions from measurements of the
global transverse energy and multiplicity.  In later sections these
estimates will be compared to densities inferred from
hydrodynamics-based models (Section \ref{Sec:therm}) and from jet
quenching evidence (Section \ref{Sec:denseN}).

\begin{figure}[tbhp]
\center
\includegraphics[width=1.0\linewidth]{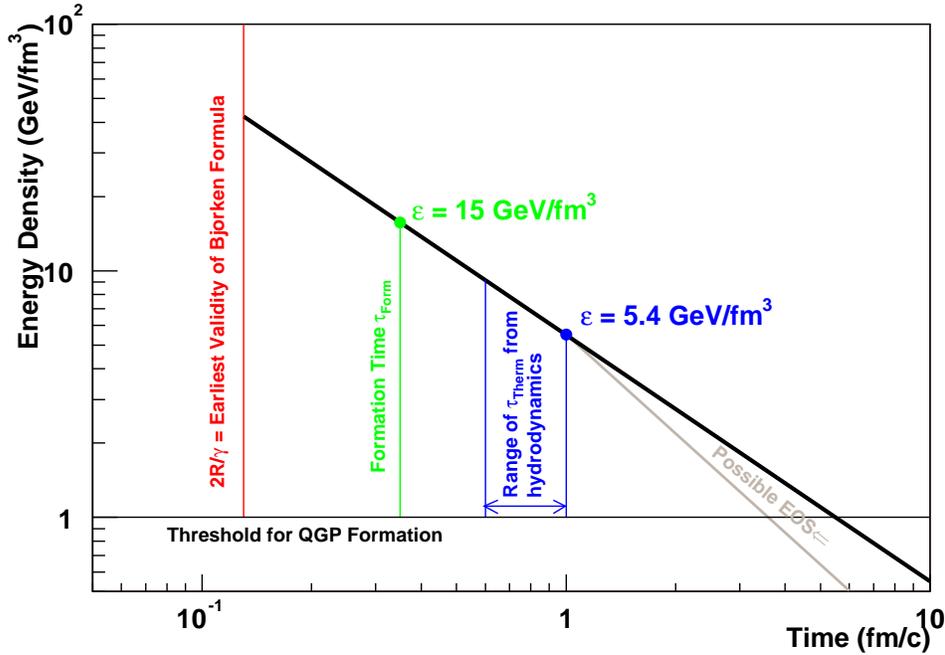}
\caption[]{\label{fig:eps_vs_time} Schematic
drawing of the time and energy density scales
derived through the Bjorken picture.}
\end{figure}

Specifically, we will address three different energy density
estimates, and introduce two distinct time scales: (i) The peak {\em
general energy density} that is achieved when the incoming nuclei
overlap; (ii) The peak {\em formed energy density} involving created
particles at proper time $\tau_{Form}$; and (iii) The peak {\em
thermalized energy density} present at proper time $\tau_{Therm}$
when local thermal equilibrium is first achieved (assuming that this
occurs).  The values and time scales for formed and thermalized
energy densities are indicated schematically in
Fig \ref{fig:eps_vs_time}; detailed explanations follow in
Sections \ref{SubSec:Realistic_EDense}
and \ref{SubSec:Thermalized_EDense}.

In this Section we will also review data on overall
particle multiplicities, and through them distinguish
between different models of the initial
particle production.

\subsection{General Energy Density}
\label{SubSec:General_EDense}

The simplest definition of ``energy density'' is the total mass-energy
within some region of space divided by the volume of that region, as
seen at some instant of time in some Lorentz frame.  However, this
definition is not satisfactory since we can ``trivially'' raise any
simple energy density by viewing the system in a different frame. For
example, a static system with constant energy density $\rho_{0}$ in
its rest frame---say, a gold nucleus---will appear to have energy
density $\gamma^{2} \rho_{0}$ when viewed in a frame boosted by
Lorentz $\gamma$.  Accordingly, we can only calculate a {\em
meaningful} energy density $\langle
\varepsilon \rangle$ as \mbox{mass-energy/volume} for some region
{\em in the case} when the total momentum in the region is zero.

Now let us imagine a symmetric RHIC $A+A$ collision at a moment when
the two original nuclei are overlapping in space, as seen in the
laboratory/center-of-mass frame.  The total momentum in any overlap 
region is zero by
symmetry, so we can calculate a meaningful---if short-lived---energy
density for such a region.  If each nucleus has energy density
$\rho_{0}$ in its rest frame then the total energy density in the
overlap region is just $\langle \varepsilon \rangle = 2 \rho_{0}
\gamma^2$.  If we take a nominal $\rho_{0}$ = 0.14 GeV/fm$^3$ for a
nucleus at rest and $\gamma = 106$ for a full-energy RHIC collision,
then the result for the peak general energy density is $\langle
\varepsilon \rangle = $3150 GeV/fm$^3$.  This is a spectacularly,
almost absurdly high number on the scale of $\sim$1 GeV/fm$^3$
associated with the familiar transition described by lattice QCD.

This energy density is of course artificial, in that it would be
temporarily present even in the case of no interactions between the
two nuclei. It is instructive to consider the (again artificial)
case where the nucleons in the two nuclei have only elastic
interactions.
Then the time during which a high energy density is present over
{\em any} volume cannot last longer than $t = 2R/\gamma$, where $R$ is
the rest-frame radius of the nucleus. With $R = $7 fm for Au this time
is only 0.13 fm/$c$ at RHIC, and after this time all energy densities
will fall precipitously back to $\rho_{0}$ if no secondary particles
are created. The scale of this interval is so short that a
scattering cannot even be said to have occurred within that volume
unless its momentum transfer scale $Q$ exceeds at least 1.5 GeV/$c$,
or more. Accordingly, we will turn our attention instead to energy
densities involving only produced particles,
as the potential source for a QCD transition.

\subsection{Formed Energy Density}
\label{SubSec:Formed_EDense}

\begin{figure}[tbhp]
\center
\includegraphics[width=1.0\linewidth]{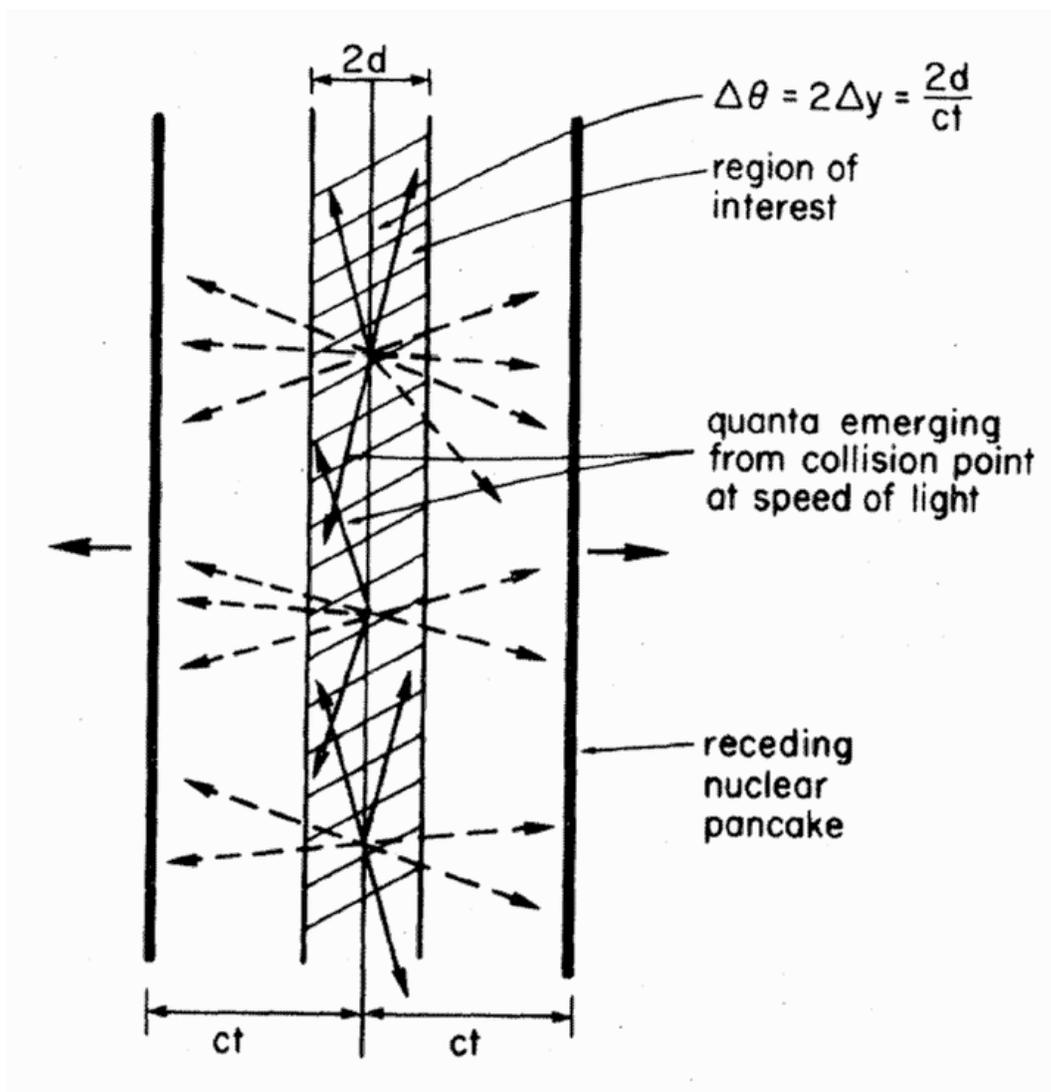}
\caption[]{\label{fig:Bj_quanta} Figure from
Bjorken \cite{Bjorken:1983qr} illustrating the geometry of initially
produced particles at a time $t$ after the overlap of the incoming
nuclei in some frame.
 The picture is valid in any
frame in which the incoming nuclei have very
high energies and so are highly Lorentz contracted.}
\end{figure}

In any frame (not just the center-of-mass frame) where the two incoming nuclei
have very high energies the region when/where the nuclei overlap
will be very thin in the longitudinal direction and very short in
duration.    In this limit, then, it is fair to describe all
secondary produced particles as having been radiated out from a very
thin ``disk'', and that they are all created at essentially the same
time. These realizations lead directly to the picture described by
Bjorken \cite{Bjorken:1983qr}, whose original diagram is reproduced
in Fig. \ref{fig:Bj_quanta} and whose derivation we retrace briefly
here.

Once the  beam ``pancakes'' recede after their initial overlap, the
region between them is occupied by secondaries at intermediate
rapidities.  We can calculate the local energy density of these
created particles if we make one further assumption: that the
secondaries can be considered ``formed'' at some proper time
$\tau_{Form}$ after they are radiated out from the thin source disk.

Our region of interest, in any frame, will be a slab perpendicular
to the beam direction, with longitudinal thickness $dz$, with one
face
 on the ``source''  plane in this frame, and transverse
extent with area $A$ covering the nuclear overlap
region\footnote{The region described here corresponds to half the
shaded region shown in Fig. \ref{fig:Bj_quanta}. Since
$\beta_{\parallel} \simeq 0$ for particles near the source location,
this is an appropriate region over which we  can calculate a
meaningful energy density.}.  At time $t = \tau_{Form}$ this volume will
contain all the (now-formed) particles with longitudinal velocities
$0 \leq \beta_{\parallel} \leq dz/\tau_{Form}$ (since we assume that
the particles cannot scatter before they are formed!). We can then
write this number of particles as $dN = (dz/\tau_{Form})
\frac{dN}{d\beta_{\parallel}}$, or equivalently $dN = (dz/\tau_{Form})
\frac{dN}{dy}$, where $y$ is longitudinal rapidity, since
$dy = d\beta_{\parallel}$ at $y = \beta_{\parallel} = 0$.  If these
particles have an average total energy $\langle m_{T} \rangle$ in
this frame ($E = m_{T}$ for particles with no longitudinal velocity),
then the total energy divided by the total volume of the slab at
$t = \tau_{Form}$ is just

\begin{eqnarray}
\langle \varepsilon(\tau_{Form}) \rangle & = & \frac{ dN \langle
m_{T} \rangle }{ dz \, A } = \frac{ dN(\tau_{Form}) }{ dy }
        \frac{\langle m_{T} \rangle}{\tau_{Form} A}
\nonumber \\
 & = &
\frac{1}{\tau_{Form} A}
\frac{ dE_{T}(\tau_{Form}) }{ dy }
\label{eq:eBj_1}
\end{eqnarray}

\noindent where we have equated $\frac{dE_{T}}{dy} = \langle m_{T}
\rangle \frac{dN}{dy}$ and emphasized that Eq. \ref{eq:eBj_1} is true
for the transverse energy density present at time $t = \tau_{Form}$.

Equation \ref{eq:eBj_1} here is essentially identical\footnote{A
(well-known) factor of 2 error appears in the original.} to
Eq. 4 of Bjorken's result \cite{Bjorken:1983qr}, and so is
usually referred to as the {\em Bjorken energy density}
$\varepsilon_{Bj}$. It should be valid as a measure of peak energy
density in created particles, on very general grounds and in all
frames, as long as two conditions are satisfied: (1) A finite
formation time $\tau_{Form}$ can meaningfully be defined for the
created secondaries; and (2) The thickness/``crossing time'' of the
source disk is small compared to $\tau_{Form}$, that is,
$\tau_{Form} \gg 2R/\gamma$.  
In particular, the validity of Eq. \ref{eq:eBj_1} is completely 
independent of the shape of the $dE_T(\tau_{Form})/dy$ distribution 
to the extent that $\beta_{\parallel}$ is infinitesimally small in 
a comoving frame; a plateau in $dE_T/dy$ is {\em not} required.
For present practical purposes, we will consider
condition (2) above to be satisfied as long as $\tau_{Form}
> 2R/\gamma$ is true, corresponding to $\tau_{Form}>$0.13 fm/$c$ for
full-energy Au+Au collisions at RHIC.

Bjorken's original motivation was to estimate, in advance of data,
the energy densities that would be reached in high-energy $A+A$
collisions, using knowledge of $p(\bar{p})+p$ collisions to estimate
$\langle m_{T} \rangle$ and $dN/dy$, and choosing 
$\tau_{Form}\sim$1 fm/$c$ 
without any particular justification other than as an
order-of-magnitude estimate.  With $A+A$ collision data in hand,
attempts have been made to use Eq. \ref{eq:eBj_1} to estimate the
energy densities that are actually achieved in the collisions.
Historically, $\varepsilon_{Bj}$ has been calculated using the
final-state $dE_{T}/dy$ and simply inserting
a nominal value of 1 fm/$c$ for $\tau_{Form}$.
In addition, fixed target experiments have been using
$dE_{T}/d\eta$ as an estimate for $dE_{T}/dy$,
which is a good approximation for these experiments;
at RHIC a correction is made for the Jacobian $dy/d\eta$ which
is important for a collider geometry.
These ``nominal Bjorken energy density'' estimates, which we term
$\varepsilon^{Nominal}_{Bj}$, range for central event
samples from about 1.5 GeV/fm$^3$ in Au+Au collisions at 
AGS energies \cite{Ahle:1994nk} ($\sqrt{s_{_{NN}}} = $5 GeV), 
to about 2.9 GeV/fm$^3$ in Pb+Pb collisions at 
SPS energies \cite{Margetis:1995tt,Adcox:2001ry} 
($\sqrt{s_{_{NN}}} = $17 GeV; and see also \cite{Adler:2004zn}) 
to about 5.4 GeV/fm$^3$ in Au+Au collisions at
full RHIC energy \cite{Adler:2004zn} ($\sqrt{s_{_{NN}}} = $200 GeV).

It has often been noted that all of these values are similar to, or
higher than, the 1 GeV/fm$^3$ scale required for the QCD transition.
However, we cannot take these $\varepsilon^{Nominal}_{Bj}$ estimates
seriously as produced energy densities without some justification
for the value of  1 fm/$c$ taken for $\tau_{Form}$. An indication of
potential problems with this choice arises immediately when
considering AGS Au+Au and SPS Pb+Pb collisions, where the center-of-mass
``crossing times'' $2R/\gamma$ are 5.3 fm/$c$ and 1.6 fm/$c$
respectively, which implies that this choice for
$\tau_{Form} = $1 fm/$c$ actually violates the validity condition
$\tau_{Form} > 2R/\gamma$ we set for the use of Eq. \ref{eq:eBj_1}.
So we will deprecate the use of $\varepsilon^{Nominal}_{Bj}$ as an
quantitative estimate of actual produced energy density, and instead
treat it only as a compact way of comparing $dE_{T}/d\eta$
measurements across different systems, centralities and beam
energies.

%
%

\subsection{Realistic $\tau_{Form}$ and
$\varepsilon_{Bj}$ estimates}
\label{SubSec:Realistic_EDense}

Can we justify a better estimate for $\tau_{Form}$?
We might say, on general quantum mechanical grounds,
that in a frame where its motion is entirely
transverse a particle of energy $m_{T}$ can be
considered to have ``formed'' after a time $t = \hbar/m_{T}$
since its creation in that frame.
%
%
%
To estimate the average transverse mass, we can use the final-state
$dE_{T}/d\eta$ to estimate $dE_{T}(\tau_{Form})/dy$ and,
correspondingly, use the final-state $dN/d\eta$ as an estimate for
$dN(\tau_{Form})/dy$ to obtain

\begin{equation}
\langle m_{T} \rangle = 
\frac{dE_{T}(\tau_{Form})/dy}{dN(\tau_{Form})/dy} \simeq
\frac{dE_{T}/d\eta}{dN/d\eta} \; (\mbox{Final state}).
\label{eq:mean_mt}
\end{equation}

\begin{figure}[tbhp]
\center
\includegraphics[width=1.0\linewidth]{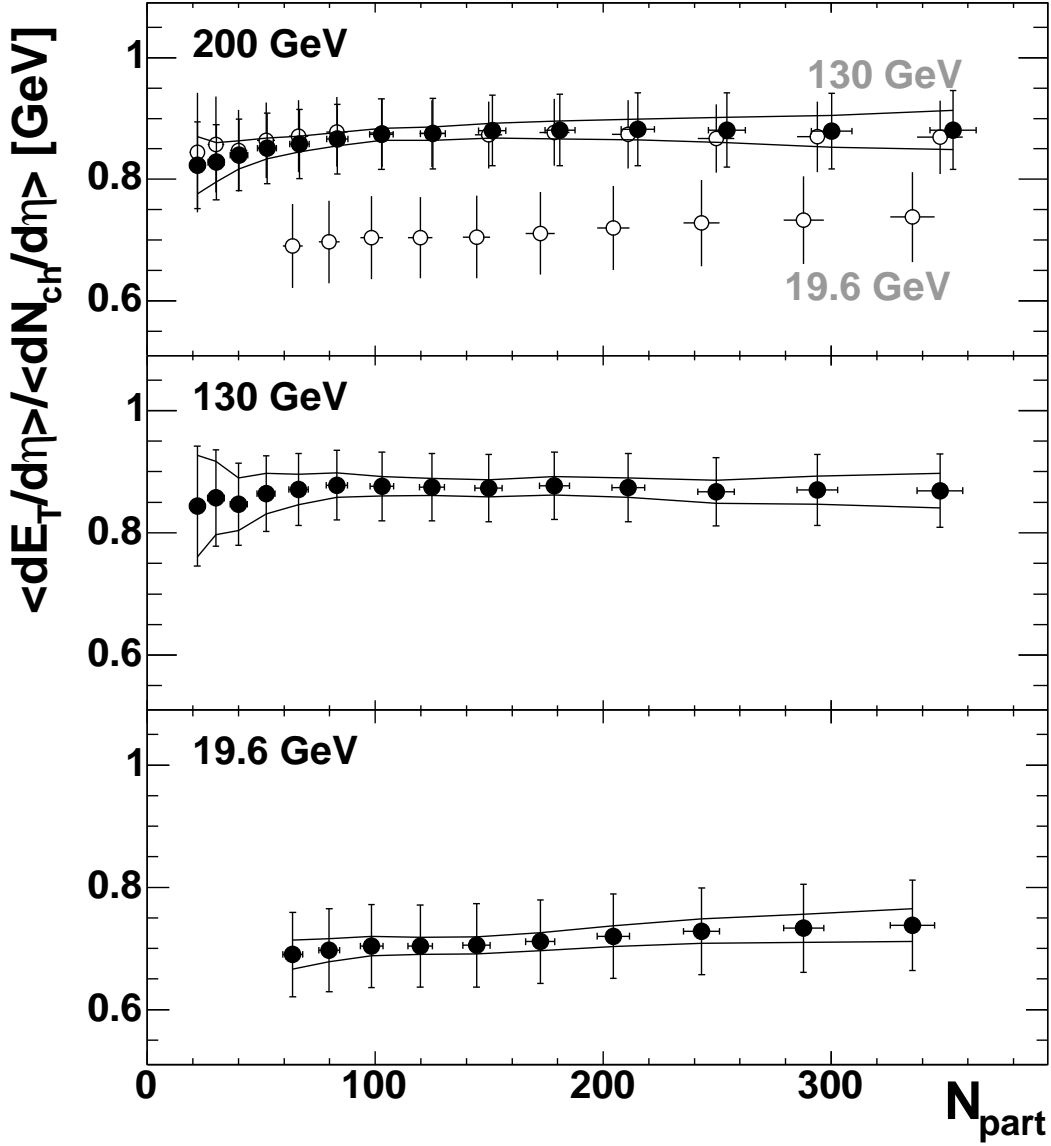}
\caption[]{\label{fig:det_over_dnch} The ratio
of transverse energy density in pseudorapidity
to charged particle density in pseudorapidity,
at mid-rapidity; shown as a function of centrality,
represented by the number of nucleons participating
in the collision, $N_{part}$,
for three different RHIC beam energies \cite{Adler:2004zn}.}
\end{figure}

\noindent PHENIX has measured the ratio of final-state transverse-energy 
density to charged-particle density, each per unit
pseudorapidity, and the results are shown in
Fig. \ref{fig:det_over_dnch}.  For a wide range of centralities the
ratio is remarkably constant at about 0.85 GeV for full-energy central
Au+Au collisions and shows very little change with beam energy,
decreasing to only 0.7 GeV when $\sqrt{s_{_{NN}}}$ is decreased by an
order of magnitude down to 19.6 GeV.

If we approximate $dN^{Ch}/d\eta = (2/3)dN/d\eta$ in the final state
then Eq. \ref{eq:mean_mt} would imply $\langle m_{T} \rangle
\simeq$ 0.57 GeV and corresponding $\tau_{Form} \simeq$ 0.35 fm/$c$, a
value shorter than the ``nominal'' 1 fm/$c$ but still long enough to
satisfy our validity condition $\tau_{Form} > 2R/\gamma$ at RHIC.
Inserting this value into Eq. \ref{eq:eBj_1}, along with the highest
$dE_{T}/d\eta$ = 600 GeV for 0--5\% central events as
measured by PHENIX \cite{Adler:2004zn}, yields a value of
$\langle \varepsilon \rangle$ = 15 GeV/fm$^3$ for the energy density
in initially produced, mid-rapidity particles in a central RHIC
Au+Au collision, that is, roughly 100 times the mass-energy density
of cold nuclear matter.



It is important to note that this large value of the energy density
as obtained from Eq. \ref{eq:eBj_1} represents a conservative {\em
lower} limit on the actual $\langle \varepsilon(\tau_{Form})
\rangle$ achieved in RHIC collisions.  This follows from two
observations: (1) The final-state measured $dE_{T}/d\eta$ is a solid
lower limit on the $dE_{T}(\tau_{Form})/dy$ present at formation
time; and (2) The final-state ratio $(dE_{T}/d\eta)/(dN/d\eta)$ is a
good lower limit on $\langle m_{T} \rangle$ at formation time, and
so yields a good {\em upper} limit on $\tau_{Form}$.  We justify
these statements as follows:

Several mechanisms are known that will decrease $dE_{T}/dy$ as the
collision system evolves after the initial particle formation, while
no mechanism is known that can cause it to increase (for $y = 0$, at
least). Therefore, its final-state value
should be a solid lower limit on its value at any earlier time.  A
partial list of the mechanisms through which $dE_{T}/dy$ will
decrease after $t = \tau_{Form}$ includes: (i) The initially formed
secondaries in any local transverse ``slab'' will, in a comoving
frame, have all their energy in transverse motion and none in
longitudinal motion; if they start to collide and thermalize, at
least some of their $E_{T}$ will be converted to longitudinal modes
in the local frame; (ii) Should rough local thermal equilibrium be
obtained while the system's expansion is still primarily
longitudinal, then each local fluid element will lose internal
energy through $pdV$ work and so its $E_{T}$ will decrease; (iii) If
there are pressure gradients during a longitudinal hydrodynamic
expansion then some fluid elements may be accelerated to higher or
lower rapidities; these effects are complicated to predict, but
we can state generally that they will always tend to {\em decrease}
$dE_{T}/dy$ where it has its maximum, namely at $y = 0$.   Given that
we have strong evidence that thermalization and hydrodynamical
evolution do occur in RHIC collisions (Section \ref{Sec:therm}), it
is likely that all these effects are present to some degree, and so
we should suspect that final-state $dE_{T}/d\eta$ is substantially
lower than $dE_{T}(\tau_{Form})/dy$ at mid-rapidity.

Turning to our estimate of $\tau_{Form}$, the assumption that
$\tau_{Form} = \hbar/\langle m_{T} \rangle$ cannot be taken as exact,
even if the produced particles' $m_{T}$'s are all identical, since
``formed'' is not an exact concept.  However, if we accept the basic
validity of this uncertainty principle argument,  then we can see
that the approximation in Eq. \ref{eq:mean_mt} provides a lower
limit on $\langle m_{T} \rangle$. First, the numerator
$dE_{T}/d\eta$ is a lower limit on $dE_{T}(\tau_{Form})/dy$, as
above. Second, the argument is often made on grounds of entropy
conservation that the local number density of particles can never
decrease \cite{Krasnitz:2002mn}, which would make the final-state
denominator in Eq. \ref{eq:mean_mt} an upper limit on its early-time
value.

%
%

With  these limits in mind, then, it is not unreasonable for us to
claim that the peak energy density of created particles reached in
central Au+Au collision at RHIC is at least 15 GeV/fm$^3$, and in
all likelihood is significantly higher.

\subsection{Thermalized Energy Density}
\label{SubSec:Thermalized_EDense}

We have arrived at a reasonably solid, lower-limit
estimate for the energy density in produced particles
in a RHIC Au+Au collision, and it is more than enough
to drive a QCD transition.  But the situation
at $t = \tau_{Form}$ pictured in Fig. \ref{fig:Bj_quanta}
looks nothing like local thermal equilibrium.  It is
an important question, then, to ask: if and when the
system evolves to a state of local thermal equilibrium,
is the energy density still sufficient to drive the
transition to a QGP?

To answer this we
begin by looking at the state of the system at
$t = \tau_{Form}$ and immediately afterward.
At the time they are formed the
particles have sorted themselves out automatically,
with all the particles on a ``sheet'' at a longitudinal
position $z$ having the same longitudinal velocity
$\beta_{\parallel} = z/t$; and so in the rest frame
of a sheet all the sheet's particles have only transverse
motion.  If the particles continue free-streaming and
never reinteract then the energy density will
continue to fall as $\varepsilon \sim 1/t$ and
the Bjorken formula in Eq. \ref{eq:eBj_1} will be
valid, with $t$ in place of $\tau_{Form}$, as
long as the expansion is primarily
longitudinal\footnote{For long times $t>R$ transverse
expansion will become significant and the energy
density will decrease as $\varepsilon \sim 1/t^3$.}.

For thermalization to occur the particles will have to start
interacting and/or radiating.  Once this happens the particles which
were originally together on one ``sheet'' will start to spread in
longitudinal velocity, though on short time scales we would expect
their group average longitudinal velocity to remain the same. If the
thermalization process is fast enough, then, we would expect that at
time $t = \tau_{Therm}$ these groups will have formed locally
equilibrated fluid elements, with a velocity profile
following $\beta^{Fluid}_{\parallel} = z/t$.  The energy density at
this time will be reduced from the energy density at formation time
$\varepsilon(\tau_{Form})$ by a factor $\tau_{Form}/\tau_{Therm}$;
{\em i.e.} the $\varepsilon_{Bj}$ of Eq. \ref{eq:eBj_1} but with
$\tau_{Therm}$ in place of $\tau_{Form}$. This evolution is
illustrated in Fig. \ref{fig:eps_vs_time}.

Once local equilibration is achieved we would then expect the system
to evolve  hydrodynamically, and the behavior of $\varepsilon(t)$
will depend on the details of the local equations of state (EOS).
Without knowing those details, though, we can say that in the limit
of low pressure, $p/\varepsilon \sim 0$, the energy density will
continue to evolve (during longitudinal expansion) as $\varepsilon
\sim 1/t$, while in the limit of high pressure,  $p/\varepsilon \sim
1/3$, the energy density will decrease somewhat more quickly,
$\varepsilon \sim 1/t^{4/3}$, within a fluid element.  This range of
possible behaviors for $t>\tau_{Therm}$ is indicated schematically
in Fig. \ref{fig:eps_vs_time}.

A direct theoretical determination of $\tau_{Therm}$ would require a
detailed description of both the parton-parton interactions and the
resulting evolution of the system density.  However, other lines of
reasoning may provide information on $\tau_{Therm}$.  For example, it
has been argued \cite{Kolb:2000sd} that the strong elliptic flow in
RHIC collisions can be taken as evidence for fast thermalization (see
Section \ref{Sec:thermFlow}).  In a hydrodynamic picture the source of
elliptic flow is the spatial anisotropy of the energy density in the
transverse plane at the time hydrodynamics becomes valid.  If local
equilibration and the onset of hydrodynamics is delayed because
interactions between the initially produced particles are weak at
first, then the spatial anisotropy which could give rise to elliptic
flow will be reduced (see Fig. \ref{fig:time_delayed_eccentricity}).
This, in effect, limits how high $\tau_{Therm}$ can be if
hydrodynamics is the mechanism for generating elliptic flow.

We can see from Table \ref{tab:hydro_v2_spectra_RHIC} in
Section \ref{Sec:thermHydro} that hydrodynamical models
typically require quite short thermalization times, in the range of
0.6--1.0 fm/$c$, in order to reproduce the magnitude of elliptic
flow which is observed at RHIC.  If we take this range as 
typical of what hydrodynamics would imply for $\tau_{Therm}$,
then we can calculate the corresponding ``typical'' implied
energy densities at thermalization time as in range of 
5.4 GeV/fm$^3$ to 9.0 GeV/fm$^3$.  These densities are well 
above that required to drive the QCD transition, so the combination 
of our transverse energy measurements and the fast thermalization 
times from hydrodynamics can be taken, to some degree, as evidence
that conditions to create the equilibrated upper phase of QCD
matter are achieved at RHIC.

\subsection{What Are the Initial Quanta?}
\label{SubSec:initial_quanta}

With our extensive use of the picture in Fig. \ref{fig:Bj_quanta} it
is only natural to ask, ``What are these initially produced
particles?''  that Bjorken referred to, nonspecifically, as
``quanta''. What models do we have for initial production, and what
can we say about them using our data on $E_T$ and multiplicity?

\begin{figure}[tbhp]
\center
\includegraphics[width=1.0\linewidth]{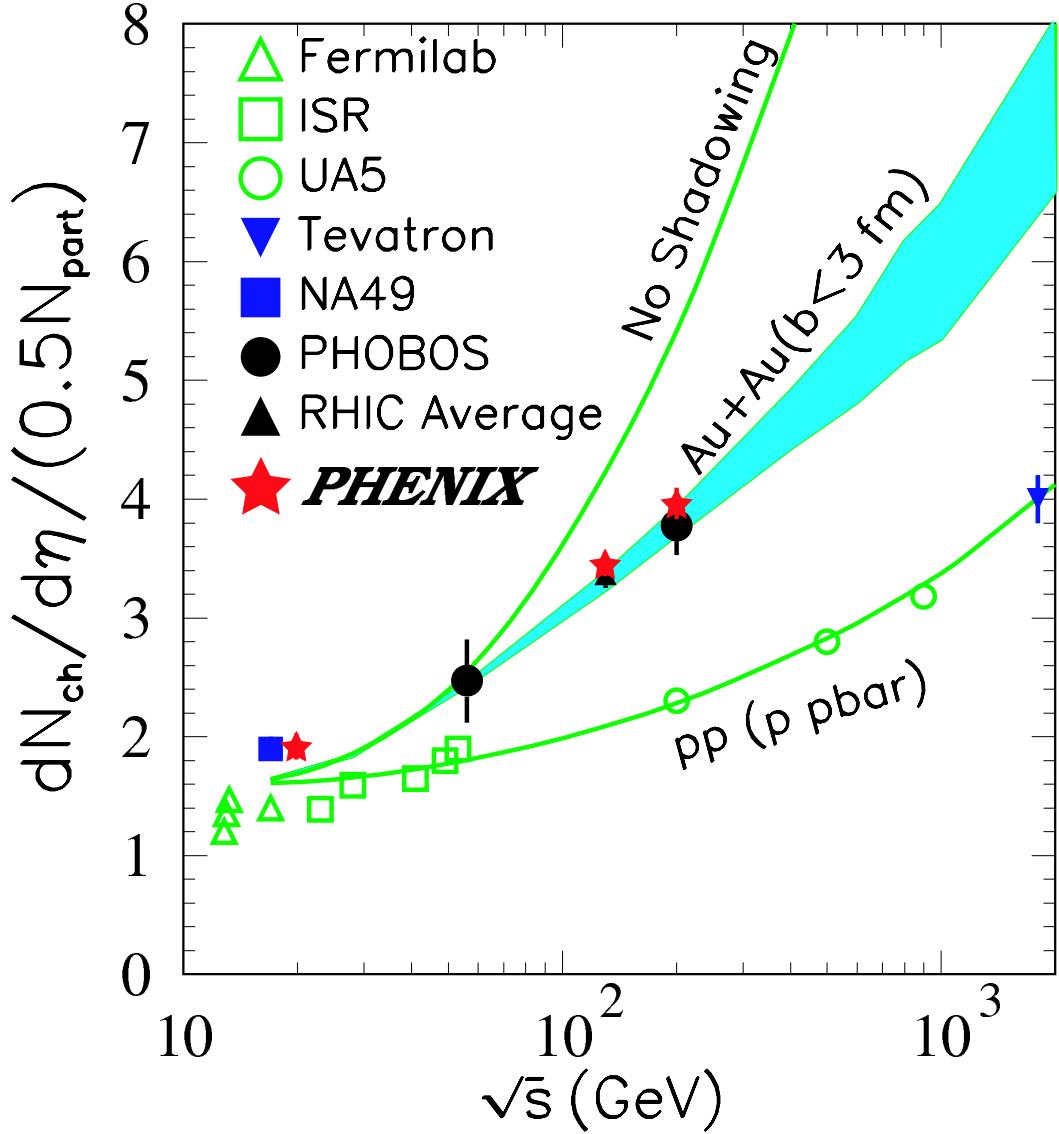}
\caption[]{\label{fig:dndeta_WangLi} Figure from Li and
Wang \cite{Li:2001xa} showing trends in final-state
charged multiplicity per participant pair vs.
(nucleon-nucleon) beam energy. (PHENIX data points\cite{Adler:2004zn}
have been added.)  The curves are the result of their
two-component ``hard/soft'' model, which reproduces
well the multiplicities from elementary $p(\bar{p})+p$
collisions at RHIC energies.  The same model
extended to nuclear collisions with no regulating
mechanism on hard processes (the ``No Shadowing'' line)
over-predicts the multiplicities in central RHIC collisions,
while the data can be matched if substantial nuclear
shadowing of gluons is invoked (shaded band).}
\end{figure}

The simplest assumption is that the initially produced particles in
a RHIC collision are scattered partons at mid- to low-$p_{T}$,
traditionally known as ``mini-jets''.  For a long period in advance
of RHIC data, it was widely expected that mini-jets would be the
dominant channel for $E_T$ and particle production, and this led to
two further, general expectations: first, that multiplicity and
$E_T$ per interacting nucleon would go up sharply at collider
energies, as compared to fixed-target energies, since jet and
mini-jet cross sections are increasing quickly with $\sqrt{s}$ (see
Fig. \ref{fig:dndeta_WangLi}); and, secondly, that $E_T$ and
multiplicity per participating nucleon would increase steeply in
more central events, since the rate of hard pQCD scatterings goes up
faster with centrality than does the number of interacting nucleons.

It was therefore quite surprising when the first RHIC
data \cite{Back:2000gw,Adcox:2000sp,Adcox:2001ry} showed lower
multiplicities than had been predicted from mini-jet models, and
only a modest increase in $E_T$ and multiplicity per participant as
functions of centrality. Compared to the sharp rise, shown in
Fig. \ref{fig:dndeta_WangLi}, predicted by straightforward
factorized pQCD, it was clear that some mechanism must be acting at
RHIC energies to restrict, or regulate, particle
production \cite{Wang:2000bf,Li:2001xa}.

\begin{figure}[tbhp]
\center
\includegraphics[width=1.0\linewidth]{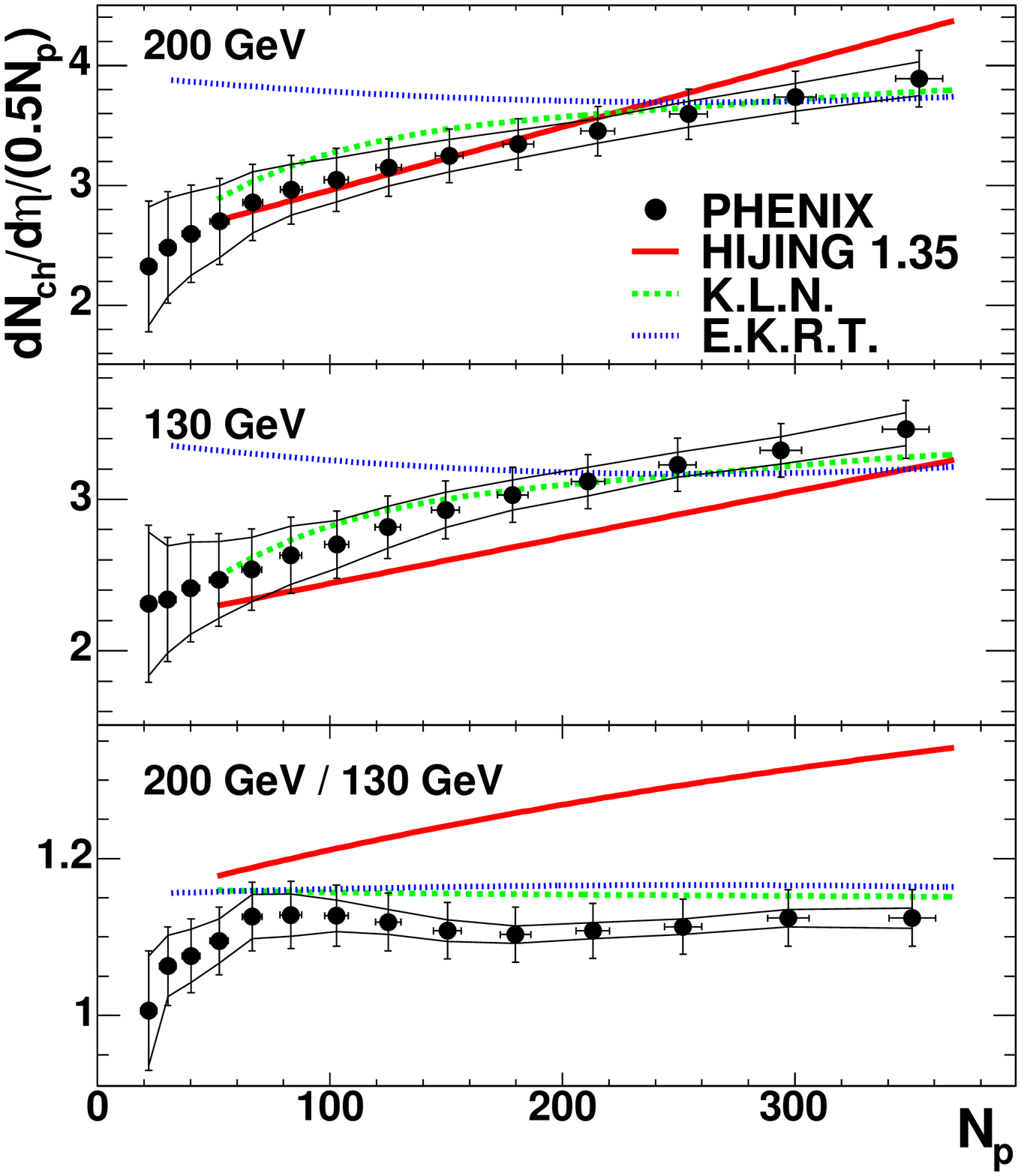}
\caption[]{\label{fig:Bazilevsky_fig3} Multiplicity per participant
nucleon pair, as a function of centrality, for
$\sqrt{s_{_{NN}}} = $130 GeV and 200 GeV Au+Au collisions as measured
in PHENIX \cite{Adler:2004zn}; compared to theoretical
predictions available in 2002. ``HIJING'' is a pQCD-based
model \cite{ToporPop:2002gf}, while ``KLN'' features gluon
saturation in the initial
state \cite{Kharzeev:2000ph,Kharzeev:2001gp}; ``EKRT'' assumes
saturation in the final state \cite{Eskola:1999fc,Eskola:2001bf}.}
\end{figure}

pQCD-based models have parameters regulating the momentum scales;
these include a lower-momentum cutoff, and the factorization and
fragmentation scales. Figure \ref{fig:Bazilevsky_fig3} shows that the
pQCD-based HIJING model, circa 2002, was able to reproduce 130 GeV
and 200 GeV $dN_{ch}/d\eta$ reasonably well.  However, in that model
jet production via hard scattering is an important mechanism for
particle production, and the combination of the $\sqrt{s}$ 
dependence of
hard-scattering cross sections with the growth of the nuclear overlap
with centrality causes the model to predict an increase in the ratio
between the two data sets with centrality.  The observed ratio is,
instead, quite constant.  Thus the authors found it necessary to
introduce a centrality-dependent shadowing to regulate the jet
growth \cite{Li:2001xa}.

An alternative to models which use collinearly factorized pQCD is found
in the ``color glass condensate'' picture, in which the gluon
population of low-$x$, low-$p_T$ states in the initial nuclear
wave function is limited by transverse overlap and fusion of these
low-$p_T$ gluons.  The phase-space density saturates because of the
competition between extra gluon radiation from higher-$x$ gluons and
non-linear fusion of the gluons at high density. Au+Au collisions are
then collisions of two sheets of colored glass, with the produced
quarks and gluons materializing at a time given by the inverse of the
saturation momentum, $\tau = 1/Q_s$. Saturation of gluons with momenta
below $Q_s$ provides a regulating mechanism that limits the rise in
gluon---and later, hadron---multiplicity with centrality and beam
energy. Models featuring this initial-state gluon saturation agree
well with essentially all RHIC data on the multiplicity density, which
is dominated by low-momentum
particles \cite{Kharzeev:2000ph,Kharzeev:2001gp}. This is seen, for
instance, in Fig. \ref{fig:Bazilevsky_fig3}.

In this picture, the total gluon multiplicity is proportional to
$1/\alpha_s \cdot Q_s^2$, which limits the number of low-momentum charged
particles produced. $Q_s$ evolves slowly with collision centrality and
beam energy. For central Au+Au collisions, it has been estimated that
the typical $m_T$ scale of the gluons ``liberated'' from the colored
glass is about 1 GeV per particle \cite{Krasnitz:2002mn}, which is
above the lower limit of 0.53 GeV per particle that we set above using
the PHENIX data. Though there are fewer predictions of $E_T$ than
total charged-particle production from gluon-saturation models, the
existing models are broadly consistent with data at
RHIC. Consequently, gluon saturation is considered to be a promising
candidate for describing the initial state of RHIC collisions.

\subsection{Conclusions}
\label{SubSec:EDense_conclusion}

Using reasoning similar to that of 
Bjorken \cite{Bjorken:1983qr},
combined with some simple formation-time arguments,
we can draw the following conclusions from the
PHENIX data on transverse energy production
and overall particle multiplicity:


\medskip
$\bullet$ The peak energy density in created secondary particles is
at least 15 GeV/fm$^3$, and this is most likely an underestimate.
This is well in excess of the $\sim$1 GeV/fm$^3$ required, according
to lattice QCD predictions, to drive a QCD transition to QGP.

\medskip
$\bullet$ We note that hydrodynamical 
calculations which reproduce the magnitude of elliptic flow 
observed at RHIC require local thermalization to occur
very quickly, typically by 1 fm/$c$ or earlier
(see Section \ref{Sec:thermHydro}).  If the system does
reach local equilibrium on this time scale then the
energy density of the first thermalized state would
be in excess of 5 GeV/fm$^3$, well above
the amount required to create the QGP.


\medskip
$\bullet$ Pre-RHIC expectations that $E_T$ and
charged particle production would be dominated by
factorized pQCD processes were contradicted by
data, which showed only very modest increases
with centrality and beam energy.  A new class of
models featuring initial-state gluon saturation
compares well with RHIC multiplicity and $E_T$
data, and are also consistent with our
Bjorken-style arguments for estimating energy
densities at early times.
 
\section{THERMALIZATION}
\label{Sec:therm}
\hyphenation{equil-i-bra-tion}
A key question is whether the matter formed at RHIC is thermalized, and if so 
when in the collision was equilibration achieved. 
If thermalization  is established early then evidence for 
strong transverse expansion can
be potentially related to the equation of state of
the dense matter produced at RHIC.
To explore these issues we review several experimental
observables from integral quantities (numbers of particles produced
and in what ratios), to differential distributions 
(measured $p_T$ and azimuthal distributions), to 
two-particle (HBT) correlations. 

\subsection{Chemical Equilibrium}
\label{Sec:thermChem}

For many years it has been known that the abundances of different hadron 
species in $e^++e^-$ and $p+\overline{p}$ reactions 
can be reproduced
by statistical models \cite{Becattini:1996if,Becattini:1997rv}. 
This success is often attributed to
hadronization statistically filling the available phase space. 
At RHIC there is also the possibility that the
strong scattering deduced from the measurements of elliptic flow
(section \ref{Sec:thermFlow})
may 
prove sufficient
to establish chemical equilibrium. 

The production of strange particles provides a means 
to check whether chemical equilibrium is achieved. For 
$e^++e^-$ and $p+\overline{p}$ reactions strange particle production is
suppressed due to the small size of the system. This canonical
suppression is largely removed for central heavy-ion collisions. 
If the measured strangeness yields are still lower than full
equilibrium predictions, then the partial equilibrium can be
quantified by a
multiplicative factor of $\gamma_s$ for each strange quark in a hadron, 
where $\gamma_s = 1$ for complete equilibration and $\gamma_s < 1$ for
partial equilibration.

\begin{figure}[tbhp]
	\includegraphics[width=1.0\linewidth]{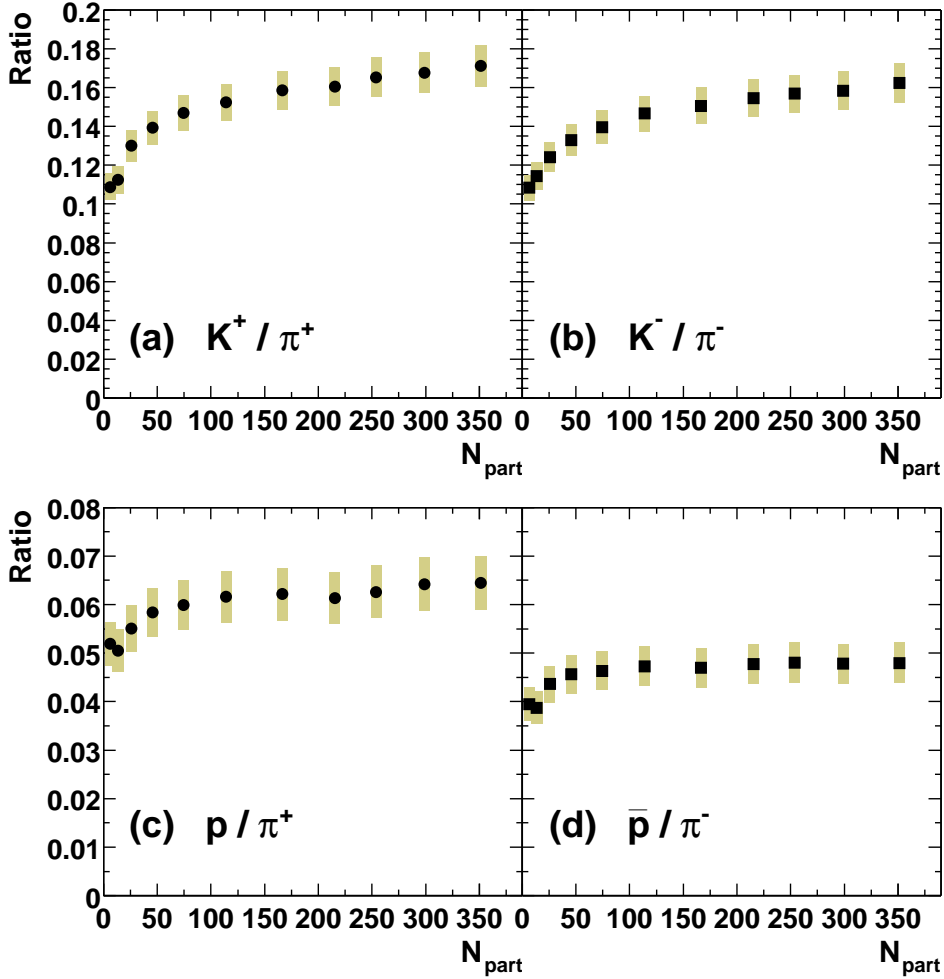}
	\caption{Centrality dependence of particle ratios for 
	(a) $K^+/\pi^+$, (b) $K^-/\pi^-$, (c) $p/\pi^+$, and (d)
	$\overline{p}/\pi^-$ in Au+Au collisions at 
        $\sqrt{s_{NN}} = 200$ GeV \cite{Adler:2003cb}.}   
	\label{fig:particleRatios}
\end{figure}

Figure \ref{fig:particleRatios} shows the centrality dependence of
$K/\pi$ and $p/\pi$ ratios in Au+Au collisions at $\sqrt{s_{NN}} = 200$ 
 GeV \cite{Adler:2003cb}.
Both $K^+/\pi^+$ and $K^-/\pi^-$
increase rapidly for peripheral collisions, and then saturate
or rise slowly from mid-central to the most central
collisions. The ratios $p/\pi^+$ and $\overline{p}/\pi^-$ also increase from peripheral
collisions but 
appear flatter than the $K/\pi$ ratios. 
Canonical statistical models \cite{Braun-Munzinger:2003zd} predict an
increase in these ratios with centrality, as the larger system-size 
effectively places less of a constraint on conserved quantities.
In addition
the chemical parameters,  T$_{chem}$ and $\mu_B$, can also vary with
centrality \cite{Kaneta:2004zr,Cleymans:2004pp}. 

\begin{figure}[tbhp]
	\includegraphics[width=1.0\linewidth]{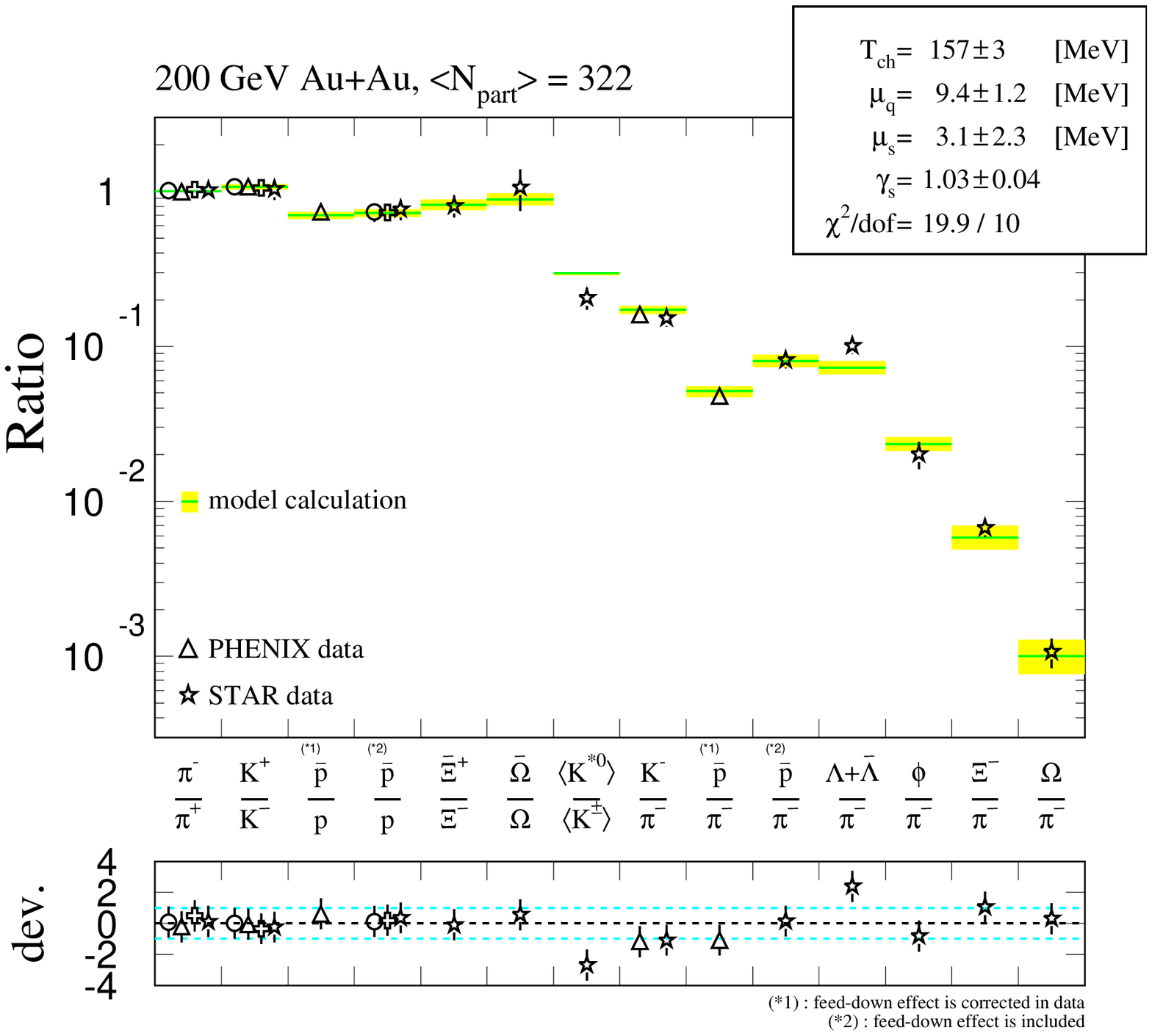}
	\caption{Comparison of PHENIX (triangles), STAR (stars), BRAHMS 
	(circles), and PHOBOS (crosses)	 
	particle ratios 
	from central Au+Au collisions at
   $\sqrt{s_{NN}} = 200$ GeV at mid-rapidity. The thermal model descriptions	
	from Kaneta \cite{Kaneta:2004zr} 
	are also shown as lines. See Kaneta \cite{Kaneta:2004zr} for the
	experimental references.}
	\label{fig:chemicalFit}
\end{figure}

Focusing on  the ratios from central collisions at $\sqrt{s_{NN}} = 200$ GeV,
the data are compared to the thermal model data analysis of
Kaneta and Xu \cite{Kaneta:2004zr} in 
Fig. \ref{fig:chemicalFit}. 
The extracted thermal parameters from this fit are T$_{chem} = 157 \pm 3$  MeV, 
$\mu_B = 23 \pm 3$  MeV, and $\gamma_s = 1.03 \pm 0.04$.
A large $\gamma_s$ is also found by STAR \cite{Adams:2003fy} who
extract $\gamma_s = 0.96 \pm 0.06$, while Cleymans
{\it et al.} \cite{Cleymans:2004pp} extract $\gamma_s$
that increases from $\gamma_s \simeq 0.85$ in peripheral collisions 
to $\gamma_s \simeq 0.95$ for central collisions at RHIC.
Similar
fits to the central RHIC data are obtained by Braun-Munzinger
{\it et al.} \cite{Braun-Munzinger:2001ip} who assume
complete chemical equilibration, {\it i.e.} $\gamma_s = 1$.

We note that there are
differences in the temperature parameter extracted by the
different authors. Kaneta 
and 
Xu \cite{Kaneta:2004zr} extract T$_{chem} = 157 \pm 3$  MeV which
is lower than that extracted by 
both Braun-Munzinger
{\it et al.} \cite{Braun-Munzinger:2003zd} of  T$_{chem} = 177 \pm 7$  
MeV and Cleymans {\it al.} \cite{Cleymans:2004pp} of  T$_{chem} = 165 \pm 
7$ MeV. However, both Braun-Munzinger {\it et al.} \cite{Braun-Munzinger:2003zd}
and Magestro \cite{Magestro:2001jz} discuss the sensitivity of the 
extracted temperature
to corrections from feed-down from decays.
Cleymans {\it et al.} \cite{Cleymans:2004pp} estimate that over 70\% of 
$\pi^+$ in the
thermal model fits come from the decay of resonances. 

At lower beam energies there is controversy over whether strangeness is in 
full chemical equilibrium.
Becattini {\it et al.} \cite{Becattini:2003wp} use data 
that is integrated over
the full rapidity and find that strangeness is in
partial equilibrium, {\it i.e.} at the 
AGS $\gamma_s = 0.65\pm0.07$ and at the SPS 
$\gamma_s = 0.84\pm0.03$.
Braun-Munzinger {\it et al.} \cite{Braun-Munzinger:2003zd} 
instead use ratios measured at mid-rapidity which 
typically have
larger strange/no-nstrange values,
and, hence, they obtain acceptable fits with $\gamma_s = 1$ at 
both AGS and SPS energies. 
At RHIC energies thermal model 
comparisons all use mid-rapidity data; a choice that is motivated in part 
by
the separation between fragmentation regions and central particle production.
  
In contrast to the controversies at lower beam energies, the 
observation that strangeness is equilibrated is common to all
thermal calculations that reproduce RHIC data. This
is consistent with chemical equilibrium being obtained before hadronization, though
does not prove that this is the case. An alternative explanation
is that scattering in the 
hadronic phase could increase $\gamma_s$ to 1,
though small interaction cross sections imply that it may be difficult to 
equilibrate the multi-strange baryons before the hadrons freezeout.

\subsection{Spectra}
\label{Sec:thermSpectra}
 
\begin{figure}[tbhp]
	\includegraphics[width=1.0\linewidth]{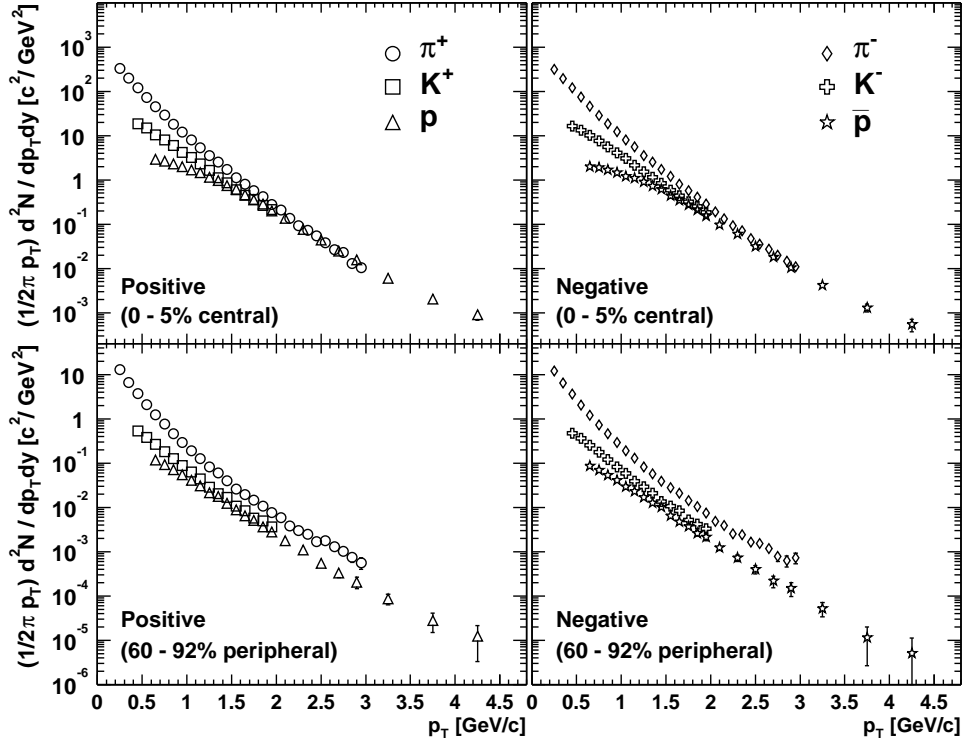}
	\caption{Transverse momentum distributions for pions, kaons, protons, and anti-protons in 
	Au+Au collisions at $\sqrt{s_{NN}}$ = 200 GeV \cite{Adler:2003cb}.}
	\label{fig:ptSpectra}
\end{figure}

Hadron spectra reflect
conditions late in the reaction, as
well as the integrated effects of expansion from the beginning
of the collision.
Figure \ref{fig:ptSpectra} shows the $p_T$ distributions for pions, kaons, protons, and anti-protons
in both central (top panel) and peripheral collisions (bottom panel) \cite{Adler:2003cb}. 
The pion spectra have a concave shape
at low p$_T$ where many of the pions
may come from the decay of resonances:
$\Delta, \rho$ etc. The kaon spectra are approximately
exponential over the full measured p$_T$ range, whereas the proton spectra
flatten at low p$_T$ for the most central collisions. A striking feature
is that the proton and anti-proton spectra in central collisions become 
comparable 
in yield to
the pion spectra above 2 GeV/$c$. This is more fully discussed in 
Section \ref{Sec:hadron}.

\begin{figure}[tbhp]
	\includegraphics[width=1.0\linewidth]{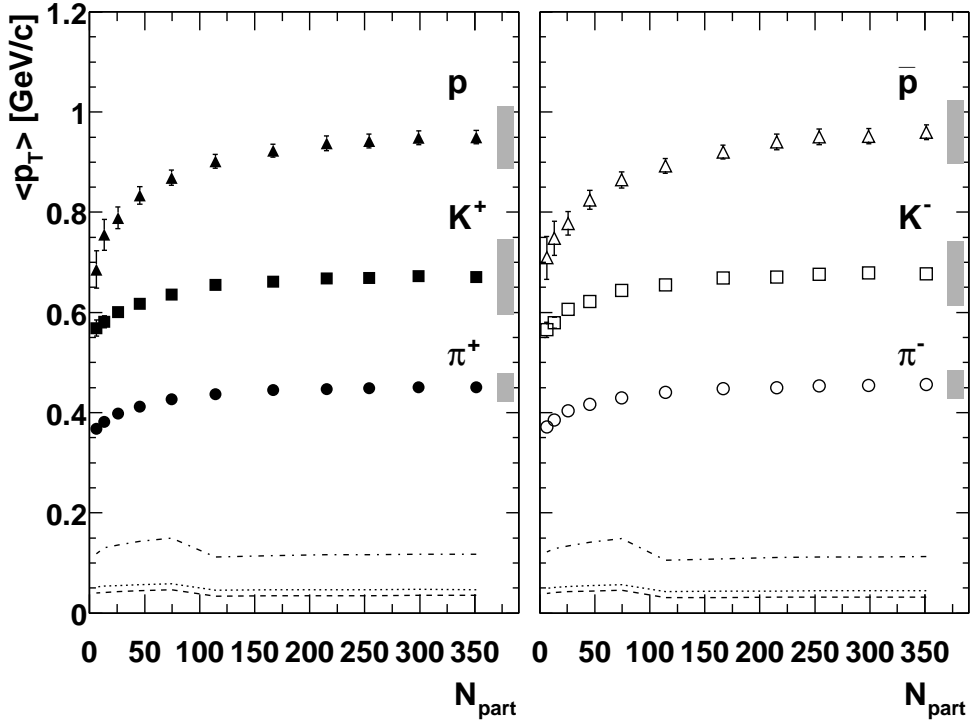}
	\caption{Mean transverse momentum as a function of $N_{part}$ for pions,
		   kaons, protons and anti-protons in Au+Au collisions at 
			$\sqrt{s_{NN}}$ = 200 GeV \cite{Adler:2003cb}.
		  The systematic errors from extrapolation, which are scaled by a factor of two
		  for clarity, are shown in the bottom for protons 
		and anti-protons (dashed-dotted lines),
		  kaons (dotted lines), and pions (dashed lines). The 
shaded bars to the right
		  represent the systematic error.}
	\label{fig:meanpt}
\end{figure}

One way to characterize the change in spectra as a function of centrality
is to calculate $\langle p_T \rangle$ for each 
spectrum \cite{Adler:2003cb} as shown in Fig. \ref{fig:meanpt}.
The $\langle p_T \rangle$ increases for all particles as a function of centrality with the
largest change occurring in peripheral collisions ($N_{part}<100$). Across
the different particles the increase 
is largest for  protons and anti-protons. This is consistent with a
collective expansion velocity that increases with centrality to produce the largest increase
in $\langle p_T \rangle$ for the heaviest particles.

\begin{figure}[tbhp]  
	\includegraphics[width=1.0\linewidth]{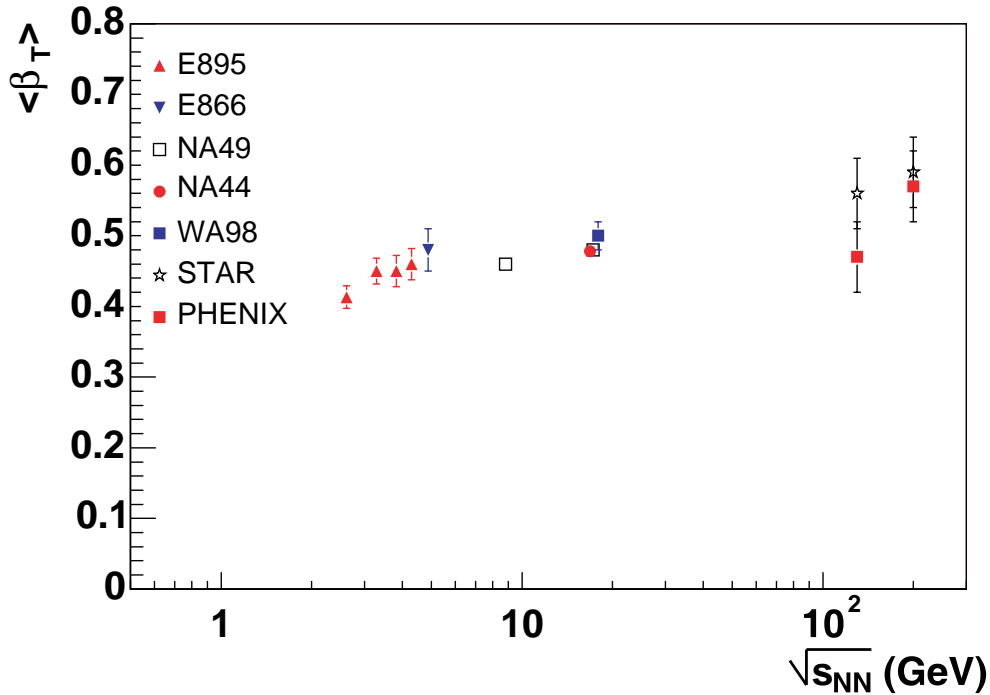}
	\caption{Beam-energy dependence of the extracted 
	mean transverse expansion velocity as a function of beam energy 
	from simultaneous fits to spectra of different mass
	 \cite{Rai:2001rv,Dobler:1999ju,Alt:2003rn,Peitzmann:2002nt,Aggarwal:2002tm,Adcox:2003nr,Adams:2003xp}. }
	\label{fig:betaSystematics}
\end{figure}

The pion, kaon, and
 proton spectra can all be fit using an ansatz of a 
thermal, expanding 
source \cite{Schnedermann:1993ws,Adcox:2003nr} to extract 
the collective transverse expansion 
velocity $\langle \beta_T \rangle$
as well as the temperature at freezeout, T$_{fo}$. Figure \ref{fig:betaSystematics} 
shows $\langle \beta_T \rangle\sim0.45$ at  
AGS energies \cite{Rai:2001rv,Dobler:1999ju},
which increases 
to $\langle \beta_T \rangle\sim 0.5$ at the
SPS \cite{Alt:2003rn,Peitzmann:2002nt,Aggarwal:2002tm} 
and RHIC \cite{Adcox:2003nr,Adams:2003xp}. All the above fits
use similar model assumptions of a linear velocity profile and
a Woods-Saxon density
profile. 
That the spectra at these beam energies can be reproduced by a thermal source
is necessary, but not sufficient, evidence for thermal equilibrium at each
of these energies.
However, it is difficult to draw strong conclusions
from the increase in $\langle \beta_T \rangle$ as a function of
beam energy since the parameters $\langle \beta_T \rangle$
and T$_{fo}$ are strongly anticorrelated
and their values depend on fit ranges and treatment of decays. 

\subsection{Elliptic Flow}
\label{Sec:thermFlow}
At the beginning of a heavy ion collision, 
the spatial distribution of the colliding
matter resembles an ellipsoid due to the incomplete
overlap of the two colliding nuclei.
Any strong scattering in this early stage converts
the spatial anisotropy to a momentum anisotropy 
which is observable as an elliptic flow of the emitted hadrons. 
Elliptic flow is a self-limiting phenomenon, which is
readily understood in the thermodynamic limit.
If strong scattering is sufficient to establish
local thermal equilibrium, then the pressure gradient 
is largest in the shortest direction of the ellipsoid. 
This gradient produces higher momenta in that direction, quickly 
reducing the spatial asymmetry. 

\begin{figure}[tbhp]
	\includegraphics[width=1.0\linewidth]{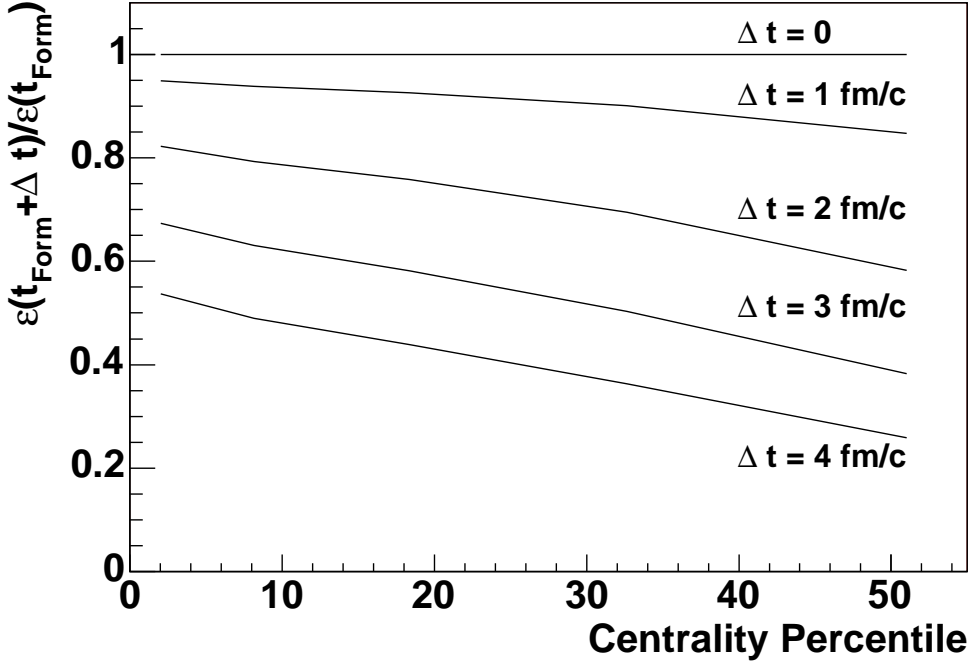}
	\caption{The ratio of the eccentricity
	after a time delay $\Delta t$ compared to its value at
	formation time, as a function of Au+Au collision centrality.
	The calculations follow the prescription of \cite{Kolb:2000sd} where
	the produced particles are allowed to free-stream at first and
	reach local equilibrium only after some time delay.}
	\label{fig:time_delayed_eccentricity}
\end{figure}

The absence of any strong scattering in the early stage of the colliison would reduce
the amount of elliptic flow that could be created.  If the
initially produced particles are allowed to initially free stream and
reach local equilibrium only after some time delay, then the spatial
anisotropy at the start of hydrodynamic evolution will be reduced;
the longer the delay, the greater the reduction.
Following the prescription of Kolb {\it et al.} \cite{Kolb:2000sd}, we plot 
in Fig. \ref{fig:time_delayed_eccentricity} the eccentricity
after a time delay $\Delta t$ compared to its value at
formation time, as a function of Au+Au collision centrality.
The eccentricity ($\varepsilon$) of the reaction zone is
\begin{equation} 
\varepsilon = \frac{\langle y^2 \rangle-\langle x^2 \rangle}{\langle y^2
\rangle+\langle x^2 \rangle}.
\label{eq:ecc_def}
\end{equation}
The eccentricity can be analytically calculated once the density
profile of the nuclei is chosen (typically a Woods-Saxon shape).
It can also be calculated using Monte Carlo techniques, where
the positions of those nucleons that participate in the 
reaction are used to calculate the averages in Eq. \ref{eq:ecc_def}. 
From Fig. \ref{fig:time_delayed_eccentricity} 
we can see that for time delays of 2 fm/$c$ or greater the
magnitude of the eccentricity is significantly reduced, and
its shape vs. centrality is also altered.

If locally equilibrated
hydrodynamics is taken as the mechanism for generating elliptic flow,
then the observation of {\em any} substantial amount of elliptic flow
can be taken as evidence that local thermal equilibrium is achieved on
a time scale before the spatial anisotropy would be completely erased.
The order of this time-scale would be $t \sim R/c$, where $R$
is the nuclear radius. However, the hydrodynamical calculations we
will examine here (see Sec. \ref{Sec:thermHydro} and
Table \ref{tab:hydro_v2_spectra_RHIC}) all require quite short
thermalization times, from 0.6--1.0 fm/$c$, in order to reproduce the
magnitude of elliptic flow observed at RHIC.

The azimuthal anisotropy of the spectra can be
characterized in terms of Fourier coefficients, which
at RHIC are dominated by the elliptic
flow, the second Fourier coefficient, 
$v_2(p_T)$, where 
\begin{equation}
\frac{d^2N}{d\phi d p_T} = N_0(1+2v_2(p_T)\cos(2\phi))\quad .
\end{equation}
Both the first Fourier coefficient, $v_1$, and higher order
coefficients have been neglected in the above expression.

\begin{figure}[tbhp]
	\includegraphics[width=1.0\linewidth]{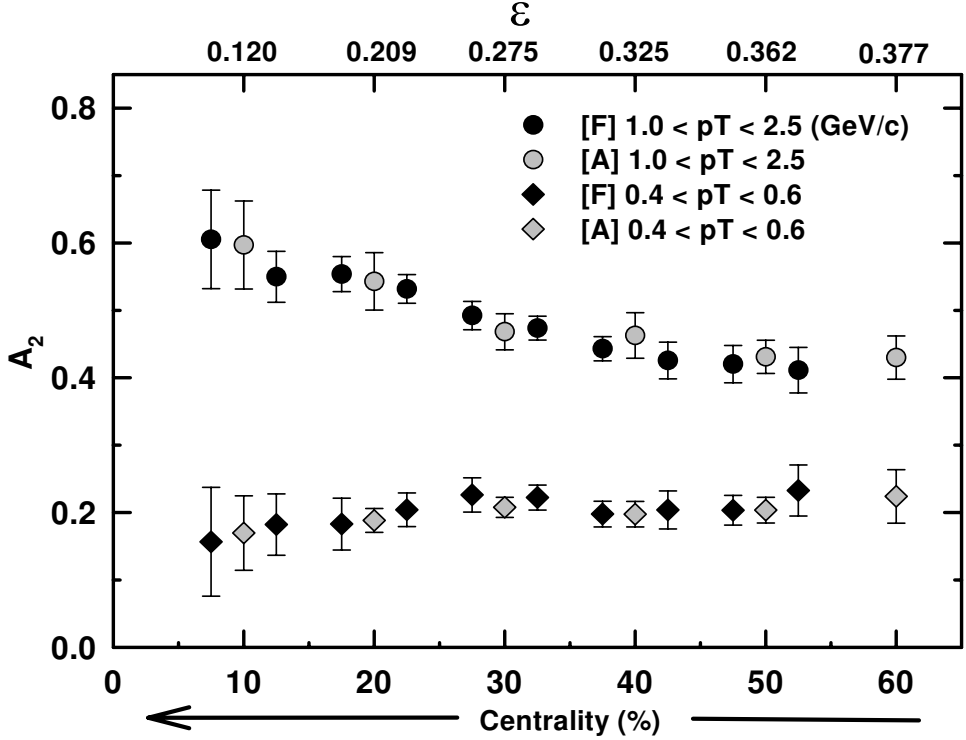}
	\caption{$A_2 = v_2/\varepsilon$ 
	vs. centrality for Au+Au 
	collisions at 
	$\sqrt{s_{NN}}$ = 130 GeV \cite{Adcox:2002ms}.
	The data points come from two different types of
	two-particle correlations: ``fixed" $p_T$ correlations when both 
particles are at the same $p_T$ 
	(points are labeled as ``F"), and ``assorted" $p_T$ correlations 
when
	the two particles have different $p_T$ (points are labeled as 
``A"). In
	this case the labeled $p_T$ range is for the higher-momentum particle of the pair.}
	\label{fig:eccScaling}
\end{figure}

The most direct evidence that $v_2$ is related to 
spatial asymmetries present early in the reaction is that 
$v_2$ at low $p_T$ approximately scales with the 
initial eccentricity ($\varepsilon$) of the reaction zone.
The measured values of 
$v_2$ normalized by $\varepsilon$ are shown in Fig. \ref{fig:eccScaling}
vs. centrality for two different $p_T$ 
ranges \cite{Adcox:2002ms}.   
At low momentum $v_2/\varepsilon$  is
independent of centrality to within 20\%. This scaling is increasingly broken at higher 
$p_T$. 

The measured values of the integrated $v_2$ at RHIC are larger than those at lower energies, 
but this is in
part due to the fact that $v_2(p_T)$ increases with $p_T$ and $\langle p_T \rangle$ increases as a 
function of beam energy. To remove this effect we will
concentrate on the differential flow, {\it i.e.} the shape
of $v_2(p_T)$.
\begin{figure}[tbhp]
	\includegraphics[width=1.0\linewidth]{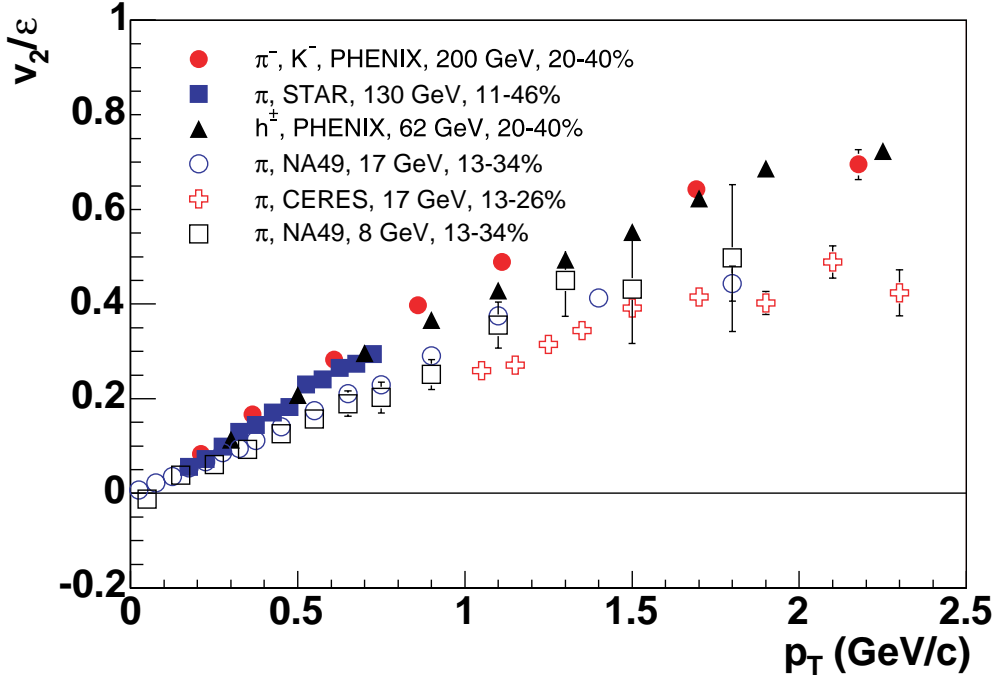}
	\caption{$v_2(p_T)/\varepsilon$ vs. $p_T$ for mid-central
	collisions at RHIC (filled symbols) and SPS (open symbols). 
Dividing by 
	eccentricity removes to first order the effect of different centrality 
	selections across the experiments
	 \cite{Adler:2003kt,Adler:2001nb,Adler:2004cj,Alt:2003ab,Agakichiev:2003gg}.}
	\label{fig:v2vspt_World}
\end{figure}

\begin{figure}[tbhp]
	\includegraphics[width=1.0\linewidth]{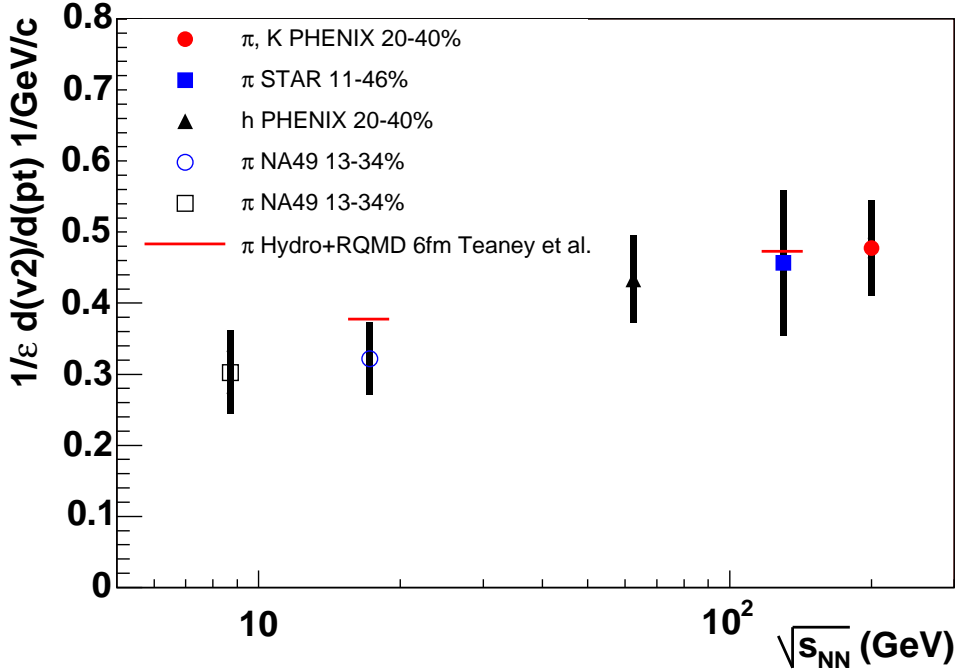}
	\caption{The slope of the scaled elliptic flow, 
	$(dv_2/dp_T)/\varepsilon$, 
	for mid-central
	collisions at RHIC (filled symbols) and the SPS (open symbols). 
	The slope is calculated from the data in Fig. \ref{fig:v2vspt_World}
        for the data $p_T<$ 1 GeV/$c$. The solid error bars
	represent the total systematic error including the systematic error on 
$v_2$ and $\varepsilon$
         \cite{Adler:2003kt,Adler:2001nb,Adler:2004cj,Alt:2003ab}.}
	\label{fig:dv2dpt_World}
\end{figure}

To make a uniform comparison between different colliding nuclei 
(Pb+Pb at SPS and Au+Au at RHIC) as well as different impact parameter 
selections from the different experiments, we normalize $v_2$ by the
eccentricity, $\varepsilon$, as shown in Fig. \ref{fig:v2vspt_World}. 
The values of $\varepsilon$ have been calculated
via a Glauber Monte Carlo using Woods-Saxon density distributions for 
the Au and Pb nuclei. The averages in Eq. \ref{eq:ecc_def} 
are over the participating nucleons,
hence $\varepsilon$ is calculated at the start of the collision.
The pion data in Fig. \ref{fig:v2vspt_World} show that $v_2(p_T)/\varepsilon$ 
increases approximately linearly with $p_T$ for low $p_T$.
The rate of increase of $v_2/\varepsilon$ as a function of $p_T$ is larger
at RHIC \cite{Adler:2003kt,Adler:2001nb}
than at SPS \cite{Alt:2003ab,Agakichiev:2003gg} as
can most easily be seen by calculating the slope of $v_2/\varepsilon$ 
below 
$p_T = $1 GeV/$c$ (Fig. \ref{fig:dv2dpt_World}).
The slope $(dv_2/dp_T)/\varepsilon$ increases from SPS to RHIC by approximately 50\%.
Hydrodynamical calculations \cite{Teaney:2001av} shown in Fig. \ref{fig:dv2dpt_World} 
reproduce the data both at RHIC and at CERN SPS within one standard 
deviation.
More extensive comparisons with hydro calculations will be discussed in 
section \ref{Sec:thermHydro}, while the 
behavior of $v_2$ at higher $p_T$, which follows a scaling with respect
to the number of quarks, is discussed in Section \ref{Sec:hadron}.

\begin{figure}[tbhp]
	\includegraphics[width=1.0\linewidth]{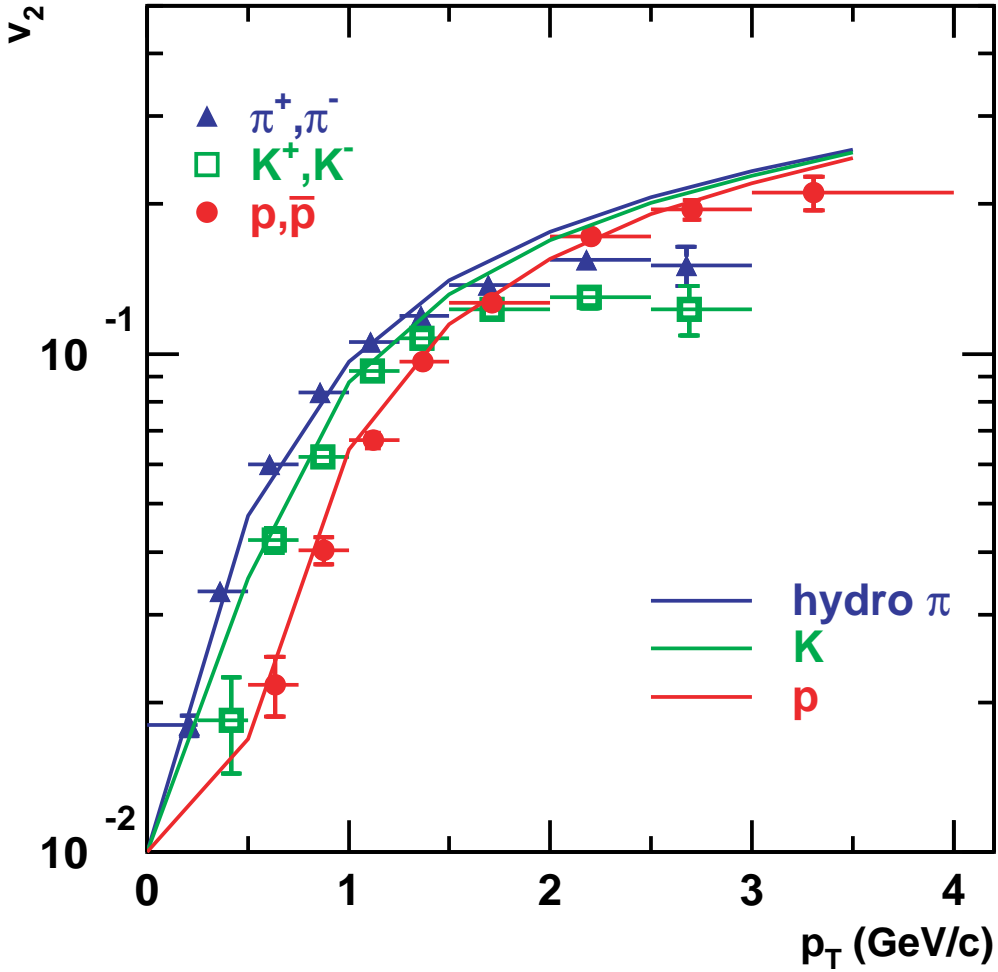}
	\caption{$v_2(p_T)$ for pions, kaons and protons
	produced in minimum-bias collisions at 
	RHIC \cite{Adler:2003kt} 
	compared
	to hydro calculations from Huovinen {\it et 
al.} \cite{Huovinen:2001cy}.}
	\label{fig:minBiasv2PID}
\end{figure}

Further insight into the expansion dynamics can be obtained from
the mass dependence of $v_2(p_T)$ shown in Fig. \ref{fig:minBiasv2PID}
for pions, kaons and protons \cite{Adler:2003kt}
along with a comparison with an early
hydrodynamic model calculation \cite{Huovinen:2001cy}.
The $v_2(p_T)$ for pions is larger than for kaons and protons at low $p_T$,
and this mass ordering has been explained as resulting from
radial expansion\cite{Huovinen:2001cy} that produces a larger
distortion of the elliplic flow induced velocity profile for larger hadron
masses.
However, as will be discussed in Section \ref{Sec:thermHydro}, this 
calculation fails to reproduce the proton spectra, and attempts
to remedy
this failure lead to calculations that 
no longer reproduce the
measured $v_2$ for pions and protons.

\subsection{HBT}
\label{Sec:thermHBT}

Bose-Einstein correlations 
between identical particles
provide a measure of the space-time extent of the source at the
end of the reaction. Because the  extracted source parameters
as measured by the HBT technique are
driven by space-time correlations, HBT results
are sensitive to expansion dynamics integrated throughout the
collision. 
HBT measurements were originally motivated by theoretical
predictions of a large source size and/or a long duration of particle
emission \cite{Rischke:1996cm,Bertsch:1989vn,Pratt:1986cc}---which would result from
the presence of a long-lived mixture of phases in the 
matter as it undergoes a first-order phase transition
from a quark-gluon plasma back to the hadronic phase.

In HBT analyses, multidimensional Gaussian fits are made to the
normalized relative momentum distributions yielding fit parameters, 
$R_{\rm long}$, $R_{\rm side}$,
$R_{\rm out}$ \cite{Pratt:1984su}, also referred to as HBT radii, where 
\begin{equation}
C_{2} = 1+\lambda\exp(-R_{\rm side}^{2}q_{\rm side}^{2}-R_{\rm
out}^{2}q_{\rm out}^{2}-R_{\rm long}^{2}q_{\rm long}^{2})
\label{equ:BPEQ}.
\end{equation}
The coordinate system is chosen so that the longitudinal direction  is
parallel to the beam axis, the out direction is in the direction of
the pair's total transverse momentum, and the side direction is in the 
transverse plane perpendicular to the out axis.
For dynamic ({\it i.e.} expanding) sources, the HBT radii depend on the 
mean
transverse momentum of the particle pairs, $k_T = |{\bf p}_{\rm 1T} +
{\bf p}_{\rm 2T}|/2$, and correspond to lengths of homogeneity:
regions of the source which emit particles of similar 
momentum \cite{Makhlin:1988gm}.
Measuring the $k_T$ dependence of HBT radii provides essential
constraints on dynamical models \cite{Pratt:1990zq}. 
In particular, the ratio $R_{\rm out}/R_{\rm side}$
is predicted to be larger than unity for 
sources which emit particles over a long time.

\begin{figure*}
	\includegraphics[width=1.0\linewidth]{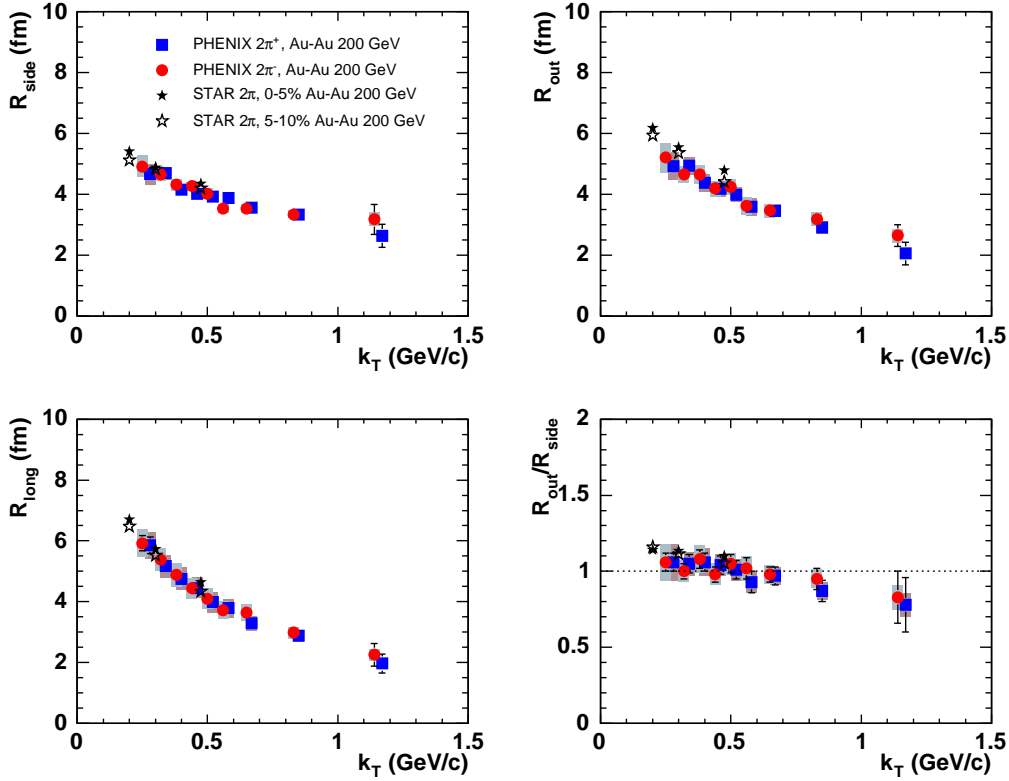}
	\caption{The $k_T$ dependence of the Bertsch-Pratt parameters for
$\pi^{+}\pi^{+}$ (blue square) and $\pi^{-}\pi^{-}$ (red circle) for
$0-30\%$ centrality with statistical error bars and systematic error
bands. Results from PHENIX \cite{Adler:2004rq} and  STAR \cite{Adams:2003ra} are overlaid.}
	\label{fig:HBTvskt}
\end{figure*}

The measured $k_T$  dependence of all 
radii \cite{Adler:2004rq} and the ratio $R_{\rm out}/R_{\rm side}$ are 
shown in
Fig. \ref{fig:HBTvskt}, along with STAR
results \cite{Adams:2003ra}. The data from PHENIX and STAR are 
in excellent agreement.
Both sets 
of data have been corrected for Coulomb 
repulsion between the detected particles.

The measured radii all decrease with increasing $k_T$ 
as expected for a rapidly expanding source.
The ratio $R_{\rm out}/R_{\rm side}$ was measured to be 1 within errors,
with a slight systematic  decrease for increasing $k_T$.
As is discussed in the next section, 
these data have excluded the validity of a large majority of
hydrodynamical models developed to describe Au+Au collisions at RHIC, 
indicating that in their present form these models do not describe well 
the space-time evolution of the Au+Au collisions. 

\subsection{Hydrodynamic Model Comparisons}
\label{Sec:thermHydro}

Many of the experimental features in the spectra and elliptic flow are consistent with
equilibrium being established early in the collision with
large pressure gradients
that drive a strong expansion. Moving from a statement of ``consistency" to
a statement that equilibrium has been ``established" is difficult. Some progress
can be made by comparing the data to hydrodynamic models that assume full
equilibrium early in the collision.

A variety of hydrodynamic models have been published.
Our approach is to confront these models with 
the following broad
set of data; $v_2(p_T)$, spectra,  and HBT. 
In this paper we will not compare the data
with hydro-inspired parameterized fits, {\it e.g.} 
blast-wave \cite{Retiere:2004wa}
or Buda-Lund \cite{Csanad:2004mm}
models, but will restrict ourselves to dynamical hydro models.

\begin{figure*}
$\begin{array}{c@{\hspace{0.01cm}}c}
	\multicolumn{1}{l}{\mbox{\bf (a)}} &
	\multicolumn{1}{l}{\mbox{\bf (b)}} \\ [-0.53cm]
	\includegraphics[width=0.51\linewidth]{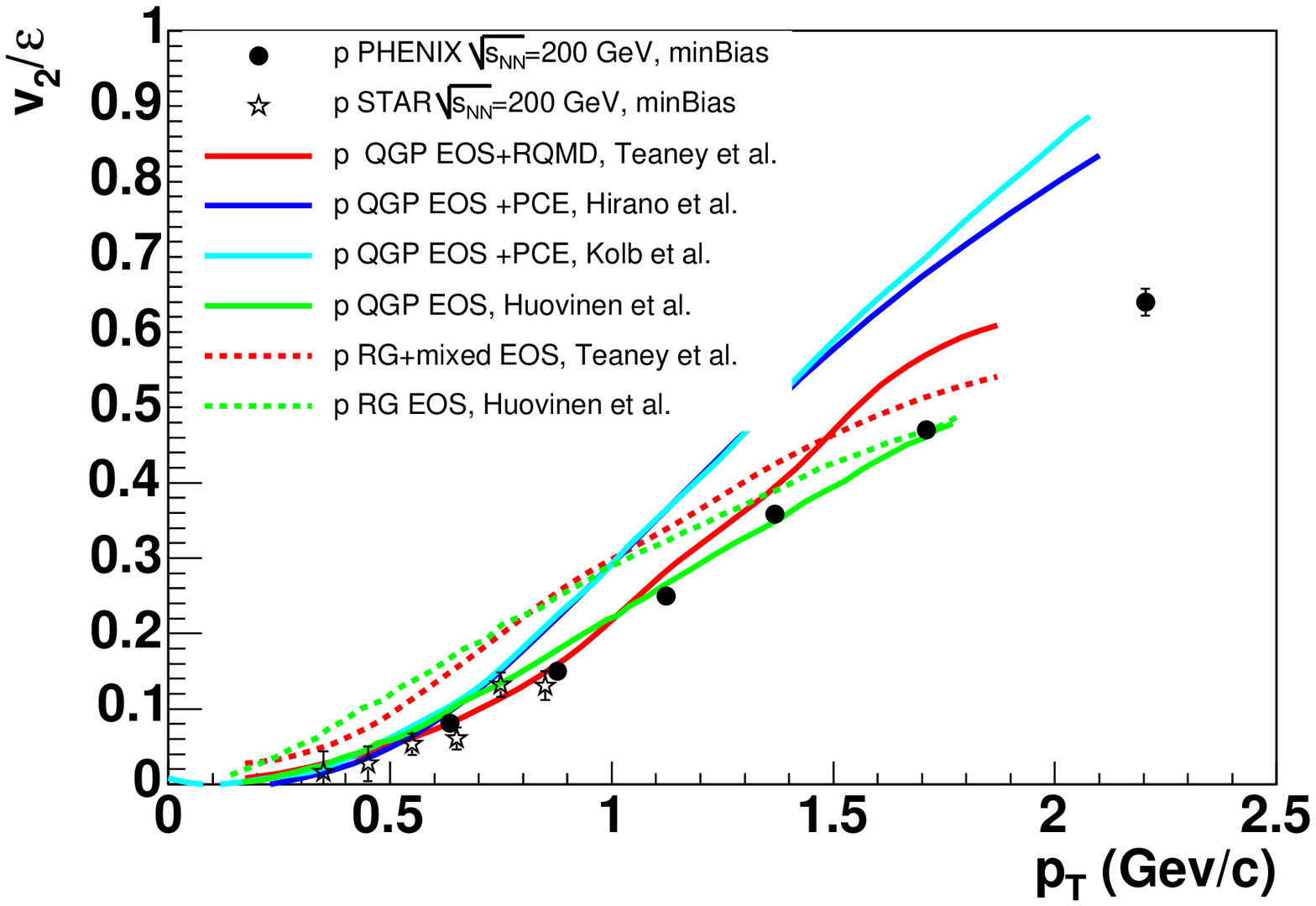} &
	\includegraphics[width=0.51\linewidth]{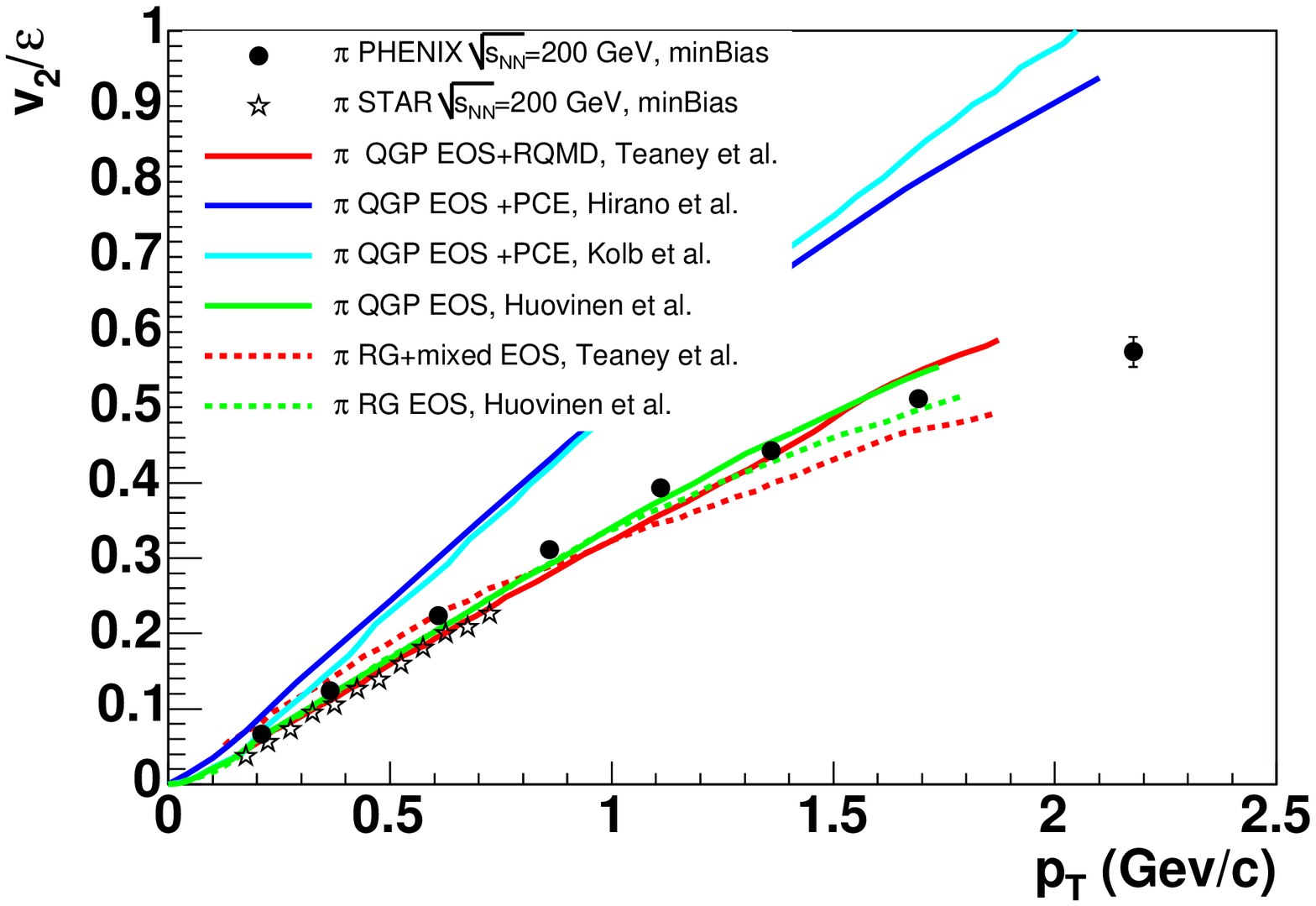} \\ [0.4cm]
	\includegraphics[width=0.51\linewidth]{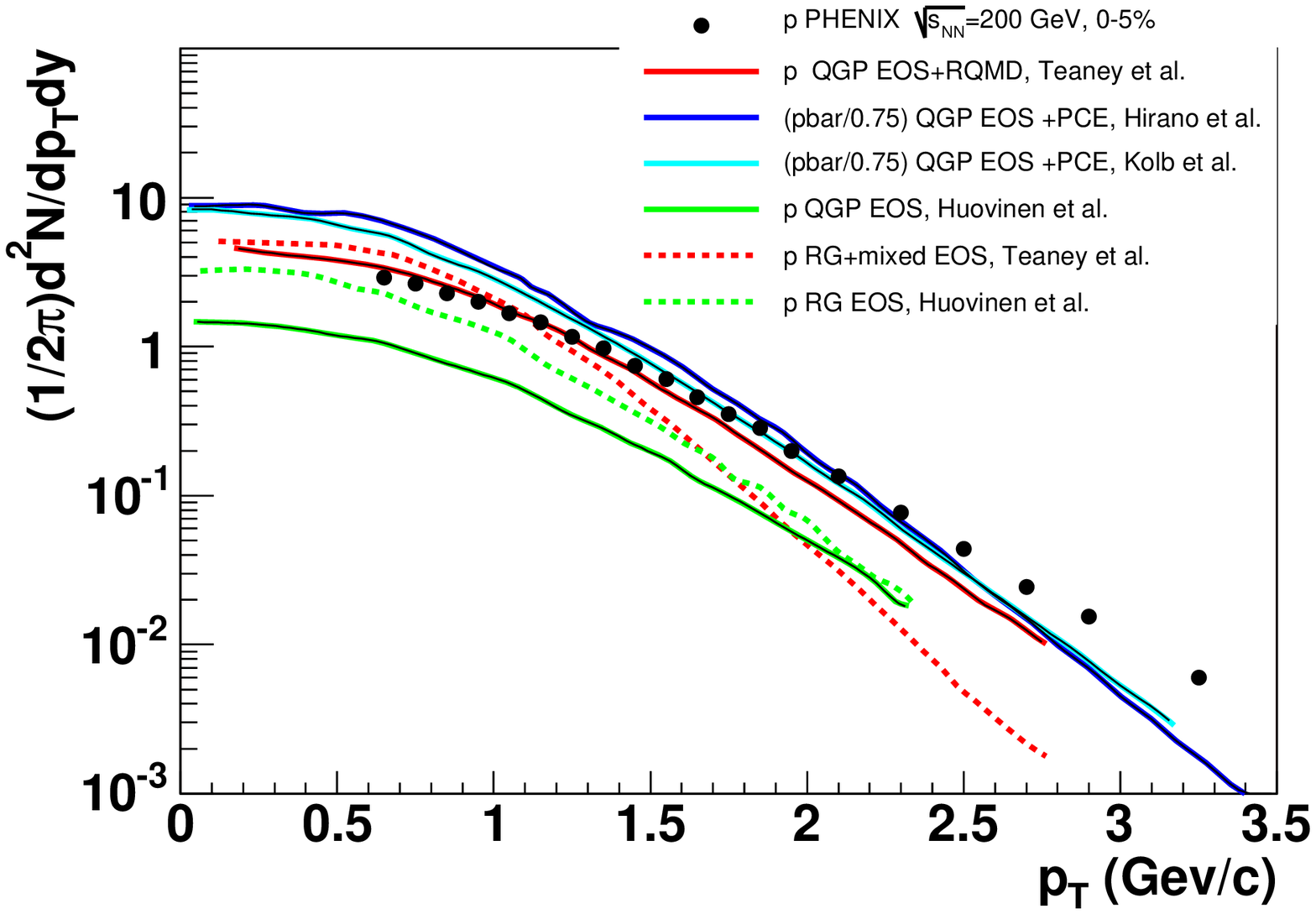} &
	\includegraphics[width=0.51\linewidth]{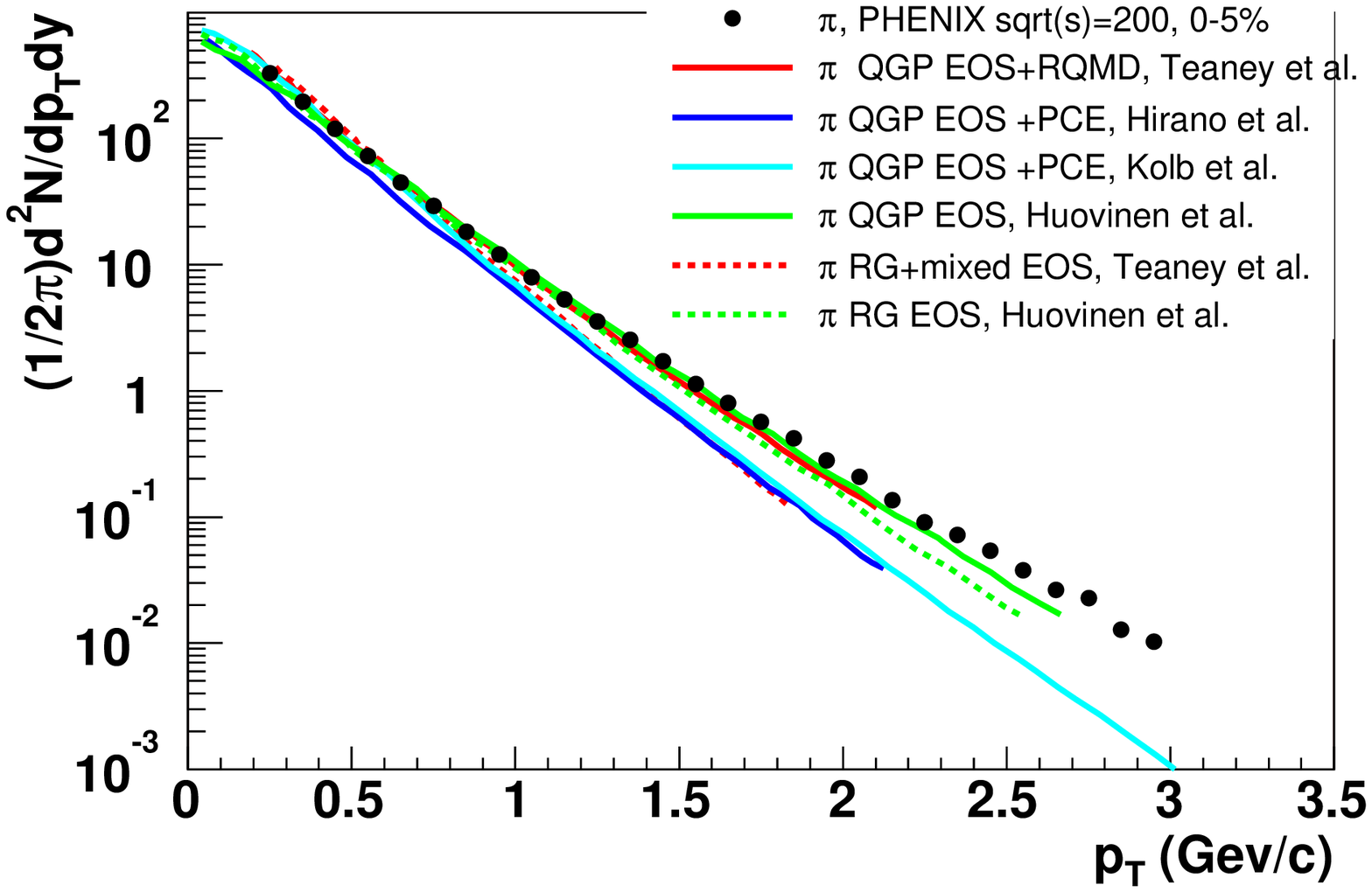} \\ [0.4cm]
\end{array}$

\caption{Top two panels: On the left, proton 
$\frac{1}{\varepsilon}v_2(p_T)$ vs. $p_T$ for 
minimum-bias collisions at RHIC \cite{Adler:2003kt,Adler:2001nb} 
are compared with hydro 
calculations \cite{Teaney:2001av,Hirano:2002ds,Kolb:2003dz,Huovinen:2001cy}, 
and on the right is the same comparison for pions. 
Bottom two panels: On the left, 
proton spectra for 0--5\% collisions at RHIC \cite{Adler:2003cb} are compared 
with the same hydro calculations and on the the right is the same comparison for pions.}

\label{fig:hydro_v2_spectra_RHIC}
\end{figure*}

\begin{figure*}
	\includegraphics[width=1.0\linewidth]{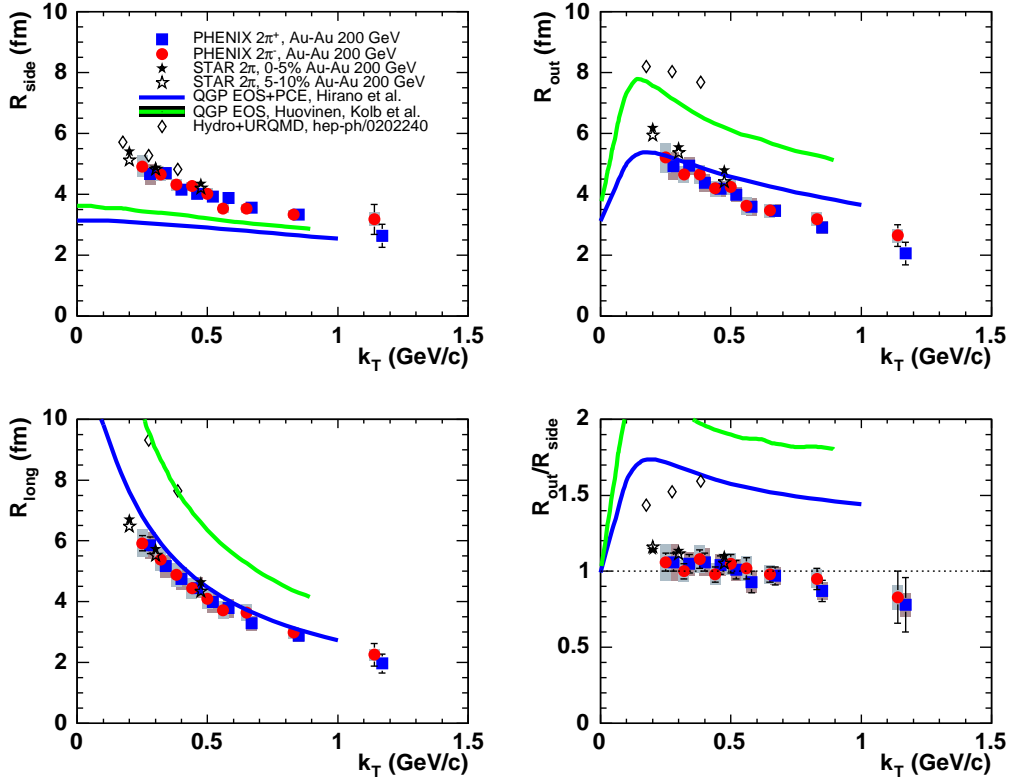}
	\caption{The $k_T$ dependence of the Bertsch-Pratt parameters for
	$\pi^{+}\pi^{+}$ (blue square) and $\pi^{-}\pi^{-}$ (red circle) for
	0--30\% centrality with statistical error bars and systematic error
	bands. Results from PHENIX \cite{Adler:2004rq}, 
	STAR \cite{Adams:2003ra} and 
	hydrodynamics models
	(Hirano \cite{Hirano:2002ds}, Kolb/Huovinen \cite{Heinz:2002un}
	and Soff \cite{Soff:2002qw},
	diamonds) are overlaid.}
	
	\label{fig:HBTvsktModel}
\end{figure*}

In Figs. \ref{fig:hydro_v2_spectra_RHIC} and \ref{fig:HBTvsktModel},
hydro calculations that include a
phase transition from the QGP phase to a hadronic phase are shown with solid lines, 
while hydro calculations that do not
include a pure QGP phase at any stage in the dynamics are drawn with dashed lines. 
The four calculations that include a QGP phase all assume
an ideal gas EOS for the QGP phase, a resonance gas for the hadronic phase
and connect the two using a first-order phase transition and a Maxwell
construction. These calculations use
latent heats that range from 0.8 GeV/fm$^3$ (Teaney 
{\it et al.} \cite{Teaney:2001av}) to
1.15 GeV/fm$^3$ (Huovinen {\it et 
al.} \cite{Huovinen:2001cy} and
Kolb {\it et al.} \cite{Kolb:2002ve}), to 
1.7 GeV/fm$^3$ (Hirano {\it et al.} \cite{Hirano:2002ds,Hirano:2004rs}).
For comparison the bag model of the nucleon with external bag pressure  
$B$ = (230 MeV)$^4$ and a $T_{crit}$ = 164 MeV produces a latent heat of  
1.15 GeV/fm$^3$ \cite{Kolb:2003dz}.
The calculations that do not include a QGP phase (dashed lines) 
either include 
a hadron phase and a phase mixture by forcing the latent heat of the 
transition to
infinity \cite{Teaney:2001av},
or use an
hadronic resonance gas equation of state, {\it i.e.} no mixed or QGP 
phases \cite{Huovinen:2001cy}.

The calculations also differ in how they solve the hydro equations and
how they treat the final hadronic phase. The work of Hirano, Tsuda, and Nara 
cited here are the only calculations in this
paper that solve the hydro equations in 
3D \cite{Hirano:2002ds,Hirano:2004rs}.
For the final
hadronic stage Teaney \cite{Teaney:2001av} uses a 
hybrid model that couples the hadronic phase to RQMD to allow 
hadrons to freezeout according to their cross section, {\it i.e.} for 
chemical equilibrium
to be broken in the hadronic phase. 
Hirano \cite{Hirano:2002ds} and Kolb \cite{Kolb:2003dz} both allow for 
partial 
chemical equilibrium by
chemically freezing out earlier than the kinetic freezeout. This
has been done in order to reproduce the large proton yield
measured at RHIC (see later in this section). In contrast,
Huovinen \cite{Huovinen:2001cy} maintains full 
chemical equilibrium throughout the hadronic
phase.

Figure \ref{fig:hydro_v2_spectra_RHIC} compares these
calculations to the measured minimun-bias proton and pion
$v_2(p_T)/\varepsilon$.  
Minimim-bias results were chosen in order
to have the broadest set of data and model calculations for 
comparison.
The four calculations that include a
phase transition from the QGP phase to a hadronic phase  
(solid lines) reproduce the low-$p_T$ proton
data better than the two hydro 
calculations that do not have a QGP phase at any stage in the dynamics (dashed lines).
The presence of the first-order QGP phase transition softens the 
EOS which reduces the elliptic
flow. 
At higher $p_T$ there is considerable variation between
the models. Part of this is due to how the final
hadronic stage is modeled. For example,
Kolb's (solid light-blue line) and Hirano's (solid dark-blue line) calculations allow for 
partial chemical equilibrium in the final stage
compared to Huovinen (solid green line) which chemically freezes out
late in the collision. The difference is observable
above $p_T\sim 1$ GeV/$c$.

The same hydro models are compared to the 
pion $v_2(p_T)/\varepsilon$ measurements from STAR and PHENIX in
Fig. \ref{fig:hydro_v2_spectra_RHIC}. 
The Kolb (solid light-blue line) and Hirano (solid dark-blue line) calculations fail completely by predicting too 
strong a $v_2$. These two models
have very similar partial chemical equilibrium assumptions in the late
hadronic stage. It is worth noting that the Kolb calculation is
the same as the Huovinen (solid green line) calculation with the exception of the 
final hadronic stage.

All the above models have assumed ideal hydrodynamics, {\it i.e.} with no 
viscosity and zero 
mean free path. Non-zero viscosity in the QGP reduces 
$v_2$ \cite{Molnar:2004yh,Teaney:2004qa} and since
the early hydro calculations from
Teaney and Huovinen reproduced the magnitude of the
pion $v_2$ data, 
it is often stated that viscosity of the 
matter at RHIC must be small \cite{Shuryak:2003xe}.
However recent
calculations from Hirano (3D) (solid dark-blue line)  and Kolb 
(solid light-blue line) overpredict the measured $v_2$.
As these results do not include
dissipative effects, such as those resulting from hadronic interactions
in the final state, their failure indicates that further work will be
necessary before a quantitative determination of the viscosity in the QGP
phase is possible.
Progress will require both theoretical
development and experimental measures that are
less sensitive to how the azimuthal asymmetry of the
energy-momentum tensor is distributed between different
particles in the final stage of the reaction, {\it e.g.} the
elliptic flow of the total transverse energy.

The same hydro models
are now compared to the measured spectra from central collisions.
The bottom right panel of
Fig. \ref{fig:hydro_v2_spectra_RHIC} shows that
all the hydro models reproduce the pion spectra below $p_T\sim$ 1 GeV/$c$;
at higher $p_T$ the particles are less likely to be equilibrated and
hydro models are not expected to work well. 
In the bottom left panel 
the calculated proton spectra from Huovinen \cite{Huovinen:2001cy}
(solid green line)  
are lower than the
data, due to the calculation maintaining chemical
equilibrium throughout the hadronic phase. The lower
temperature chemical freezeout suppresses the final calculated
yield of heavier particles such 
as protons. Of the two calculations from Teaney \cite{Teaney:2001av}
the calculation that includes the QGP phase (solid red line) reproduces the proton spectra,
presumably because of the increased transverse flow from the 
stronger early pressure gradients. 
Hirano's and Kolb's (solid  dark and light-blue lines) calculations break chemical equilibrium during
the hadronic phase and overpredict the proton spectra at low $p_T$.

One difficulty is that the spectra comparison with hydrodynamic models
is for central collisions while the $v_2$ comparison is for minimum-bias
collisions. It is dificult to use central
collisions for the $v_2$ comparison since the collisions
are nearly symmetric and hence $v_2$ is small.
In addition, hydrodyamic calculations that reproduce $v_2$ values over a broad
range of centrality (from 0-45\% in Ref. \cite{Adler:2001nb}) 
tend to overpredict the
data for more peripheral collisions by approximately 25\%,
presumably because of a breakdown in the hydrodynamic assumptions. 
Hence when comparing to minimum-bias data sets, an overprediction 
of $v_2$ from the hydro models of less than  20\% should be acceptable.
  
\begin{table*}
\begin{tabular}{|p{1.7cm}|p{1.9cm}|p{1.9cm}|p{1.9cm}|p{1.9cm}|p{1.9cm}|p{1.9cm}|} \hline
  & \multicolumn{4}{|c|}{\it QGP+mixed+RG} 
  & {\it mixed+RG} & {\it RG} \\ \cline{2-7}
  & {\it Teaney} & {\it Hirano} & {\it Kolb} &  {\it Huovinen} 
  & {\it Teaney} &  {\it Huovinen} \\   \hline
Reference & \cite{Teaney:2001av} & \cite{Hirano:2002ds} & 
\cite{Kolb:2002ve,Heinz:2002un} & 
	\cite{Huovinen:2001cy} & \cite{Teaney:2001av} &	 \cite{Huovinen:2001cy} \\ \hline \hline
latent heat (GeV/fm$^3$)& 0.8 & 1.7 & 1.15 & 1.15 & 0.8 &  \\ \hline
init. $\varepsilon_{max}$ (GeV/fm$^3$)& 16.7 &  & 23 & 23 & 16.7  & 23 \\ 
\hline
init. $\langle \varepsilon \rangle$ (GeV/fm$^3$)& 11.0 & 13.5 &  &  &11.0 &  \\ \hline
$\tau_0$ fm/$c$ & 1.0 & 0.6 & 0.6 & 0.6 & 1.0 & 0.6 \\ \hline
hadronic stage& RQMD & partial chemical equil. &  partial chemical equil. 
& full equil. & RQMD & full equil. \\ \hline
proton v2 & yes &  $<$ 0.7 GeV/$c$ &  $<$ 0.7 GeV/$c$  & yes & no  & no \\ \hline
pion v2 & yes & no &  no  & yes & yes  & yes \\ \hline
proton spectra & yes & overpredict & overpredict  & no & no  & no \\ \hline
pion spectra & yes &  $<$ 1 GeV/$c$ &  $<$ 1 GeV/$c$  & yes &  $<$ 0.7 GeV/$c$   & yes \\ \hline
HBT & Not available & No & No & No & Not available  & Not available\\
\hline \hline
\end{tabular}
\caption{Summary of various hydro model assumptions
and a comparison between measurements and hydro calculations. Two initial energies
are tabulated, either the maximum energy density at the center of the collision
or the energy density averaged over the transverse profile.}
\label{tab:hydro_v2_spectra_RHIC}
\end{table*}

These comparisons between data and hydro  models
are summarized in Table \ref{tab:hydro_v2_spectra_RHIC}
and in the following conclusions;
\begin{itemize}{} {}

\item $v_2(p_T, PID)$ is sensitive to all
stages of the reaction. Elliptic flow is produced by strong 
scattering in the initial phase, while
the detailed shape of $v_2(p_T)$ and how the momentum
asymmetry is distributed to different particles is
affected by the transition from a QGP to
hadronic phase and scattering in the final hadronic stage.
\item The  hydro models that reproduce the low-$p_T$ proton $v_2$ are 
those that include
both a QGP and hadronic phase.
\item The hadronic phase critically affects the final values of $v_2(p_T, 
PID)$. 
Models (Hirano, Kolb)
that include partial chemical equilibrium to reproduce the baryon yield, completely fail
on the pion $v_2$. 
\item The only model that survives this comparison with measured
$v_2$ and spectra is Teaney's (solid red line) which includes
 a strong expansion in a QGP phase, a phase transition to a mixed phase, and 
then a hadronic cascade in the final hadronic state. There are
open questions in this hybrid model, {\it e.g.} the sensitivity of the
results to the matching conditions between hydro and RQMD. All other models
fail in at least one $v_2$ or spectra comparison, partially due to
differences in modeling the final hadronic state.
\item Until the model uncertainty in the final state is reduced, it
is not yet possible to use the measured splitting between
proton and pion $v_2(p_T)$ to extract quantitative information on the EOS 
during the reaction, including the possible
softening of the EOS due to 
the presence of a mixed phase.
\end{itemize}

A comparison with the HBT data and
some of the hydro models is shown in Fig. \ref{fig:HBTvsktModel}.
It is unfortunate that not all hydro models have been compared to HBT 
data, {\it e.g.} the hydro+RQMD model from Teaney \cite{Teaney:2001av}
has not been confronted with this 
observable. 
The hydro calculation from Kolb, Heinz and Huovinen \cite{Heinz:2002un}
(solid green line) includes a first-order phase transition which leads to a long lifetime
for the system. The source parameter $R_{\rm long}$ is considered most sensitive to the 
duration of the whole collision, {\it i.e.}
from initial overlap to final particle emission, and the Kolb/Huovinen hydro 
calculation \cite{Heinz:2002un}
(solid green line) overpredicts the 
measured $R_{\rm long}$ data. Changing to partial chemical equilibrium in 
the
hadronic stage \cite{Hirano:2002ds}, indicated with the dark blue line, reduces 
the lifetime of the
collision which improves the agreement with $R_{\rm long}$. However
the ratio $R_{\rm out}/R_{\rm side}$, which is sensitive to the duration over which particles are
emitted,
is still overpredicted. 

There have been many attempts to understand what may be causing the
disagreement with data (known collectively as the HBT puzzle);
\begin{itemize}
\item Sinyukov {\it et al.} \cite{Sinyukov:2002if} and Grassi et 
al. \cite{Grassi:2000ke}
have suggested that the sharp Cooper-Frye freezeout
condition \cite{Cooper:1974mv} should be replaced by an emission function that decouples hadrons
depending on their hadronic cross section.
\item However when this has been effectively implemented by using a hadronic 
cascade (URQMD) for the final hadronic stage, the predicted ratio $R_{\rm out}/R_{\rm side}$ increases
and diverges further from the data \cite{Soff:2000eh}. 
Modeling the final stage with 
a hadronic cascade effectively includes dissipative effects which should
increase the duration of emission and produce a larger ratio $R_{\rm out}/R_{\rm side}$ .
\item One method to reduce the lifetime of the reaction is to change the QGP
EOS. Using a crossover instead of a first-order transition reduces the
ratio $R_{\rm out}/R_{\rm side}$ 
by  about 50\% to $R_{\rm out}/R_{\rm side} \sim 1.5$ \cite{Zschiesche:2001dx}
which is still larger than the data.
Because the calculation was restricted to $\eta = 0$
Zschiesche {\it et al.} were unable to compare with the measured values of
$R_{\rm long}$.
\end{itemize}

In summary, model comparisons seem to 
be closer to the HBT data when the lifetime of the collision
is made smaller than the long time resulting from a 
first-order phase transition. The small values of
$R_{\rm out}/R_{\rm side}$ may indicate that there is little to no mixed phase present in the reactions. 
One possible direction for future comparisons with data 
is to include a more realistic
EOS into the hydro models, {\it e.g.}
to take the EOS from lattice QCD
calculations \cite{Karsch:2003jg}.
Such a calculation needs to be compared with all the available data,
including spectra and $v_2$, as well as HBT.

\subsection{Conclusions}

In summary we can make the following conclusions

\begin{itemize}{} {}

\item The measured yields and spectra of hadrons are consistent with 
thermal emission from a strongly expanding source.
\item Strangeness is fully saturated at RHIC, consistent with full chemical
equilibrium.
\item The scaling of $v_2$ with eccentricity shows that collective behavior is
established early in the collision. 
\item Elliptic flow is stronger at RHIC than at the SPS, since
the measured slope of $v_2(p_T)$ for pions is 50\% larger at RHIC.
\item The measured proton $v_2(p_T)$ is less than that for pions at low
$p_T$; the small
magnitude of the proton $v_2$ at low $p_T$ is reproduced by hydro models 
that include
both a QGP and hadronic phase.
\item However several of the hydro models that reproduce
the proton $v_2(p_T)$ fail for the pion $v_2(p_T)$. 
\item The HBT source parameters, especially the small value  of $R_{\rm long}$ and
the ratio $R_{\rm out}/R_{\rm side}$, suggest that
the mixed phase is too long-lived in the current hydro calculations.
\end{itemize}

Hence we currently do not have a consistent picture of the 
space-time dynamics of reactions at RHIC as revealed by spectra, $v_2$, and HBT. 
The lack of a consistent picture of the dynamics means that 
it is not yet possible to extract quantitative properties
of the QGP or mixed phase using the observables $v_2$, spectra, or HBT.
 
\section{FLUCTUATIONS}
\label{Sec:fluc}
\subsection{Net-Charge Fluctuations}
In the study of the fluctuations of multiplicity as a means to
understand the dynamics of charged particle production, one important
realization was to use small regions of phase space, where
energy-momentum conservation constraints would not be
significant \cite{Fowler:1977gx,Fowler:1981ks,VanHove:1987kb}. 
Such studies led to the important observation that the distribution
of multiplicity, even in small intervals near mid-rapidity, was
Negative Binomial rather than Poisson, which indicated large
multiplicity correlations even in small $\delta\eta$
intervals \cite{Alner:1985zc}. No such studies are yet available at
RHIC.

Based on predictions that event-by-event fluctuations of the net
charge in local phase space regions would show a large decrease as
a signature of the QGP \cite{Jeon:2000wg,Asakawa:2000wh,Heiselberg:2000ti}, 
net-charge fluctuations were measured in
PHENIX \cite{Adcox:2002mm}. The idea is that in a QGP composed of
fractionally charged quarks, the larger number of fractionally
charged particles compared to unit-charged hadrons would result in
smaller 
relative 
net-charge fluctuations in a QGP than for a pure gas of
hadrons and that this original fluctuation would survive the
transition back to ordinary hadrons.
   
It is important to realize that the study of net-charge
fluctuations represents the study of fluctuations in a
quantity that is conserved over all phase space.  Consider
$N = N_+ +N_-$ charged particles produced in the full phase
space.  By charge conservation $N_+ = N_- = N/2$, and the net
charge $Q\equiv N_+ - N_-$ is identically zero so that there
are no net-charge fluctuations---the variance $V(Q) = 0$,
where
\begin{equation}
V(Q)\equiv \langle Q^2\rangle- \langle Q\rangle^2 \:.
\end{equation}
In a smaller region of 
phase space,  where $p$ is the fraction of $N$ observed in a stochastic 
scenario, the mean and 
variance of the number of positive $n_+$ and negative $n_-$ particles are 
equal, but the variance of $Q$ is no longer identically zero: 
\begin{equation}
\langle n_+\rangle = \langle n_-\rangle = p N/2 \: ,
\end{equation}
\begin{equation}
V(n_+) = V(n_-) = p (1-p) N/2 \: , 
\end{equation}
from which it follows that 
\begin{equation}
V(Q) = (1-p) n_{ch}\:,   
\end{equation}
where $n_{ch} = pN$ is the expected number of charged particles on the interval.  
Thus the normalized variance in $Q$ (normalized to Poisson statistics) is defined as: 
\begin{equation}
v(Q) = {V(Q)\over n_{ch}} = (1-p) \: .
\end{equation} 
In the limit $n_{ch}\gg 0$, the variance of the charge ratio
$R = n_+/n_-$ approaches $V(R) = 4(1-p)/n_{ch}$. However, it is well known
in mathematical statistics that moments of the inverse of a stochastic
variable, e.g. $1/n_-$, diverge if there is any finite probability, no
matter how small, for $n_{-} = 0$. Thus, the charge ratio is not a
stable measure of fluctuations.

The previous arguments are based on fixed $N$. The results
where $N$ varies according to a specified distribution are
also interesting. 
If $n_-$ is Poisson distributed, with mean value
$\mu = N/2$ over the whole phase space, then in the region of
phase space with probability $p$ the distribution is also
Poisson, with mean $\langle n_-\rangle|_{p} = \mu_{p} = p
N/2$. If, on the other hand $n_-$ is Negative Binomial
distributed, with mean value $\mu = N/2$ and NBD parameter
$\sigma^2/\mu^2-1/\mu = 1/k$ for the whole phase space, then in
the region of phase space with probability $p$, the
distribution is Negative Binomial with mean $\langle
n_-\rangle|_{p} = \mu_{p} = p N/2$ and the same value of $1/k$.

Actually, the binomial division preserves
$\sigma_{p}^2/\mu_{p}^2-1/\mu_{p} = 1/k$, for any
distribution \cite{Albrecht:1989kh}. This appears to indicate
that smaller intervals, which tend to have larger values of
$\sigma_{p}^2/\mu_{p}^2$ would be less sensitive to the global
$1/k$, the long-range correlation. This would be true except
for the fact that there are short-range correlations which are
better seen on small intervals of phase space. Another
important thing to note regarding a binomial split of a
Negative Binomial distribution is that the two subintervals
are not statistically independent. The conditional probability
distribution on the interval $(1-p)$ depends upon the outcome on
the interval $p$ \cite{Carruthers:1985jd}. It is unfortunate
that these elegant arguments can not be applied to the
net-charge fluctuations since $\langle Q\rangle = 0$.
     
\begin{figure}[tbhp]
\includegraphics[width=1.0\linewidth]{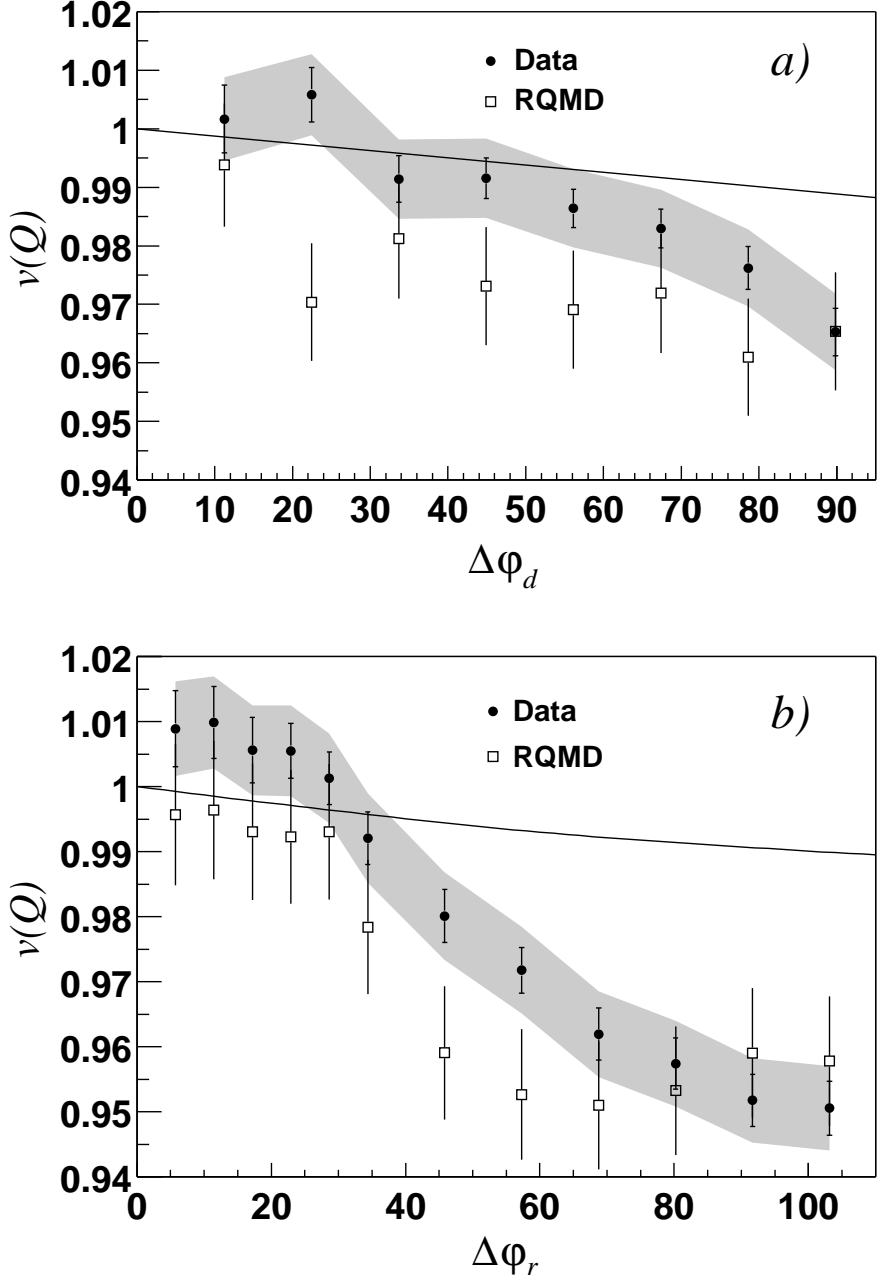}
\caption[]{$v(Q)$ for the 10\% most central events in data and
RQMD, as a function of the azimuthal interval in degrees of reconstructed tracks,  either a) at the
detector, $\Delta\phi_d$, or b) at the vertex, $\Delta\phi_r$,
chosen symmetrically around the detector acceptance. For data, the error 
band shows the total statistical error, whereas the error bars indicate 
the uncorrelated part. The solid line 
shows the expected reduction in $v(Q)$ in the stochastic 
scenario when global charge conservation is taken into 
account \cite{Adcox:2002mm}.
}
\label{fig:vQ}
\end{figure}

The PHENIX measurement \cite{Adcox:2002mm} of the normalized variance $v(Q)$ of
net-charge fluctuations is shown in Fig. \ref{fig:vQ} in the
interval $-0.35\leq \delta\eta\leq +0.35$ as a function of the
azimuthal angular interval of reconstructed tracks, either at the
detector, $\Delta\phi_d$, or at the vertex, $\Delta\phi_r$,
chosen symmetrically around the detector acceptance.
For smaller $\Delta\phi_r$ the data agree with the purely
stochastic $(1-p)$ dependence shown as the solid line, but
deviate from the stochastic prediction at larger values due to
correlations from resonance decay, such as $\rho^0\rightarrow
\pi^+ + \pi^-$ as nicely explained by RQMD \cite{Sorge:1995dp}.
     
Absent new theoretical insight, it is difficult to understand
how quark-level net-charge fluctuations in a QGP can be
related to net-charge fluctuations of hadrons, where, by
definition, strong correlations exist, e.g., in the formation of a meson
from a $q-\bar{q}$ pair. Also, the study of the fluctuations
of net charge, which is conserved, may not be as useful to
detect interesting fluctuations as the study of fluctuations of the
total charged multiplicity, which is much less constrained by conservation laws.   
This has yet to be tried at RHIC.


\subsection {Event-by-Event Average-$p_T$ Fluctuations} 
   Fluctuations in the event-by-event average $p_T$, denoted
$M_{p_T}$, have been measured and provide a severely small limit on
possible fluctuations from a sharp phase transition.
For events with $n$ detected charged particles with magnitudes of
transverse momenta, $p_{T_i}$, the event-by-event average $p_T$,
denoted $M_{p_T}$ is defined as:   
\begin{equation}
M_{p_T} = \overline{p_T} = {1\over n} \sum_{i = 1}^n p_{T_i}
\: .\label{eq:defMpT}
\end{equation}
Mixed events are used to define the baseline for random fluctuations
of $M_{p_T}$ in PHENIX \cite{Adcox:2002pa,Adler:2003xq}. This has the
advantage of effectively removing any residual detector-dependent
effects. The event-by-event average distributions are very sensitive
to the number of tracks in the event (denoted $n$ or $N_{tracks}$), so
the mixed event sample is produced with the {\em identical}
$N_{tracks}$ distribution as the data. Additionally, no two tracks
from the same data event are placed in the same mixed event in order
to remove any intra-event correlations in $p_T$. Finally, $\langle
M_{p_T}\rangle$ must exactly match the semi-inclusive $\langle
p_T\rangle$.

For the case of statistical independent emission, where the
fluctuations are purely random, an analytical formula for the
distribution in $M_{p_T}$ can be obtained assuming Negative Binomial 
distributed event-by-event multiplicity, with Gamma distributed
semi-inclusive $p_T$ spectra \cite{Tannenbaum:2001gs}. The formula
depends on the four semi-inclusive parameters $\langle n\rangle$, $1/k$,
$b$ and $p$ which are derived from the means and standard deviations
of the semi-inclusive $p_T$ and multiplicity distributions, $\langle
n\rangle$, $\sigma_n$, $\langle p_T\rangle$, $\sigma_{p_T}$:
\begin{equation}
       f(y) = 
\sum_{n = {n_{\rm min}}}^{n_{\rm max}} f_{\rm NBD}(n,1/k,\langle 
n\rangle) 
\, f_{\Gamma}(y,np,nb) \: ,   
\label{eq:nbdgamma}
\end{equation} 
where $y = M_{p_T}$. For fixed $n$, and purely random fluctuations, the
mean and standard deviation of $M_{p_T}$ follow the expected behavior,
$\langle M_{p_T}\rangle = \langle p_T\rangle$,
$\sigma_{M_{p_T}} = \sigma_{p_T}/\sqrt{n}$.  In PHENIX,
Eq. \ref{eq:nbdgamma} is used to confirm the randomness of
mixed events which are used to define the baseline for random
fluctuations of $M_{p_T}$ \cite{Adcox:2002pa,Adler:2003xq}.

\begin{figure}[tbhp]
\begin{center}
\begin{tabular}{c}
\includegraphics[width=0.7\linewidth]{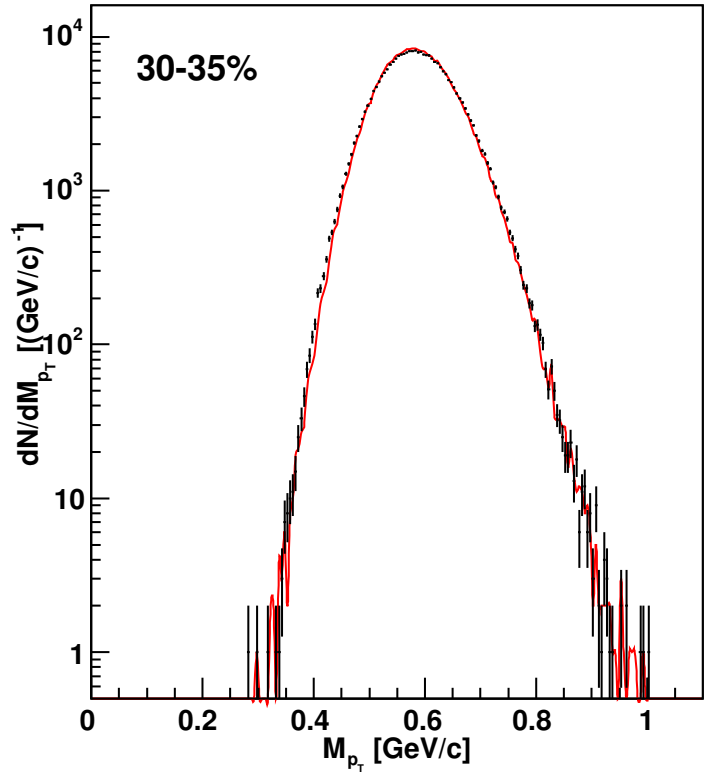}\\
\includegraphics[width=0.7\linewidth]{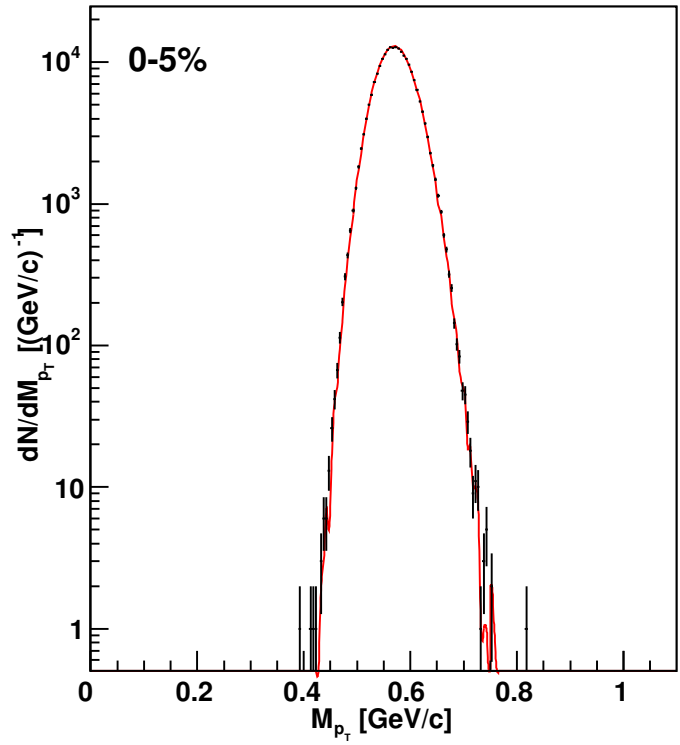} 
\end{tabular}
\end{center}
\caption[]{$M_{p_T}$ for 30--35\% and 0--5\% centrality classes: data (points) mixed events (histogram) \cite{Adler:2003xq}. 
\label{fig:MpT}} 
\end{figure}

\begin{figure}[tbhp]
\includegraphics[width=1.0\linewidth]{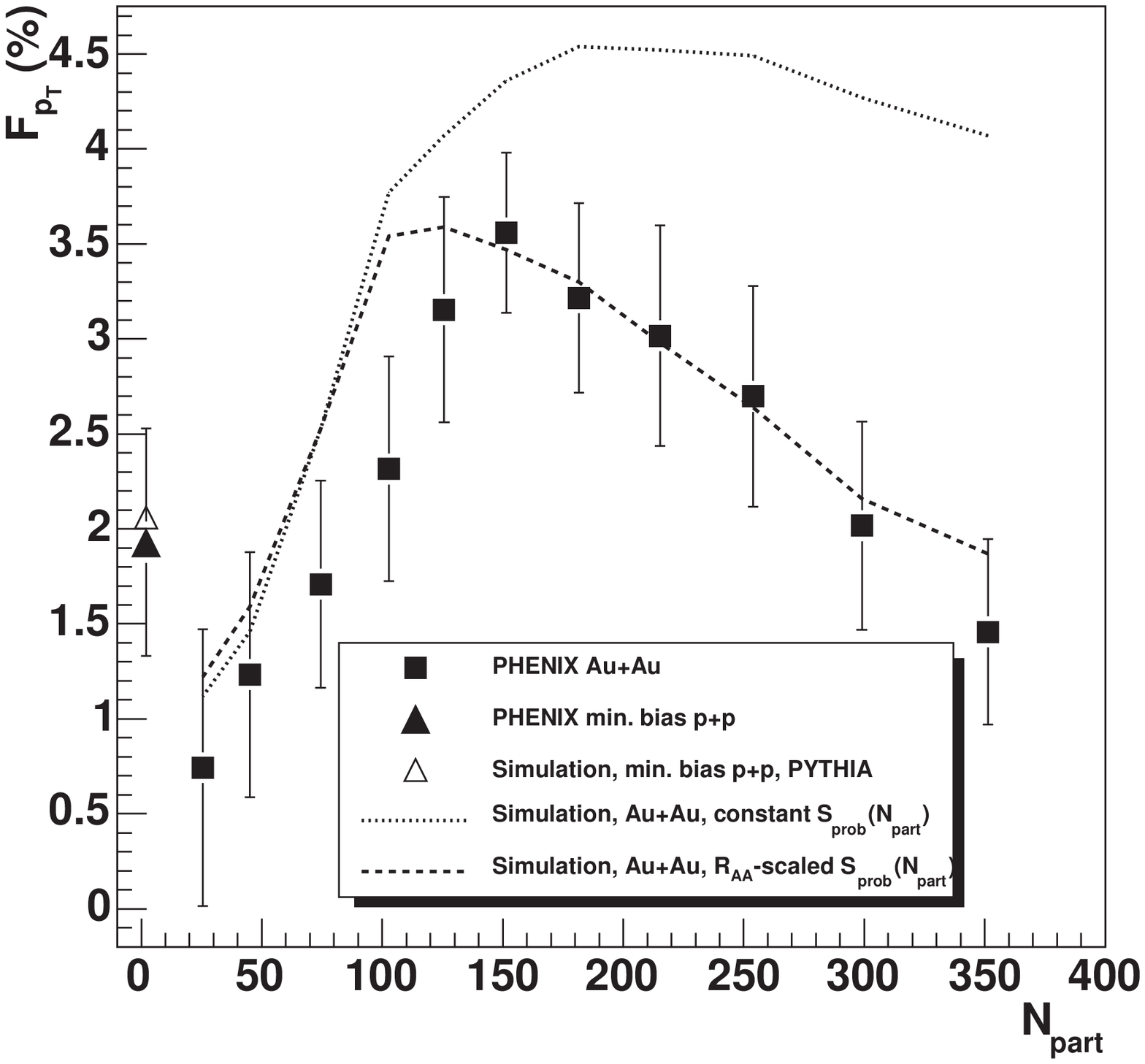}
\caption[]{$F_{p_T}$ vs. centrality, represented as the average number of participants $(N_{part})$ in a centrality class, compared to 
jet simulation \cite{Adler:2003xq}.}
\label{fig:FpTcent}
\end{figure}

The measured $M_{p_T}$ distributions for the data in two
centrality classes for $\sqrt{s_{NN}} = 200$ GeV Au+Au
collisions in PHENIX \cite{Adler:2003xq} are shown in Fig. \ref{fig:MpT} (data
points) compared to the mixed-event distributions
(histograms).
The non-Gaussian, Gamma-distribution shape of the $M_{p_T}$
distributions is evident.  The difference between the data and the
mixed-event random baseline distributions is not visible to the
naked eye. The nonrandom fluctuation is quantified by the fractional
difference of $\omega_{p_T}$, the normalized standard deviation of
$M_{p_T}$, for the data and the mixed-event (random) samples:
\begin{equation}
\omega_{p_T} = \frac{\sigma_{M_{p_T}}}{\langle M_{p_T} \rangle} \: ,
\label{eq:defomega}
\end{equation}
	\begin{equation}
	F_{p_T} = \frac{\omega_{p_T,{\rm data}} -\omega_{p_T,{\rm mixed}}} {\omega_{p_T,{\rm mixed}}} \: .
	\label{eq:def:FpT}
	\end{equation}
The results are shown as a function of centrality, represented by $N_{part}$ in Fig. \ref{fig:FpTcent}. 

\begin{figure}[tbhp]
\includegraphics[width=1.0\linewidth]{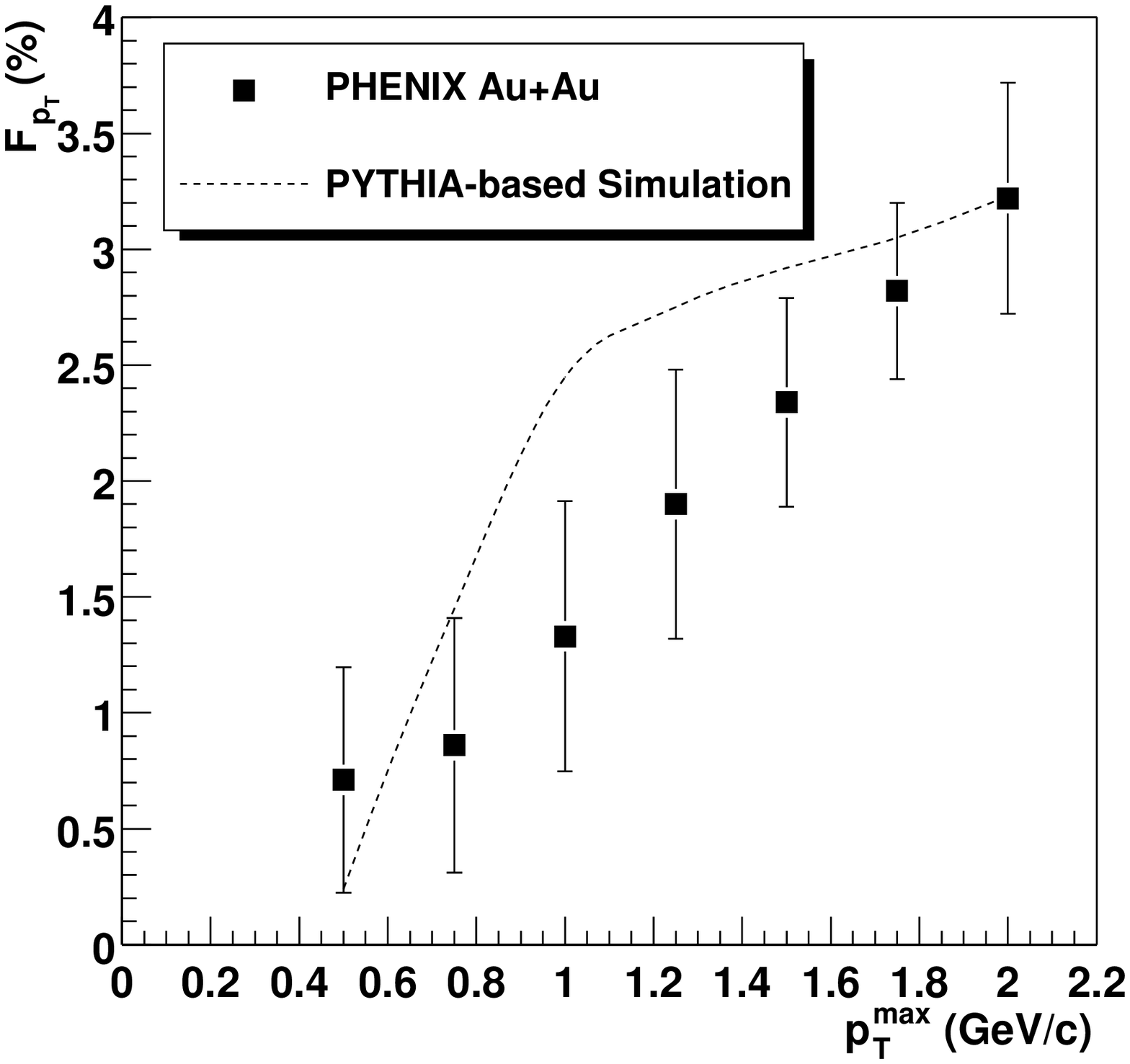}
\caption[]{$F_{p_T}$ vs. $p_T^{\rm max}$ compared to 
$R_{AA}$-scaled 
jet simulation
for the 20--25\% centrality class ($N_{part} = 181.6$) \cite{Adler:2003xq}.}
\label{fig:FpTmax}
\end{figure}

The dependence of $F_{p_T}$ on $N_{part}$ is striking. To
further understand this dependence and the source of these
nonrandom fluctuations, $F_{p_T}$ was measured over a varying
$p_T$ range, \mbox{0.2 GeV/$c$ $\leq p_T\leq p_T^{\rm max}$}
(Fig. \ref{fig:FpTmax}), where $p_T^{\rm max} = 2.0$ GeV/$c$ for
the $N_{part}$ dependence.
	
The increase of $F_{p_T}$ with $p_T^{\rm max}$ suggests
elliptic flow or jet origin. This was investigated using a
Monte Carlo simulation of correlations due to elliptic flow
and jets in the PHENIX acceptance. The flow was significant
only in the lowest centrality bin and negligible ($F_{p_T}
<0.1$\%) at higher centralities. Jets were simulated by
embedding (at a uniform rate per generated particle,
$S_{prob}(N_{part})$) $p+p$ hard-scattering events from the
PYTHIA event generator into simulated Au+Au events assembled
at random according to the measured $N_{tracks}$ and
semi-inclusive $p_T$ distributions. This changed $\langle
p_T\rangle$ and $\sigma_{p_T}$ by less than 0.1{\%}.  $S_{prob}(N_{part})$ was either constant for all centrality
classes, or scaled by the measured hard-scattering suppression
factor $R_{AA}(N_{part})$ for $p_T > 4.5$
 GeV/$c$ \cite{Adler:2003qi}. A value $F_{p_T} = 2.06$\% for $p+p$
collisions was extracted from pure PYTHIA events in the PHENIX
acceptance in agreement with the $p+p$ measurement 
(Fig. \ref{fig:FpTcent}). The value of
$S_{prob}(N_{part})$ was chosen so that the simulation with
$S_{prob}(N_{part})\times R_{AA}(N_{part})$ agreed with the
data at $N_{part} = 182$. The centrality and $p_T^{\rm max}$
dependences of the measured $F_{p_T}$ match the simulation
very well, but only when the $R_{AA}$ scaling is included.

	A less experiment-dependent method to compare
nonrandom fluctuations is to assume that the entire $F_{p_T}$
is due to temperature fluctuations of the initial state, with
RMS variation $\sigma_{T}/\langle
T\rangle$ \cite{Korus:2001au,Adcox:2002pa}. Then,
\begin{equation} {\omega^2_{p_T,{\rm data}}} -
{\omega^2_{p_T,{\rm mixed}}} = (1-{1\over {\langle n\rangle}})
\frac{\sigma^2_{T}}{\langle T\rangle^2} = 2
F_{p_T}{\omega^2_{p_T,{\rm mixed}}} \: , \label{eq:sigmaT}
\end{equation} This yields $\sigma_{T}/\langle T\rangle$ = 1.8\%
for central collisions in PHENIX with similarly small values
for the other 
Relativistic Heavy Ion  
experiments \cite{Mitchell:2004xz}, 1.7\% in
STAR, 1.3\% in CERES, and 0.6\% in NA49.  These results put
severely small limits on the critical fluctuations that were
expected for a sharp phase transition, both at SPS
energies and at RHIC,
but are consistent with the expectation from lattice QCD that the transition is a smooth crossover \cite{Rajagopal:2000uu}.

	Other proposed explanations of the centrality and $p_T^{\rm max}$ dependences of $F_{p_T}$ include: overlapping color strings which form clusters so that the number of sources and $\langle p_T\rangle$ per source is modified as a function of centrality \cite{Ferreiro:2003dw}; and near equilibrium $p_T$ correlations induced by spatial inhomogeniety \cite{Gavin:2003cb}.
\subsection{Conclusions}
Critical behavior near the phase boundary can produce nonrandom
fluctuations in observables such as the net-charge distribution and
the average transverse momentum distribution. Our search for
net-charge fluctuations has ruled out the most naive model of charge
fluctuations in a QGP, but it is unclear whether the charge
fluctuation signature can survive hadronization. Our measurement of
the event-by-event average $p_T$ distribution shows a nonrandom
fluctuation that is consistent with the effect expected from
high-$p_T$ jets. This puts a severe constraint on the critical
fluctuations that were expected for a sharp phase transition 
but is 
consistent with the expectation from lattice QCD that the transition is a smooth crossover \cite{Rajagopal:2000uu}.
 
\section{BINARY SCALING}
\label{Sec:scale}
\subsection{Hard Scattering and pQCD}
One way to get a partonic probe into the midst of an $A+A$ collision is
to use the high-$p_T$ partons produced by hard scattering. For $p+p$ collisions in
the RHIC energy range, hard scattering is considered to be the dominant process
of particle production with $p_T \geq 2$ GeV/$c$ at mid-rapidity. Typically, particles
with $p_T\geq 2$ GeV/$c$ are produced from states with two roughly back-to-back jets
which are the result of scattering of constituents of the nucleons (partons)
as described by pQCD \cite{Owens:1987mp}. 
        
The overall $p+p$ hard-scattering cross section in ``leading logarithm" pQCD   
is the sum over parton reactions $a+b\rightarrow c +d$ 
(e.g. $g+q\rightarrow g+q$) at parton-parton center-of-mass (c.m.) energy $\sqrt{\hat{s}}$:   
\begin{equation}
\frac{d^3\sigma}{dx_1 dx_2 d\cos\theta^*} = 
\frac{1}{s}\sum_{ab} f_a(x_1) f_b(x_2) 
\frac{\pi\alpha_s^2(Q^2)}{2x_1 x_2} \Sigma^{ab}(\cos\theta^*) \: ,
\label{eq:QCDabscat}
\end{equation} 
where $f_a(x_1)$, $f_b(x_2)$, are parton distribution functions, 
the differential probabilities for partons
$a$ and $b$ to carry momentum fractions $x_1$ and $x_2$ of their respective 
protons (e.g. $u(x_2)$), and where $\theta^*$ is the scattering angle in the parton-parton c.m. system. 
The parton-parton c.m. energy squared is $\hat{s} = x_1 x_2 s$,
where $\sqrt{s}$ is the c.m. energy of the $p+p$ collision. The parton-parton 
c.m. system moves with rapidity $y = 1/2 \ln (x_1/x_2)$ in the $p+p$ c.m. system. 

Equation \ref{eq:QCDabscat} gives the $p_T$ spectrum of outgoing parton $c$, which then
fragments into hadrons, e.g. $\pi^0$.  The fragmentation function
$D^{\pi^0}_{c}(z,\mu^2)$ is the probability for a $\pi^0$ to carry a fraction
$z = p^{\pi^0}/p^{c}$ of the momentum of outgoing parton $c$. Equation \ref{eq:QCDabscat}
must be summed over all subprocesses leading to a $\pi^0$ in the final state.
The parameter $\mu^2$ is an unphysical ``factorization" scale introduced to account
for collinear singularities in the structure and fragmentation
functions \cite{Owens:1987mp,Bunce:2000uv}.

In this formulation, $f_a(x_1,\mu^2)$, $f_b(x_2,\mu^2)$ and $D^C_c (z,\mu^2)$ 
represent the ``long-distance phenomena" to be determined by experiment;
while the characteristic subprocess angular distributions,
{\bf $\Sigma^{ab}(\cos\theta^*)$},
and the coupling constant,
$\alpha_s(Q^2) = \frac{12\pi}{25} \ln(Q^2/\Lambda^2)$,
are fundamental predictions of QCD \cite{Cutler:1978qm,Cutler:1977mw,Combridge:1977dm}
for the short-distance, large-$Q^2$, phenomena.
The momentum scale $Q^2\sim p_T^2$ for the scattering
subprocess, while $Q^2\sim\hat{s}$ for a Compton or annihilation subprocess,
but the exact meaning of $Q^2$ tends  to be treated as a parameter rather than a dynamical
quantity. 
The transverse momentum of a scattered constituent is:
\begin{equation}
p_T = p_T^* = { \sqrt{\hat{s}} \over 2 } \; \sin\theta^* \: .
\label{eq:cpT}
\end{equation}

Equation \ref{eq:QCDabscat} leads to a general `$x_T$-scaling' form for the invariant cross
section of high-$p_T$ particle production: 
\begin{equation}
E \frac{d^3\sigma}{d^3p} = \frac{1}{p_T^{n}} F({x_T}) = 
 \frac{1}{\sqrt{s}^{n}} G({x_T}) \: ,
 \label{eq:bbg}
 \end{equation} 
where $x_T = 2p_T/\sqrt{s}$. 
The cross section has two factors, a function $F({x_T})$ ($G({x_T})$) which `scales',
i.e. depends only on the ratio of momenta, and a dimensioned factor,
${1/p_T^{n}}$ ($1/\sqrt{s}^{\, n}$),   
where $n$ equals 4  in lowest-order (LO) calculations, analogous to the $1/q^4$ form
of Rutherford Scattering in QED. The structure and fragmentation
functions   
are all in the
$F(x_T)$ ($G(x_T)$) term. Due to higher-order effects such as the running of
the coupling constant, $\alpha_s(Q^2)$, the evolution of the
structure and fragmentation functions, and the initial-state
transverse momentum
$k_T$, $n$ is not a constant but is a function of $x_T$, $\sqrt{s}$.
Measured values of ${\,n(x_T,\sqrt{s})}$ in $p+p$ 
collisions are between 5 and 8.

\subsection{Mid-Rapidity $p_T$ Spectra from $p+p$ Collisions}   

\begin{figure}[tbhp]
\begin{center}
\begin{tabular}{cc}
\hspace*{-0.12in}
\includegraphics[width=0.50\linewidth]{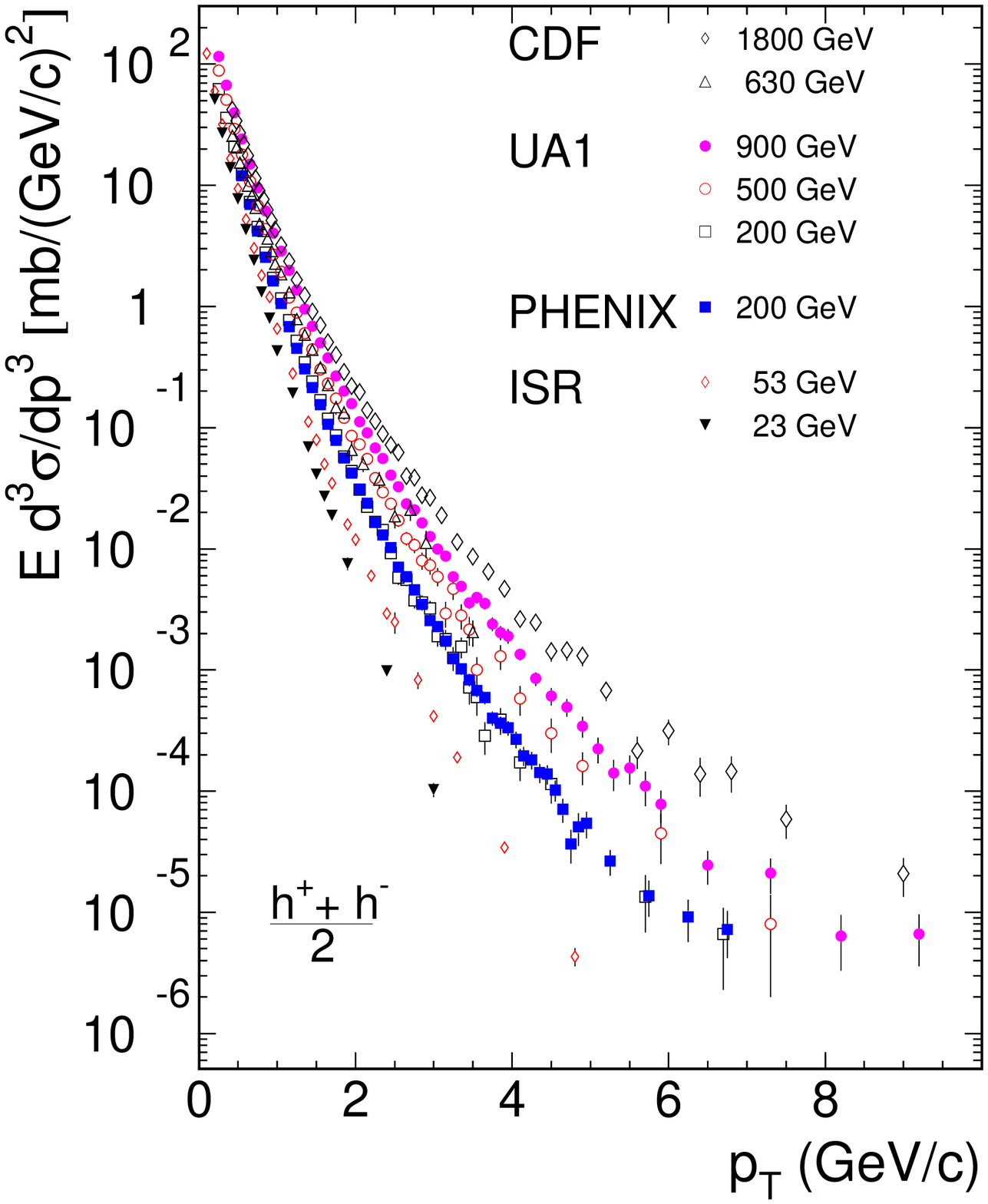} &
\includegraphics[width=0.50\linewidth]{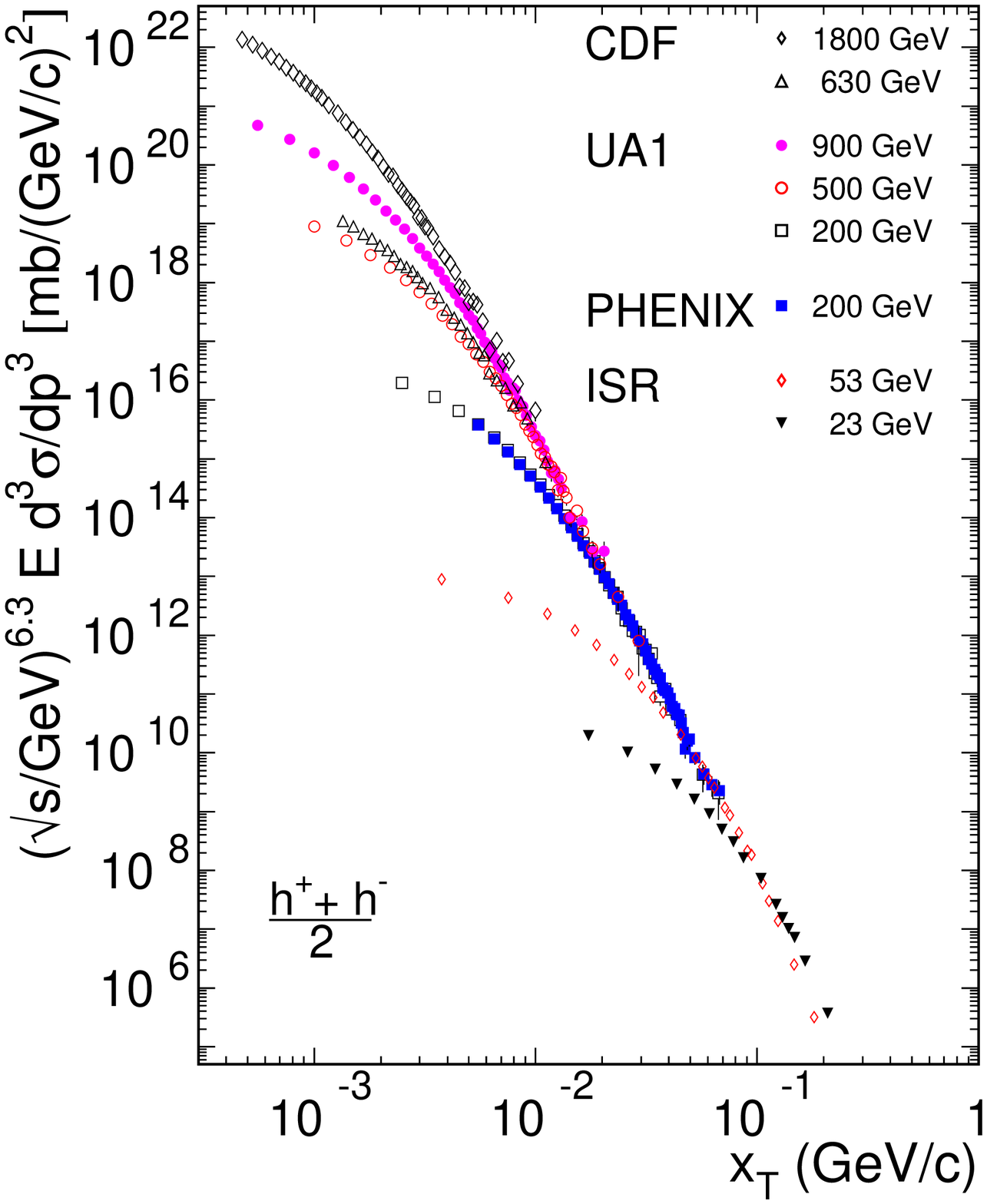}
\end{tabular}
\end{center}
\caption[]{(left) $E {d^3\sigma}(p_T)/{d^3p}$ at mid-rapidity as a function of
$\sqrt{s}$ in $p+p$ collisions.
(right) $\sqrt{s}({\rm GeV})^{6.3}\times Ed^3\sigma/d^3p$ vs.
$x_T = 2{p_T}/\sqrt{s}$ \cite{Adler:2003au} (and references therein). }
\label{fig:hpTxT}
\end{figure}

The scaling and power-law behavior of hard scattering are evident from the $\sqrt{s}$
dependence of the $p_T$ dependence of the $p+p$ invariant cross sections.  This is
shown for nonidentified charged hadrons,
$(h^+ + h^-)/2$, in Fig. \ref{fig:hpTxT}a. 
At low $p_T\leq 1$ GeV/$c$ the cross sections exhibit a ``thermal" 
$\exp {(-6 p_T)}$ dependence, which is largely independent of $\sqrt{s}$, while at high $p_T$
there is a power-law tail, due to hard scattering, which depends strongly on $\sqrt{s}$. 
The characteristic variation with $\sqrt{s}$ at high $p_T$ is produced by the fundamental
power-law and scaling dependence of Eqs. \ref{eq:QCDabscat}, \ref{eq:bbg}. This is best
illustrated by a plot of 
\begin{equation}
\sqrt{s}^{{\,n(x_T,\sqrt{s})}} \times E \frac{d^3\sigma}{d^3p} = G(x_T) \: ,
\label{eq:xTscaling}
\end{equation}
as a function of $x_T$, with ${{\,n(x_T,\sqrt{s})}} = 6.3$, which is valid for the $x_T$
range of the present RHIC measurements (Fig. \ref{fig:hpTxT}b).  The data show an
asymptotic power law with increasing $x_T$. Data at a given $\sqrt{s}$ fall
below the asymptote at successively lower values of $x_T$ with
increasing $\sqrt{s}$, corresponding to the transition region from
hard to soft physics in the $p_T$ region of about 2 GeV/$c$. 

\begin{figure}[tbhp]
\includegraphics[width=1.0\linewidth]{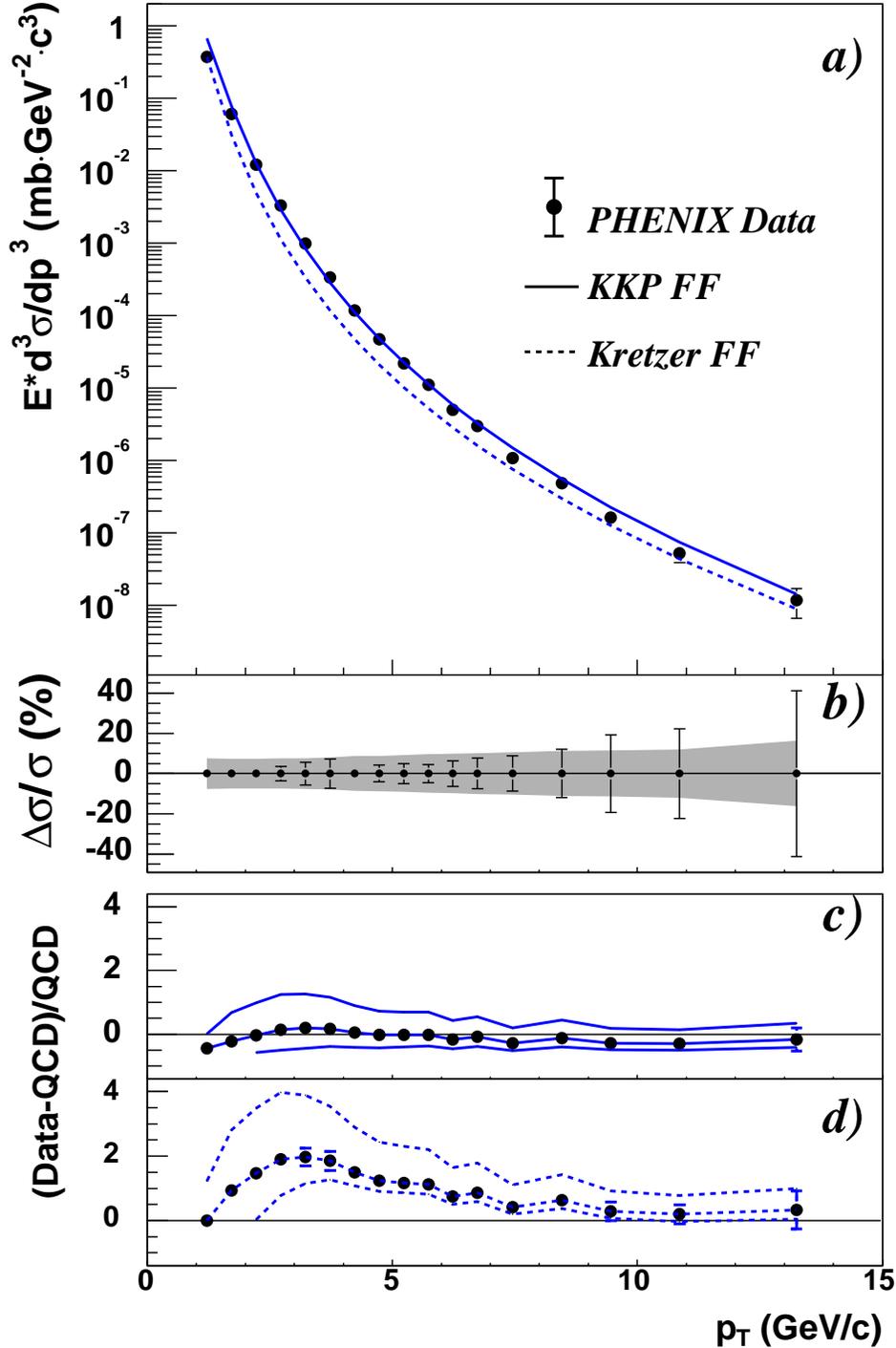}
\caption[]{PHENIX $\pi^0$ invariant cross section at mid-rapidity from $p+p$ collisions
at $\sqrt{s} = 200$ GeV, together with NLO pQCD predictions from Vogelsang \cite{Aversa:1988vb,Jager:2002xm}. 
a) The invariant differential cross section for inclusive $\pi^{\circ}$ 
production (points) and the results from 
NLO pQCD calculations with equal renormalization and factorization scales of $p_T$ 
using the ``Kniehl-Kramer-P\"{o}tter'' (solid line)  and ``Kretzer'' (dashed line) 
sets of fragmentation functions.  
b) The relative statistical (points) and point-to-point systematic (band) errors.  
c,d)  The relative difference between
the data and the theory using KKP (c) and Kretzer (d) fragmentation functions with
scales of $p_T$/2 (lower curve), $p_T$, and 2$p_T$
(upper curve).  In all figures, 
the normalization error of 9.6\% is not shown \cite{Adler:2003pb}.} 
\label{fig:pizpp200}
\end{figure}

The PHENIX measurement of the invariant cross section for $\pi^0$ production in $p+p$
collisions at $\sqrt{s} = 200$ GeV \cite{Adler:2003pb} agrees with NLO pQCD predictions
over the range $2.0\leq p_T\leq 15$ GeV/$c$ (Fig. \ref{fig:pizpp200}). 

\subsection{Scaling Hard Scattering from $p+p$ to $p+A$ and $A+B$ Collisions} 
Since hard scattering is point like, with distance scale $1/p_T \leq 0.1$ fm,
and the hard-scattering cross section factorizes as shown in Eq. \ref{eq:QCDabscat},
the cross section in $p+A$ or $A+B$ collisions, compared to $p+p$, is proportional
to the relative number of possible point-like encounters.
The number of encounters of point-like constituents of nucleons is then proportional to
$A$ $(AB)$, for $p+A$ ($A+B$)
minimum-bias collisions. For $A+B$ collisions at 
impact parameter $b$, it is proportional to 
$T_{AB}(b)$, the nuclear thickness function, which
is the integral of the product of nuclear thickness over
the geometrical overlap region of the two nuclei.
In detail, the semi-inclusive
invariant yield of 
e.g. high-$p_T$ $\pi^0$'s for $A+B$ inelastic collisions, with centrality $f$, is related
to the $p+p$ cross section by:
\begin{equation}
\left . \frac{1}{N^{\rm evt}_{AB}} \frac {d^2 N^{\pi^0}_{AB}}{dp_T dy} \right |_{f} 
 = {\langle T_{AB} \rangle_{f}} \times \frac{d^2 \sigma^{\pi^0}_{pp}}{dp_T dy }\: .
 \label{eq:TABscaling}   
 \end{equation}
Note that  
\begin{equation}
\langle T_{AB}\rangle_{f} = \frac{\int_{f} T_{AB}(b)\,d^2b}{\int_{f} (1- e^ {-\sigma_{NN}\,T_{AB}(b)})\, d^2 b} = \frac{\langle N_{coll}\rangle_f}{\sigma_{NN}} \:, 
\label{eq:TABf}
\end{equation}
where $\langle N_{coll}\rangle_f$ is the average number of binary nucleon-nucleon
inelastic collisions, with cross section $\sigma_{NN}$, in the centrality class $f$.
This leads to the description of the scaling for point-like processes as binary-collision
(or $N_{coll}$)  scaling. 

Nuclear medium effects, either in the initial or final state, can modify the expected scaling. 
These modifications can be quantitatively studied by measurement of the
{\it Nuclear modification factor} $R_{AB}$, which is defined as
\begin{equation}
R_{AB} = \frac{dN_{AB}^P}{\langle T_{AB} \rangle_{f} \times d\sigma_{NN}^P}
      = \frac{dN_{AB}^P}{\langle N_{coll} \rangle_{f} \times dN_{NN}^P} \: ,
\end{equation}
where $dN_{AB}^P$ is the differential yield of a point-like process $P$
in a $A+B$ collision and $d\sigma_{NN}^P$ is the cross section of $P$ in $N+N$ collision.
If there are no initial- or final-state effects that modify the yield of $P$
in $A+B$ collisions, the process $P$ scales with $\langle T_{AB}\rangle_{f}$ and
$R_{AB} = 1$. Sometimes, the central to peripheral ratio, $R_{CP}$, is used as
an alternative to
$R_{AB}$. The central to peripheral ratio is defined as
\begin{equation}
R_{CP} = \frac{dN^{Central}/\langle N_{coll}^{Central}\rangle}
              {dN^{Peripheral}/\langle N_{coll}^{Peripheral}\rangle} \: ,
\end{equation}
where $dN^{Central}$ and $dN^{Peripheral}$ are the differential yield per event of
the studied process in a central and peripheral collision, respectively. If the yield of
the process scales with the number of binary collisions, $R_{CP} = 1$.

\subsection{Binary Scaling in $l+A$, $p+A$, and Low-Energy $A+A$ \mbox{ }}
\begin{figure}[tbhp]
\includegraphics[width=1.0\linewidth]{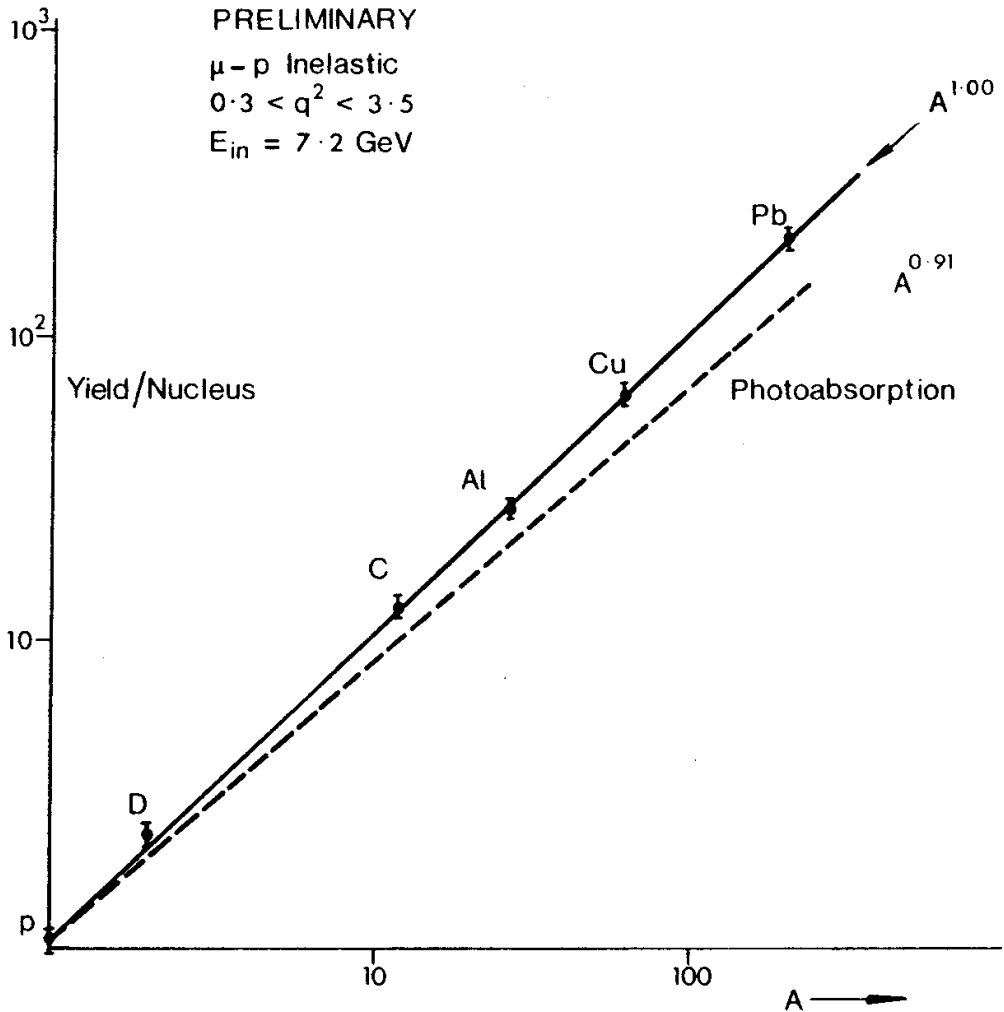}
\caption[]{$\mu$-$A$ cross section vs. $A$ \cite{May:1975ju}.}
\label{fig:muA}
\end{figure}

In 
deeply inelastic 
lepton scattering, where hard scattering was
discovered \cite{Bloom:1969kc,Breidenbach:1969kd,Bjorken:1969dy}, the cross section for
$\mu$-A collisions is indeed proportional to $A^{1.00}$ (Fig. \ref{fig:muA}). 
This indicates that the structure function of a nucleus of mass $A$ is simply $A$ times the
structure function of a nucleon (with only minor deviations, $\leq 10$\% for
$0.02\leq x\leq 0.50$ \cite{Arneodo:1996rv}), which means that the nucleus acts like
an incoherent superposition of nucleons for hard scattering of leptons. 

\begin{figure}[tbhp]
\includegraphics[width=1.0\linewidth]{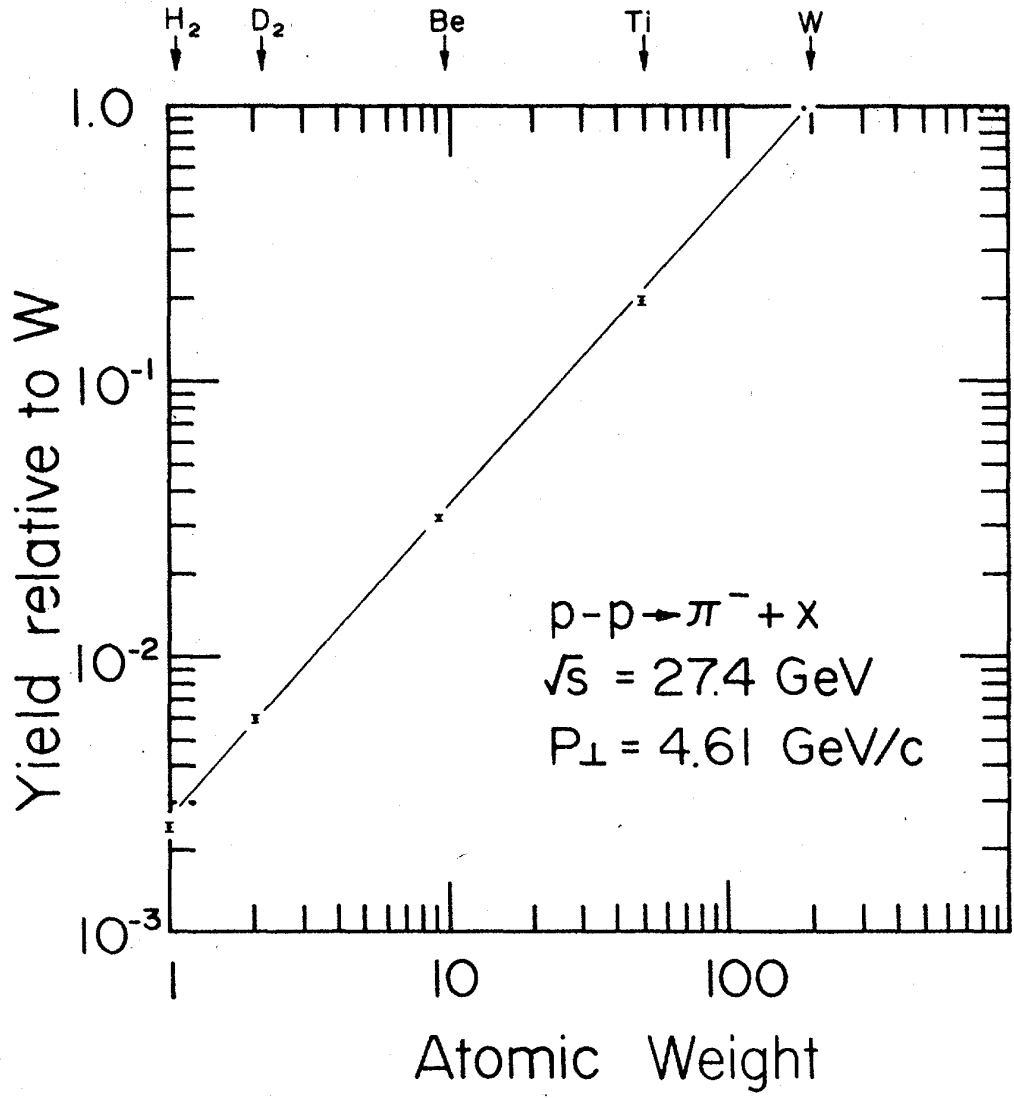}
\caption[]{Cronin Effect in $p+A$, for $\pi^-$ with $p_T = 4.61$ GeV/$c$.
$\alpha(p_T) = 1.148\pm 0.010$ \cite{Antreasyan:1979cw}.}
\label{fig:pA}
\end{figure}

\begin{figure}[tbhp]
\includegraphics[width=1.0\linewidth]{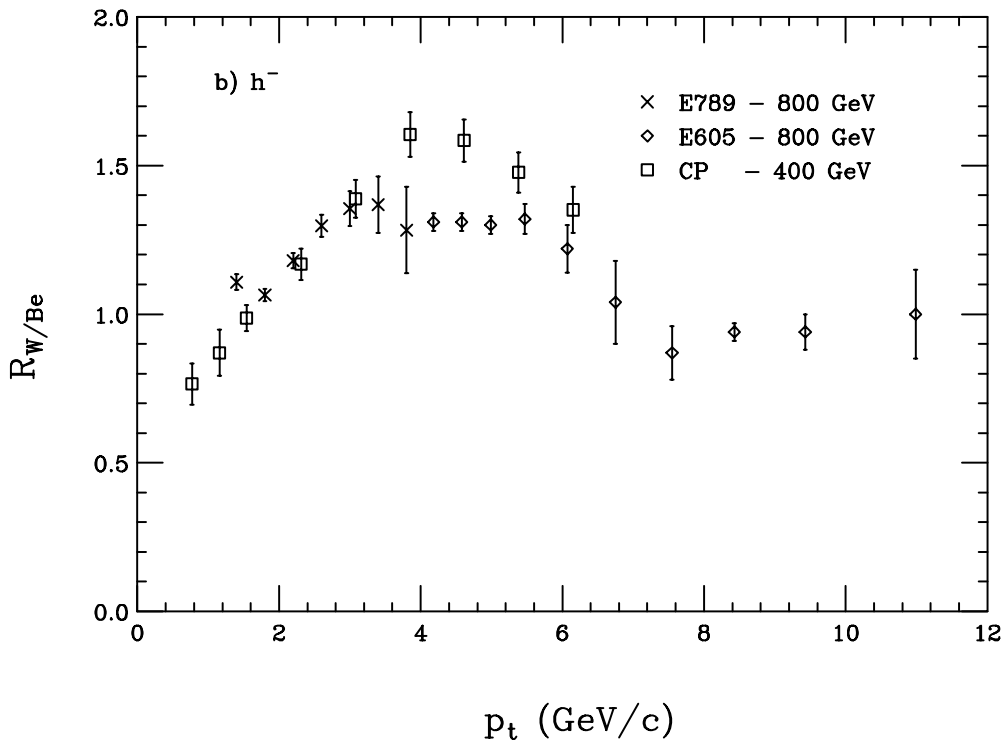}
\caption[]{Cronin effect at fixed target energies expressed as $R_{\rm W/Be}$, the ratio
of the point-like scaled cross sections in $p+W$ and $p+Be$ collisions vs.
$p_T$ \cite{Brown:1996fd}.}
\label{fig:CBrown}
\end{figure}

The situation is rather different in $p+A$ collisions: the cross section at a given
$p_T$ also scales as a power law, $A^{\alpha (p_T)}$ (Fig. \ref{fig:pA}),
but the power ${\alpha (p_T)}$ is greater than 1. 
This is called the ``Cronin Effect" \cite{Antreasyan:1979cw}. 
The enhancement (relative to $A^{1.00}$) is thought to be due to the multiple scattering
of the incident partons while passing through the nucleus $A$ before the
collision \cite{Krzywicki:1979gv,Lev:1983hh}, which smears the axis of the
hard scattering relative to the axis of the incident beam, 
leading to the characteristic ``Cronin Effect'' shape for $R_A(p_T)$ (Fig. \ref{fig:CBrown}). At low $p_T <1$ GeV/$c$, the cross-section is no longer point like, so the scattering is shadowed ($\propto A^{2/3}$), thus $R_A <1$. At larger $p_T >2$ GeV/$c$, as the hard-scattering, power-law $p_T$ spectrum begins to dominate,  the multiple scattering smears the spectrum to larger $p_T$ leading to an enhancement relative to binary-scaling which dissipates with increasing $p_T$  as the influence of the multiple scattering diminishes.

\begin{figure}[tbhp]
\includegraphics[width=1.0\linewidth]{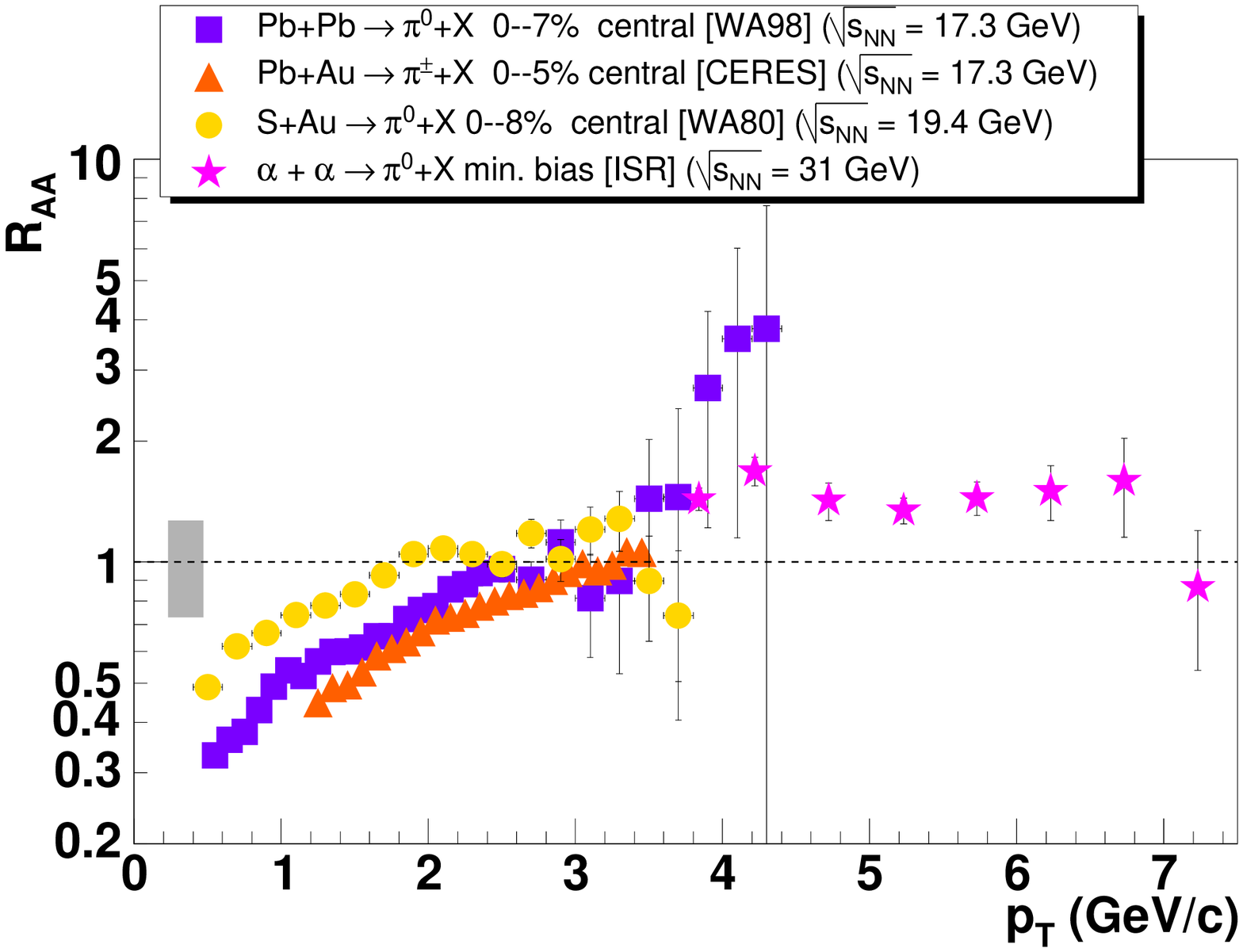}
\caption[]{Nuclear modification factors for $\pi^0$ production at the CERN-ISR in minimum-bias $\alpha+\alpha$ reactions at $\sqrt{s_{NN}}$ = 31  
 GeV \protect\cite{Angelis:1985fk} and for pion production at the CERN-SPS  
in central Pb+Pb \protect\cite{Aggarwal:2001gn}, Pb+Au \protect\cite{ceres:THESIS},
and S+Au \protect\cite{Albrecht:1998yc} reactions at  
$\sqrt{s_{NN}}\approx$ 20 GeV.  
 The $R_{AA}$ from SPS are obtained using  
the $p+p$ parametrization proposed in ref. \protect\cite{d'Enterria:2004ig}. The shaded  
band around $R_{AA}$ = 1 represents the overall fractional uncertainty of the SPS data (including in quadrature the 25\% uncertainty of the $p+p$ reference and  
the 10\% error of the Glauber calculation of $N_{coll}$). There is an additional overall  
uncertainty of $\pm$15\% for the CERES data not shown in the plot \protect\cite{ceres:THESIS}.}
\label{fig:AAlows}
\end{figure}

\begin{figure}[tbhp]
\includegraphics[width=1.0\linewidth]{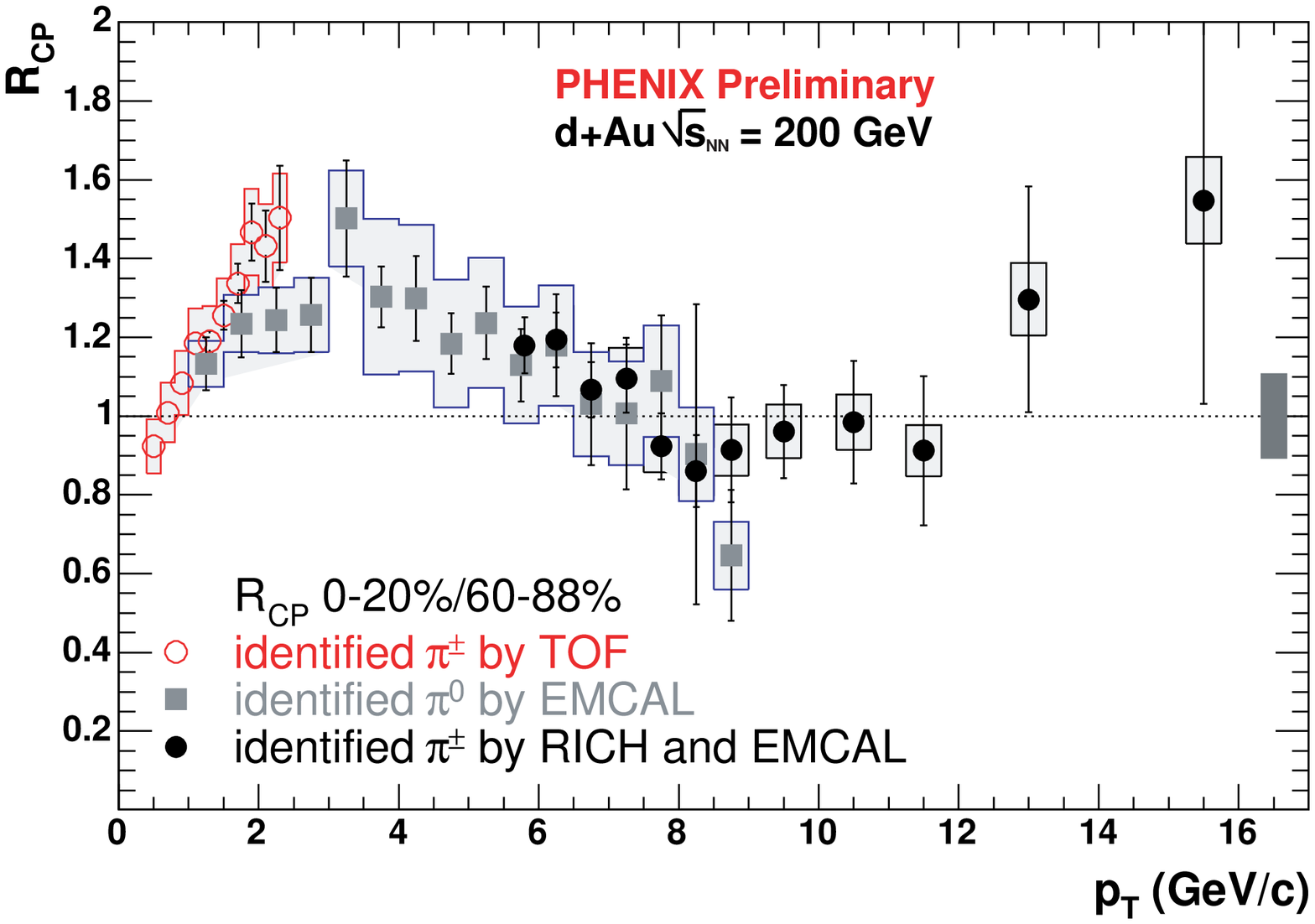}
\caption[]{Cronin effect in $R_{CP}$, the ratio of point-like scaled central to
peripheral collisions for pions in $d+$Au at $\sqrt{s_{NN}} = 200$ GeV\cite{Busching:2004fr}. Data points for low $p_T$ are $\pi^{\pm}$ identified by Time of Flight (TOF). Data at medium $p_T$ are for $\pi^0$ identified by reconstruction in the Electromagnetic Calorimeter (EMCAL). Highest $p_T$ data are for $\pi^{\pm}$ identified by a count in the Ring Imaging Cerenkov Counter (RICH) and a deposited energy/momentum and shower shape in the EMCAL inconsistent with those of a photon or electron. The shaded band on the right represents the overall fractional systematic uncertainty due to $N_{coll}$. }
\label{fig:dA}
\end{figure}

	Previous measurements of high-$p_T$ particle production in $A+A$ collisions at 
$\sqrt{s_{NN}}\leq 31$ GeV (Fig. \ref{fig:AAlows})
and in $p+A$ (or $d+A$)
collisions (Fig. \ref{fig:CBrown}) including measurements
at RHIC \cite{Adler:2003ii} at mid-rapidity (Fig. \ref{fig:dA})
all show binary scaling or a Cronin effect.

This establishes that the initial condition for hard scattering at RHIC at mid-rapidity is an incoherent
superposition of nucleon structure functions, including gluons, where multiple scattering
before the hard collision smears the $p_T$ spectrum of scattered particles to be somewhat
above the simple point-like binary ($N_{coll}$) scaling.

	An alternative view of the initial state of a nucleus at RHIC is provided by the color glass condensate (CGC), in which the gluon population at low 
$x$ is not an incoherent superposition of nucleon structure functions 
but is limited with increasing $A$ by non-linear gluon-gluon fusion 
resulting from the overlap of gluons from several nucleons in the plane 
of the nucleus transverse to the collision axis \cite{Kharzeev:2003wz}. 
A Cronin effect in $d+A$ collisions, as shown in Fig. \ref{fig:dA}, can 
be reproduced in the CGC with a suitable choice of initial state 
parameters, which must also reproduce quantitatively the observed 
binary scaling of the direct photon production and total charm 
production in Au+Au collisions to be shown below (Figs. \ref{fig:dirphoton}, 
\ref{fig:charm}). However, at this writing, no detailed quantitative 
description of the CGC initial state which satisfies these three 
conditions has been published.

\subsection{Binary Scaling in Au+Au Collisions at RHIC---Direct Photons and Charm Yield} 
\begin{figure}[tbhp]
\includegraphics[width=1.0\linewidth]{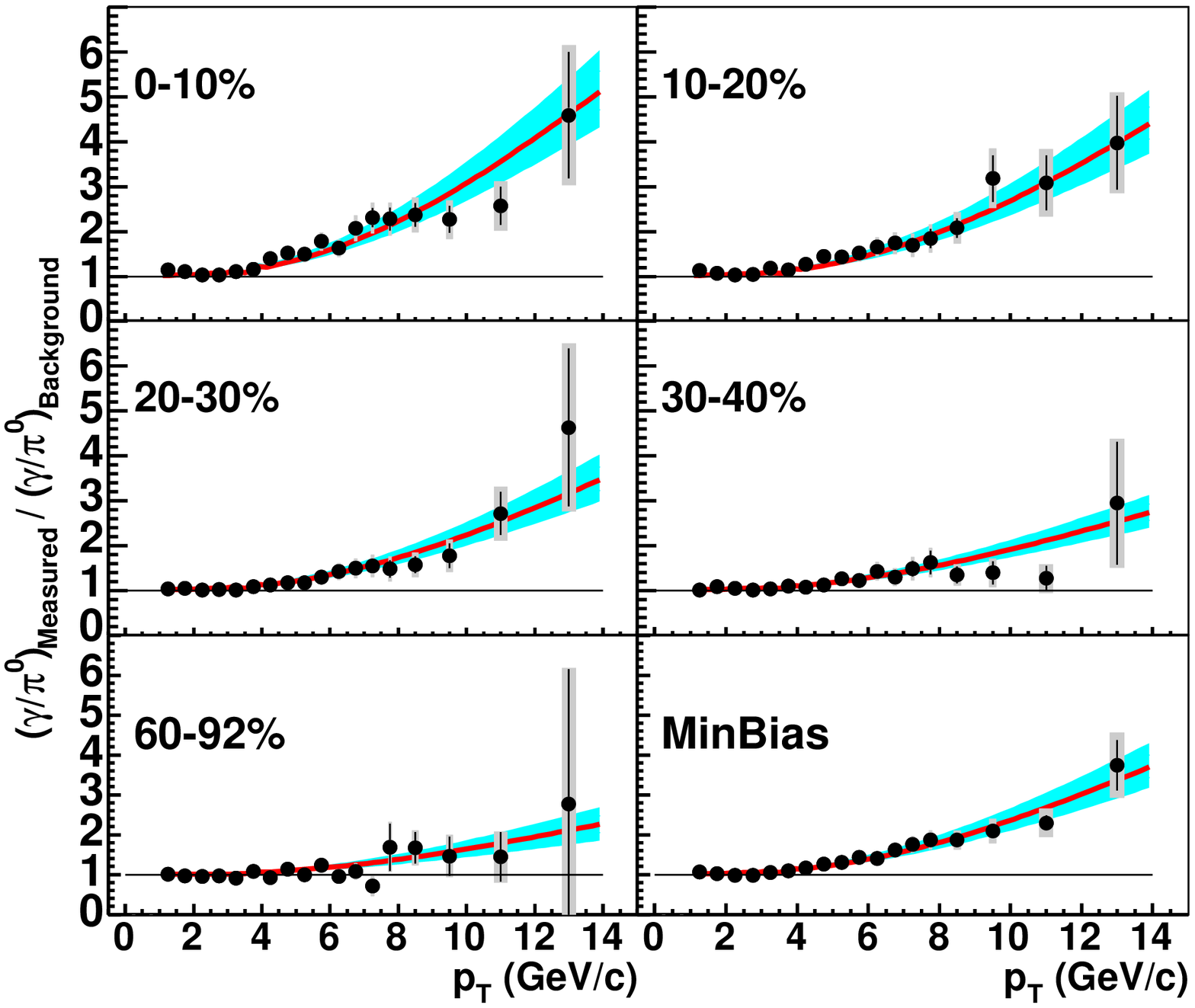}
\caption[]{PHENIX direct photon measurements relative to the background for 
for minimum bias
and for five centralities of Au+Au
collisions at $\sqrt{s_{_{NN}}}$ = 200 GeV (0-10\% is the most
central). Statistical and total errors are indicated separately on each
data point by the vertical bar and shaded region, respectively. The
curves represent a pQCD calculation of direct photons in $p+p$ collisions 
from Vogelsang \cite{Baer:1990ra,Aurenche:1983ws,Aurenche:1987fs,Gordon:1993qc} scaled to
Au+Au assuming pure point-like ($N_{\rm coll}$) scaling, with no suppression. The shaded region around the curves indicate the variation of the pQCD calculation for scale changes from $p_T/2$ to $2p_T$, plus the $\langle N_{coll}\rangle$ uncertainty \cite{Adler:2005ig}.}
\label{fig:dirphoton}
\end{figure}

The production of hard photons in Au+Au collisions at RHIC via the
constituent reactions
(e.g. $g+q\rightarrow \gamma +q$)
is a very important test of QCD and the
initial state, because the photons only interact electromagnetically, hence hardly at
all, with any final-state medium  produced. The direct-photon cross section and
centrality dependence should then reflect only the properties of the initial state,
notably the product of the gluon and quark structure functions of the Au nuclei.

The first measurement of direct photon production in Au+Au collisions at RHIC has
been reported by the PHENIX collaboration (Fig. \ref{fig:dirphoton}) \cite{Adler:2005ig}.
The data exhibit pure point-like
($N_{coll}$) scaling as a function of centrality relative to a  pQCD calculation for
$p+p$ collisions. 
The statistical and systematic errors still leave some
room for a small Cronin effect and/or some thermal photon production. The observation
of direct photon production establishes the importance of gluon degrees of freedom at
RHIC.

\begin{figure}[tbhp]
\includegraphics[width=1.0\linewidth]{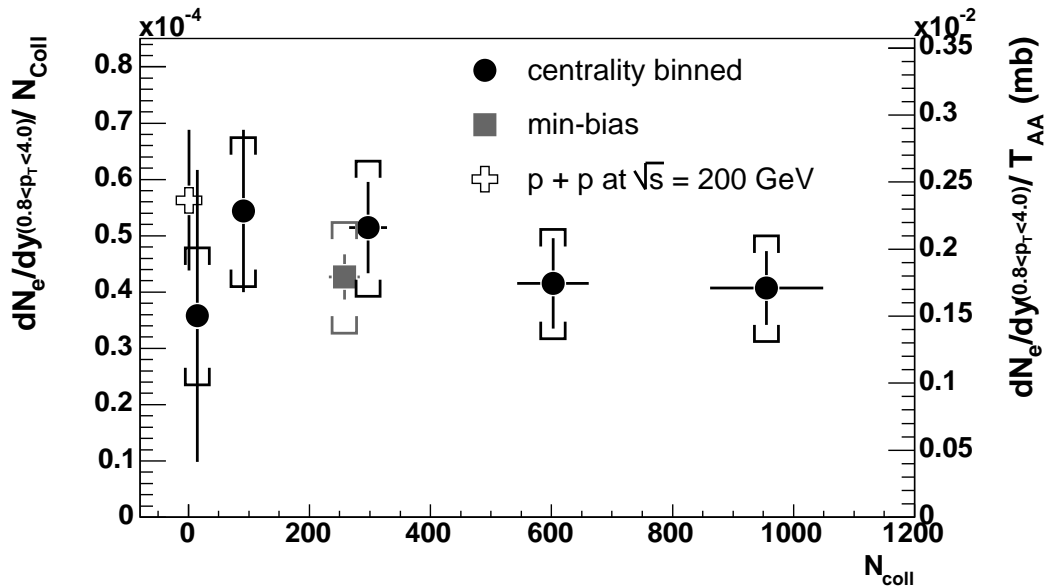}
\caption[]{Non-photonic electron yield ($0.8<p_T<4.0$ GeV/$c$),
dominated by semi-leptonic charm decays,
measured in Au+Au collisions at $\sqrt{s_{NN}} = 200$ GeV
scaled by $N_{coll}$ as a function of $N_{coll}$. The right-hand scale
shows the corresponding electron cross section per $NN$ collision in
the above $p_T$ range. The yield in $p+p$ collision at 200 GeV is also shown \cite{Adler:2004ta}.}
\label{fig:charm}
\end{figure}

PHENIX measured the single-electron yield from nonphotonic sources in
Au+Au collision at 
130 GeV \cite{Adcox:2002cg} and 200
 GeV \cite{Adler:2004ta}.  Since semi-leptonic decay of charm is the
dominant source of the non-photonic electrons at low $p_T$ ($p_T \leq$
3 GeV/$c$), the total yield of charm can be determined from the
integrated yield of non-photonic electrons in the low-$p_T$
region. Figure \ref{fig:charm} shows the yield of non-photonic
electrons ($0.8 < p_T < 4.0$ GeV/$c$) per $NN$ collision in Au+Au
reactions at $\sqrt{s_{NN}}$ = 200 GeV as a function of
$N_{coll}$ \cite{Adler:2004ta}.  The $N_{coll}$ dependence of the
yield is fit to $N_{coll}^\alpha$, where $\alpha$ = 1 is the
expectation for binary scaling. We find 
$\alpha = 0.938 \pm 0.075$(stat.)$\pm 0.018$(sys.), showing that the total
yield of charm-decay electrons is consistent with binary scaling. It should
be noted that medium effects, such as energy loss of charm in the dense hot
medium, can only influence the momentum distribution of charm, and have little
effect on the total yield of charm. Initial-state effects, such as
shadowing, and other effects, such as thermal production of charm, are
believed to be very small for charm production at RHIC energy.
Therefore, the observation of binary scaling of the total charm yield
in Au+Au collisions at RHIC may also be considered as an experimental
verification of the binary scaling of a point-like pQCD process.

\subsection{Conclusions}
    In this section evidence has been presented to show that the initial
condition for hard-scattering at RHIC at mid-rapidity is an incoherent
superposition of nucleon structure functions, including gluons, where
multiple scattering before the hard collision can smear the  $p_T$
spectrum of scattered particles to be somewhat above the simple
point-like binary  ($N_{coll}$) scaling. This was demonstrated using
the reactions: pion production in $d$+Au collisions,
where there is no final-state medium, and direct photon production in
Au+Au collisions, where the outgoing photons
interact electromagnetically, hence hardly at all, with any final-state
medium produced. The total charm yield in Au+Au, a reaction dominated
by the subprocess  $g+g\rightarrow c +\bar{c}$, and which is not
sensitive to final-state medium effects for the total yield of $c
+\bar{c}$ pairs, also exhibits binary scaling.
The latter two measurements provide experimental evidence for the
binary scaling of point-like pQCD processes in Au+Au collisions.

        The color glass condensate (CGC) provides an alternative view of the
initial state of a nucleus at RHIC in which coherence of gluons due to
non-linear gluon-gluon fusion can produce a Cronin-like effect,
depending on the initial conditions and the kinematic range covered.
However, at the present writing, there is no CGC description of the
initial state nuclear structure function which reproduces the observed
Cronin effect for pions in $d$+Au collisions and the observed binary scaling
for both direct photon production and the total charm yield in Au+Au
collisions.

 
\section{NUMBER DENSITY AND HIGH $p_T$ SUPPRESSION}
\label{Sec:denseN}

To study the initial properties of the matter created in heavy ion
collisions we need a probe that is already present at earliest times
and that is directly sensitive to the properties of the
medium. Partons resulting from hard scatterings during the initial
crossing of the two nuclei in \mbox{A+A} collisions provide such a
probe. Energetic partons propagating through a dense medium are
predicted to lose energy
\cite{Bjorken:1982tu,Gyulassy:1990ye,Wang:1992xy,Wang:1995fx,Baier:1995bd,Baier:1997kr,Baier:1998yf,Gyulassy:2000fs,Gyulassy:2000er}
thus producing a suppression in the yield of high-\pperp\ hadrons
produced from the fragmentation of these partons. Initial measurements
from RHIC Run 1 \cite{Adcox:2001jp,Adcox:2002pe,Adler:2002xw} and Run 2
\cite{Adler:2003qi,Adler:2003au,Adams:2003kv,Back:2003qr}
demonstrated such a suppression, and the results of \dAu\ measurements
\cite{Adler:2003ii,Adams:2003im,Back:2003ns,Arsene:2003yk} showed that
the suppression was not due to initial-state effects. Further
measurements have indicated a modification of di-jet angular
correlations \cite{Adams:2003im} that has also been attributed to
in-medium parton energy loss \cite{Wang:2003mm,Qiu:2003pm}.  

While the energy loss of hard-scattered partons was originally
proposed as a signature of the quark-gluon plasma and deconfinement,
it has been argued recently that the energy loss is sensitive only to
the density of unscreened color charges and not directly to
deconfinement \cite{Baier:1995bd,Baier:1997kr,Baier:1998yf,Gyulassy:2000fs,Gyulassy:2000er,Wang:2001cs,Wang:2002ri}.
Ideally, a measurement of initial parton
densities together with constraints on initial energy density might
allow an estimate of the temperature of the medium. As will be seen
below, the current high-\pperp\ measurements and theoretical tools for
interpreting the experimental data are not yet
sufficient to take such a step. Instead, the energy loss results are
currently being used to provide estimates of the initial energy
density. The remainder of this section summarizes PHENIX experimental
data related to high-\pperp\ suppression, discusses the current state
of theoretical understanding of the energy loss process and concludes
with a statement of estimates for initial parton number and energy
densities that currently can be made.

\subsection{Single particle spectra, \RAA}
\label{SubSec:denseN:hptsingle}
\begin{figure}[tbhp]
\includegraphics[width=1.0\linewidth,bb = 0 0 470 530]{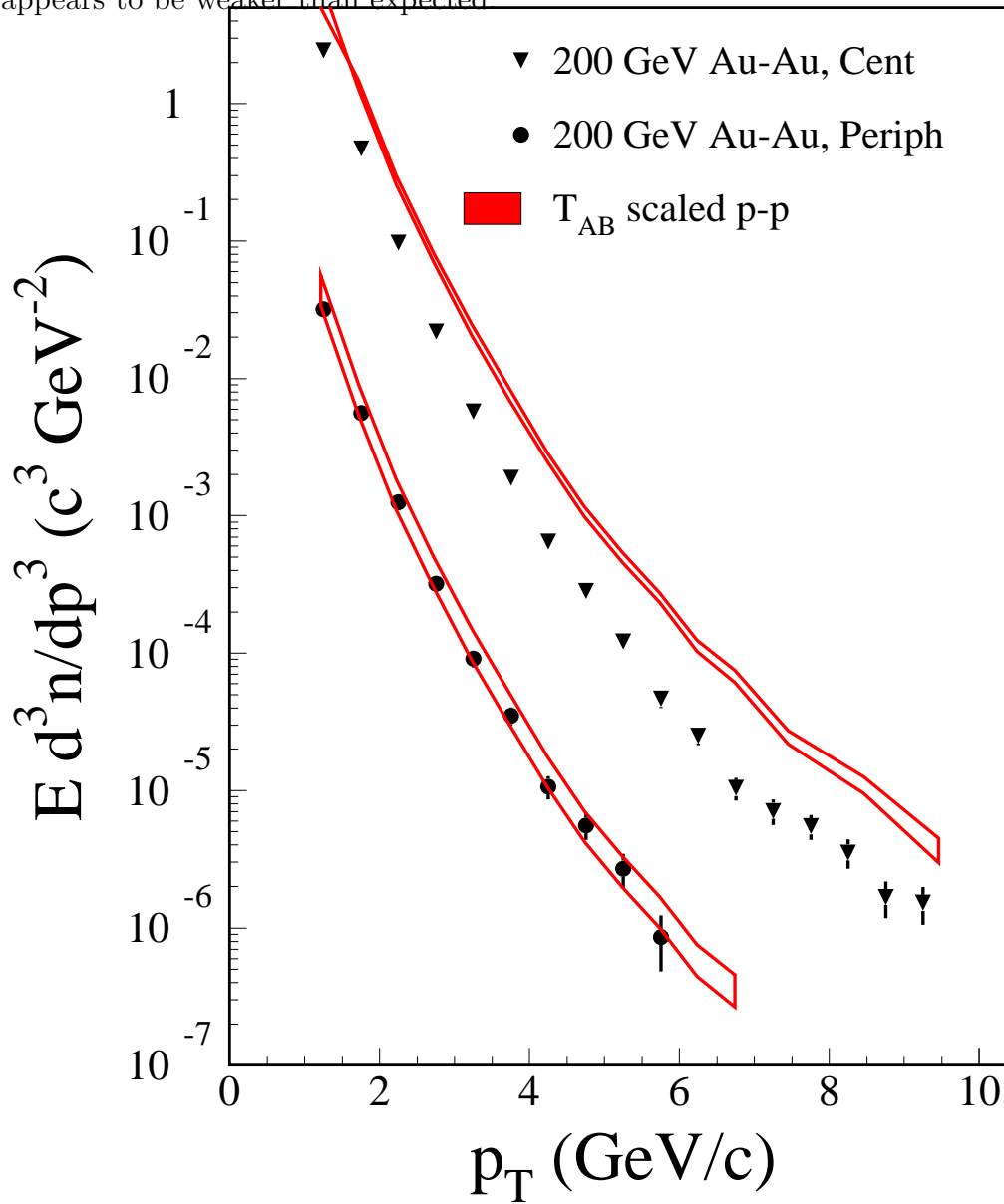}
\caption{\piz\ \pperp\ spectra in 200 GeV \AuAu\ collisions 
  \cite{Adler:2003qi}
  compared to a \TAB\ scaling of the 200 GeV \pp\ \piz\ differential
  cross section \cite{Adler:2003pb}. The central data were obtained with a 0--10\% centrality
  cut while the peripheral data were obtained with an 80--92\% cut.
} 
\label{fig:hpt:spectra}
\end{figure}

As described in Section \ref{Sec:scale}, in the absence of
modifications due to initial-state or final-state effects, the rate
for the production of particles through hard-scattering processes in
nucleus-nucleus collisions is expected to be given by the equivalent
\pp\ hard-scattering cross section multiplied by \TAB.
Figure \ref{fig:hpt:spectra} shows PHENIX \piz\ spectra,
$d^2N/d\pperp dy$, measured in 200 GeV \cite{Adler:2003qi} peripheral
(80--92\%) and central (0--10\%) \AuAu\ collisions compared to measured
\cite{Adler:2003pb} \pp\ cross sections multiplied by the peripheral and 
central \TAB\ values estimated using the procedure described in
Section \ref{Sec:scale}. The error bands on the \pp\ data points reflect
both the systematic errors on the \pp\ cross sections and the
uncertainties in the \TAB\ values. As the figure clearly demonstrates,
the central \AuAu\ \piz\ yields are strongly suppressed relative to
the ``expected'' yields over the entire measured \pperp\ range. In
contrast, the peripheral yields compared to the \TAB-scaled \pp\
cross sections show little or no suppression. The results 
incontrovertibly demonstrate that there is
a strong and centrality-dependent suppression of the production of
high-\pperp\ pions relative to pQCD-motivated expectations. This is
quite different from measurements of $R_{AA}$ in Pb+Pb collisions
at $\sqrt{s_{NN}} = 17.3$ GeV where in semi-peripheral Pb+Pb collisions
there is a nuclear enhancement increasing with $p_T$ similar to the
well-known Cronin effect, while in central collisions the Cronin
enhancement appears to be weaker than expected.

\begin{figure}[tbhp]
\includegraphics[width=1.0\linewidth,bb = 50 0 567 567]{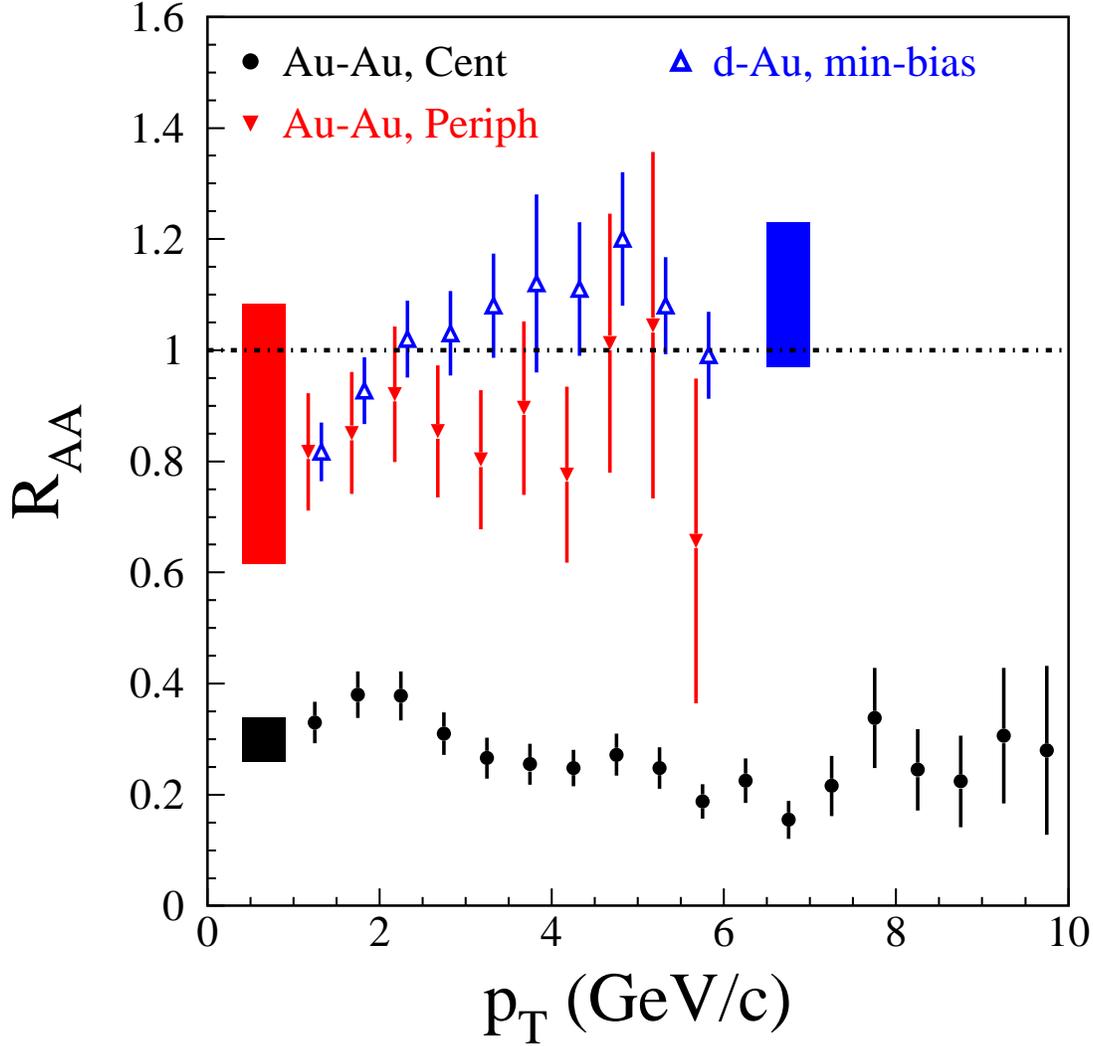}
\caption{\piz\ \RAApt\ for central (0--10 \%) and
peripheral (80--92 \%) \AuAu\ collisions 
\cite{Adler:2003qi} and minimum-bias \dAu\ collisions
\cite{Adler:2003ii}.  The shaded boxes on the left show the 
systematic errors for the \AuAu\
\RAA\ values resulting from overall normalization of spectra and
uncertainties in \TAB. The shaded box on the right shows the same
systematic error for the \dAu\ points.}
\label{fig:hpt:RAApiz}
\end{figure}

To better demonstrate quantitatively the suppression in central
collisions indicated in Fig. \ref{fig:hpt:spectra}, we show in
Fig. \ref{fig:hpt:RAApiz} 
\RAApt\ for mid-rapidity \piz's in central and peripheral
200 GeV \AuAu\ collisions. We also show the 
values obtained from minimum-bias 200 GeV \dAu\ collisions
\cite{Adler:2003ii} which provide a stringent test of the possible
contribution of initial-state nuclear effects to the observed
suppression in \AuAu\ collisions.  The error bands on the data
indicate combined statistical and point-to-point systematic errors and
the bars shown next to the different data sets indicate common
systematic errors due to uncertainties in the \pp\ cross section
normalization and
\TAB. 

Figure \ref{fig:hpt:RAApiz} shows that the central \AuAu\ \piz\
suppression changes only slightly over the measured \pperp\ range and
reaches an approximately \pperp-independent factor of 5 ($\RAA \approx
0.2$) for $\pperp > 4-5$ GeV/$c$. The peripheral \AuAu\ \RAA\ values are
consistent with one after taking into account systematic errors but we
cannot rule out a slight suppression suggested by the peripheral \RAA\
values. In all of the data sets \RAA\ decreases with decreasing \pperp\ 
for $\pperp < 2$ GeV/$c$. This decrease, known since the original
measurements of the $A$ dependence of particle production in $pA$
collisions is due to contributions of soft hadronic
processes at low \pperp\ that are expected to increase more slowly
than proportional to \TAB.
The \dAu\ \RdA\ values are also consistent with one within
systematic uncertainties, but in
contrast to the \AuAu\ results, the data suggest a slight
enhancement. The \dAu\ \RdA\ values above 2 GeV/$c$
exceed one for nearly the entire experimentally covered
\pperp\ range.
As shown previously
in Fig. \ref{fig:dA}, only for $\pperp \gsim
6$ GeV/$c$ does the \dAu\ pion yield return to the \TAB-scaling
expectation.  
Such a small enhancement is
consistent with expectations based on prior measurements of the Cronin
effect \cite{Cronin:1975zm,Antreasyan:1977va},
and it is also quantitatively
consistent with calculations incorporating the initial-state multiple
scattering that is thought to produce the Cronin effect
\cite{Gyulassy:1998nc,Wang:1998ww,Papp:1999ra,Vitev:2002pf,Vitev:2003xu,Wang:2003vy,Levai:2003at}. Therefore
the Cronin effect at RHIC cannot mask a strong initial-state
suppression of the parton distributions in the Au nucleus
\cite{Kharzeev:2002pc}. 

\begin{figure}[tbhp]
\centerline{\includegraphics[width=0.8\linewidth]{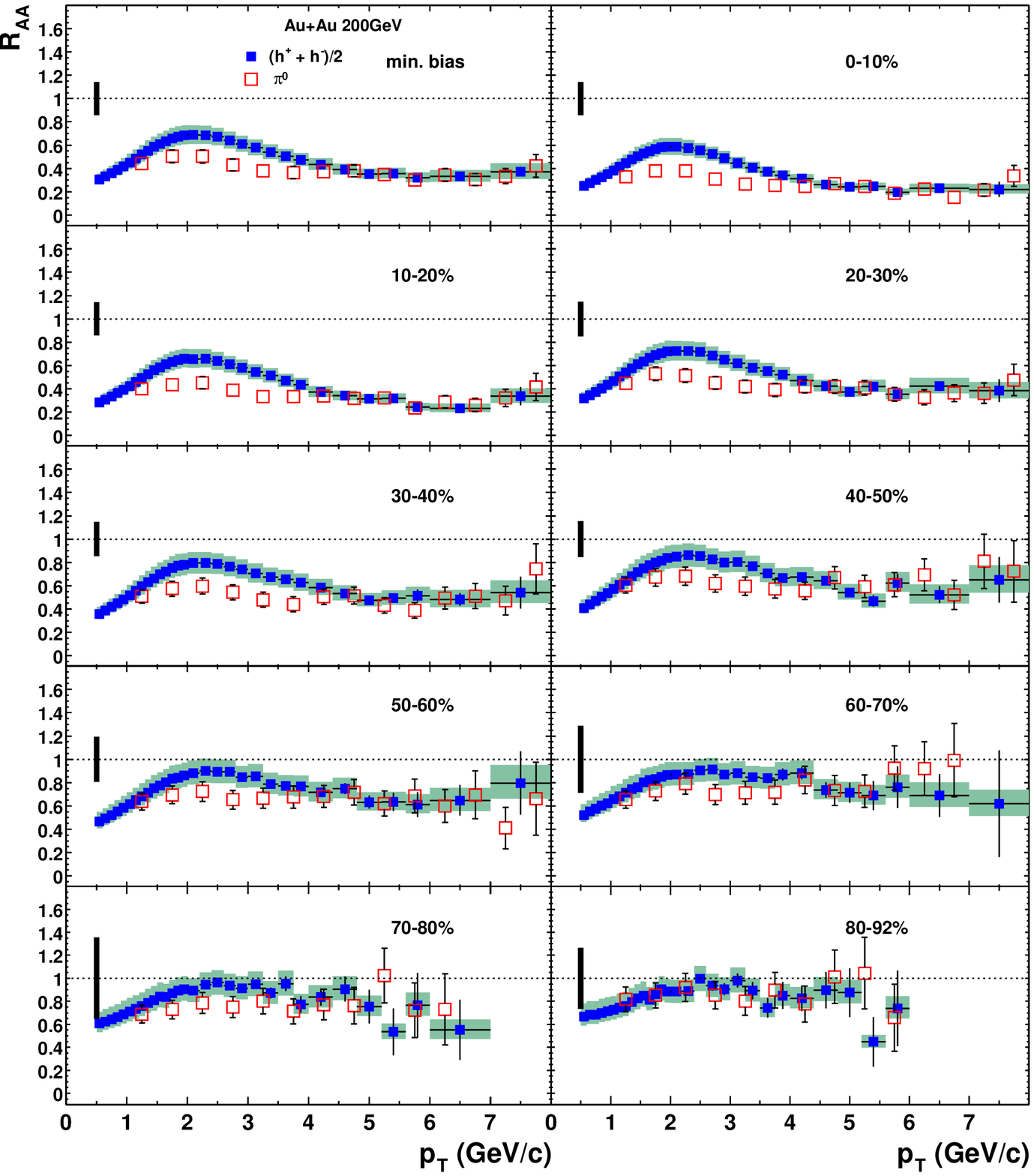}}
\centerline{\includegraphics[width=0.8\linewidth]{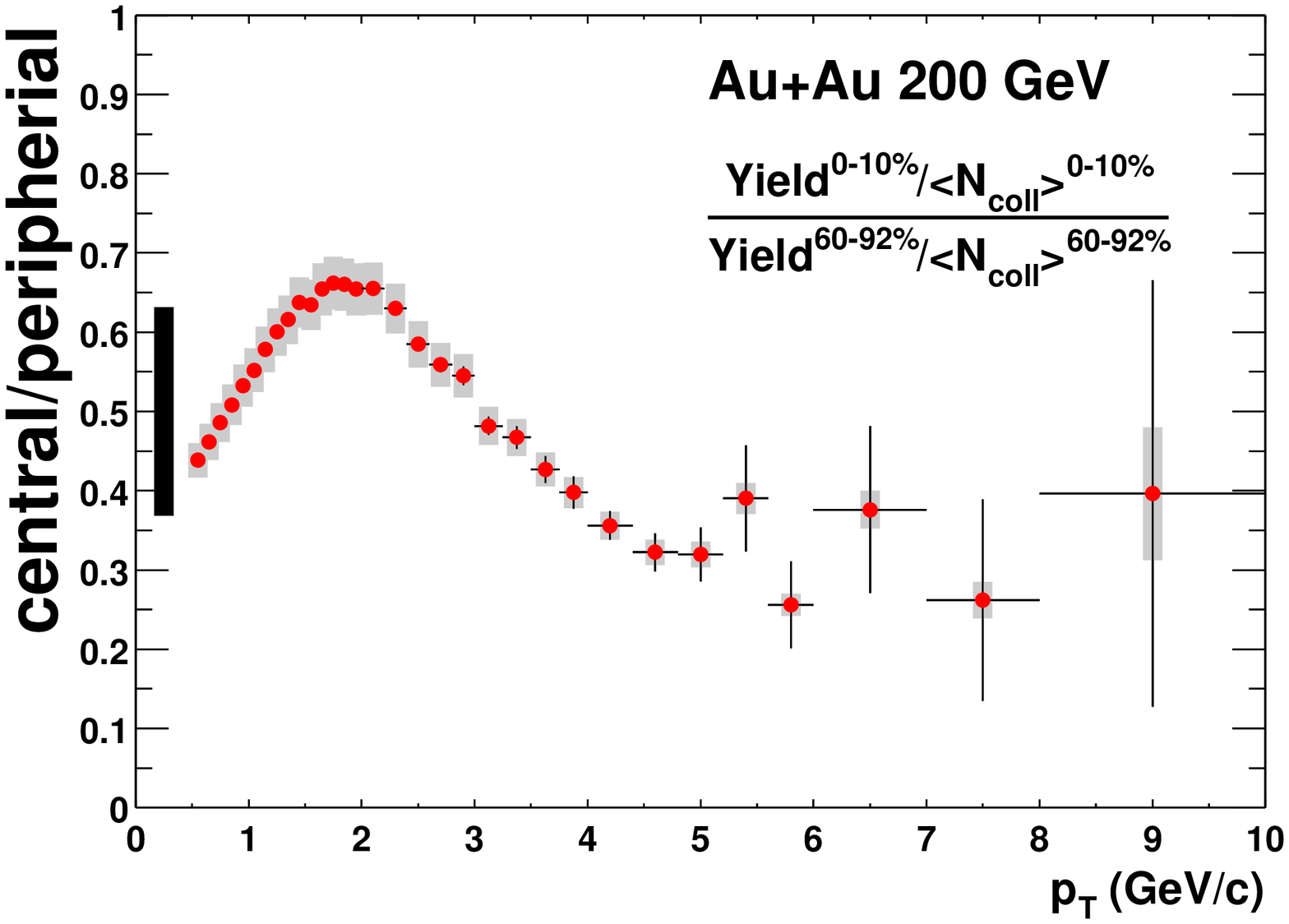}}
\caption{Centrality and \pperp\ dependence of nuclear modification
factors in 200 GeV \AuAu\ collisions\cite{Adler:2003au}. Top panel: \piz\
and charged particle \RAApt\ for ten 
centrality bins. Bottom panel: charged particle
\Rcp\ vs. \pperp.} 
\label{fig:hpt:RAAcent}
\end{figure}
\begin{figure}[tbhp]
\includegraphics[width=1.0\linewidth]{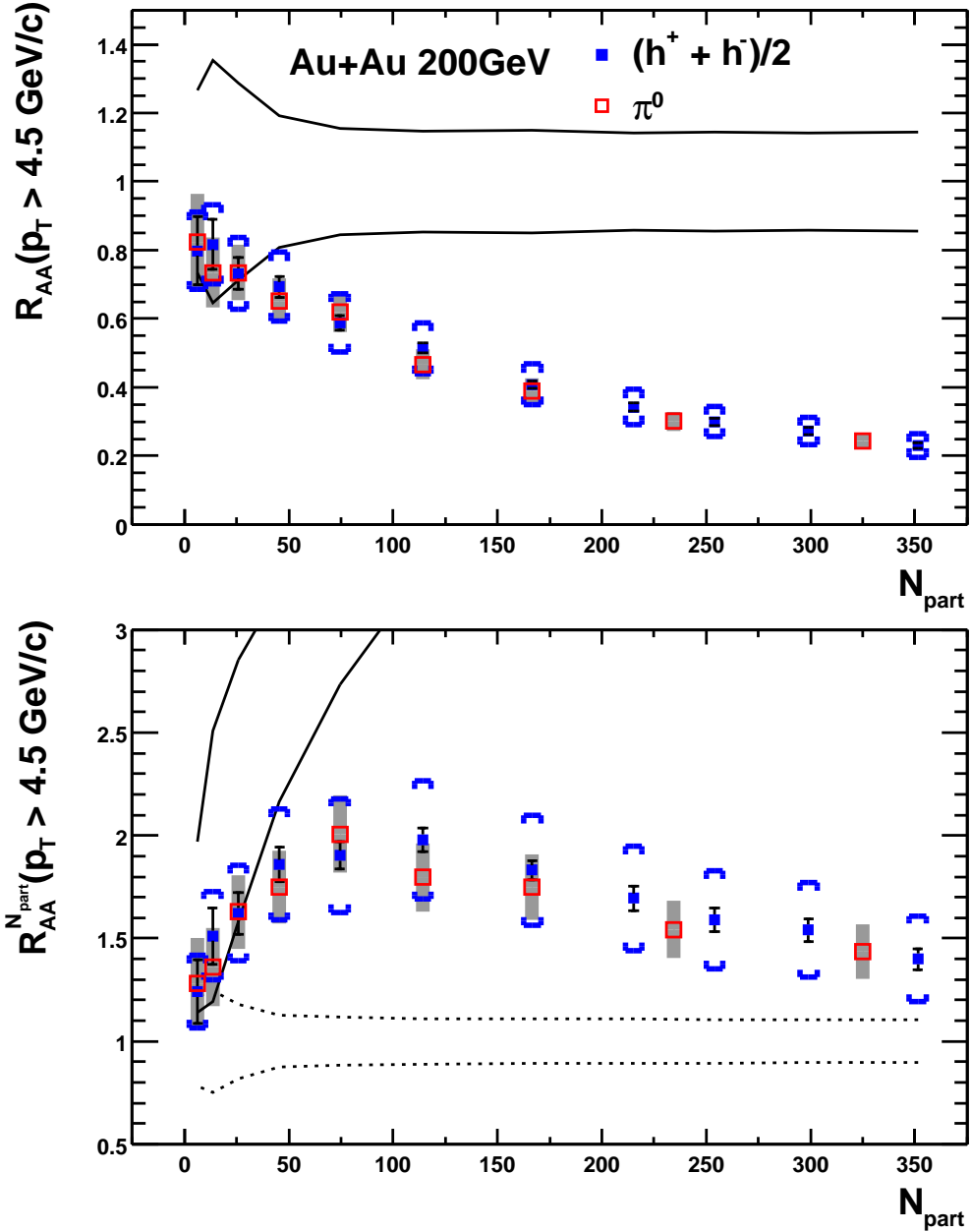}
\caption{Top panel: \RAA\ vs. \Npart\ obtained from \pperp-integrated
  ($\pperp > 4.5$ GeV/$c$) \AuAu\ \piz\ and charged-hadron
  spectra. The
  band indicates the systematic error bands on a hypothetical \TAB\
  scaling of the \pp\ \pperp-integrated cross section. Bottom panel:
  \piz\ and charged hadron yield per participant vs. \Npart\ divided by
  the same quantity in \pp\ collisions (\RAAnpart). The solid band shows the same 
  band as in the top panel expressed in terms of yield per participant
  pair while the dashed band indicates the systematic error bands around
  a hypothetical \Npart\ scaling.
  Both plots are from \cite{Adler:2003au}.
}
\label{fig:hpt:avgRAAnpart}
\label{fig:hpt:hptYieldpart}
\end{figure}

To better demonstrate the systematic behavior of the high-\pperp\
suppression we show in Fig. \ref{fig:hpt:RAAcent} \piz\ \cite{Adler:2003qi} and
unidentified charged particle \RAA\ values \cite{Adler:2003au} as a function of
\pperp\ for various centrality bins. While for moderate \pperp\ values
($2 < \pperp < 5$ GeV/$c$) total charged particle production is suppressed
less than pion production, the charged particle and \piz\ \RAA\ values
become equal, within errors, at high \pperp. This evolution in the
charged particle suppression is related to contributions from the
(anti)protons that will be discussed further below. Despite
the differences resulting from the protons, the charged particles and
\piz's exhibit very similar trends in the suppression vs. \pperp\ and
vs. centrality. The suppression increases smoothly with centrality
though the change in \RAA\ values at high \pperp\ is most rapid in the
middle of the centrality range. Figure \ref{fig:hpt:RAAcent} also shows
that the suppression is approximately constant as a function of
\pperp\ for $\pperp > 4.5 $ GeV/$c$ in all centrality bins. We take
advantage of this feature of the data to better illustrate the
centrality dependence of the suppression by integrating both the
\AuAu\ spectra and the reference \pp\ cross sections over $\pperp >
4.5$ GeV/$c$ and using these integrated quantities to determine an
average suppression factor, \RAAAvgnpart\ for $\pperp > 4.5$ GeV/$c$.  We
plot the charged particle and \piz\ \RAAAvgnpart\ values vs. \Npart\ in
Fig. \ref{fig:hpt:avgRAAnpart}(top). This figure suggests that the
suppression evolves smoothly with \Npart, showing no abrupt onset of
suppression. The charged particles and \piz's exhibit similar
evolution of suppression with \Npart. In the most central collisions
we obtain \RAAAvgnpart\ values of $0.24 \pm 0.04(total)$ and $0.23 \pm
0.05(total)$  for charged particles and \piz's respectively. In
peripheral collisions, \RAAAvgnpart\ approaches one, but the systematic
errors on the most peripheral \TAB\ values are sufficiently large that
we cannot rule out $\sim 20\%$ deviations of the peripheral \AuAu\
hard-scattering yields from the \TAB-scaled \pp\ cross sections.

An alternative method for evaluating the evolution of the high-\pperp\
suppression with centrality is provided in
Fig. \ref{fig:hpt:hptYieldpart}(bottom) which presents the charged and
\piz\ yields per participant integrated over $\pperp > 4.5$ GeV/$c$ as a 
function of \Npart\ \cite{Adler:2003au} divided by the same quantity
in \pp\ collisions.  Also shown in the figure are curves demonstrating
the \Npart\ dependence that would result if the \piz\ and charged
particle yields exactly \TAB\ scaled and what an \Npart\ scaling  from
$p+p$ collisions would imply. As Fig. \ref{fig:hpt:hptYieldpart}
demonstrates, the high-\pperp\ yields of both charged hadrons and
\piz's {\em per participant} increase proportional to \TAB\ for small
\Npart\ but level off and then decrease with increasing \Npart\ in
more central collisions. The PHENIX measurements do not naturally
support an approximate \Npart\ scaling of high-\pperp\ particle
production suggested in an analysis of PHOBOS data. The PHENIX \RAAnpart\
values decrease from mid-peripheral ($\Npart \approx 75$) to central
collisions by an amount larger than the systematic errors in the
measurement. For more peripheral collisions, \RAAnpart\ increases with
\Npart\ consistent with the modest suppression of high-\pperp\ production
shown for peripheral collisions in the top panel of
Fig. \ref{fig:hpt:avgRAAnpart}. The initial rise and subsequent decrease of
\RAAnpart\ with increasing \Npart\ suggests that the high-\pperp\
hadron yield in \AuAu\ collisions has no simple dependence on \Npart.
The observation that the
high-\pperp\ yields initially increase proportional to \TAB\
demonstrates that in the most peripheral \AuAu\ collisions the
hard-scattering yields are consistent with point-like scaling. 
However, the deviation from \TAB\ scaling sets in rapidly, becoming
significant by $\Npart = 50$. By $\Npart = 100$ the high-\pperp\
suppression is so strong that high-\pperp\ yields grow even more
slowly than proportional to \Npart.

 \subsection{$x_T$ scaling in Au+Au collisions at RHIC} 
\label{SubSec:denseN:xtscale}
\begin{figure}[tbhp]
\includegraphics[width=1.0\linewidth]{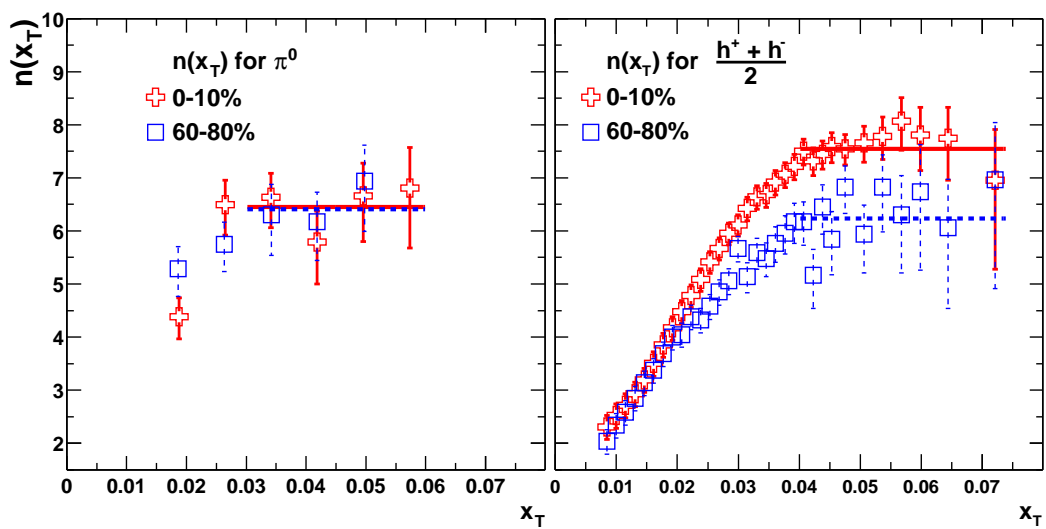}
\caption[]{Power-law exponent $n(x_T)$ for $\pi^0$ and $h$ spectra in central
and peripheral Au+Au collisions at $\sqrt{s_{NN}} = 130$ and 200 GeV \cite{Adler:2003au}. } 
\label{fig:nxTAA}
\end{figure}
If the production of high-$p_T$ particles in Au+Au collisions is the
result of hard scattering according to pQCD, then $x_T$ scaling should
work just as well in Au+Au collisions as in $p+p$ collisions and should
yield the same value of the exponent $n(x_T,\sqrt{s})$.  The only assumption
required is that the structure and fragmentation functions in Au+Au
collisions should scale, in which case Eq. \ref{eq:xTscaling} still
applies, albeit with a $G(x_T)$ appropriate for Au+Au. In
Fig. \ref{fig:nxTAA}, $n(x_T,\sqrt{s_{NN}})$ in Au+Au is derived from
Eq. \ref{eq:xTscaling}, for peripheral and central collisions, by
taking the ratio of $E d^3\sigma/dp^3$ at a given $x_T$ for
$\sqrt{s_{NN}} = 130$ and 200 GeV, in each case. The $\pi^0$'s exhibit
$x_T$ scaling, with the same value of $n = 6.3$ as in $p+p$ collisions,
for both Au+Au peripheral and central collisions, while the
non-identified charged hadrons $x_T$-scale with $n = 6.3$ for peripheral
collisions only. Notably, the $h^{\pm}$ in Au+Au central collisions
exhibit a significantly larger value of $n$, indicating different
physics, which will be discussed below.  The $x_T$ scaling establishes
that high-$p_T$ $\pi^0$ production in peripheral and central Au+Au
collisions and $h^{\pm}$ production in peripheral Au+Au collisions
follow pQCD as in $p+p$ collisions, with parton distributions and 
fragmentation functions that scale with $x_T$, at least within
the experimental sensitivity of the data.

\subsection{Two-hadron azimuthal-angle correlations}
\label{SubSec:denseNa:azimcorr}
\begin{figure}[tbhp]
\includegraphics[width=1.0\linewidth]{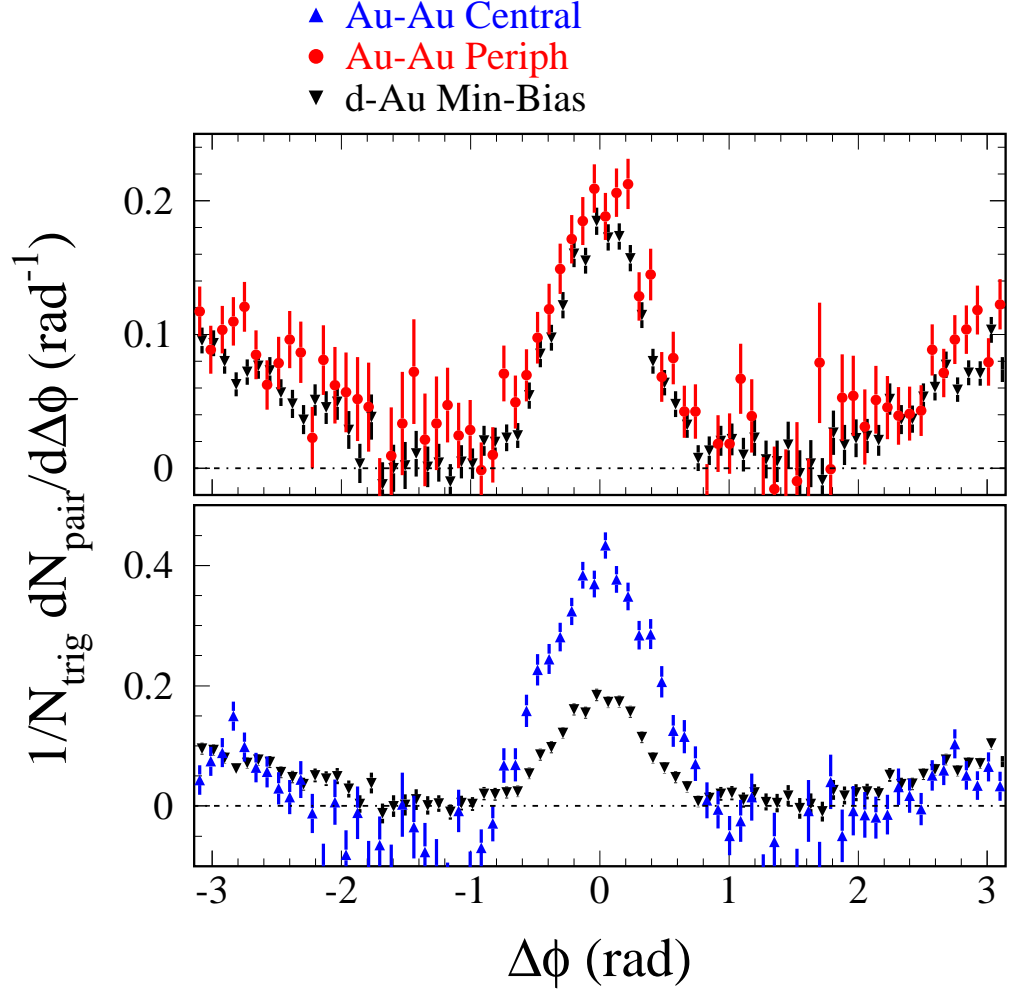}
\caption{Differential yields per \dphi\ and per trigger particle of
pairs of charged hadrons in \dAu, peripheral \AuAu\ and
central \AuAu\ collisions at $\sqrt{s_{NN}}$ = 200 GeV\cite{Rak:2004gk}.
The pairs were selected with the higher-momentum
``trigger'' particle in the range $2.5 < \pperp < 4.0$ GeV/$c$ and the
lower-momentum 
``associated'' particle in the range $ 1.0 < \pperp <
2.5$ GeV/$c$. A constant background has been subtracted for all three
distributions.} 
\label{fig:hpt:correlation}
\end{figure}

We argued in Sec. \ref{Sec:scale} that the production of hadrons at
high-\pperp\ results predominantly from hard scattering followed by
fragmentation of the outgoing parton(s). While this result is
well established in $p(\bar{p})+p$ collisions, it might not
be true in \AuAu\ collisions when the yield of high-\pperp\
particles is modified so dramatically compared to expectations. Since
a hard-scattered parton fragments into multiple particles within a
restricted angular region (i.e. a jet) a reasonable way to check the
assumption that high-\pperp\ hadron production in \AuAu\ collisions is
due to hard scattering is to directly observe the angular correlations
between hadrons in the jets. None of the experiments at RHIC are
currently capable of reconstructing jets in the presence of the large
soft background of a \AuAu\ collision. However, both STAR
\cite{Adler:2002ct,Adler:2002tq} and PHENIX \cite{Rak:2004gk,Chiu:2002ma}
have directly observed the presence of jets by studying two-hadron
azimuthal-angle correlations. Figure \ref{fig:hpt:correlation} shows
preliminary distributions \cite{Rak:2004gk} of the relative azimuthal angle (\dphi)
between pairs of charged particles detected within the PHENIX
acceptance in \dAu\ collisions and peripheral (60--90\%) and central
(0--10\%) \AuAu\ collisions after the subtraction of combinatoric background. The pairs of particles are chosen such
that one particle lies within a ``trigger'' \pperp\ range ($ 2.5 <
\pperp_{\rm trig} < 4$ GeV/$c$) while the other ``associated'' particle
falls within a lower \pperp\ window $ 1.0 < \pperp < 2.5$ GeV/$c$. The
distributions show the differential
yield per \dphi\ of associated particles per detected trigger particle
within the given \pperp\ ranges and within the \psy\ acceptance of the
PHENIX central arms ($-0.35 < \psy < 0.35$).  

The peaks observed at
$\dphi = 0$ (near side) reflect the correlation between hadrons produced
within the same jet while the broader peaks observed at $\dphi = \pi$
(away side) reflect the correlations between hadrons produced in one jet and
hadrons produced in the ``balance'' jet.  In the \AuAu\ cases, a
$\cos{2\dphi}$ modulation underlies the jet angular correlations due
to the elliptic flow of particles in the combinatoric background and
possibly also in part due to azimuthal anisotropies in the jets
themselves (see below).  Nonetheless, the \cosmod\ contribution has
little effect on the narrow same-jet (near-side) peak in the \dphi\ distribution.

\begin{figure}[tbhp]
\includegraphics[width=1.0\linewidth]{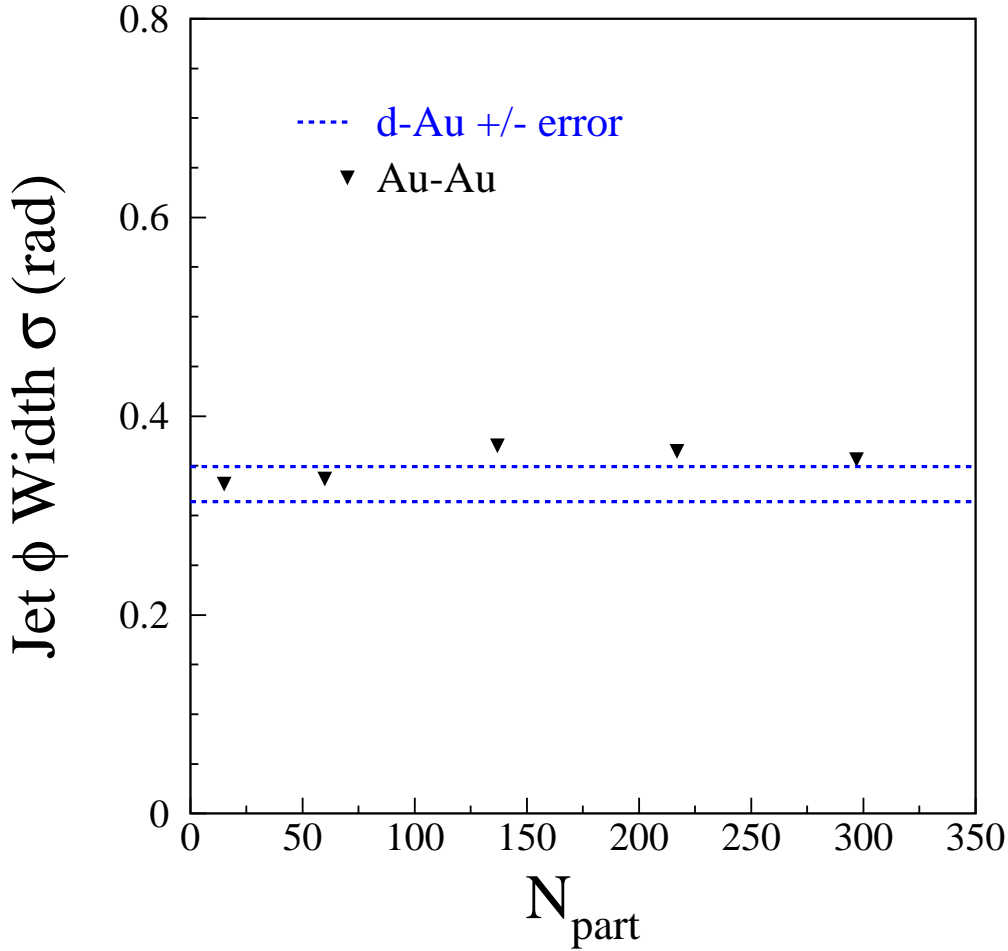}
\caption{The azimuthal angle width of jets in 200 GeV \AuAu\
collisions extracted as the $\sigma$'s of Gaussian fits to the
$0^\circ$ peak in the two-charged-hadron azimuthal-angle (\dphi)
correlation functions \cite{Rak:2004gk}. The correlation functions were formed from
pairs with trigger hadron in the \pperp\ range $2.5 < 
\pperp < 4.0$ GeV/$c$ and the associated hadron in the range $1.0 <
\pperp < 2.5$ GeV/$c$. 
The dashed lines show the $\pm
1\sigma$ range of the jet widths in \dAu\ collisions using the same
momentum bins. 
In the Au+Au data, the effect of the elliptic flow has
been subtracted in the extraction of the jet width.}
\label{fig:hpt:jetwidths}
\end{figure}

We observe that the angular widths of the same-jet correlations are
the same within errors in all three data sets in spite of the factor
of two larger yield of associated hadrons in central \AuAu\ collisions
compared to \dAu\ and peripheral \AuAu\ collisions. This result is
demonstrated more quantitatively in Fig. \ref{fig:hpt:jetwidths} which
shows the centrality dependence of the Gaussian widths of the same-jet
peaks in the \AuAu\ \dphi\ compared to the jet widths extracted from
\dAu\ collisions \cite{Rak:2004gk}.  We see that the \AuAu\ two-hadron
correlation functions show peaks with the same jet width as 
\dAu\ collisions. Since this width is a unique characteristic of the
parton  fragmentation process, we conclude that high-\pperp\ hadrons
in \AuAu\ collisions result from hard scattering followed by jet
fragmentation regardless of any medium modifications of the
fragmentation multiplicity.

\subsection{High-\pperp\ suppression and energy loss}
\label{Sec:denseN:eloss}
The suppression of the production of high-\pperp\ hadrons in heavy ion
collisions at RHIC had been predicted long before RHIC started running
\cite{Bjorken:1982tu,Gyulassy:1990ye,Wang:1992xy,Wang:1995fx,Baier:1995bd,Baier:1997kr,Baier:1998yf,Bass:1999zq}.  It is now generally accepted that partons
propagating in colored matter lose energy predominantly through
medium-induced emission of gluon radiation
\cite{Kovner:2003zj,Gyulassy:2003mc}.  An energetic parton scatters
off color charges in the high-parton-density medium and radiates gluon
bremsstrahlung.  The reduction in the parton energy translates to a
reduction in the average momentum of the fragmentation hadrons, which,
in turn, produces a suppression in the yield of high-\pperp\ hadrons
relative to the corresponding yield in \pp\ collisions. The
power-law spectrum for $p_T\geq 3$ GeV/$c$ implies that a modest
reduction in fragmenting parton energy can produce a significant
decrease in the yield of hadrons at a given $p_T$. Thus, the
suppression of the yield of high-\pperp\ hadrons is generally believed
to provide a direct experimental probe of the density of color charges in
the medium through which the parton passes
\cite{Gyulassy:1992xb,Wang:2002ri,Vitev:2002pf}.  However, before
proceeding to an interpretation of our results, we briefly discuss the
theoretical understanding of the radiative energy loss mechanism and
limitations in that understanding.

\begin{figure}[tbhp]
\centerline{\includegraphics[width=1.0\linewidth,bb = 0 0 567 420]{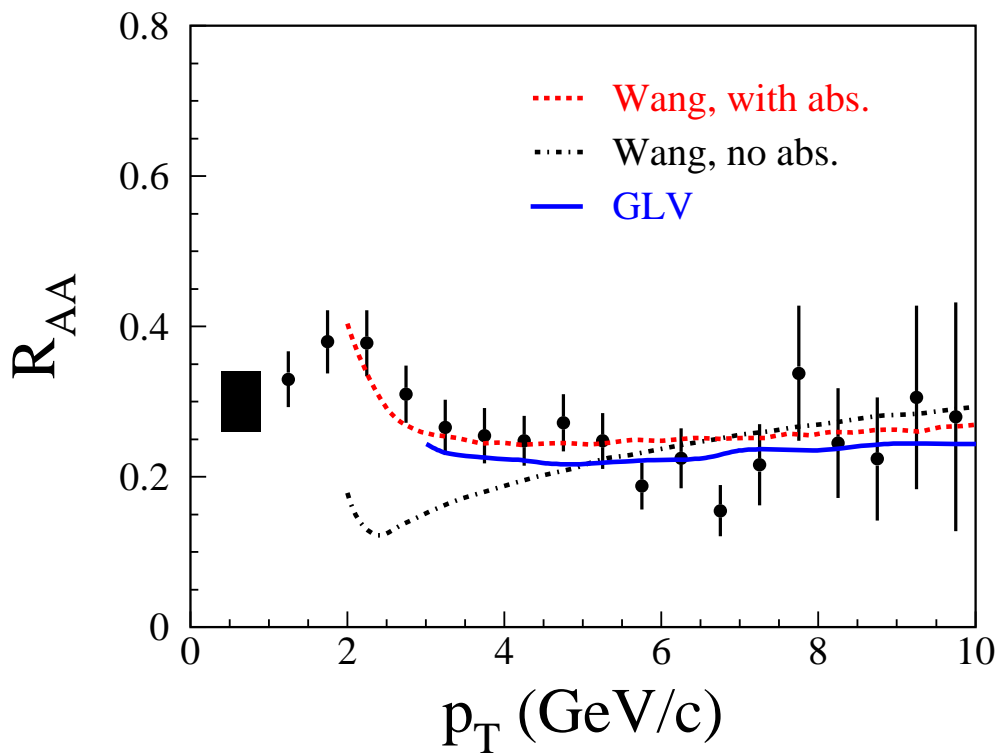}}
\caption{ Comparisons of energy loss calculations \cite{Vitev:2004bh,Wang:2003mm} used to extract
  estimates for the initial parton number or energy density (see text
  for details) to the central 200 GeV \AuAu\ \piz\
  \RAApt\  measured by PHENIX. The Wang curves compare results with and without
  energy absorption from the medium.}
\label{fig:hpt:models}
\label{fig:hpt:wangraa}
\end{figure}

The dominant role of radiative gluon emission was identified early on
\cite{Wang:1992xy}, but it took several years and much effort before
rigorous calculations of the energy loss taking into account
Landau-Pomeranchuk-Migdal suppression \cite{Wang:1995fx} and the
time evolution of the medium were available. Initial estimates of the
radiative energy loss suggested an approximately constant \dEdx\
\cite{Gyulassy:1990ye,Wang:1995fx}, but later calculations
\cite{Baier:1995bd,Zakharov:1997uu,Gyulassy:2000fs,Gyulassy:2000er} showed that the
quantum interference can produce a loss of energy that grows faster
than linearly with the propagation path length, $L$, of the parton in
the medium. However, this ideal growth of \dEdx\ with increasing path
length is never realized in heavy ion collisions due to the rapid
decrease of the energy density and the corresponding color charge
density with time \cite{Baier:1998yf,Wang:2002ri,Vitev:2002pf,Wang:2003mm}. 
Generally, all energy loss calculations
predict that the fractional energy loss of a propagating parton
decreases with increasing parton energy. However, the precise
evolution with parton energy depends on the assumptions in the energy
loss models and on the treatment of details like kinematic limits and
non-leading terms in the radiation spectrum \cite{Gyulassy:2003mc,Kovner:2003zj}. There are many
different calculations of medium-induced energy loss currently
available based on a variety of assumptions about the thickness of the
medium, the energy of the radiating parton, and the coherence in the
radiation process itself (see
\cite{Baier:2000mf,Gyulassy:2003mc,Kovner:2003zj} for recent
reviews). The \pperp\ dependence of the PHENIX
\piz\ \RAA\ values has ruled out the possibility of a constant (energy
independent) \dEdx\ \cite{Bass:1999zq} and the original BDMS energy
loss formulation (which the authors argued should not be applied at
RHIC energies). In fact, the only detailed energy loss model that
{\it predicted} the flat \pperp\ dependence of \RAA\ over the \pperp\ range
covered by RHIC data was the GLV prescription
\cite{Gyulassy:2000er,Gyulassy:1999zd,Levai:2001dc,Vitev:2001td,Vitev:2002pf}. In the GLV formulation,
the fractional energy loss for large jet energies varies approximately
as $\log(E)/E$ but the authors observe that below $20$ GeV the full
numerical calculation of the energy loss produces a nearly constant
$\Delta E/E$ \cite{Gyulassy:2003mc}. However, the same authors argue
that the flat \RAApt\ observed at high \pperp\ at 200 GeV also
requires an accidental cancellation of several different contributions
including the separate \pperp\ dependences of the quark and gluon jet
contributions, the \pperp\ dependence of the Cronin enhancement, and
shadowing/EMC effect. A comparison of the GLV results for the \pperp\
dependence of the  \piz\ suppression to the PHENIX data is shown in
Fig. \ref{fig:hpt:models}. 

One of the most critical issues in the energy loss calculation is the
treatment of the time evolution of the energy density of the matter
through which the radiating parton is propagating. Even if transverse
expansion of the created matter is ignored, the longitudinal expansion
produces a rapid reduction in the energy density as a function of
time.  Most energy loss calculations assume that the color charge
density decreases as a function of proper time as $\rho(\tau) = \rho_0
\tau_0/\tau$ in which case the measured \RAA\ can be used to infer the
product $\rho_0 \tau_0$. Here $\tau_0$ represents the formation time
of the partons from which the medium is composed and $\rho_0$ the
initial number density of those partons. Since the gluons have the
largest cross section for scattering with other partons, the initial
color-charge density is interpreted as the gluon density. Making the
usual assumption that the produced partons are spread over a
longitudinal spatial width $\delta z = \tau_0 \delta y$, the GLV
authors relate the product $\rho_0 \tau_0$ 
to the initial \dngdy\ and
obtain $\dngdy = 1000 \pm 200$ from the PHENIX \piz\ \RAA\ values
\cite{Vitev:2002pf}.  The sensitivity of the GLV calculations to the
details of the description of the transverse parton density and the
transverse expansion of the matter has been tested by using the
results of hydrodynamic calculations of the energy density as a
function of position and time \cite{Gyulassy:2001kr}. The average
energy loss for partons in central \AuAu\ collisions evaluated under
dramatically different assumptions was shown to be remarkably
insensitive to details of the description of the parton density. The
GLV results are also potentially sensitive to a ``screening mass''
that determines both the transverse momentum distributions of the
virtual gluons absorbed from the medium in the bremsstrahlung process
and an energy cutoff for the radiated gluons. This mass is related to
the local energy density using lattice QCD calculations of the plasma
screening mass \cite{Vitev:2002pf}.  However, it was shown by the
authors that a factor of two change in the screening mass produces
only a 15\% change in the \dngdy\ needed to describe the data.

An alternative analysis of parton energy loss \cite{Guo:2000nz} starts
from explicit calculation of higher-twist matrix elements for $e+A$
collisions that account for coherent rescattering of the struck quark
in the nucleus. The contributions of these
higher-twist terms can be incorporated into modified jet fragmentation
functions, producing an effective energy loss. This calculation can 
reproduce \cite{Wang:2002ri} the HERMES measurements of modified jet
fragmentation in nuclear deep-inelastic scattering
\cite{Airapetian:2000ks}. By relating the modified fragmentation
functions from the higher-twist calculation to energy-loss results
obtained from the leading term in an opacity expansion calculation
(e.g. GLV) of medium-induced energy loss the parameters describing the
rescattering in the nucleus in $e+A$ collisions can be related to the
parameters describing the medium in an explicit energy-loss
calculation. By relating the two sets of parameters, the parton
density in the hot medium can be related to the parton density in a
cold nucleus \cite{Wang:2002ri}. Results of this analysis are shown in
Fig. \ref{fig:hpt:models} for parameters that give an initial
energy loss per unit length of $13.8 \pm 3.9$ GeV/fm when the HIJING
\cite{Li:2001xa} parameterization of shadowing is used
\cite{Wang:2003mm} (Note: this result is a factor of two larger that
in \cite{Wang:2002ri} which was based on analysis of the 130 GeV
results). However, an alternative (EKS) \cite{Eskola:1998df}
shadowing description results in an initial energy loss of $16.1 \pm
3.9$ GeV/fm \cite{Wang:2003mm} in the same calculation indicating at
least a 25\% systematic error in the energy loss estimates due to
uncertainties in the description of nuclear shadowing.  Nonetheless,
these initial-energy-loss values are much larger than the
time-averaged energy loss extracted from the calculation, $0.85
\pm 0.24$ GeV/fm for HIJING shadowing \cite{Wang:2003mm}, due to the assumed
$1/\tau$ decrease in the color-charged density. In fact, the average
energy loss per unit path length in central \AuAu\ collisions
\cite{Wang:2002ri} is comparable to the value for cold nuclear matter
extracted from HERMES data
\cite{Wang:2002ri}. However, the initial energy loss is estimated
by Wang to be a factor of $\sim 30$ larger than that in a cold nucleus
\cite{Wang:2003mm} implying that the initial \AuAu\ parton density is larger
by a factor $>30$ than in cold nuclear matter \cite{Wang:2004dn}. 

As shown in Fig. \ref{fig:hpt:models} the Wang higher-twist calculation
predicts a suppression that varies strongly with \pperp\ over the range
where the experimental \RAApt\ 
values are flat. However, Wang and Wang have argued that absorption of
energy from the medium needs to be accounted for in calculating the
energy loss of moderate-\pperp\ partons \cite{Wang:2001cs}. They
provide a formula which incorporates both parton energy loss and
``feedback'' from the medium that can reproduce the shape of the
observed high-\pperp\ suppression as shown by the lower curve in
Fig. \ref{fig:hpt:wangraa}. This formula, then, provides the energy
loss estimate given above. This explanation for the observed \pperp\
independence of \RAA, a crucial feature of the experimental data, is
disquieting, however, because it contradicts the explanation provided
by the GLV model which provides a consistent estimate of the initial
energy density. The feedback of energy from the medium is {\em not}
included in the GLV calculations and if this contribution is
significant, then the agreement of the GLV predictions with the \piz\
\RAApt\ over the entire \pperp\ range would have to be considered
``accidental''. Also, the variation of the suppression in the Wang
higher-twist calculation with \pperp\ reflects the $\Delta E \propto
\log{E}$ variation of parton energy loss naturally obtained from
approximations to the full opacity expansion \cite{Gyulassy:2003mc}.
As noted above, the GLV approach finds that incorporating non-leading
terms in the opacity expansion produces $\Delta E \propto E$. Thus,
while the absorption of energy from the medium in the Wang \etal\
approach may only be significant below $\pperp = 5$ GeV/$c$, the
differences between the variation of energy loss with parton energy in
the two approaches will not be confined to low \pperp.

One source of uncertainty in the interpretation of the high-\pperp\
suppression is the role of possible inelastic scattering of hadrons
after fragmentation. It was originally argued that final-state
inelastic scattering of hadrons could produce all of the observed
suppression \cite{Gallmeister:2002us}. The persistence of the jet
signal with the correct width in \AuAu\ collisions would be difficult
to reconcile with this hypothesis. Indeed, more recent analyses
\cite{Cassing:2003sb} discount the possibility that hadronic
re-interaction could account for the observed high-\pperp\ suppression
and indicate  that only $\sim 1/3$ of fragmentation hadrons undergo
final-state inelastic scattering \cite{Cassing:2003sb}. 
Wang has also argued \cite{Wang:2003aw}
that the complete pattern of high-\pperp\ phenomena observed in the
RHIC data cannot be explained by hadronic rescattering. However, this
leaves open the question of whether hadronic re-interactions after
jet fragmentation can be partially responsible for the observed 
high-\pperp\ suppression.
There are a number of other open issues with the quantitative
interpretation of the observed high-\pperp\ suppression. The calculations
all assume that the jets radiate by scattering off static color
charges while the typical initial gluon \pperp\ is often assumed to be
$\sim 1$ GeV. Also the radiated gluons are assumed to be massless
though a plasmon cutoff equal to the screening mass is applied. The
systematic errors introduced by these and other assumptions made in
the current energy loss calculations have not yet been evaluated
though the gluon screening mass is being included in analyses of 
heavy-quark energy loss.

\subsection{Empirical Energy Loss Estimate}
\label{SubSec:denseN:empirical}

\begin{figure}[tbhp]
\includegraphics[width=1.0\linewidth]{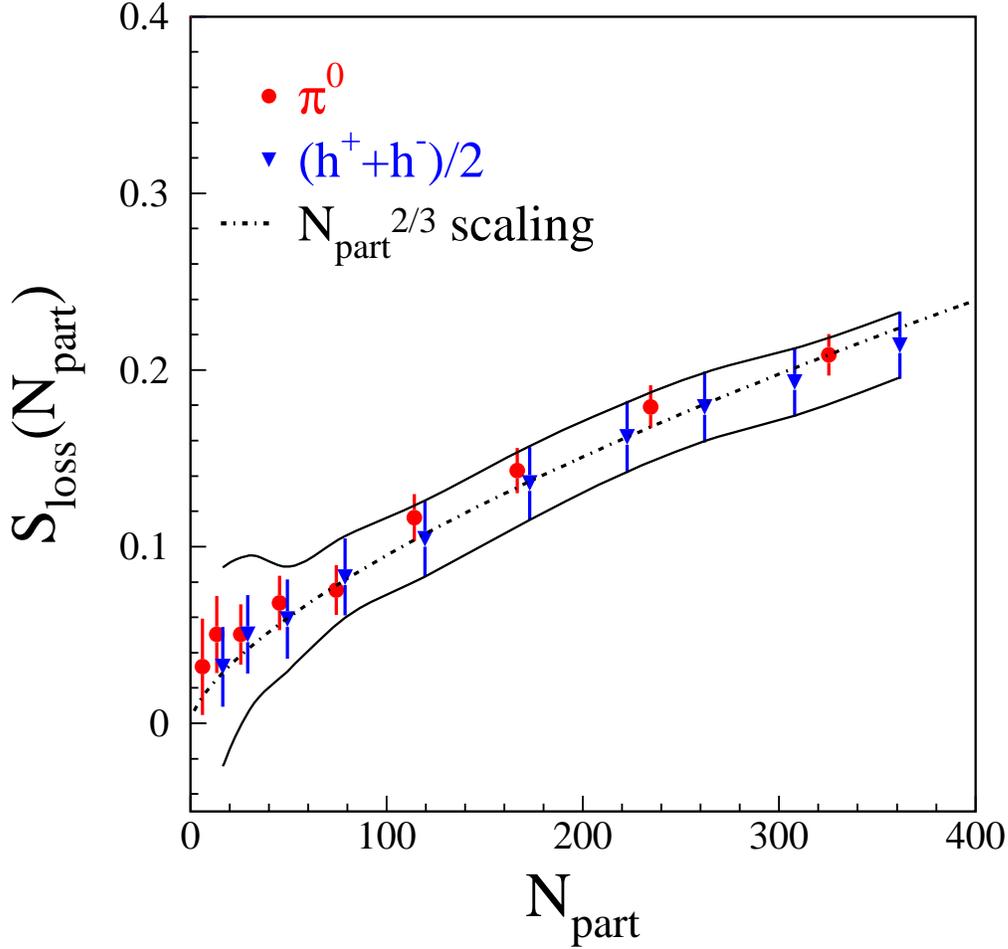}
\caption{Calculated energy loss shift factor, \Sloss\ vs. \Npart\ for \piz\
and charged hadron production in 200 GeV \AuAu\ collisions. The band
around the values indicates systematic errors resulting from
uncertainties in \TAB\ and the normalization of the \pp\ spectrum. 
The dot-dashed curve shows an $\Npart^{2/3}$ scaling of \Sloss\ using
the most central bin to fix the proportionality constant.}
\label{fig:hpt:eloss}
\end{figure}

The observation that the suppression of high-\pperp\ particle
production is approximately independent of \pperp\ above 4 GeV/$c$ and
that the \pp\ \pperp\ spectra are well described by a pure power-law
function in the same \pperp\ range allows a simple empirical estimate
of the energy loss of hard-scattered partons in the medium. 
The \piz\ invariant cross section measured by PHENIX in \pp\
collisions \cite{Adler:2003pb} is found to be well described by a
power law 
\begin{equation}
E \frac{d^3n}{dp^3} = \frac{1}{2\pi} \frac{d^2 n}{\pperp d\pperp dy} = 
\frac{A}{\pperp^n}
\end{equation}
for $\pperp > 3.0$ GeV/$c$ with an exponent $n = 8.1 \pm {0.1}$. 
If we assume that none of the hard-scattered 
partons escape from the medium without losing energy, then the
approximately \pperp-independent suppression above 4.5 GeV/$c$ can be
interpreted as resulting from an average fractional shift in the
momentum of the final-state hadrons due to energy loss of the parent
parton. The suppressed spectrum can be evaluated from the unsuppressed
(\pp) spectrum by noting that hadrons produced in \AuAu\ collisions at
a particular \pperp\ value, would have been produced at a larger \pperp\
value $\pperp' = \pperp + S(\pperp)$ in \pp\ collisions. If the
energy loss is proportional to \pperp\ then we can write 
$S(\pperp) = S_0 \pperp$ so  $\pperp' = (1+S_0)\pperp$
Then, the number of particles observed after suppression in a given 
$\Delta \pperp$ interval is given by 
\begin{equation}
\frac{dn}{d\pperp} = \frac{dn}{d\pperp'} \frac{d\pperp'}{d\pperp} = 
\frac{A}{(1+S_0)^{(n-2)} \, \pperp^{(n-1)}}.
\end{equation}
We note that the factor $\frac{d\pperp'}{d\pperp}$ accounts for the
larger relative density of particles per measured \pperp\ interval
due to the effective compression of the \pperp\ scale caused by the
induced energy loss; this factor is necessary for the total number of
particles to be conserved.
The nuclear modification factor then can be expressed in terms of $S_0$,
\begin{equation}
\RAApt = \frac{1}{(1+S_0)^{(n-2)}}.
\label{eq:raas}
\end{equation}
Using this very simple picture, we can estimate the fraction of
energy lost by hard-scattered partons in the medium from our measured
\RAA\ values. First we obtain $S_0$ from Eq. \ref{eq:raas}
\begin{equation}
S_0 = \frac{1}{\RAA^{1/(n-2)}} - 1.
\end{equation}
Then we observe that the hadrons that would have been produced in \pp\
collisions at a momentum $(1+S_0)\pperp$ were actually produced at
$\pperp$, implying a fractional energy loss 
\begin{equation}
S_{\rm loss} = 1 - 1/(1+S_0) = 1 - \RAA^{1/(n-2)}.
\end{equation}
Figure \ref{fig:hpt:eloss} shows the centrality dependence of 
\Sloss\ obtained from the \pperp-averaged
\RAA\ values shown in Fig. \ref{fig:hpt:avgRAAnpart}. 
For the most central \AuAu\ collisions at 200 GeV we obtain $\Sloss = 0.2$,
which naively implies
that an average 20\% reduction in the energy of partons in the medium
will produce the suppression observed in the \piz\ spectra above
4.5 GeV/$c$. The extracted \Sloss\ values are well described by an
$\Npart^{2/3}$ dependence using the most central bin to fix the
proportionality constant. This result agrees with the GLV prediction
for the centrality dependence of the medium-induced energy loss.

It has been shown previously \cite{Gyulassy:2001nm,Baier:2001yt} that
fluctuations in the radiation process can distort an estimate of
parton energy loss using the procedure described above. Because of the
steeply falling \pperp\ spectrum, the partons that lose 
less energy dominate the yield at a given \pperp\ so our determination
of \Sloss\ will significantly underestimate the true energy loss. However,
it has also been observed that this distortion
can largely be compensated by a single multiplicative factor of value  
$\sim 1.5-2$ \cite{Gyulassy:2001nm}.
While we cannot use the empirically extracted energy loss to estimate
an initial gluon density, we can evaluate the consistency of our
results with estimates of $\langle dE/dx \rangle$ in the medium.  If
we take into account the factor of $1.5-2.0$ renormalization of the
\Sloss, we estimate that 10 GeV partons lose $\sim 3-4$ GeV of
energy. If the typical path length of these partons is on the order of
the nuclear radius then we can infer a $\Delta E/\Delta x \sim
0.5$ GeV/fm which is in good agreement with the estimate from Wang
\cite{Wang:2002ri}. 
We can also use the above empirical energy loss
approach to evaluate possible systematic errors in the estimate of the
initial gluon density. For example, if one third of the observed
suppression were a result of final-state hadronic interactions in the
medium, then the suppression due to energy loss would be a
factor of 1.5 smaller than that implied by the measured \RAA\
values, assuming that every fragmentation hadron
that interacts effectively ``disappears'' by being shifted to much
lower momentum. As a result, $S_0$ in central \AuAu\ collisions would be
reduced from 0.25 to 0.17, implying 30\% reduction in the estimated
energy loss. If the energy loss is indeed proportional to the initial
gluon density then the uncertainty in the effect of the final-state
hadronic interactions would introduce a 30\% systematic error in
\dngdy.

\subsection{Conclusions}
\label{Subsec:denseN:conclusions}

The observed suppression of high-\pperp\ particle production at RHIC
is a unique phenomenon that has not been previously observed in any
hadronic or heavy ion collisions at any energy. The suppression
provides direct evidence that \AuAu\ collisions at RHIC have produced
matter at extreme densities, greater than ten times the energy density
of normal nuclear matter and the highest energy densities
ever achieved in the laboratory. Medium-induced energy loss,
predominantly via gluon bremsstrahlung emission, is the only currently
known physical mechanism that can fully explain the magnitude and
\pperp\ dependence of the
observed high-\pperp\ suppression. This conclusion is based on
evidence provided above that we summarize here:

\begin{itemize}
\item Observation of the \xt\ scaling of the high-\pperp\ hadron spectra
  and measurements of two-hadron azimuthal-angle correlations at high
  \pperp\ confirm the dominant role of hard scattering and subsequent
  jet fragmentation in the production of high-\pperp\ hadrons.  
\item \dAu\ measurements demonstrate that any initial-state
  modification of nuclear-parton distributions has little effect on
  the production of hadrons with $\pperp > 2$ GeV/$c$ at mid-rapidity.  
\item This conclusion is further strengthened by preliminary PHENIX
  measurements showing that the yield of direct photons with $\pperp >
  5$ GeV/$c$ is consistent with a \TAB\ scaling of a pQCD-calculated
  \pp\ direct-photon spectrum.   
\item Analyses described above indicate that final-state hadronic
  interactions can only account for a small fraction of the observed
  high-\pperp\ suppression.
\end{itemize}

Interpreted in the context of in-medium energy loss, the high-\pperp\
suppression data rule out the simplest energy loss prescription---a
jet energy independent \dEdx. The approximately flat \RAApt\ was
predicted by the GLV energy loss model from which the most explicit
estimates of the initial gluon-number density, $dn_g/dy = 1000 \pm
200$ and a corresponding initial energy density 
$\varepsilon_0 \approx
15 {\rm GeV/fm^3}$ \cite{Vitev:2002pf}, have been obtained. An alternative estimate from
the analysis of Wang \etal\ \cite{Wang:2002ri} yields a path-length-averaged energy
loss of 0.5 GeV/fm. Assuming a $1/\tau$ time evolution of the energy
density a much larger initial energy loss of 13--16 GeV/fm is
obtained. That estimate combined with the estimated 0.5 GeV/fm energy
loss of partons in cold nuclear matter yields an initial \AuAu\ gluon
density $> 30$ times larger than that in nuclei \cite{Wang:2004dn}. 
From this result, Wang concludes that the initial energy density
is a factor of $\sim 100$ times larger than that of a nucleus which
would correspond to $16 {\rm GeV/fm^3}$ \cite{Wang:2004dn}. While this
conclusion is consistent with the independent estimate from GLV, we
note that the two models provide completely different explanations for
the nearly \pperp-independent \RAA\ -- the most unique feature of the
single-particle high-\pperp\ suppression -- and the differences
between the approaches may not be confined to low \pperp.
An empirical analysis of the parton energy
loss suggests that the Wang estimate of $> 0.5$ GeV/fm for the average
parton \dEdx\ is consistent with the measured \RAA\ values in central
\AuAu\ collisions. However, some outstanding issues with current
energy loss calculations and the interpretation of high-\pperp\
suppression were noted above. Most notably, rescattering of hadrons
after parton fragmentation could affect the observed
high-\pperp\ suppression even if such rescattering cannot explain the
pattern of jet quenching observations. Using results from
\cite{Cassing:2003sb} and our empirical energy loss analysis, we estimated that
hadronic interactions {\it could} modify extracted values for initial parton
densities by only 30\%. However, we cannot evaluate the potential
systematic error in extracted parton densities due to other
untested assumptions of the energy loss calculations. Therefore, to be
conservative we interpret the extracted initial gluon
number and energy densities as order-of-magnitude estimates. Even
then, the $15 {\rm GeV/fm^3}$ estimated by Gyulassy and Vitev from the
central 200 GeV \AuAu\ \piz\ \RAApt\ measurements indicates that the
matter produced in central \AuAu\ collision has an energy density $>
10$ times normal nuclear matter density.

\section{HADRON PRODUCTION}
\label{Sec:hadron}
%

Descriptions of heavy ion collisions have provided an understanding of
early energy densities, production rates and medium effects of hard
partons, and collective flow of matter.  However, hadronization---the
process by which partons are converted into hadrons---is not well
understood.  The process of hadronization is particularly important
since it includes both the dressing of the quarks from their bare
masses, i.e. the breaking of approximate chiral symmetry, and the
confinement of quarks into colorless hadrons. One could conclude that
a quark-gluon plasma had been formed if one had conclusive evidence of
hadronization occurring from a thermal distribution of quarks and
gluons.

Hadronization processes have been studied over many years in
proton-proton and electron-positron reactions.  Hadron formation, by
its very nature a nonperturbative process, has often been
parameterized from data (e.g. fragmentation functions $D(z)$) or
phenomenologically described (e.g. string models) \cite{Webber:1999ui}.
From QCD one expects that hadron production at high transverse momentum is
dominated by hard scattering of partons followed by fragmentation into
``jets'' or ``mini-jets'' of hadrons.  Following the assumptions of
collinear factorization, the fragmentation functions should be
universal. This universality has proved a powerful tool in comparing
$e^{+}e^{-}$ annihilation to hadron-hadron reactions. One 
feature of jet fragmentation is that baryons and antibaryons are
always suppressed relative to mesons at a given
$p_{T}$ \cite{Alper:1975jm,Abreu:2000nw}.  Phenomenologically this can be thought of
as a large penalty for creating a diquark-antidiquark pair for baryon
formation vs. a quark-antiquark pair for meson formation.

In hadron-hadron reactions, hard scattering followed by fragmentation
is considered to be the dominant process of hadron production for
particles with $p_{T} \ge 2$ GeV/$c$ at mid-rapidity.  At low transverse
momentum, where particles have  $p_{T}<2$ GeV/$c$, particle
interactions are often referred to as ``soft''.  In small momentum
transfer reactions the effective wavelength of interactions is longer
than the spacing of individual partons in a nucleon or nucleus.  Thus
coherence effects are expected to result in large violations of
factorization and universality of fragmentation functions.  Hadron
formation mechanisms in this ``soft'' regime are poorly understood.
We are particularly interested in the study of hadron formation in the
region of $p_{T} \approx$ 2--5 GeV/$c$, where production is expected to
make the transition from ``soft'' to ``hard'' mechanisms.

\subsection{Baryons and Antibaryons}

\begin{figure}[tbhp]
\includegraphics[width=1.0\linewidth]{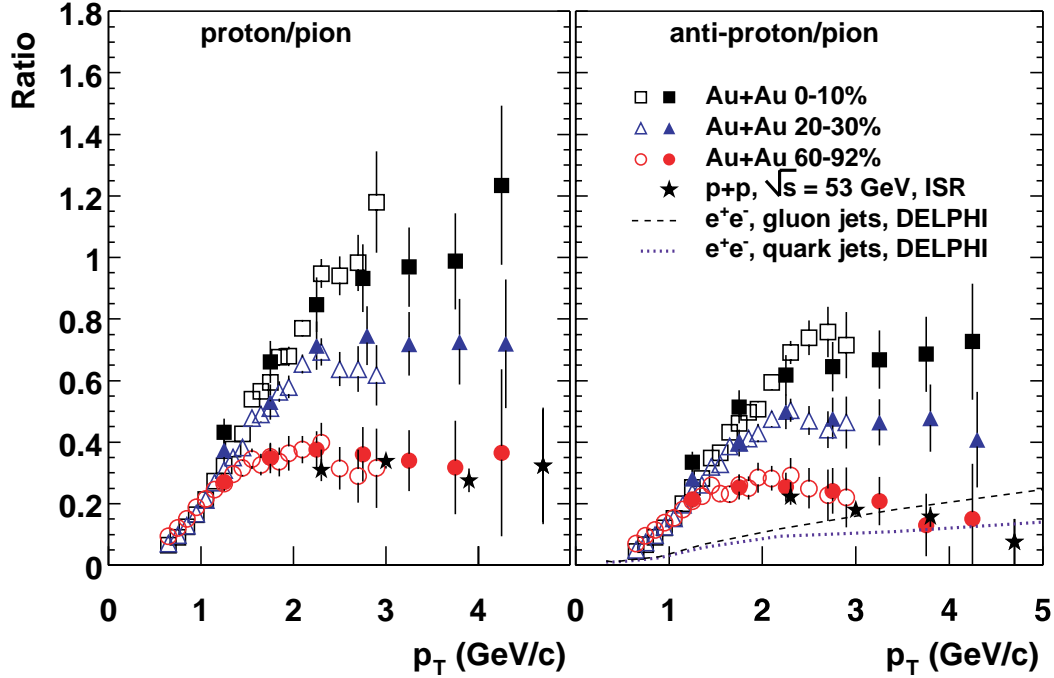}
\caption{
$p/\pi$ (left) and $\overline{p}/\pi$ (right) ratios for central 
(0--10\%), mid-central (20--30\%) and peripheral (60--92\%)
Au+Au collisions at $\sqrt{s_{NN}} = 200 $ GeV\cite{Adler:2003kg}.  Open (filled)
points are for $\pi^{+/-}$ ($\pi^{0}$), respectively.
Data from $\sqrt{s} = 53 $ GeV $p+p$ collisions \cite{Alper:1975jm}
are shown with stars.  The dashed and dotted lines are
($\overline{p}+p$)/($\pi^{+}+\pi^{-}$) ratio in
gluon and in quark jets \cite{Abreu:2000nw}.}
\label{fig:ptopi}
\end{figure}

One of the most striking and unexpected observations in heavy ion
reactions at RHIC is the large enhancement of baryons and antibaryons
relative to pions at intermediate $p_{T} \approx$ 2--5 GeV/$c$.  As
shown in Fig. \ref{fig:ptopi}, the (anti)proton to pion ratio is
enhanced by almost a factor of three when one compares peripheral
reactions to the most central gold-gold
reactions \cite{Adler:2003kg}\footnote{All PHENIX (anti)proton spectra
shown in this section are corrected for feed down from heavier
resonances.}.  This of course is in sharp contrast to the suppression
of pions in this region.

\begin{figure}[tbhp]
\includegraphics[width=1.0\linewidth]{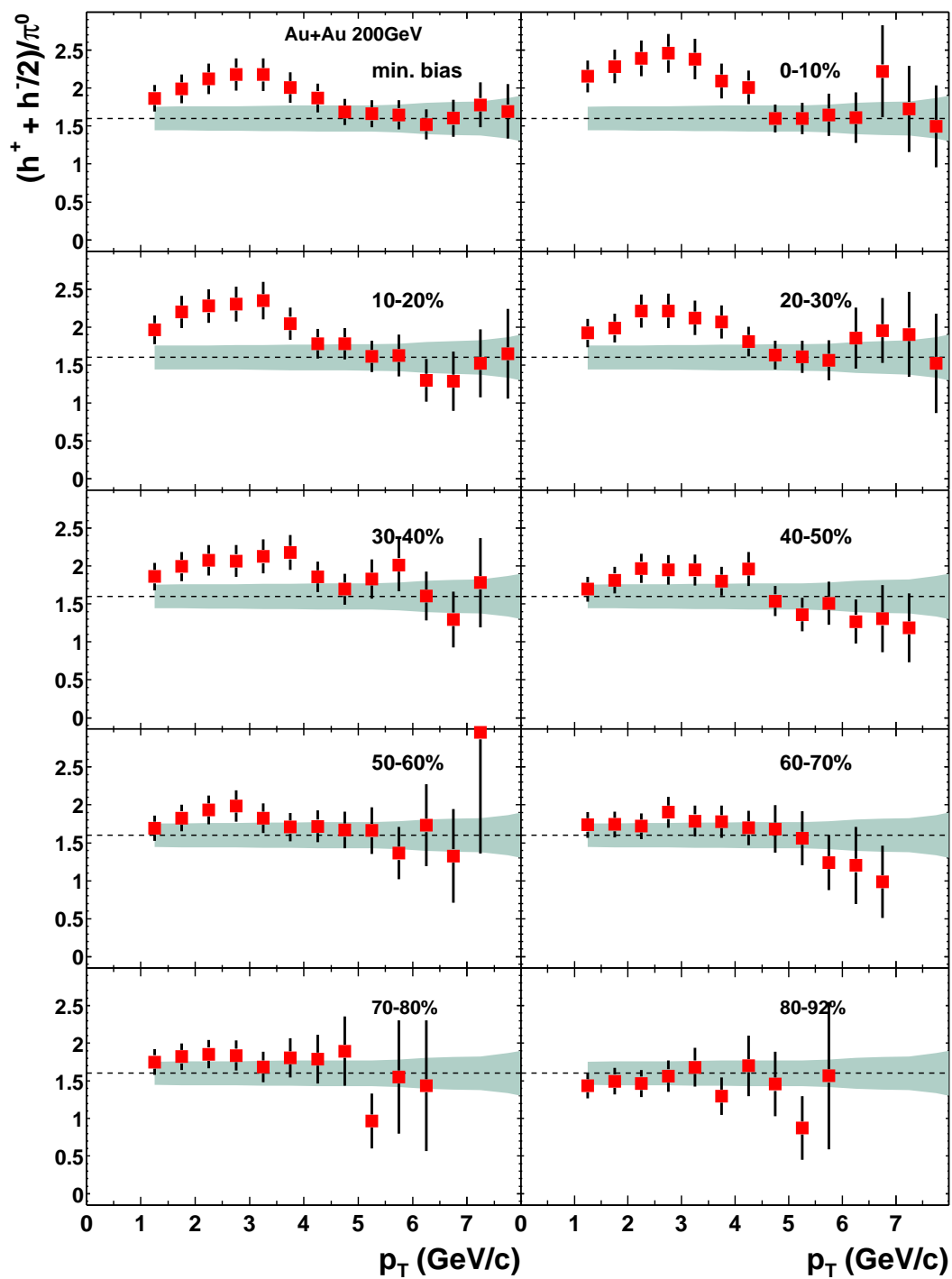}
\caption{Charged hadron to $\pi^{0}$ ratio for different centrality classes for
Au+Au collisions at $\sqrt{s_{NN}} = 200 $ GeV\cite{Adler:2003au}. 
Error bars represent the quadratic sum of statistical and point to
point systematic errors. The shaded band shows the
normalization error common to all centrality classes. 
The line at 1.6 is the
$h/\pi$ ratio measured in $p+p$ collisions at $\sqrt{s} = 53 $ GeV \cite{Alper:1975jm}
and e+e- collisions \cite{Abreu:2000nw}.}
\label{fig:htopi}
\end{figure}

We can investigate this (anti)baryon excess to much higher $p_{T}$ by
comparing our inclusive charged spectra (primarily pions, kaons and
protons) with our neutral pion 
measurements \cite{Adler:2003kg}. Shown in Fig. \ref{fig:htopi} is
the charged hadron to $\pi^{0}$ ratio as a function of transverse
momentum in ten centrality bins.  We observe a significant increase of
the $(h^{+}+h^{-})/\pi^{0}$ ratio above 1.6 in the $p_{T}$ range
1--5 GeV/$c$ that increases as a function of collision centrality.  The
ratio of $h/\pi = 1.6$ is the value measured in $p+p$
reactions \cite{Alper:1975jm}, and is thought to arise from jet
fragmentation.  In Au+Au central reactions, above $p_{T} \approx
5 $ GeV/$c$, $h/\pi$ returns to the $p+p$ measured baseline. 
This implies that the (anti)baryon excess occurs only in the
limited $p_{T}$ window $\approx$2--5 GeV/$c$, and then returns to the
universal fragmentation function expectation.

\begin{figure}[tbhp]
\includegraphics[width=1.0\linewidth]{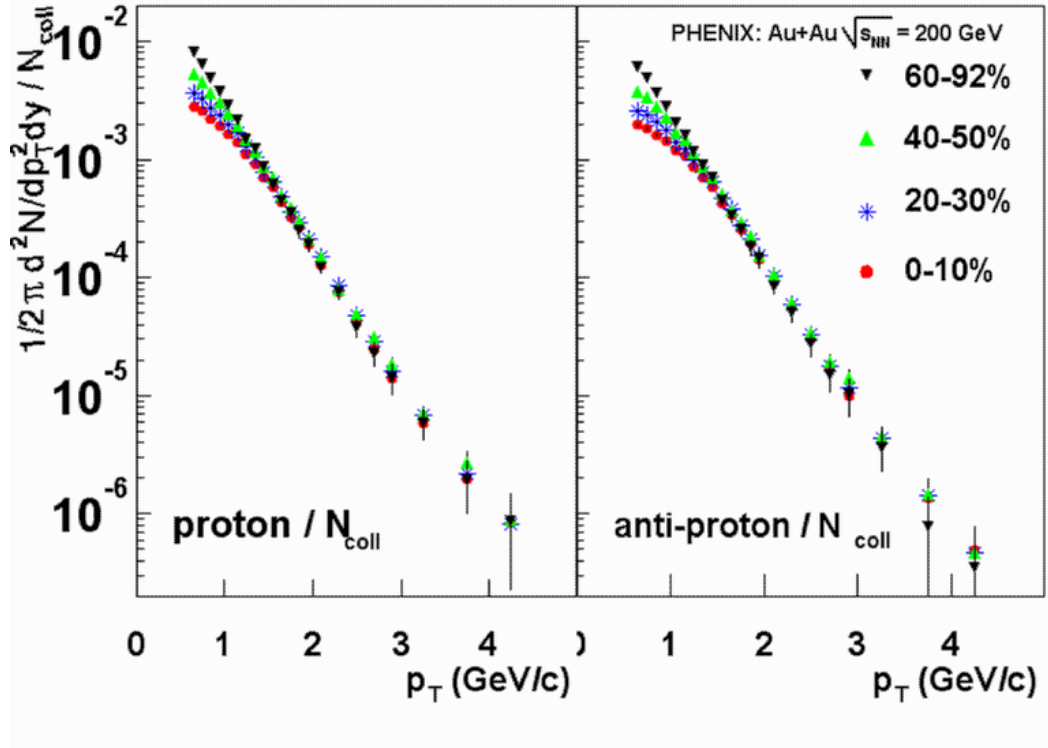}
\caption{$p$ and $\overline{p}$ invariant yields scaled by $N_{coll}$
in Au+Au collision at $\sqrt{s_{NN}}$ = 200 GeV\cite{Adler:2003kg}.  Error
bars are statistical.  Systematic errors on $N_{coll}$ range from
$\approx$10\% for central to $\approx$28\% for 60--92\% centrality.
Multiplicity dependent normalization errors are $\approx$3\%.}
\label{fig:pncoll}
\end{figure}

As discussed in section \ref{Sec:denseN}, pions in this $p_{T}$ range
are suppressed by almost a factor of five relative to binary collision
scaling for central Au+Au reactions.  Thus, one possible
interpretation of the large (anti)proton to pion ratio is that somehow
the baryons are not suppressed in a manner similar to the pions.
Figure \ref{fig:pncoll} shows that in fact (anti)proton production
appears to follow binary collision scaling over the transverse
momentum range $p_{T} = $ 2--5 GeV/$c$ \cite{Adler:2003kg}.  However, 
the $h/\pi^0$ ratios shown in Fig. \ref{fig:htopi} imply
that above $p_{T} > 5$ GeV/$c$, the (anti)protons must be as suppressed as
the pions. 
  
Characteristics of the intermediate $p_T$ (anti)protons are:
\begin{itemize}
\item A large enhancement of the $p/\pi$ and $\overline{p}/\pi$ ratios in central Au+Au collisions.
\item A ratio in peripheral collisions which is in agreement with that from $p+p$ collisions.
\item A smooth increase from peripheral to central Au+Au collisions.
\item A similar effect for protons and antiprotons.
\item Approximate scaling of (anti)proton production at $p_T \approx$ 2--4 GeV/$c$ with the
number of binary nucleon-nucleon collisions.
\item Suppression relative to binary collision scaling similar for (anti)protons and 
pions for $p_{T} > 5$ GeV/$c$.
\end{itemize}

\begin{figure}[tbhp]
\includegraphics[width=1.0\linewidth]{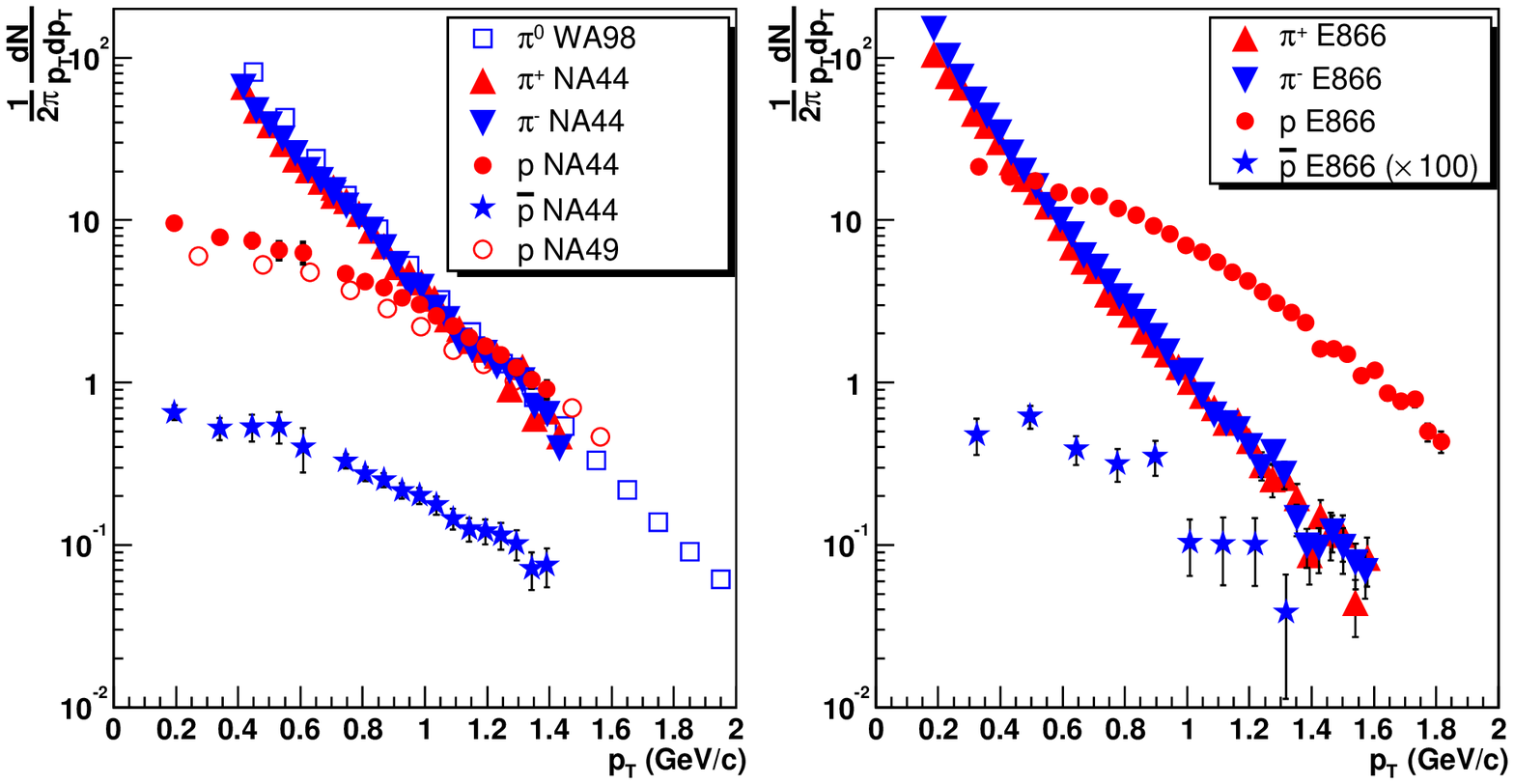}
\caption{Invariant yields of $p$, $\bar{p}$, and $\pi$ as function
of $p_T$ in central Pb+Pb collisions at the SPS ($\sqrt{s_{NN}} = 17$ GeV)
(left panel) and in central Au+Au collisions
at the AGS ($\sqrt{s_{NN}} = 5$ GeV) (right panel).
The $\bar{p}$ spectrum from the AGS is scaled up by a factor 100.
All data are at mid-rapidity ($y - y_{cm} \approx 0$) and are
from W98 \cite{Aggarwal:2001gn}, NA44 \cite{Bearden:2002ym},
NA49 \cite{Anticic:2004yj}, and E866 \cite{Ahle:1997vx,Ahle:1998wk}.}
\label{fig:ags_sps_p_pi}
\end{figure}

Large proton to pion ratios have also been observed in heavy ion collisions
at lower energies. Figure \ref{fig:ags_sps_p_pi} shows $p_T$ distributions
of protons, antiprotons, and pions in central Pb+Pb collisions at the SPS and
in central Au+Au collisions at the AGS. The 
$p/\pi$ ratio in central Pb+Pb collisions at the SPS is
greater than unity for $p_T \ge$ 1.3 GeV/$c$. At the AGS,
the proton spectrum crosses pion spectra at $p_T$ $\sim$ 0.5 GeV/$c$,
and the $p/\pi$ ratio is about 20 at $p_T$ = 1.6 GeV/$c$. The $p/\pi$
ratios in the low-energy heavy-ion collisions are also enhanced
compared with $p+p$ collisions at the same energy.

Most of the protons in these lower-energy heavy-ion collisions are not
produced in the collision. Rather they are protons from the beam or
target nucleus (Pb or Au) that are transported to large $p_T$ at
mid-rapidity.  As discussed in section \ref{Sec:therm}, a strong
radial flow with velocity $\beta_T \sim 0.5$ is produced in heavy ion
collisions at AGS and SPS energies. The large $p/\pi$ ratio can be
interpreted as a result of this radial flow.  Since the proton is
heavier, a fixed velocity boost results is a  larger momentum boost than for 
pions,
and thus enhances $p/\pi$ ratio at higher $p_T$.  In contrast, at RHIC
energies, most of protons are produced particles \cite{Adcox:2001mf}.  
The anomalously
large antibaryon-to-meson ratio $\bar{p}/\pi \sim 1$ at high $p_T \ge
$ 2 GeV/$c$ is a unique result from RHIC. Such a large $\bar{p}/\pi$
ratio has not been observed in any other collision system.
Figure \ref{fig:ags_sps_p_pi} shows that $\bar{p}/\pi$ is less than
$\sim 0.1$ at the SPS, and it is less than 1/100 at the AGS. 
It should also be noted that the measurements from the AGS/SPS
are limited to lower $p_T$ ($p_T < $ 2 GeV/$c$), where soft physics is
still dominant, while at RHIC we observe a large $p(\bar{p})/\pi$
ratio in $ p_T \approx $ 2--5 GeV/$c$ where hard processes are
expected to be the dominant mechanism of particle production.

\subsection{The $\phi$ Meson}

We have extended our identified hadron studies to include the $\phi$
vector meson as measured in the $K^{+}K^{-}$ decay channel.  
The $\phi$ is a meson, and is in that sense similar to the pion
with a valence quark and antiquark, and yet its mass is comparable to
that of the proton.

\begin{figure}[tbhp]
\includegraphics[width=1.0\linewidth]{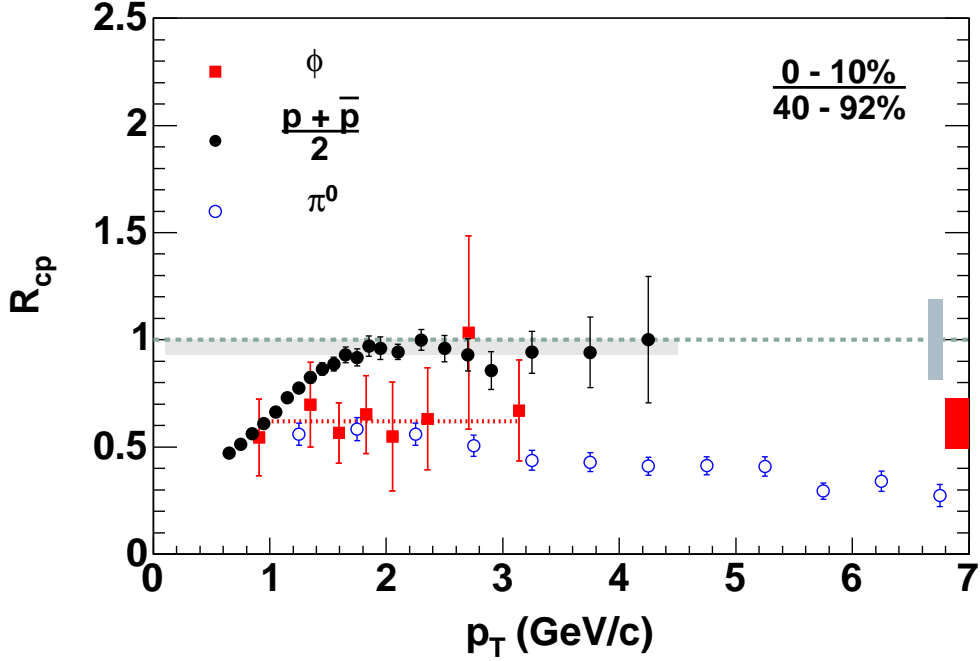}
\caption{The $R_{CP}$ of the $\phi$ as measured in the $KK$ channel, compared
to the protons and pions for Au+Au collisions at $\sqrt{s_{NN}} = 200 $ GeV\cite{Adler:2004hv}.}
\label{fig:phircp}
\end{figure}

\begin{figure}[tbhp]
\includegraphics[width=1.0\linewidth]{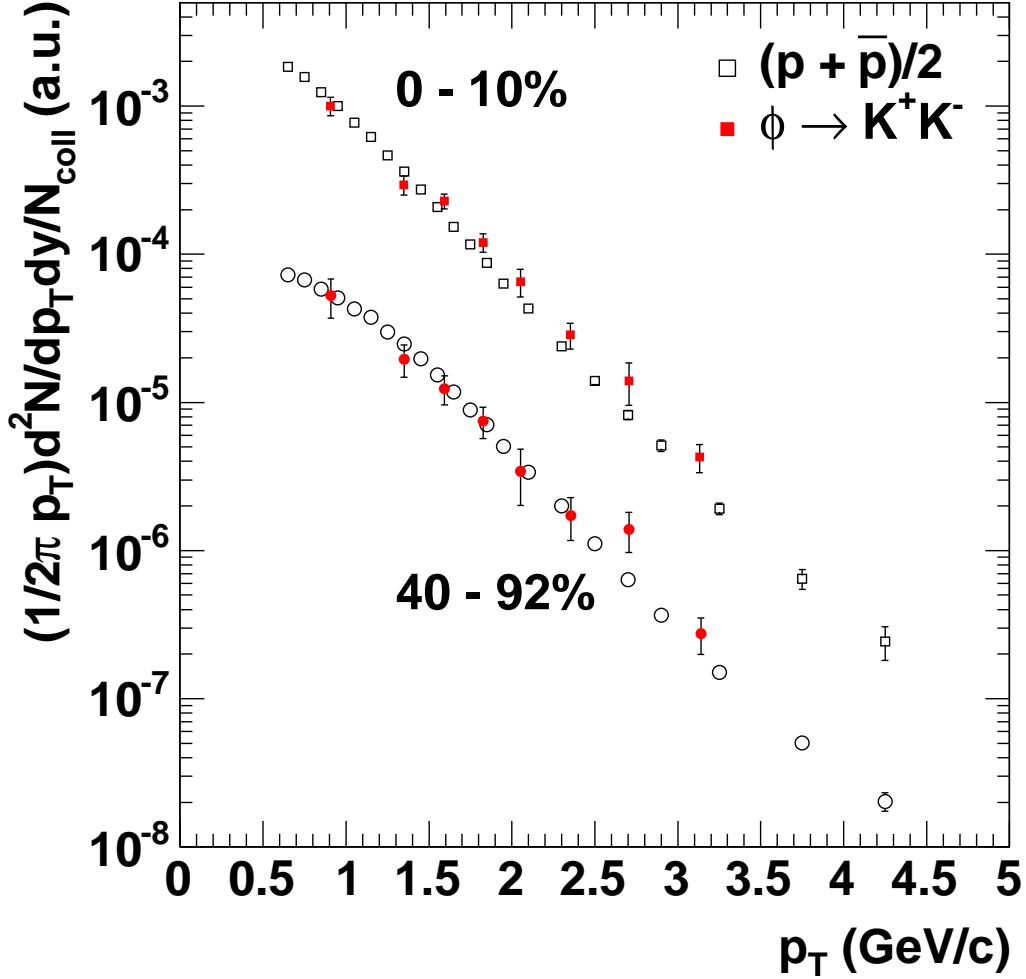}
\caption{$(p+\overline{p})$ and $\phi$ invariant yield as a function
of transverse momentum for central 0--10\% and peripheral 40--92\% Au+Au reactions at
$\sqrt{s_{NN}} = 200 $ GeV.  The two distributions for each centrality
class are given an
arbitrary relative normalization to allow for comparison of the
$p_{T}$ dependent shapes. Data are from \cite{Adler:2004hv}.}
\label{fig:phislopenorm}
\end{figure}

Figure \ref{fig:phircp} shows $R_{CP}$, the ratio of production in
central to peripheral Au+Au collisions scaled by binary collisions,
for protons, pions and $\phi$ mesons detected via its $KK$ decay
channel \cite{Adler:2004hv} in Au+Au collisions at
$\sqrt{s_{NN}} = 200 $ GeV.  A large suppression of pions at
$p_{T}>2 $ GeV/$c$ is observed (as detailed in Section \ref{Sec:denseN}),
and a lack of suppression for the protons and antiprotons as expected
from Fig. \ref{fig:pncoll}.  The $\phi$ follows the suppression
pattern of the pions within errors, indicating that the surprising
behavior of the protons is not followed by the $\phi$.
Figure \ref{fig:phislopenorm} shows a comparison between the $p_{T}$
spectral shape for protons and the $\phi$ in central and peripheral Au+Au
reactions. The two spectra agree with each other within errors for the
most central events.  Thus,
although the yields are evolving differently with
collision centrality, giving rise to the deviation from unity of $R_{CP}$, the
$p_{T}$ distributions appear quite similar.

\subsection{Jet Correlations}

\begin{figure}[tbhp]
\includegraphics[width=1.0\linewidth]{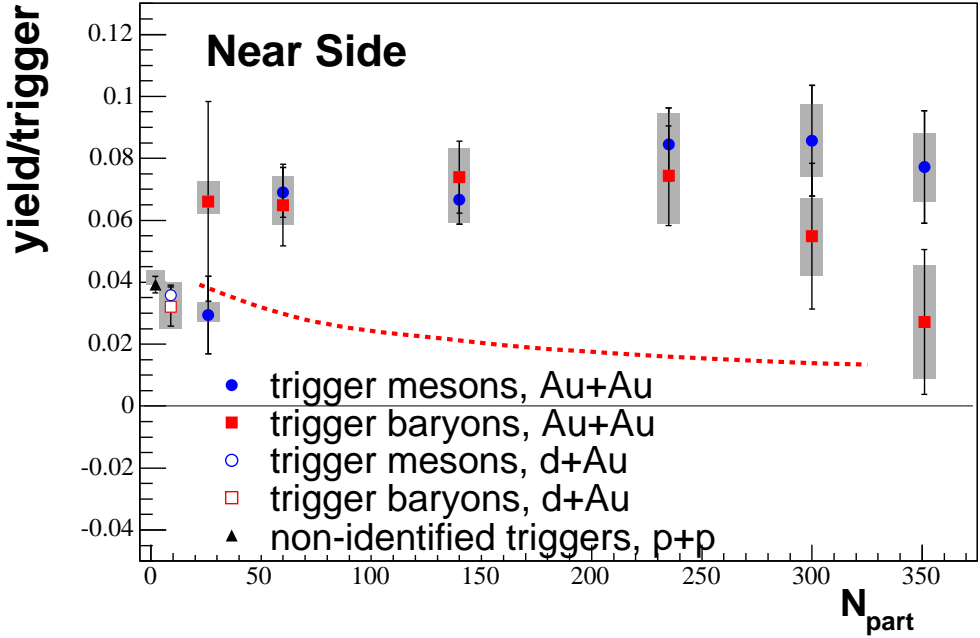}
\caption{ Centrality dependence of associated charged hadron yield 
($1.7<p_T<2.5$ GeV/$c$) above combinatorial background for trigger baryons and
 trigger mesons in the p$_T$ range 2.5-4.0 GeV/$c$ in a $54^\circ$ cone
 around the trigger particle in Au+Au collisions at
 $\sqrt{s_{NN}} = 200 $ GeV\cite{Adler:2004zd}.  The error bars are statistical errors and
 the gray boxes are systematic errors.  The dashed line represents
an upper limit of the centrality dependence of the near-side partner
yield from thermal recombination (see text).
}
\label{fig:jetcorr}
\end{figure}

A crucial test of the origin for the enhanced (anti)proton to pion
ratio is to see if baryons in this intermediate $p_T$ regime exhibit
correlations characteristic of the structure of jets from
hard-scattered partons. Particles which exhibit these correlations are
termed ``jet-like''.  Figure \ref{fig:jetcorr} shows the associated
partner particle yield within the relative angular range $0.0 < \phi <
0.94$ radians on the same side as trigger baryons and
mesons \cite{Adler:2004zd}. Correlated pairs are then formed between
the trigger particle and other particles within the above mentioned
angular range. Mixed events are used to determine the combinatorial
(i.e. non-jet-like) background distribution, which is subtracted after
modulation according to the measured $v_2$.

The partner yield increases for both trigger baryons and mesons by
almost a factor of two from deuteron-gold to peripheral and
mid-central Au+Au reactions.  We then observe
a decrease in the jet-like correlations for baryons
relative to mesons for the most central collisions.  
It is notable that this observation is of limited significance within
our current statistical and systematic errors.  
Over a broad range of centrality 10-60\% the partner yield is the same for protons
and pions within errors.  This is notable since the (anti)proton to pion
ratio has already increased by a factor of two for mid-central Au+Au relative
to proton-proton reactions, with the implication that the increase in
the $p/\pi$ ratio is inclusive of the particles with jet-like correlations.

The dashed line in Figure \ref{fig:jetcorr} shows the expected centrality dependence of
partners per baryon if all the ``extra'' baryons which increase the $p/\pi$
over that in p+p collisions were to arise solely from soft processes.
Baryons from thermal quark recombination should have no jet-like
partner hadrons and would dilute the per-trigger conditional yield.
Because this simple estimate does not allow for meson production by
recombination, which must also occur along with baryon production,
it represents an upper limit to the centrality dependence of jet
partner yield from thermal recombination.  The data clearly disagree
with both the centrality dependence and also the absolute yields of
this estimation, indicating that the baryon excess has the same jet-like
origin as the mesons, except perhaps in the highest centrality bin.

\begin{figure}[tbhp]
\includegraphics[width=1.0\linewidth]{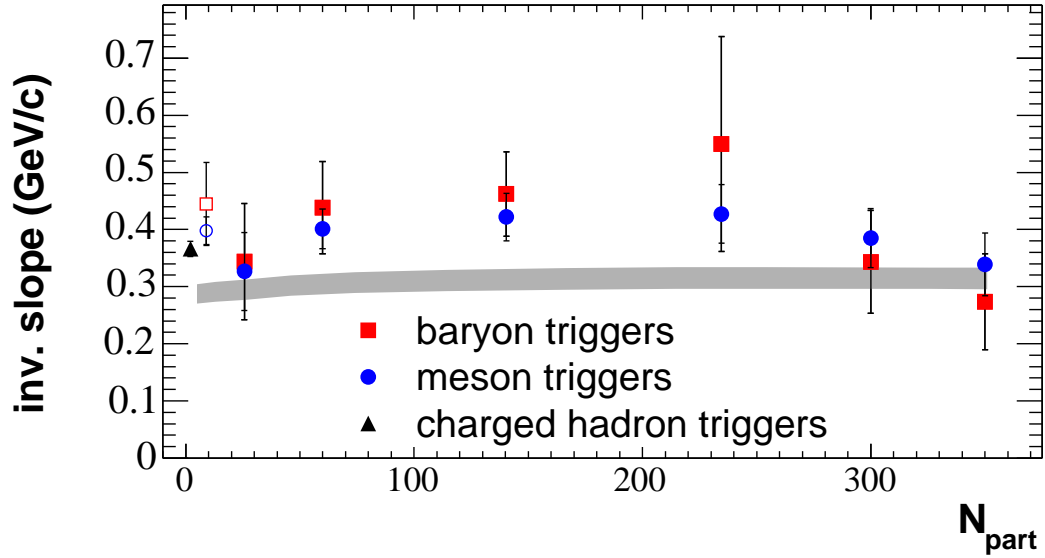}
\caption{The inverse slopes for the momentum distributions of the associated particles shown in Fig. \ref{fig:jetcorr}.  The gray band is the inverse slope
of the momentum distribution of the inclusive hadrons from Au+Au
collisions at $\sqrt{s_{NN}} = 200 $ GeV\cite{Adler:2004zd}.}
\label{fig:jetcorrslope}
\end{figure}

The characteristics of the jet-like particles are compared to
inclusive hadrons in Fig. \ref{fig:jetcorrslope}, which shows the
centrality dependence of the $p_T$ distributions of jet-like partners
and inclusive hadrons.  One can see that, within the statistics
available, the slopes of the associated particle spectra in $p+p$,
$d+$Au, peripheral and mid-central Au+Au collisions are very similar
for both trigger mesons and trigger baryons. The partner spectra are
harder than the inclusive hadron spectra, as expected from jet
fragmentation. In the most central collisions, the number of particles
associated with trigger baryons is very small, resulting in large
statistical error bars. However, the inverse slopes of the jet-like
partners and inclusive hadron distributions agree better in central
collisions than in peripheral collisions.

We can then make the following general observations:
\begin{itemize}
\item Trigger (anti)protons and mesons have comparable near-side 
associated-particle yields over a broad range in centrality, indicating 
a significant jet-like component for both.
\item There is an indication
that the proton partner yield tends to diminish for
the most central collisions, unlike for leading mesons. 
\item Within the limited statistics available for the measurement, 
the inverse slopes of the associated particles are similar for both mesons
and baryons. These are harder than for the inclusive spectra.
\item Trigger particles in
Au+Au collisions appear to have \begin{it}more\end{it} associated
particles than in $d+$Au collisions. This is true for all centralities
aside from the most peripheral, and except for leading baryons in
central collisions.
\end{itemize}

\subsection{Soft Physics}

In hadron-hadron reactions, hard scattering followed by fragmentation
is considered to be the dominant process of hadron production with
$p_{T} \ge 2$ GeV/$c$ at mid-rapidity.  However, as detailed in
section \ref{Sec:therm}, there is strong evidence for explosive
collective motion of particles in the medium.  If the mean free path
for particles in the medium is small, then all particles must move
with a common local velocity as described by hydrodynamics.
Therefore, heavier particles receive a larger momentum boost than
lighter particles.  This effective shifting of particles to higher p$_T$
 results
in a ``shoulder-arm'' shape for the (anti)proton $p_{T}$ spectra,
visible in Fig. \ref{fig:phislopenorm}.

\subsubsection{Hydrodynamics}

Is it possible that this soft hadron production extends to higher
$p_{T}$ for baryons than mesons?  Hydrodynamic boosting of ``soft''
physics for heavier particles into the $p_{T}>2 $ GeV/$c$ offers a
natural explanation for the enhanced $p/\pi$ and $\overline{p}/\pi$
ratios \cite{Teaney:2001av}.

\begin{figure}[tbhp]
\includegraphics[width=1.0\linewidth]{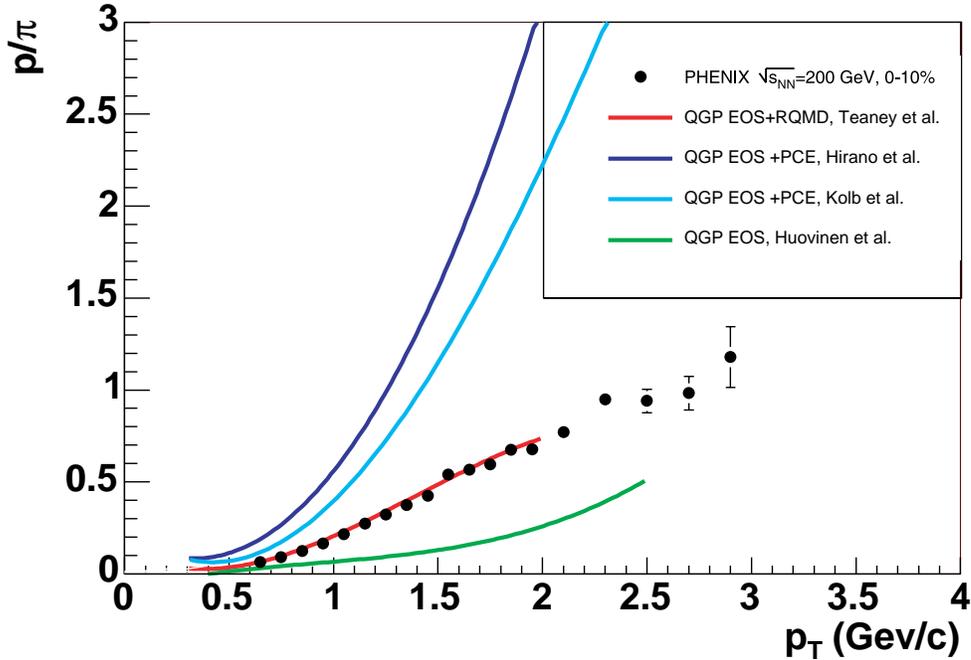}
\caption{
$p/\pi$ ratios for central (0--10\%) Au+Au collisions at
$\sqrt{s_{NN}} = 200 $ GeV\cite{Adler:2003kg} compared to hydrodynamic
models \cite{Teaney:2001av,Huovinen:2001cy,Kolb:2002ve,Hirano:2002ds,Hirano:2004rs}.  }
\label{fig:ppiratioCent}
\end{figure}

As seen in Section \ref{Sec:therm}, some hydrodynamical models can describe
both the proton and the pion spectra. Consequently, the $p/\pi$
ratio is also reproduced (Fig. \ref{fig:ppiratioCent}).  It is clear that
the description of the $p/\pi$ ratio is not unique and different calculations 
yield quite different results.  
Above some $p_{T}$, hydrodynamics should fail to describe the data and
fragmentation should dominate.  Pure hydrodynamics predicts that this
ratio would continue to increase essentially up to $p_{T} \rightarrow
\infty$.  However, these particles cannot have a zero mean free path 
in the medium. Any finite mean free path and
a finite volume will limit the number of $p_{T}$ ``kicks''
a particle can receive.  For this reason many of the hydrodynamic
calculations are not extended into the $p_{T}$ region 2--5 GeV/$c$ in
which we are interested.

Hydrodynamic calculations do not specify the quanta that flow; rather
they assume an equation of state.  When applied at RHIC,
most calculations start with a quark-gluon-plasma equation of state
and transition to a resonance gas.  The mapping of the fluid onto
hadrons is somewhat \it ad hoc, \rm and often uses the Cooper-Frye
freezeout \cite{Cooper:1974mv}, giving the typical hierarchies of
momenta one sees where heavier particles receive a larger boost.

As mentioned previously, this generic feature of a transverse velocity
boost yielding an increase in the baryon to meson ratio relative to
proton-proton reactions is not unique to RHIC as shown in
Fig. \ref{fig:ags_sps_p_pi}.  However, a major difference between
lower-energy results and those at $\sqrt{s_{NN}} = 200$ GeV is that at these
highest energies there is a significant hard-process contribution.  If
the source of the excess baryons is the transport of soft baryons to
the intermediate $p_{T}$ range, then it is purely coincidental that
the baryons scale with binary collisions.  More importantly, we should
expect a significant decrease in the jet-like partner yield for
baryons relative to mesons.  Although there may be a hint of this for
the most central reactions, one expects this decrease to follow the
centrality dependence of the increase in $p/\pi$ ratio.  Thus, this
effect should already reduce the partner yield by a factor of two in
mid-central Au+Au reactions. This is ruled out by the data.

\subsubsection{Recombination Models}

The quark recombination or coalescence model is a different physics
framework in which baryons receive a larger $p_{T}$ boost than mesons.
These models were frequently invoked in the
1970's \cite{Das:1977cp,Roberts:1979ku} in an attempt to describe the
rapidity distribution of various hadronic species in $hadron-hadron$
reactions.  More recently, these models have been applied to describe
the forward charm hadron production in $hadron-nucleus$ reactions at
Fermilab \cite{Braaten:2002yt}.  In this case they calculate a
significant probability for $D$ meson formation from a
hard-scattering-created charm quark with a light valence quark in the
projectile. The quark coalescence mechanisms have some similarities to
light nuclei coalescence. However, wave functions are relatively well
determined for light nuclei, whereas the hadron wave functions are
neither easily described by partons nor directly calculable from QCD.

Recently, quark recombination has been successfully applied to
describe a number of features of heavy ion
collisions \cite{Fries:2003vb,Fries:2003kq} (Duke model).  In this
picture, quarks in a densely populated phase space combine to form the
final-state hadrons.  This model uses
the simplifying assumption that the mass is small relative to the
momentum giving a prediction largely independent of the final hadron
wave function\footnote{The recombination model prediction of these
models is independent of the final hadron wave function with an
accuracy of about  20\% for protons and  10\% for pions}.  The
coalescing parton distribution was assumed to be exponential,
i.e. thermal, and recombination applied for hadrons where
$m^2/p_T^2<<1$.  At very high $p_T$ particles are assumed to arise
from fragmentation of hard partons with a standard power law
distribution; the relative normalization of the thermal source with
respect to this process is an important external parameter to the
model.  A crucial component of recombination models is the assumption
that the partons which recombine carry a mass which is essentially 
equal to the mass of the dressed constituent quarks\footnote{The actual
source of this mass is under discussion. It may be that the chiral
phase transition is slightly above the deconfinement transition. In
this case, the mass would be from the dressing of the quarks.  Another
possibility is that the mass is a thermal mass which happens to be
similar to the constituent quark mass.}.  If all observables of
intermediate $p_T$ hadrons can be explained by recombination of only
thermal quarks, this would essentially prove the existence of a
quark-gluon plasma in the early stage of the collisions.

Three essential features are predicted by recombination models. First,
baryons at moderate $p_T$ are greatly enhanced relative to mesons as
their transverse momentum is the sum of 3 quarks rather than
2. Recombination dominates over parton fragmentation in this region,
because, for an exponential spectrum recombination is a more efficient
means of producing particles at a particular $p_T$. This enhancement
should return to its fragmentation values at higher $p_T$. In the
intermediate range, all mesons should behave in a similar manner
regardless of mass, as should all baryons.  Secondly, recombination
predicts that the collective flow of the final-state hadrons should
follow the collective flow of their constituent quarks. Finally,
recombination
causes thermal features to extend to
higher transverse momentum, $p_T >>T_C$ than one
 might naively expect since the underlying thermal spectrum of the
 constituents gets a multiplication factor of essentially 3 for 
 baryons and 2 for mesons. A last general feature which is true
 for the simplest of the models, but may not necessarily be true for
 more complex models, is that at intermediate $p_T$, recombination is
 the dominant mechanism for the production of hadrons---particularly
 of baryons.

Other recombination calculations have relaxed the assumptions
previously described, at the cost of much more dependence on the
particular form of the hadronic wave function used. One such
calculation \cite{Hwa:2004ng,Hwa:2002tu} (Oregon model) uses a
description of hadronization which assumes that all
hadrons---including those from fragmentation---arise from
recombination. Hard partons are allowed to fragment into a shower of
partons, which can in turn recombine---both with other partons in the
shower and partons in the thermal background. Another
model \cite{Greco:2003xt} (TAMU model) uses a Monte-Carlo method to
model the production of hadrons allowing
recombination of hard partons with thermal partons, and 
includes particle decays, such as $\rho \rightarrow 2\pi$ which
produces low-$p_T$ pions.

\begin{figure}[tbhp]
\includegraphics[width=1.0\linewidth]{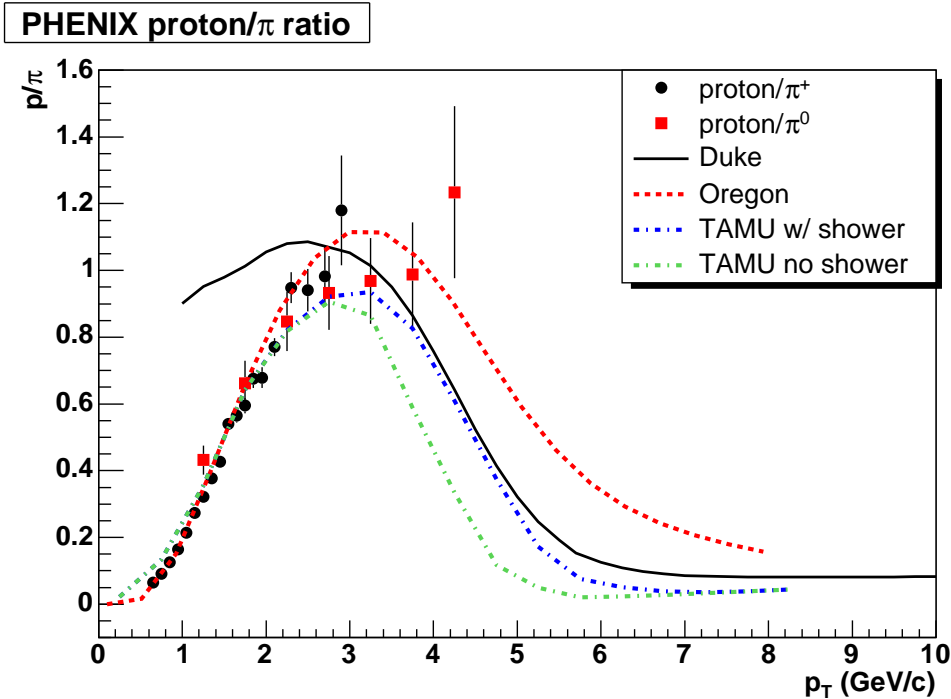}
\caption{The proton to pion ratio measured by PHENIX for Au+Au collisions at $\sqrt{s_{NN}} = 200 $ GeV\cite{Adler:2003kg}. Several comparisons to recombination
	models as mentioned in the text are shown.}
\label{fig:ppivsmodel}
\end{figure}

Figure \ref{fig:ppivsmodel} shows several recombination model calculations
compared to the $p/\pi$ ratio from PHENIX. 
The general features at $p_T>3$ GeV/$c$ are reasonably reproduced---that
is the protons show a strong enhancement at moderate $p_T$ which
disappears at $p_T > 5$ GeV/$c$ consistent with the measured h/$\pi$
ratio shown in Fig. \ref{fig:htopi}. The more complicated models do a
better job, as one might expect in the $p_T< 3$ GeV/$c$ region, where
the assumptions made by the Duke model begin to break down. Since the
recombination model's essential ingredient is the number of
constituent quarks in a hadron, the similarity of $R_{CP}$ for the
$\phi$ and pions is nicely explained.

\begin{figure}[tbhp]
\includegraphics[width=1.0\linewidth]{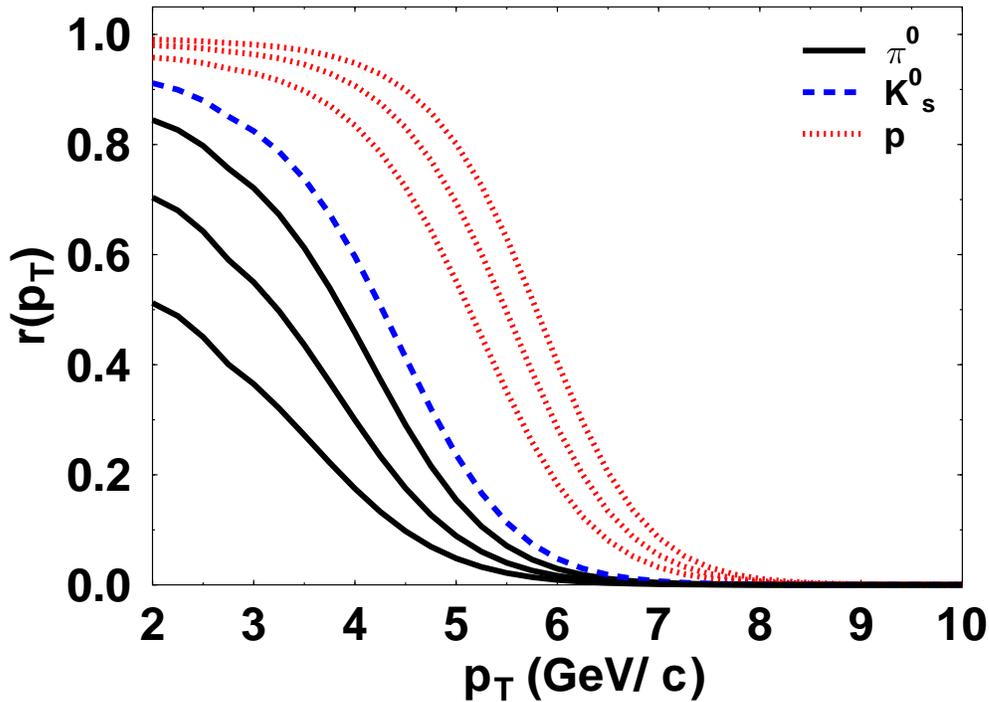}
\caption{The ratio $r(P_T)$ = $R/(R+F)$ 
of recombined hadrons to the sum of recombination ($R$) and
fragmentation ($F$) for pions (solid), $K^0$s (dashed) and $p$ (dotted
lines) \cite{Fries:2003kq} in Au+Au collisions at
$\sqrt{s_{NN}} = 200 $ GeV. For protons and pions different impact
parameters $b$ = 0, 7.5 and 12 fm (from top to bottom) are
shown. $K^0$s is for $b$ = 0 fm only.}
\label{fig:recomfragratio}
\end{figure}

Figure \ref{fig:recomfragratio} shows the fraction of hadrons arising
from recombination of only thermal quarks, as a function of $p_T$. 
For $p_T$ between
2.5 and 4 GeV/$c$ the fraction of protons from recombination is greater
than 90\% for all impact parameters, and is essentially 100\% for the most
central collisions. For pions the value is between 40 and 80\%,
depending on the centrality. This is contradicted by the
data in Fig. \ref{fig:jetcorr} which clearly shows jet-like correlations 
for both pions and protons in mid-central collisions.
It should be noted that the yield of
particles associated with baryons in very central collisions appears 
to decrease, indicating a possible condition where the simple picture
of recombination of purely thermal quarks may apply. 

\begin{figure}[tbhp]
\includegraphics[width=1.0\linewidth]{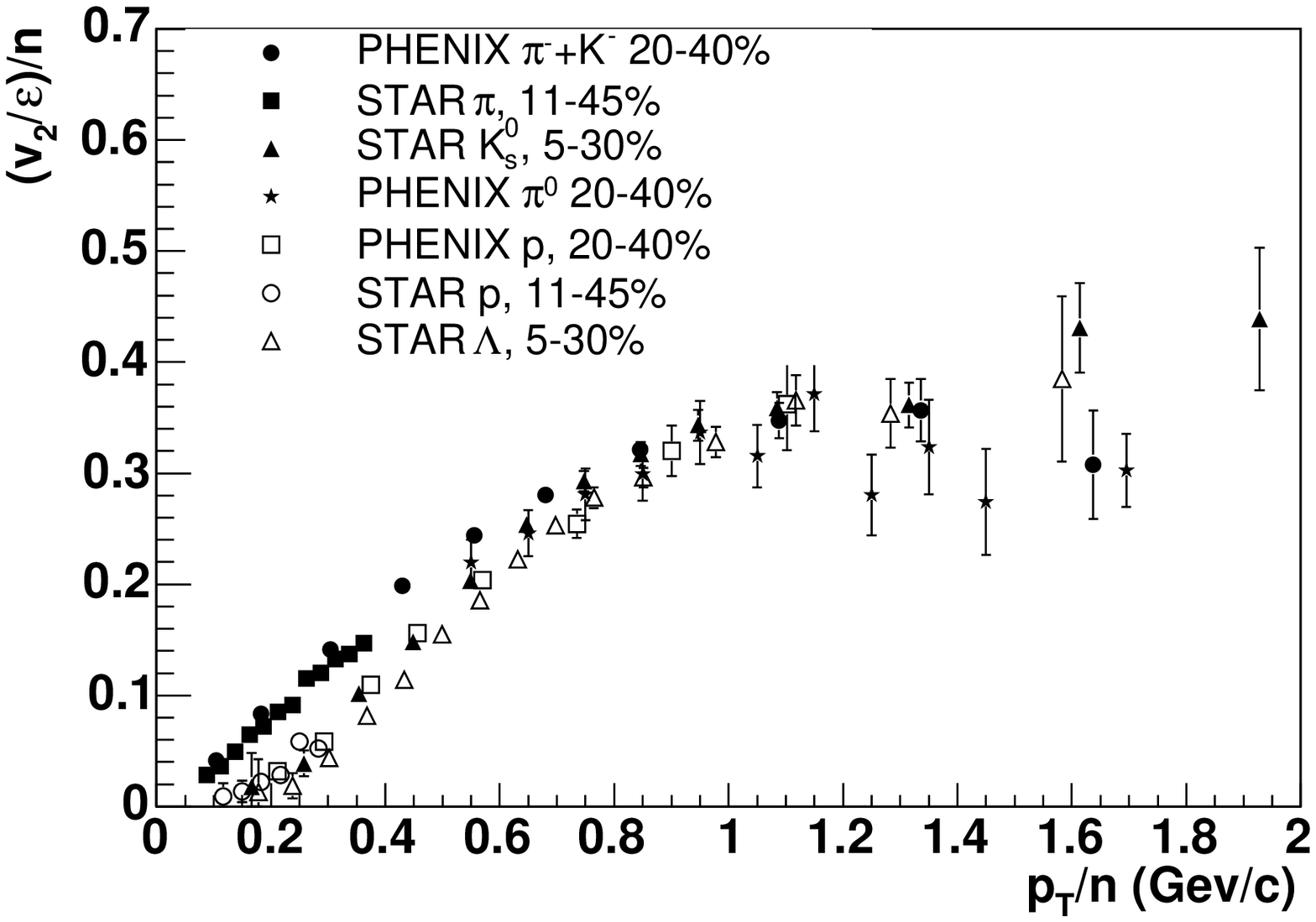}
\caption{
$v_2$ as a function of transverse momentum for a variety of particles
for Au + Au collisions
where both $v_2$ and
$p_T$ have been scaled by the number of constituent quarks in the
particle.
The meson data are shown with filled symbols;
$\pi^-+K^-$ from PHENIX at 
$\sqrt{s_{NN}} = 200 $ GeV \cite{Adler:2003kt} (filled circles), 
charged $\pi$ from STAR at 
$\sqrt{s_{NN}} = 130 $ GeV \cite{Adler:2001nb} (filled squares),
$K^0_s$ from STAR at 
$\sqrt{s_{NN}} = 200 $ GeV \cite{Adams:2003am} (filled triangles),
and $\pi^0$ from PHENIX at 
$\sqrt{s_{NN}} = 200 $ GeV \cite{Kaneta:2004sb} (filled stars).
While the baryons are shown with open symbols;
$p$ from PHENIX at 
$\sqrt{s_{NN}} = 200 $ GeV \cite{Adler:2003kt} (open squares),
$p$ from STAR at 
$\sqrt{s_{NN}} = 130 $ GeV \cite{Adler:2001nb} (open circles), and
$\Lambda$ from STAR at 
$\sqrt{s_{NN}} = 200 $ GeV \cite{Adams:2003am} (open triangles).
}
\label{fig:v2}
\end{figure}

One can examine the general prediction for the elliptic flow of
identified particles by rescaling both the $v_2$ and the transverse
momentum by the number of constituent quarks as shown in
Fig. \ref{fig:v2}. This scaling was first suggested by 
Voloshin \cite{Voloshin:2003zzz}. Above $p_T/n$ of 1 GeV/$c$ 
(corresponding to 3 GeV/$c$ in
the proton transverse momentum) all particles essentially plateau at a
value of about 0.35 presumably reflecting the elliptic flow of the
underlying partons. Interestingly, even at lower values of the
transverse momentum, all particles also fall on the same curve aside
from pions.

\begin{figure}[tbhp]
\includegraphics[width=1.0\linewidth]{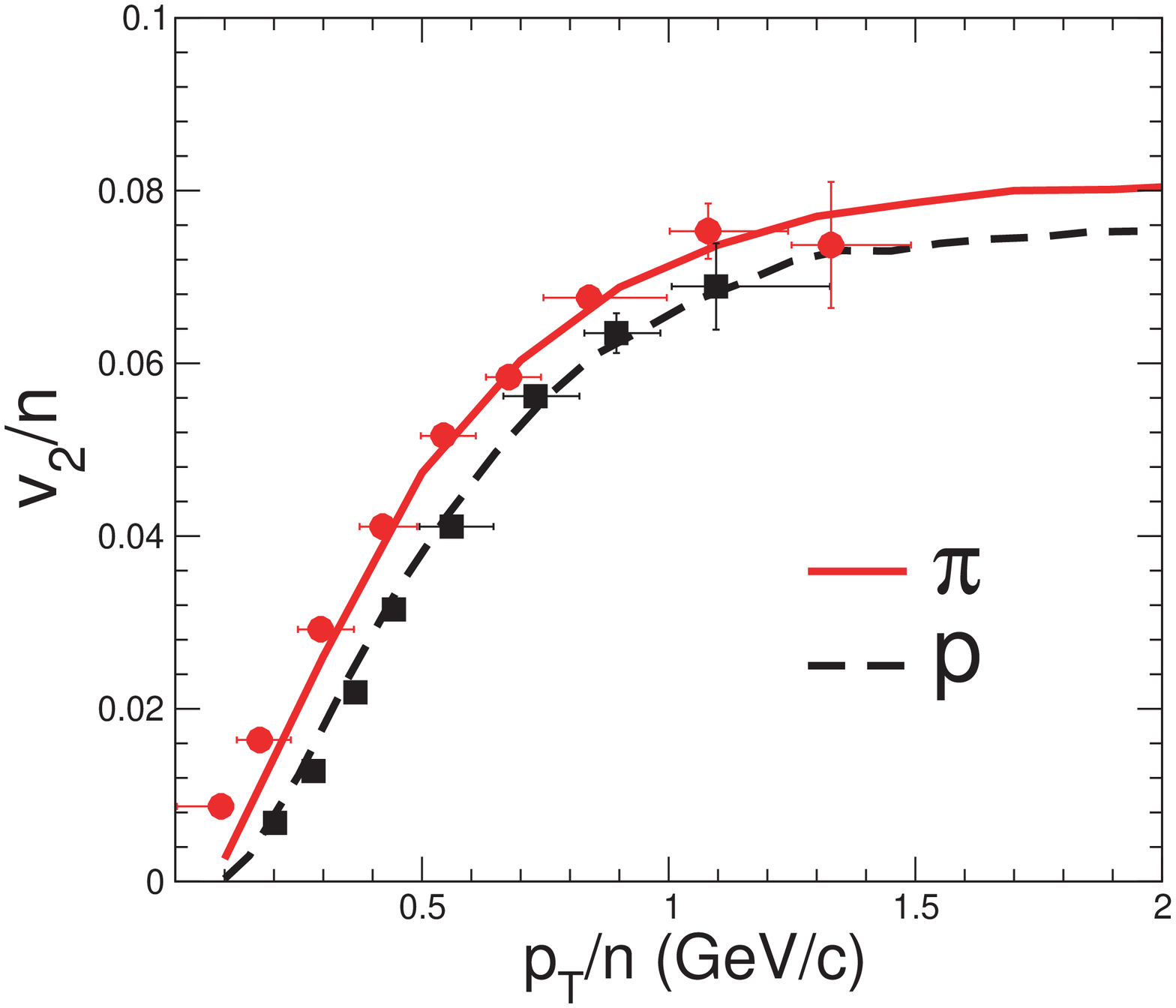}
\caption{ $v_2/n$ in the TAMU model, where n is the number of constituent quarks 
in a particle for protons and pions\cite{Greco:2004ex}. 
Scaled pion (dashed line) and proton (dotted line) results from the TAMU
model are shown in addition to charged pion 
and proton measurements from the
PHENIX experiment from minimum bias Au+Au reactions
at $\sqrt{s_{NN}} = 200 $ GeV \cite{Adler:2003kt}.
This model allows for the recombination of hard partons and soft
partons, as well as the decay of resonances such as the two pion decay
of the $\rho$ meson.  One sees that, at least in this calculation, the
addition of processes which mix hard and soft partons do not destroy
the agreement for the model with $v_2/n$ which is presumably a soft
process.}
\label{fig:kov2}
\end{figure}

It is clear from the jet correlations observed that the majority
of moderate $p_T$ baryons in peripheral and mid-central collisions
cannot arise from a purely thermal source, as that would dilute the 
per-trigger partner yield. The jet structure and collision scaling
indicate that at least some of the baryon excess is jet-like in origin. 
The relatively short formation time for baryons of such momenta 
suggests that allowing recombination of fragmentation partons with 
those from the medium may solve the problem and better reproduce the 
data. Both the Oregon and TAMU models have mechanisms to do this.
However, such modification
of the jet fragmentation function must also modify the elliptic
flow, and could break the quark scaling needed to reproduce the
observed $v_2$ trends. Hence, the jet structure of hadrons at 2--5 
 GeV/$c$ $p_T$ presents a challenge to models of the hadron formation. 

Figure \ref{fig:kov2} shows a comparison 
of the elliptic flow calculated by the TAMU model \cite{Greco:2004ex} 
with PHENIX data from Au+Au collisions at $\sqrt{s_{NN}} = 200 $ GeV. The
model includes the recombination of hard and soft partons, as well as
the decay of resonances such as the $\rho$. In this model, at least,
the agreement of $v_2$ with the data is preserved---in addition a
simple explanation is given for the excess of pion $v_2$ at low
$p_T$. A similar conclusion in shown in \cite{Dong:2004ve}. 
This would seemingly attribute all the elliptic flow to the
partonic phase leaving no room for additional flow to be produced in
the later, hadronic stage---which may be in contradiction to
hydrodynamic interpretations of the hadronic state as demanded by a
variety of signatures such as the $p_T$ spectra of the protons and
pions (see Section \ref{Sec:therm}). It is clear that a more
comprehensive comparison of observables should be undertaken to check
the validity of these models. Higher-statistics jet studies with
different identified particles by PHENIX in Run-4 will help clarify
the situation.



\subsection{Hadron Formation Time}

In the discussion of the suppression of pions for $p_{T}>2 $ GeV/$c$, we
treat the pions as resulting from the fragmentation of hard-scattered 
quarks and gluons.  The explanation of this suppression in
terms of partonic energy loss assumes that the hadronic wave function
only becomes coherent outside the medium.
Protons have a different hadronic structure and larger mass, and
so may have a different, shorter time scale for coherence. 

Following \cite{Wang:2003aw}, we can estimate the formation time for the
different mass hadrons at moderate $p_T$ in two different ways.
According to the uncertainty principle, the formation time in the
rest frame of the hadron can be related to the hadron size, $R_h$.
In the laboratory frame, the hadron formation time is then given by
\begin{equation}
\tau_f \approx R_h \frac{E_h}{m_h}
\end{equation}
where $R_h$ is taken to be 0.5--1 fm. For a 10 GeV/$c$ pion, this gives
a formation time of 35--70 fm/$c$. For the $p_T = $2.5 GeV/$c$ pions 
considered in this section, the formation time is 9--18 fm/$c$, well
outside the collision region. However for $p_T = $2.5 GeV/$c$ protons,
the corresponding formation time is only 2.7 fm/$c$ in the vacuum, 
suggesting the possibility that the hadronization process may begin inside 
the medium. However the formation of such heavy particles would presumably
be delayed in a deconfined medium until the entire system began to hadronize.

If quarks and antiquarks from gluon splitting are assumed to combine
into dipole color singlets leading to the final hadrons, the formation
time may be estimated from the gluon emission time. Then the formation time
for a hadron carrying a fraction $z$ of the parton energy is given by
\begin{equation}
\tau_f  \approx \frac{2E_h(1-z)}{k_T^2+m_h^2}.
\end{equation}
If $z$ is 0.6--0.8 and $k_T \approx \Lambda_{QCD}$, proton formation
times in the range of 1--4 fm result \cite{Wang:2003aw}. Such values
again imply formation of the proton within the medium. Thus, it is
possible that differing (and perhaps complicated) interactions with 
the medium may produce different scalings of proton and pion production
and result in modified fragmentation functions in Au+Au collisions.
However, most expectations are that this should lead to greater suppression
rather than less.  In fact, modified fragmentation functions measured
in electron deep-inelastic scattering on nuclei by the HERMES experiment
are often interpreted in terms of additional suppression for hadrons
forming in the nuclear material.

\subsection{Hard-Scattering Physics}

If the dominant source of (anti)protons at intermediate $p_{T}$ is not
soft physics, is the explanation a medium-modified hard-process
source?  The near-side partner yields indicate that a significant
fraction of the baryons have jet-like partners.  However, in the
parton energy loss scenario as described in
Section \ref{Sec:denseN}, hard-scattered partons lose energy in medium
prior to hadronization.  Thus one would expect the same suppression
for baryons and mesons.  Furthermore, we know that the (anti)protons
are as suppressed above $p_{T} = 5 $ GeV/$c$ in a manner similar to
pions . Hence for this explanation to be correct, there must exist
a mechanism by which only partons leading to baryons between 2 and 5 GeV/$c$ 
in $p_T$ escape suppression.

Another key piece of information is that the elliptic flow $v_{2}$ for
protons is large for $p_{T}$ in the range 2-4 GeV/$c$.  At low $p_{T}$
this collective motion is attributed to different pressure gradients
along and perpendicular to the impact parameter direction in
semi-central collisions.  At higher $p_{T}$ it has been hypothesized
that one could observe a $v_{2}$ due to smaller partonic energy loss
for partons traveling along the impact parameter direction (shorter
path in the medium) as opposed to larger partonic energy loss in the
perpendicular direction (larger path in the medium). However, the data
suggest that the pions have a large energy loss (a factor of five
suppression in central Au+Au reactions) , while the protons do not.
In this case one might expect that if the source of proton $v_{2}$
were energy loss, then proton $v_{2}$ would be significantly less than
the $v_{2}$ for the pions. In fact, the opposite is experimentally
observed: for $p_T > 2$ GeV/$c$, the proton $v_{2}$ is always larger than
the pion $v_{2}$.

The contradictions the data create for both the ``soft''- and
``hard''-physics explanations may indicate that the correct physics
involves an interplay between the two.

\subsection{Conclusions}

The anomalous enhancement of (anti)protons relative to pions at
intermediate $p_T$ = 2--5 GeV remains a puzzle.
At lower transverse momentum particle production is a long-wavelength
``soft'' process and the transport of these hadrons and their precursor
partons is reasonably described by hydrodynamics.  As observed at
lower energies, soft particles emitted from an
expanding system receive a collective velocity boost to higher $p_T$
resulting in an enhanced $p/\pi$ and $\overline{p}/\pi$ ratio relative
to proton-proton reactions at the same energy. We observe a similar
phenomena at RHIC, for which the (anti) proton spectra and $v_{2}$ are
roughly described in some hydrodynamic models up to approximately 2
 GeV/$c$.  Another class of calculations, referred to as recombination
models, also boosts soft physics to higher $p_{T}$ by coalescence of
``dressed'' partons.  In the hydrodynamic models the quanta which are
flowing are initially partons and then hadrons.  The recombination 
models describe comoving valence partons which coalesce into hadrons, and
do not reinteract.
These two points of view may not be entirely contradictory, since both
include a flowing partonic phase. If fact, it may be that the recombination
models provide a mechanism by which hydrodynamics works to a much
higher $p_T$ than one might expect. The simplifying assumption of
hadrons which do not interact is most probably an oversimplification
and further refinement of the models will include this, though
it may be that the hadronic phase will not modify the spectra as much
as the hydrodynamic models might predict. 

In both models, the (anti)proton enhancement as a function of
centrality can be tuned to reproduce the apparent binary collision
scaling observed in the data.  An important distinction between the
two is that in one case this enhancement is mass dependent and in the
other it comes from the combination of quark momenta and thus
distinguishes between baryons and mesons\footnote{ A caveat to this
fact is that in the recombination models, it is the constituent-quark
mass that is important, thereby giving a slightly larger mass to the
strange quark. }.  $R_{CP}$ for the $\phi$ is similar to other mesons
despite the fact that they are more massive than protons. This scaling
with quark content, as opposed to mass, favors recombination models.


Further investigations into these intermediate $p_T$ baryons reveals a
near-angle correlation between particles, in a fashion characteristic
of jet fragmentation.  The near-angle associated particle yield increases by
almost a factor of two in going from proton-proton and deuteron-gold
reactions to gold-gold peripheral collisions.  In addition, the
partner yield is similar for trigger pions and protons, except in the
most central gold-gold reactions. This appears to indicate a
hard process source for a significant fraction of these baryons in
contrast to the previous mentioned physics scenarios.  Quantifying the
precise contribution is an important goal for future measurements.


The large (anti) baryon to pion excess relative to expectations from
parton fragmentation functions at intermediate $p_T = 2-5$ GeV/$c$ remains
one of the most striking unpredicted experimental observations at RHIC.
The data clearly indicate a new mechanism other than universal parton
fragmentation as the dominant source of baryons and anti-baryons at
intermediate $p_T$ in heavy ion collisions.  The boosting of soft physics,
that dominates hadron production at low $p_T$, to higher transverse
momentum has been explored with the context of hydrodynamic and
recombination models.  However, investigations into these intermediate
$p_T$ baryons reveals a near-angle correlation between particles,
in a fashion characteristic of jet fragmentation.  If instead these
baryons have a partonic hard scattering followed by fragmentation source,
this fragmentation process must be significantly modified.  It is
truly remarkable that these baryons have a large $v_2$ (typically 20\%)
indicative of strong collective motion and also a large ``jet-like''
near-side partner yield.  At present, no theoretical framework provides
a complete understanding of hadron formation in the intermediate
$p_T$ region.
 
\section{FUTURE MEASUREMENTS}
\label{Sec:future}
The previous sections have documented the breadth and depth of the
PHENIX data from the first three years of RHIC operations, along
with the physics implications of those results. Here we describe
those measurements required to further define and characterize the
state of matter formed at RHIC. In particular, we note that the
study of penetrating probes, which are the most sensitive tools in
this endeavor, is just beginning. The PHENIX experiment was
specifically designed to address these probes with capabilities that
are unique within the RHIC program and unprecedented in the field of
relativistic heavy ion physics.

One can distinguish two broad classes of penetrating probes:
\begin{enumerate}
\item Hard probes created at the very early stage of the collision
which propagate through, and could be modified by, the medium. These
are the QCD hard-scattering probes and the main observables are
high-$p_T$ particles coming from the fragmentation of jets, hidden charm
($J/\psi$ production), open charm and eventually also bottom quark
and $\Upsilon$ production.

\item Electromagnetic probes (either real or virtual photons)
which are created by the medium. Due to their large mean free path
these probes can leave the medium without final-state interaction
thus carrying direct information about the medium's conditions and
properties. The main observables here are low-mass $e^+e^-$ pairs
and the thermal radiation of the medium.
\end{enumerate}

By their very nature, penetrating probes are also rare probes and
consequently depend on the development of large values of the
integrated luminosity. In the present data set the reach for high-$p_T$
particles in PHENIX extends to roughly 10 GeV/$c$, and
lower-cross-section measurements such as charmonium are severely
limited. The dramatic improvement of the machine performance in
the year 2004 run
provides confidence that both this data set and those from
future RHIC runs will dramatically extend our reach in the rare
probes sector.

As part of a decadal planning of the RHIC operation,
PHENIX has prepared a
comprehensive document that outlines in great detail its scientific
goals and priorities for the next 10 years together with the
associated detector upgrade program needed to achieve them. The
decadal plan \cite{PHENIX:DecadalPlan} is centered around the
systematic study of the penetrating probes listed above.
The program is broad and can accommodate additions or modifications 
provided that a compelling physics case can be made.
Measurements are mainly planned in Au+Au collisions at the full RHIC
energy but they will be supplemented by other measurements varying
the energy and/or the species and by the necessary reference
measurements of $p+p$ and $p+A$ collisions. A short summary is given
below.

\subsection {High-$p_T$ Suppression and Jet Physics}
The most exciting results to date at RHIC are the discovery  of
high-$p_T$ suppression of mesons, interpreted in terms of energy loss of
quarks in a high-density medium, and the nonsuppression of baryons
or equivalently, the anomalously high $p/\pi$ ratio which still
awaits a clear explanation. These two topics were extensively
discussed in Sections \ref{Sec:denseN} and \ref{Sec:hadron},
respectively.

The data collected so far are superb. However, they suffer from
limited reach in transverse momentum, limited particle
identification capabilities and limited statistics in particular for
detailed studies of jet correlations. PHENIX has a program for
further studies of the high-$p_T$-suppression phenomena and jet
physics which aims at overcoming these limitations.

It will be necessary to trace the suppression pattern to much higher
$p_T$  to determine whether (and if so, when) the suppression
disappears and normal perturbative QCD behavior sets in.
High-luminosity runs will be needed, with at least a factor of 50
more statistics. PHENIX is particularly able to perform these
measurements with its excellent capability of triggering on
high-momentum $\pi^0$'s.

PHENIX has performed several particle correlation analyses and has
demonstrated that the experiment's aperture at mid-rapidity is
sufficient to conduct these studies. Currently, these analyses are
limited by the available statistics. Again, increasing the data
sample by a factor of 50--100 will allow a variety of correlation
studies using trigger particles with much-higher-momentum than
studied to date. A particularly interesting case is the study of
high-momentum $\gamma$-jet correlations, which have vastly reduced
trigger bias, since the trigger photons propagate through the
medium with a very long mean free path.

To further elucidate the baryon puzzle, additional data is
required with better separation between baryons and mesons. An
upgrade consisting of an aerogel Cerenkov counter and a
high-resolution TOF detector 
is expected to be
completed in time for the year 2006.
A portion of this aerogel counter was
already installed prior of the year 2004 run and performed
according to expectations. Once completed, this high-$p_T$ detector
will allow identification of $\pi,K/p$ to beyond 8 GeV/$c$ in $p_T$.

\subsection {$J/\psi$ Production}
Suppression of heavy quarkonia is one of the earliest and most striking
proposed signatures of deconfinement. The suppression mechanism follows
directly from the Debye screening
expected in the medium, which reduces the range of the potential between
charm quark and anti-quark pairs \cite{Matsui:1986dk}.
The NA38 and NA50 experiments have carried out a systematic study of
$J/\psi$ and $\psi'$ at the CERN-SPS in $p+p$, $p+A$, light ion, and
Pb+Pb collisions providing some of the most intriguing
results of the relativistic heavy ion program for more than ten years.
The NA50 experiment observed an anomalous suppression of $J/\psi$ in
central
Pb+Pb collisions at $\sqrt{s_{NN}}$ = 17.2 GeV \cite{Abreu:2000ni}.
The suppression, which is of the order of 25\% with respect to the normal
suppression in nuclear matter, has been interpreted by the NA50 authors as
evidence for deconfinement of quarks and gluons. Although this
interpretation is not universally shared \cite{Capella:2000zp,Capella:2001vb}, the
results
of NA38 and NA50 demonstrate the utility and great interest in
understanding
the fate of charmonium in dense nuclear matter.

The theoretical expectations at RHIC energies are not at all clear. They
range from total suppression in the traditional Debye screening
scenario to enhancement in coalescence 
models \cite{Thews:2000rj,Thews:2003da,Kostyuk:2003kt} 
and in statistical hadronization 
models \cite{Braun-Munzinger:2000px,Andronic:2003zv},
of $c$ and $\overline{c}$ quarks. Although some versions of the coalescence 
model seem disfavored from our very limited data set \cite{Adler:2003rc},
a more conclusive statement on these models has to await the
much larger data set of the year 2004 run.

PHENIX has unprecedented capabilities for the study of the
$J/\psi$ in Au+Au collisions. The $J/\psi$ can be measured via its
$\mu^+\mu^-$ decay channel at forward and backward rapidities in
the muon spectrometers and via its $e^+e^-$ decay channel at
mid-rapidity in the central arm spectrometers. From the recorded
luminosity of the year 2004 run,
we expect several thousand and $\sim$500
$J/\psi$ in the muon and central arms, respectively. This data set
will allow us a first look at the $J/\psi$ production pattern at
RHIC. However, it could well be marginal for a complete
characterization as a function of centrality and $p_T$, so that it
is likely that further higher-luminosity runs will be required.
Also the $p+p$ and $d$+Au baseline measurements performed in the year
2001 -- 2003 runs
have large statistical uncertainties, and higher-statistics versions for
these colliding species will be needed. A high-luminosity $p+p$ run
is planned in the year 2005 and high-luminosity $d+A$ or $p+A$
are still to be scheduled in the next years.

\subsection{Charm Production}
Charm quarks are expected to be produced in the initial hard
collisions between the incoming partons. The dominant mechanism is
gluon fusion and thus the production cross section is sensitive to
the gluon density in the initial state. The $c\overline{c}$
production cross section is sizable at RHIC energies with a few
$c\overline{c}$ pairs and therefore several open charm mesons per
unit of rapidity in central Au+Au collisions. As a result, charm
observables become readily accessible at RHIC and offer additional and
extraordinarily valuable diagnostic tools. For example, it is
vitally important to perform measurements of charm flow and to
determine the energy loss of charm quarks in the medium. Such
measurements will determine if the bulk dynamics observed for light
quarks extend to charm quarks, which could in fact have very
different behavior due to their much larger mass. Again the
potential of PHENIX in this domain is unique with its capability of
measuring open charm in a broad rapidity range, in the central and
muon arms, via both the electron and muon decay channels. An
additional unique feature is the possibility to measure correlated
semileptonic charm decays by detecting $e-\mu$ coincidences from
correlated $D\overline{D}$ decays. Such a measurement is
particularly interesting for the study of charm-quark energy loss
which may differ significantly from that observed for lighter
quarks \cite{Mustafa:1997pm,Lin:1997cn,Dokshitzer:2001zm}. A first 
study of $e-\mu$ coincidences should be feasible with the year 2004 
data.

To date PHENIX has measured charm production cross section in an
indirect way through high-$p_T$ single electrons \cite{Adcox:2002cg,Adler:2004ta}
assuming that all electrons (after measuring and subtracting the
contributions from light hadrons and photon conversions) originate
from the semileptonic decays of charm quarks. Although the charm
cross section has large uncertainties, the centrality dependence of
the charm rapidity density demonstrates that charm production
follows binary scaling as shown in Fig. \ref{fig:charm}.
Improvements and additional information are expected from the much
higher statistic of the year 2004 data.

A qualitatively new advance for PHENIX in the charm and also the
beauty sector will be provided by the implementation of the silicon
vertex detector. An upgrade project is underway to install in the
next five years a silicon vertex tracker, including a central arm
barrel and two end caps in front of the two muon spectrometers. The
vertex tracker will allow us to resolve displaced vertices and
therefore to directly identify open charm mesons via hadronic, e.g.
$D \rightarrow K \pi$, as well as semi-leptonic decays. The
heavy-quark physics topics accessible with the vertex tracker
include production cross section and energy loss of open charm and
open beauty, and spectroscopy of charmonium and bottomonium states,
each of which should provide incisive new details on the properties
of the created medium.

\subsection {Low-Mass Dileptons}
Low-mass dileptons are considered the most sensitive probe of
chiral symmetry restoration primarily through $\rho$ meson decays.
Due to its very short lifetime ($\tau = 1.3$ fm/$c$) compared to
that of the typical fireball of $\sim$ 10 fm/$c$, most of the $\rho$
mesons decay inside the medium providing an unique tool to observe
in-medium modifications of its properties (mass and/or width)
which could be linked to chiral symmetry restoration. The
situation is somewhat different but still interesting for the
$\omega$ and $\phi$ mesons. Because of their much longer lifetimes
($\tau = 23$ fm/$c$ and 46 fm/$c$ for the $\omega$ and $\phi$,
respectively ) they predominantly decay outside the medium, after
regaining their vacuum properties, with only a small fraction
decaying inside the medium. Since the measurement integrates over
the history of the collision, this may result in a small
modification of the line shape of these two mesons which PHENIX
might be able to observe with its excellent mass resolution.
PHENIX also has the unprecedented capability of simultaneously
measuring within the same apparatus the $\phi$ meson decay through
$e^+e^-$ and $K^+K^-$ channels. The comparison of the branching
ratios to these two channels provides a very sensitive tool for
in-medium modifications of the $\phi$ and $K$ mesons.

The CERES experiment at CERN has confirmed the unique physics
potential of low-mass
dileptons \cite{Agakishiev:1995xb,Agakishiev:1998au,Adamova:2002kf}.
An enhancement of electron pairs was observed in the mass region
$m = 0.2$--$0.6$ GeV$/c^2$ in Pb+Au collisions at
$\sqrt{s_{NN}}$ = 17.2 GeV with respect to $p+p$ collisions. The
results have triggered a wealth of theoretical activity and can be
explained by models which invoke in-medium modification of the
$\rho$ meson (dropping of its mass and/or broadening of its
width) \cite{Rapp:1999ej}. The precision of the CERES data has
been so far insufficient to distinguish between the different
models. Results with higher statistics and better mass resolution 
are expected from the NA60 experiment that is studying the production
of low-mass dimuons in In+In collisions \cite{Arnaldi:2005bx}.  
Theoretical calculations \cite{Rapp:2002mm} show that the
enhancement should persist at RHIC energies and that PHENIX with
its excellent mass resolution has an unique opportunity to do
precise spectroscopy of the light vector mesons and to shed more
light on the origin of the enhancement of the low-mass-pair
continuum.

The measurement of low-mass electron pairs is however a very
challenging one. The main difficulty stems from the huge
combinatorial background created by the pairing of $e^+$ and $e^-$
tracks from unrecognized $\pi^0$ Dalitz decays and $\gamma$
conversions. PHENIX is developing a novel Cerenkov detector that,
in combination with the recently installed coil which makes the
magnetic field zero close to the beam axis, will effectively
reduce this combinatorial background by almost two orders of
magnitude \cite{HBDTechnicalNote}. The detector, operated in pure
CF$_4$, consists of a 50-cm-long radiator directly coupled, in a
windowless configuration, to a triple GEM detector which has a CsI
photocathode evaporated on the top face of the first GEM foil and
pad read out at the bottom of the GEM stack \cite{Kozlov:2003zr}.
The R\&D phase to demonstrate the validity of the concept is
nearing completion. The detector construction phase is starting
now with installation foreseen in time for the year 2006 --- 2007.
With this detector PHENIX will have the unprecedented ability to
perform high-quality measurements over the whole dilepton mass
range from the $\pi^0$ Dalitz decay up to the charmonium states.

\subsection {Thermal Radiation}
A prominent topic of interest in the field of relativistic
heavy-ion collisions is the identification of the thermal
radiation emitted by the system and in particular the thermal
radiation emitted by the quark-gluon plasma via $q\overline{q}$
annihilation. Such radiation is a direct fingerprint of the matter
formed and is regarded as a very strong signal of deconfinement.
Its spectral shape should provide a direct measurement of the
plasma temperature.

In principle the thermal radiation can be studied through real
photons or dileptons, since real and virtual photons carry basically
the same physics message. In practice the measurements are extremely
challenging. The thermal radiation is expected to be a small signal
compared to the large background from competing processes, hadron
decays for real photons and Dalitz decays and $\gamma$ conversions
for dileptons, the former being larger by orders of magnitude
compared to the latter. But in both cases, a very precise knowledge
of all these sources is an absolutely necessary prerequisite. After
subtracting these sources, one still needs to disentangle other
contributions which might be comparable or even stronger, mainly the
contributions of initial hard-parton scattering to direct photons
and of semileptonic decays of charm mesons to dileptons.

Theoretical calculations have singled out the dilepton mass range
$m = 1$--$3$ GeV$/c^2$ as the most appropriate window where the QGP
radiation could dominate over other
contributions \cite{Rapp:2000pe,Ruuskanen:1992au}. Measurements in
this intermediate mass range carried out at the CERN SPS by HELIOS
and NA50 have revealed an excess of dileptons, but this excess
could be explained by hadronic contributions \cite{Li:1998xn}.

There is no conclusive evidence for QGP thermal photons from the
CERN experiments (for a recent review see \cite{Turbide:2003si}).
From the theoretical point of view it is clear that in the
low-$p_T$ region ($p_T < 2$ GeV/$c$) the real photon spectrum is
dominated by hadronic sources and the thermal radiation from the
hadron gas. It is only in the high-$p_T$ region where one might
have a chance to observe the thermal radiation from the QGP.

PHENIX has measured direct real
photons at $p_T > 4$ GeV/$c$ from the initial hard 
scatterings\cite{Adler:2005ig}. The
errors are relatively large leaving room for a comparable
contribution of thermal photons. The high statistics of the year 2004
run will provide the first real opportunity to search for the QGP thermal
radiation in PHENIX both in the dilepton and real photon channels.
However, the search for this elusive signal might take some time as
it will probably require equally-high-statistics runs of reference
data in $p+p$ and $p+A$ collisions for a precise mapping of all the
other contributions (hadronic + pQCD for real photons and hadronic +
charm for dileptons).
 
\section{SUMMARY AND CONCLUSIONS}
\label{Sec:conclusion}

The PHENIX data set from the first three years of RHIC operation
provides an extensive set
of measurements, from global variables to hadron spectra to high-$p_T$ 
physics to heavy-flavor production. From this rich menu we
have reviewed those aspects of the present data that address the
broad features of the matter created in Au+Au collisions at RHIC,
namely, energy and number density, thermalization, critical
behavior, hadronization, and possible deconfinement.

We first investigated whether the transverse energy and multiplicity
measurements of PHENIX demonstrate that a state of high-energy-density 
matter is formed in Au+Au collision at RHIC. We estimated
from our $dE_T/d\eta$ measurement that the peak energy density in
the form of created secondary particles is at least 15 GeV/fm$^3$.
If we use a thermalization time of
1 fm/$c$ provided by the hydrodynamic models from the elliptic flow,
then the value of the energy density of the first thermalized
state would be in excess of 5 GeV/fm$^3$.
These values are well in excess of the
$\sim$1 GeV/fm$^3$ obtained in lattice QCD as the energy density
needed to form a deconfined phase. Na\"ive expectations prior to
RHIC turn-on that $dE_T/d\eta$ and $dN_{ch}/d\eta$ could be
factorized into a ``soft'' and a pQCD jet component are not
supported by the data. Results from a new class of models featuring
initial-state gluon saturation compare well with RHIC multiplicity
and $E_T$ data.

We then examined our data and various theoretical models to
investigate the degree to which the matter formed at RHIC appears to
be thermalized. The measured yields and spectra of hadrons are
consistent with thermal emission from a strongly expanding source,
and the observed strangeness production is consistent with
predictions based on complete chemical equilibrium. The scaling of
the strength of the elliptic flow $v_2$ with eccentricity shows that
a high degree of collectivity is built up at a very early stage of
the collision. The hydro models which include both hadronic and QGP
phases reproduce the qualitative features of the measured $v_2(p_T)$
of pions, kaons, and protons. These hydro models require early
thermalization ($\tau_{_{therm}} \leq 1$ fm/$c$) and high initial
energy density $\varepsilon \geq$ 10 GeV/fm$^3$. These points of
agreement between the data and the hydrodynamic and thermal models
can be interpreted as strong evidence for formation of high-density
matter that thermalizes very rapidly.

However several of the hydro models fail to reproduce the $v_2(p_T)$
of pions, protons, and spectra of pions and protons simultaneously.
Given this disagreement it is not yet possible to make an
unequivocal statement regarding the presence of a QGP phase
based on comparisons to hydrodynamic calculations. The
experimentally measured HBT source parameters, especially the small
value  of $R_{\rm long}$ and the ratio $R_{\rm out}/R_{\rm side} \approx 1$, are not
reproduced by the hydrodynamic calculations. Hence we currently do
not have a consistent picture of the space-time dynamics of
reactions at RHIC as revealed by spectra, $v_2$ and HBT. These
inconsistencies prevent us from drawing firm conclusions on
properties of the matter such as the equation of state and the
presence of a mixed phase.

Critical behavior near the phase boundary can produce
nonstatistical fluctuations in observables such as the net-charge
distribution and the average transverse momentum. Our search for
charge fluctuations has ruled out the most na\"ive model of charge
fluctuations in a QGP, but it is unclear if the charge fluctuation
signature can survive hadronization. Our measurement of $\langle p_T
\rangle$ fluctuations is consistent with the effect expected  of
high-$p_T$ jets, and it gives a severe constraint on the
fluctuations that were expected for a sharp phase transition.

Many of these observables---for instance, large $dE/d\eta$ and
$dN_{ch}/d\eta$, strangeness enhancement, strong radial flow, and
elliptic flow---have been observed in heavy ion collisions at lower
energies. We have found smooth changes in these observables as a
function of $\sqrt{s_{NN}}$ from AGS energies to SPS energies to
RHIC energy. The $dE_T/d\eta$ increases by about 100\% and the
strength of the elliptic flow increase by about 50\% from SPS to
RHIC. The strangeness suppression factor $\gamma_s$ and the radial
expansion velocity $\langle \beta_T \rangle$ vary smoothly from AGS
to RHIC energies. No sudden change with collision energy has been
observed.

The strong suppression of high-$p_T$ particle production at RHIC is
a unique phenomenon that has not been previously observed.
Measurements of two-hadron azimuthal-angle correlations at high
$p_T$ and the $x_T$ scaling in Au+Au collisions confirm the dominant
role of hard scattering and subsequent jet fragmentation in the
production of high-$p_T$ hadrons. Measurements in deuteron-gold
collisions demonstrate that any initial-state modification of
nuclear parton distributions causes little or no suppression of hadron
production for $p_T > 2$ GeV/$c$ at mid-rapidity. This conclusion is further
strengthened by the observed binary scaling of direct photon and
open charm yields in Au+Au. Combined together, these observations
provide direct evidence that Au+Au collisions at RHIC have produced
matter at extreme densities.

Medium-induced energy loss, predominantly via gluon bremsstrahlung
emission, is the only currently known physical mechanism that can
fully explain the magnitude of the observed high-$p_T$ suppression.
The approximately flat suppression factor $R_{AA}(p_T)$ observed in
the data, which was predicted by the GLV energy loss model, rules
out the simplest energy loss models which predicted a constant
energy loss per unit length. However, the model by Wang {\it et al.}
obtains the same flat $R_{AA}(p_T)$ from apparently different
physics. From the GLV model, the initial gluon number density,
$dn_g/dy \approx 1000$ and initial energy density, $\varepsilon_0
\approx 15 {\rm GeV/fm^3}$, have been obtained. These values are
consistent with the energy density obtained from our $dE_T/d\eta$
measurement as well as ones from the hydro models.

The large (anti)baryon to pion excess relative to expectations from
parton fragmentation functions at
intermediate $p_T$ (2 --- 5 GeV/$c$) is both
an unpredicted and one of the most striking
experimental observation at RHIC. The data clearly indicates that a
mechanism other than universal parton fragmentation is the
dominant source of (anti-)baryons in the intermediate
$p_T$ range in heavy ion collisions.
The boosting of soft physics to higher transverse momentum
has been explored within the context of hydrodynamics and
recombination models.
Hydrodynamic models can readily
explain the baryon to meson ratio as a consequence of strong radial
flow, but these models have difficulties reproducing the difference
in $v_2$ between protons and mesons above 2 GeV/$c$. Recombination
models provide a natural explanation for the large baryon to meson
ratio as well as the apparent quark-number scaling of the elliptic
flow.
However, investigations into these intermediate $p_T$ baryons
reveal a near-angle correlation between particles, in a fashion
characteristic of jet fragmentation. If instead these
baryons have a partonic hard scattering followed by
fragmentation, this fragmentation process must be
significantly modified. It is truly remarkable
that these baryons have a large $v_2$ of $\approx 20$ \%
typically indicative of strong collective motion and also a 
large jet-like near-side partner yield.
At present, no model provides a complete understanding
of hadron formation in the intermediate $p_T$ regime.

The initial operation of RHIC has produced the impressive quantity
of significant results described above. These striking findings call
for additional efforts to define, clarify and characterize the state
of matter formed at RHIC. Further study of the collisions using hard
probes such as high-$p_T$ particles, open charm, and $J/\psi$, and
electromagnetic probes such as direct photons, thermal photons,
thermal dileptons, and low-mass lepton pairs are particularly
important. The utilization of these penetrating probes is just
beginning, and we expect these crucial measurements based on the
very-high-statistics data of the year 2004 run
will provide essential results towards understanding of the dense
matter created at RHIC.

Advances in the theoretical understanding of
relativistic heavy ion collisions is vital for the quantitative
study of the dense matter formed at RHIC. While there is rapid and
significant progress in this area, a coherent and consistent picture
of heavy ion collisions at RHIC, from the initial formation of the
dense matter to the thermalization of the system to the
hadronization to the freezeout, remains elusive. With such a
consistent model, it will become possible to draw definitive
conclusions on the nature of the matter and to quantitatively
determine its properties. The comprehensive data sets from global
variables to penetrating probes provided by PHENIX at present and in
the future will prove essential in  constructing and constraining a
consistent model of heavy ion collisions to determine the precise
nature of the matter created at RHIC.

In conclusion, there is compelling experimental evidence that
heavy-ion collisions at RHIC produce a state of matter characterized
by very high energy densities, 
density of unscreened color charges ten times that of a nucleon,
large cross sections for the
interaction between strongly interacting particles, strong collective
flow, and early thermalization.  Measurements indicate that this
matter modifies jet fragmentation and has opacity that is too large to
be explained by any known hadronic processes.  This state of matter is
not describable in terms of ordinary color-neutral hadrons, because
there is no known self-consistent theory of matter composed of
ordinary hadrons at the measured densities.  The most economical
description is in terms of the underlying quark and gluon degrees of
freedom.  Models taking this approach have scored impressive successes
in explaining many, but not all, of the striking features measured to
date.  There is not yet irrefutable evidence that this state of matter
is characterized by quark deconfinement or chiral symmetry
restoration, which would be a direct indication of quark-gluon plasma
formation.  The anticipated program of additional incisive
experimental measurements combined with continued refinement of the
theoretical description is needed to achieve a complete understanding
of the state of matter created at RHIC.
 
\section*{ACKNOWLEDGEMENTS}
\label{Sec:acknowledge}

We thank the staff of the Collider-Accelerator and Physics
Departments at Brookhaven National Laboratory and the staff of
the other PHENIX participating institutions for their vital
contributions.  We acknowledge support from the Department of
Energy, Office of Science, Nuclear Physics Division, the
National Science Foundation, Abilene Christian University
Research Council, Research Foundation of SUNY, and Dean of the
College of Arts and Sciences, Vanderbilt University (U.S.A),
Ministry of Education, Culture, Sports, Science, and Technology
and the Japan Society for the Promotion of Science (Japan),
Conselho Nacional de Desenvolvimento Cient\'{\i}fico e
Tecnol{\'o}gico and Funda\c c{\ a}o de Amparo {\`a} Pesquisa do
Estado de S{\ a}o Paulo (Brazil),
Natural Science Foundation of China (People's Republic of
China),
Centre National de la Recherche Scientifique, Commissariat
{\`a} l'{\'E}nergie Atomique, Institut National de Physique
Nucl{\'e}aire et de Physique des Particules, and Association
pour la Recherche et le D{\'e}veloppement des M{\'e}thodes et
Processus Industriels (France),
Ministry of Industry, Science and Tekhnologies,
Bundesministerium f\"ur Bildung und Forschung, Deutscher
Akademischer Austausch Dienst, and Alexander von Humboldt
Stiftung (Germany),
Hungarian National Science Fund, OTKA (Hungary),
Department of Atomic Energy and Department of Science and
Technology (India),
Israel Science Foundation (Israel),
Korea Research Foundation and Center for High Energy Physics
(Korea),
Russian Ministry of Industry, Science and Tekhnologies, Russian
Academy of Science, Russian Ministry of Atomic Energy (Russia),
VR and the Wallenberg Foundation (Sweden),
the U.S. Civilian Research and Development Foundation for the
Independent States of the Former Soviet Union, the US-Hungarian
NSF-OTKA-MTA, and the US-Israel Binational Science Foundation.

\bibliographystyle{elsart-num}
\bibliography{ppg048a1}
 
\end{document}